\documentclass[12pt]{article}
\usepackage[width=16.00cm, height=22.00cm]{geometry}
\usepackage[utf8]{inputenc}
\usepackage{hyperref}
\usepackage{cite}

\usepackage{tikz}
\usepackage{graphicx}
\usepackage{dcolumn}
\usepackage{arydshln} 
\usepackage{bm}
\usepackage{color}			
\usepackage[T1]{fontenc}
\usepackage{mathrsfs,amsfonts,dsfont}
\usepackage{amssymb}
\usepackage{amsthm}
\usepackage{amstext}
\usepackage{mathtools}
\usepackage{enumitem} 
\usepackage{rotating} 
\usepackage{stmaryrd} 

\makeatletter 
 \hypersetup{pdftitle = {pdf title},
	     pdfauthor = {Daniel Miller},
	     pdfsubject = {Multipartite Quantum Entanglement},
	     pdfkeywords = {entanglement, semiseparability,  noise threshold, local noise, 
			    PPT criterion, reduction criterion, entropy criterion, entanglement witness,
			    higher-dimensional, qudit, arbitrary dimension, composite dimension, 
			    integers modulo, invertible, non-invertible, 
			    graph state, stabilizer state,
			    puzzle, star graph, dandelion graph, line graph, ring graph, 
			    Greenberger-Horne-Zeilinger, GHZ, absolutely-maximally-entangled, AME,
			    uniform state, proof, show that,
			    sector length,  full-body sector length, of noisy state 
			    }
	    }
\makeatother

 \makeatletter
\def\moverlay{\mathpalette\mov@rlay}
\def\mov@rlay#1#2{\leavevmode\vtop{%
   \baselineskip\z@skip \lineskiplimit-\maxdimen
   \ialign{\hfil$\m@th#1##$\hfil\cr#2\crcr}}}
\newcommand{\charfusion}[3][\mathord]{
    #1{\ifx#1\mathop\vphantom{#2}\fi
        \mathpalette\mov@rlay{#2\cr#3}
      }
    \ifx#1\mathop\expandafter\displaylimits\fi}
\makeatother

\newcommand{\bigcupdot}{\charfusion[\mathop]{\bigcup}{\cdot}}

\newcounter{counter}
\newcommand{\counter}{\refstepcounter{counter}\hspace{-.5em}\textbf{\arabic{counter}.}}
\newcommand{\theorem}{\paragraph{Theorem}\counter}
\newcommand{\lemma}{\paragraph{Lemma}\counter}
\newcommand{\proposition}{\paragraph{Proposistion}\counter}
\newcommand{\corollary}{\paragraph{Corollary}\counter}

\usepackage{braket} 
\usepackage{bbm}  
\usepackage{framed}
\newcommand{\e}{\mathrm{e}} 
\newcommand{\ord}{\mathrm{ord}}

\newcommand{\diag}{\mathrm{diag}} 
\newcommand{\Tr}{\mathrm{Tr}}  
\newcommand{\wt}{\mathrm{wt}}  
\newcommand{\swt}{\mathrm{swt}}
\newcommand{\norm}[1]{\left\vert\left\vert #1 \right\vert\right\vert }
\newcommand{\ZZ}{\mathbb{Z}}
\newcommand{\FF}{\mathbb{F}}

\newcommand{\cliff}{\mathcal{C} \hspace{-.2em} \ell}

\newcommand{\ZDZ}{\mathbb Z/D\mathbb Z}
\newcommand{\ZdZ}{\mathbb Z/d\mathbb Z}
\newcommand{\GHZp}{\mathrm{GHZ}_D^n(p)}
\newcommand{\TA}{^\mathrm{T_A}}  
\newcommand{\nchoosem}{\genfrac{(}{)}{0pt}{1}{n}{m}}
\newcommand{\nchoosej}{\genfrac{(}{)}{0pt}{1}{n}{j}}
\newcommand{\mchoosek}{\genfrac{(}{)}{0pt}{1}{m}{k}}
\newcommand{\Span}{\mathrm{span}} 

\newcommand{\rhoGlobGammaP}{\rho_{\mathrm{glob},\Gamma}(p)}

\newcommand{\rhoGlobPsiP}{\rho_{\mathrm{glob},\psi}(p)}
\newcommand{\rhoLocPsiP}{\rho_{\mathrm{loc},\psi}(p)}

\newcommand{\gray}{\textcolor{gray}}

\begin{document} 
\title{Small quantum networks in the qudit stabilizer formalism}	
\author{Daniel Miller\\ \normalsize Institut f\"ur Theoretische  Physik III, Heinrich-Heine-Universit\"at D\"usseldorf \\ \vspace*{2cm} Supervised by Prof. Dr. Dagmar Bru\ss}
\date{\today.}

\thispagestyle{empty}
\begin{center} 
\vspace*{2em}
 \Huge \textbf{Small quantum networks in the \\ qudit stabilizer formalism } \\
 \vspace*{2em}
 \large A thesis submitted in partial fulfillment of the requirements for the degree of \\
 \vspace*{1em}
 \Large \textbf{Master of Science} \\ \textbf{Physics} \\
 \vspace*{1em} 
 \large by \\
 \vspace*{1em} 
 \Large  \textbf{Daniel Miller}\\
 \vspace*{2em} 
 \large at the \\
 \vspace*{1em} 
 \textbf{Institut f\"ur Theoretische Physik III}\\
 \textbf{Quanteninformation} \\
 \textbf{Heinrich-Heine-Universit\"at D\"usseldorf }\\
 \vspace*{1em} 
  \includegraphics[width = .6\textwidth]{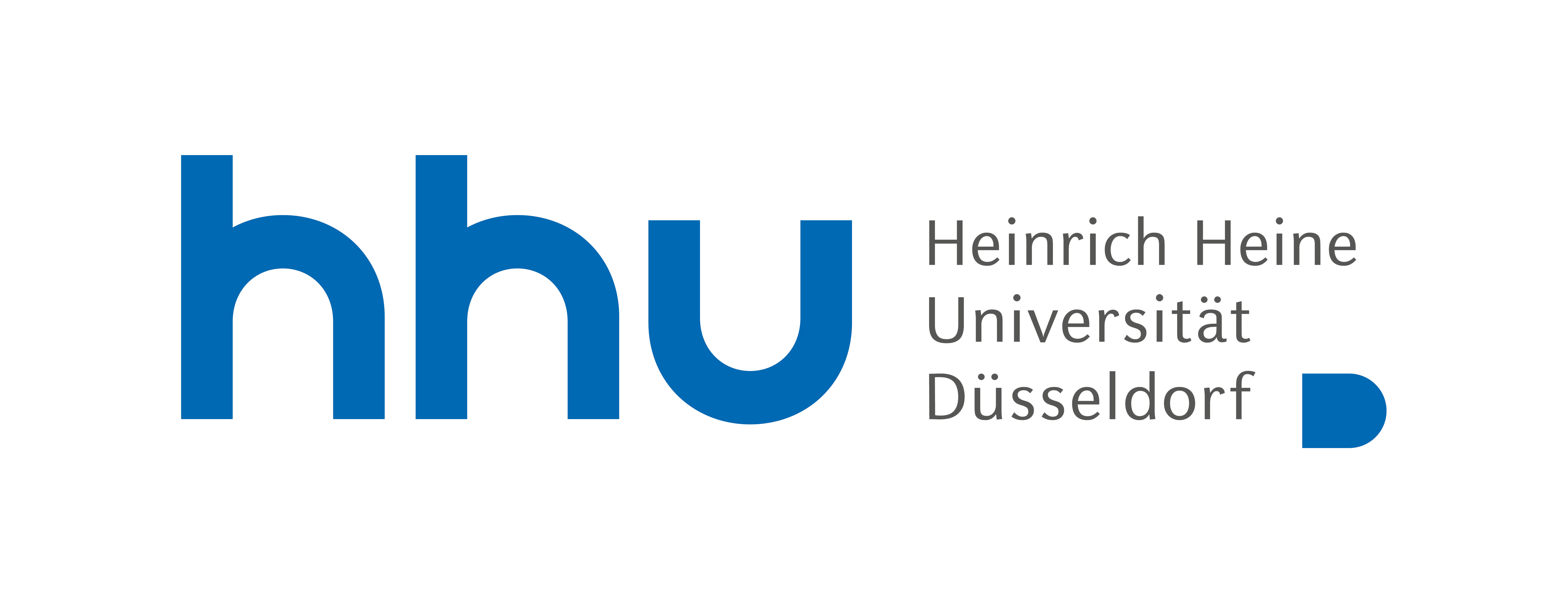}\\
 \vspace*{1em}   
 supervised by Prof. Dr. Dagmar Bru\ss.\\
 \vspace*{2em}
 2019
\end{center}
 
 
\newpage

\setcounter{page}{1} 

\section*{Abstract}

How much noise can a given quantum state tolerate without losing its entanglement? 
For qudits of arbitrary dimension, I investigate this question for two noise models: Global white noise, where a depolarizing channel is applied to all qudits simultaneously, and local white noise, where a single qudit depolarizing channel is applied to every qudit individually. Using a unitary generalization of the Pauli group, I derive noise thresholds for stabilizer states, with an emphasis on graph states, and compare different entanglement criteria. The PPT and reduction criteria generally provide high noise thresholds, however, it is difficult to apply them in the case of local white noise. Entanglement criteria based on so-called sector lengths, on the other hand, provide coarse noise thresholds for both noise models. 
The only thing one has to know about a state to compute this threshold is the number of its full-weight stabilizers.
In the special case of qubit graph states, I relate this question to a graph-theoretical puzzle and solve it for four important families of states.
For Greenberger-Horne-Zeilinger states under local white noise, I obtain for the first time a noise threshold below which so-called semiseparability is ruled out.

\vspace{10em}

\section*{Master's thesis statement of originality}
I hereby confirm that I have written the accompanying thesis by myself, without contributions from any other sources other than those cited in the text and acknowledgements. This applies also to all graphics, drawings, maps and images included in the thesis.
For an in-depth clarification about originality of the proven lemmata, propositions, theorems, and corollaries, see Section~\ref{sec:9}.

D\"usseldorf, October 21, 2019.
\vspace{10em}

Daniel Miller

\newpage
\tableofcontents

\newpage

\section{\label{sec:1}\protect Introduction}  
The prospect of an eventual world-spanning quantum internet and, more generally, quantum technologies 
has created great interest and motivates tremendous investments~\cite{RBTC17, ABBCEEEGGJKLRSTWWW18, WEH18}. 
A quantum internet offers---among an increasing number of other applications \cite{WEH18, GLM01, ChrWeh05, GJC12, KBdGL18}---the possibility of quantum key distribution,
a cryptographic procedure whose security is not based on computational hardness assumptions but on the laws of quantum mechanics~\cite{BB84, Ekert91, Bruss98}.
The crucial feature of quantum mechanics which enables these applications is called quantum entanglement~\cite{Schroedinger35, Horodecki09, WaGrEi16}. 
As quantum entanglement is a phenomenon with many facets, it is difficult to characterize and there are still many unanswered questions it has raised.  
In particular, multipartite entanglement and entangled multi-level quantum systems are not understood in full depth~\cite{WaGrEi16}.

The stabilizer formalism, originally introduced by Gottesman to study quantum error-correcting codes~\cite{GottesmanPhD}, provides an efficient description of a certain class of quantum states.
It has been used to introduce so-called graph states which are particularly suited for quantum network applications~\cite{HeinEisBri04}:
If one interprets the vertices of a graph as nodes of a quantum network, the edges of the graph correspond to optical links through which an exchange of quantum information encoded into photons is possible.
From an experimental point of view, quantum networks are difficult to realize since losses of photons limit the transmission distance of photons. 
Furthermore, operational errors deteriorate the overall performance of any quantum communication protocol.
This necessitates the investigation of quantum networks in the presence of noise.

In my Bachelor's thesis~\cite{Bachelor} and Refs.~\cite{MHKB18, MHKB19}, we have investigated the impact of physical noise on the entangled state distributed by an error-corrected, one-way quantum repeater based on higher-dimensional qudits.
In this Master's thesis, we consider noisy quantum states that have already been distributed within a quantum network.
For different noise models, we will apply several entanglement criteria to establish critical noise thresholds from which one can gain information about the entanglement of a given noisy state.
For any implementer of a quantum network it is crucial to know how much noise a given target state can tolerate without losing its entanglement. In particular, such noise thresholds provide benchmarks to the performance of such quantum networks.

This thesis is structured as follows.
In Sec.~\ref{sec:2}, we introduce the notion of quantum entanglement and discuss several criteria (entropy, PPT, reduction, witnesses) for certifying that a given state is entangled.
In Sec.~\ref{sec:3}, we formally define qudit graph states and introduce the so-called qudit stabilizer formalism which is very useful for studying these states. 
In Sec.~\ref{sec:4}, we will apply some of the entanglement criteria (entropy, PPT, reduction) to general   qudit graph states to establish noise thresholds. 
As this first approach is only easily applicable for very simple noise models,
we introduce the concept of so-called sector lengths of a quantum state in Sec.~\ref{sec:5}.
If such a sector length exceeds a certain bound, one can infer certain information about the entanglement of a given state. 
One of the main results of this thesis is a formula for how sector lengths of a pure stabilizer state get diminished for two noise models of global and local white noise, respectively.
This leads to new noise thresholds which we numerically investigate and compare to other known thresholds in 
Sec.~\ref{sec:6} for important examples of small qubit networks. 
In Sec.~\ref{sec:7}, we conclude and give an outlook.
The acknowledgements are in Sec.~\ref{sec:8}.
In Sec.~\ref{sec:9}, we provide a statement of originality where we clarify to which extend the results in this thesis were already known before.

\section{\label{sec:2}\protect Entanglement}  
If a quantum system is composed of $n\ge2$ parties, the structure of its density operator $\rho$ can be used to classify correlations between the different parties~\cite{WaGrEi16}.
Completely uncorrelated systems are in a so-called \emph{product state}, i.e., their density operator is of the form $\rho = \rho^{(1)} \otimes \ldots \otimes \rho ^{(n)}$.
More generally, $\rho$ is called \emph{fully separable} if it can be written as the convex combination of such product states, i.e., 
\begin{align}\label{eq:separable}
 \rho &= \sum_{j=1}^N p_j \rho_j^{(1)} \otimes \ldots \otimes \rho_j^{(n)},
\end{align}
where $\{p_j\}$ is a probability distribution according to which the product states have been mixed.
Separable states constitute the broadest class of physical states whose correlations of local measurement statistics can be explained without quantum mechanics. 
Any state $\rho$ which does not admit a decomposition as in Eq.~\eqref{eq:separable} is called \emph{entangled}.

Quantum entanglement manifests itself in the phenomenon that a multipartite state can contain more information than the combination of its marginals~\cite{Schroedinger35, Horodecki09}.
This is formalized by the \emph{von Neumann entropy} $S[\rho]:= -\Tr\left[\rho \log_2 [\rho ]\right]$:
Only an entangled quantum states $\rho$ can fulfill 
\begin{align}\label{eq:entropic_inequality}
 S[\rho] < S[ \Tr_J [\rho ] ], 
\end{align}
where the partial trace  $\Tr_J [\rho]$ yields the reduced state of the parties remaining after discarding a suited subset of parties $J\subset I:=\{1,\ldots,n\}$. 
Formally, it is defined as 
\begin{align}
\Tr_J [\rho]  := \sum_{j\in J} \sum_{k=1}^{\dim (\mathcal{H}_j)} \bra{b_k^{(j)}} \rho \ket{b_k^{(j)}} ,
 \end{align}
 where $\vert{b_1^{(j)}}\rangle,\ldots,\vert{b_{\dim(\mathcal{H}_j)}^{(j)}}\rangle \in \mathcal{H}_j$ is a orthonormal basis of the Hilbert space associated to party $j$. 

Consider, for example, the bipartite qubit \emph{Werner state}~\cite{Werner89}
\begin{align} \label{eq:Werner_state}
 \rho_\mathrm{W}(p) := (1-p) \ket{\Phi^+}\bra{\Phi^+} + p\frac{\mathbbm 1}{4}
\end{align}
which is a mixture of the maximally entangled qubit Bell state $\ket{\Phi^+}=\frac{1}{\sqrt 2}(\ket{00} + \ket{11})$ and the maximally mixed state on two qubits, where $0\le p \le 1$.
Regardless of $p$, both one-party marginals of $\rho_\mathrm{W}(p)$ are maximally mixed, i.e., 
$\rho^{(1)}=\Tr_2 [\rho_\mathrm{W} (p)] =\mathbbm 1/2$ and likewise for $\rho^{(2)}$.
Therefore, also the von Neumann entropies $S[\rho^{(1)}]= S[\rho^{(2)}] =1$ are independent of $p$.
The entropy of the total state, however, is given by 
\begin{align}
 S[\rho_\mathrm{W}(p)] &=  -3 \left( \frac{p}{4} \right) \log_2\left(\frac{p}{4} \right) - \left( 1-\frac{3p}{4}\right) \log_2 \left(1-\frac{3p}{4} \right) 
\end{align}
as the eigenvalues of $\rho_\mathrm{W}(p)$ are just $\lambda_1=\lambda_2=\lambda_3= p/4$ and $\lambda_4 =1-3p/4$. 
Note that the function $p \mapsto S[\rho_\mathrm{W}(p)]$ is strictly monotonically increasing on the domain $[0,1]$ and takes values $S[\rho_\mathrm{W}(0)] =0$, $S[\rho_\mathrm{W}(1)] =2$, and
$S[\rho_\mathrm{W}(p_\mathrm{crit}^\mathrm{Entr} )]=1 $ for 
$p_\mathrm{crit}^\mathrm{Entr} \approx 0.2524$. 
Hence, the Werner state $\rho_\mathrm{W}(p)$ contains more information than its marginals iff $p < p_\mathrm{crit}^\mathrm{Entr}$. 
As argued above, the Werner state is necessarily entangled in this case. Note that this noise threshold is not tight.

In the subsequent subsections, we will dive into the theory of quantum entanglement.
First, in Sec.~\ref{sec:2.1}, further criteria 
for the verification of entanglement in the bipartite setting are presented.
Afterwards, in Sec.~\ref{sec:2.2} and~\ref{sec:2.3}, we review in more detail how the notions of separability and entanglement generalize to the multipartite setting.

\subsection{\label{sec:2.1}\protect Entanglement criteria based on positive maps} 
Positive maps can provide much stronger entanglement criteria than Inequality~\eqref{eq:entropic_inequality}. 
Hereby, a linear map $\Lambda: \mathcal{B}(\mathcal{H}_\mathrm{A})\rightarrow \mathcal{B}(\mathcal{H}_\mathrm{A'}), \rho \mapsto \Lambda [\rho]$ is called \emph{positive} if $\rho_\mathrm{A} \ge 0$ implies $\Lambda[\rho_\mathrm{A}]\ge0$, where $\mathcal{H}_\mathrm{A}$ and $\mathcal{H}_\mathrm{A'}$ are two Hilbert spaces and 
 $\mathcal{B}(\mathcal{H}_\mathrm{A})$ denotes the Hilbert space of bounded operators on $\mathcal{H}_\mathrm{A}$ and likewise for $\mathcal{B}(\mathcal{H}_\mathrm{A'})$. 
 In words: $\Lambda$ is positive if it maps every positive semidefinite operator to a positive semidefinite operator.

 As it has been shown in Ref.~\cite{Horodecki96}, a bipartite state $\rho_\mathrm{AB}$ acting on $\mathcal{H}_\mathrm{A}\otimes \mathcal{H}_\mathrm{B}$ is separable iff for every positive map $\Lambda: \mathcal{B}(\mathcal{H}_\mathrm{A})\rightarrow \mathcal{B}(\mathcal{H}_\mathrm{A'})$ the operator $(\Lambda \otimes \mathbbm{1}_\mathrm{B})[\rho_\mathrm{AB}]\in \mathcal{B(\mathcal{H}_\mathrm{A'}\otimes \mathcal{H}_\mathrm{B})} $ is positive semidefinite.
If, for a given state  $\rho_\mathrm{AB}$, one finds a positive map $\Lambda$ for which $(\Lambda \otimes \mathbbm{1}_\mathrm{B})[\rho_\mathrm{AB}]$ possesses at least one negative eigenvalue, one has proven that $\rho_\mathrm{AB}$ is entangled.
Obviously, a positive map $\Lambda$ with the property that also $\Lambda\otimes \mathrm{id}_{\mathbb{C}^n}$ is positive for all $n\ge 1$ cannot provide a nontrivial entanglement criterion. Such maps are called \emph{completely positive}.
In Secs.~\ref{sec:PPT_criterion} and~\ref{sec:reduction_criterion} we will discuss two positive but not completely positive maps and their corresponding entanglement criteria.

\subsubsection{Peres-Horodecki criterion}
\label{sec:PPT_criterion}

The first positive map that was recognized to be useful for the verification of quantum entanglement is the transposition map $\Lambda^\mathrm{trans}_\mathrm{A}: \mathcal{B}(\mathcal{H}_\mathrm{A})\rightarrow \mathcal{B}(\mathcal{H}_\mathrm{A}), \rho_\mathrm{A} \mapsto \rho_\mathrm{A}^\mathrm{T}$~\cite{Peres96, Horodecki96}.
In fact, the eigenvalues of an operator are invariant under transposition, in particular, $\rho_\mathrm{A} \ge 0$ iff $\rho_\mathrm{A}^\mathrm{T}\ge 0$. 
However, when this map is extended to a second party (with $\dim( \mathcal{H}_\mathrm{B})\ge 2$), the resulting map
\begin{align}
 \Lambda^\mathrm{trans}_\mathrm{A} \otimes \mathrm{id}_\mathrm{B}: \mathcal{B(\mathcal{H}_\mathrm{A}\otimes \mathcal{H}_\mathrm{B})} \longrightarrow \mathcal{B(\mathcal{H}_\mathrm{A}\otimes \mathcal{H}_\mathrm{B})},
 \hspace{2em}
 \rho_\mathrm{AB} \longmapsto \rho_\mathrm{AB}\TA,
\end{align}
called the \emph{partial transpose} of A is not positive, where for a fixed product basis $\rho_\mathrm{AB}\TA$ is given by
\begin{align}
 \bra{i_\mathrm{A}, j_\mathrm{B}} \rho_\mathrm{AB}\TA,\ket{k_\mathrm{A}, l_\mathrm{B}} =
 \bra{k_\mathrm{A}, j_\mathrm{B}} \rho_\mathrm{AB} \ket{i_\mathrm{A}, l_\mathrm{B}}.
\end{align}
The resulting entanglement criterion (which is independent of the choice above) is given by 
\begin{align}\label{eq:PPT_criterion}
  \rho_\mathrm{AB}\TA \ge 0 & \Longleftarrow   \rho_\mathrm{AB} \text{ is separable},
\end{align}
or by its logically equivalent contrapositive, 
\begin{align}\label{eq:NPT_criterion}
  \rho_\mathrm{AB}\TA  \not \ge 0
  &\Longrightarrow  \rho_\mathrm{AB} \text{ is entangled}.
\end{align}
This is called the \emph{Peres-Horodecki criterion} or \emph{PPT-criterion} as, by Eq.~\eqref{eq:PPT_criterion}, every separable state is PPT, i.e., it has a positive partial transpose. Entangled states for which entanglement can be verified by means of Eq.~\eqref{eq:NPT_criterion} are called NPT for negative partial transpose. 
For quantum systems of combined dimension $\dim(\mathcal{H}_\mathrm{A})\dim(\mathcal{H}_\mathrm{B}) \le 6$ 
a state is entangled iff it is NPT~\cite{Horodecki96}. 
In general, however, there exist so-called \emph{bound entangled} states which are PPT but still entangled~\cite{Horodecki09}.

For the example of the Werner state $\rho_\mathrm{W}(p)$ from Eq.~\eqref{eq:Werner_state}, however, the combined dimension is small enough that the Peres-Horodecki criterion is sufficient to characterize entanglement completely. Let us derive the corresponding (tight) critical noise threshold.
The partial transpose of the Werner state is given by the block diagonal matrix
\begin{align} \label{eq:PPT_Werner}
 \rho_\mathrm{W}(p)\TA =   \left(
\def\arraystretch{1.3}
 \begin{array}{cccc} 
\frac{2-p }{4}& \gray 0 & \gray 0 & \gray 0 \\
\gray 0 & \frac{p}{4} & \frac{1-p }{2} & \gray 0\\
\gray 0 & \frac{1-p }{2}& \frac{p}{4} & \gray 0 \\
\gray 0 & \gray 0 & \gray 0 & \frac{2-p }{4}
\end{array}
  \right)
\end{align}
from which one can easily read off the eigenvalues.
The eigenvalues of the $1\times 1$-blocks, $(2-p)/4$, are positive for all $p\in[0,1]$, i.e., they cannot be used to apply the Peres-Horodecki criterion.
The eigenvalues of the $2\times 2$-block are given by $p/4 \pm (1-p)/2$.~\footnote{Note that the eigenvalues of a $2\times 2$-matrix of the form $\begin{pmatrix} x& y \\y&x \end{pmatrix}$ are simply given by $x\pm y$.}
While the ``$+$''-eigenvalue is also equal to $(2-p)/4$, the ``$-$''-eigenvalue, $(3p-2)/4$, is negative for all $p<2/3$. 
That is, the critical noise threshold for the Werner state is given by $p_\mathrm{crit}^\mathrm{PPT}=2/3$
\cite{Peres96}.

Note that the reason why the entropic inequality~\eqref{eq:entropic_inequality} can only be used to detect entanglement in Werner states for $p < p^\mathrm{Entr}_\mathrm{crit} \approx 0.2524$ is because the von Neumann entropy incorporates both classical and quantum correlations. 
Only if the correlations are so large that they cannot be explained classically, one can conclude that a given state is entangled. 

\subsubsection{Reduction criterion}
\label{sec:reduction_criterion}

Another, similarly constructed entanglement criterion is the reduction criterion~\cite{CeAdGi99, Horodecki99}.
It is based on the positive map, $\Lambda^\mathrm{red}_\mathrm{A} : \mathcal{B}(\mathcal{H}_\mathrm{A})\rightarrow \mathcal{B}(\mathcal{H}_\mathrm{A}),  \rho_\mathrm{A} \mapsto   \mathbbm{1}_\mathrm{A}\Tr[\rho_\mathrm{A}] -\rho_\mathrm{A}$.
Its extension to a second quantum system~B, 
\begin{align}
 \Lambda^\mathrm{red}_\mathrm{A} \otimes \mathrm{id}_\mathrm{B}: \mathcal{B(\mathcal{H}_\mathrm{A}\otimes \mathcal{H}_\mathrm{B})} \longrightarrow \mathcal{B(\mathcal{H}_\mathrm{A}\otimes \mathcal{H}_\mathrm{B})},
 \hspace{2em}
 \rho_\mathrm{AB} \longmapsto  \mathbbm{1}_\mathrm{A} \otimes \Tr_\mathrm{A}[\rho_\mathrm{AB}] \hspace{.2em}-\rho_\mathrm{AB},
\end{align}
is not a positive map. 
Similarly to the partial transpose one obtains the \emph{reduction criterion}
\begin{align} 
\label{eq:red_criterion_ent}
  \mathbbm{1}_\mathrm{A} \otimes \Tr_\mathrm{A}[\rho_\mathrm{AB}] \hspace{.2em}-\rho_\mathrm{AB}  \not \ge 0
  &\Longrightarrow  \rho_\mathrm{AB} \text{ is entangled}.
\end{align}
In general, this criterion is not stronger than the Peres-Horodecki criterion in the sense that 
$\mathbbm{1}_\mathrm{A} \otimes \Tr_\mathrm{A}[\rho_\mathrm{AB}] \hspace{.2em}-\rho_\mathrm{AB}  \not \ge 0$ implies $\rho_\mathrm{AB}\TA\not \ge 0$, however, it is sometimes easier work with Eq.~\eqref{eq:red_criterion_ent} rather than Eq.~\eqref{eq:NPT_criterion}.
In fact, in Sec.~\ref{sec:4}, where we will establish tolerable noise thresholds for graph states, it will turn out that both criteria lead to the same noise threshold while the result is more readily established with the reduction criterion.

Let us also illustrate this criterion at the example of the Werner state.
As we have mentioned already,  the reduced state $ \Tr_\mathrm{A}[\rho_\mathrm{W}(p)]$ is maximally mixed, regardless of $p$. 
Thus, the operator of interest has the matrix form
\begin{align}
   \mathbbm{1}_\mathrm{A} \otimes \Tr_\mathrm{A}[\rho_\mathrm{W}(p)] \hspace{.2em}-\rho_\mathrm{W}(p)
   = 
\left( 
\def\arraystretch{1.3}
\begin{array}{cccc}
    \frac{1}{2}- \frac{2-p}{4} & \gray 0 & \gray 0 & -\frac{1-p}{2} \\
    \gray 0 &  \frac{1}{2}-\frac{p}{4} & \gray 0  & \gray 0  \\
    \gray 0 &  \gray 0  & \frac{1}{2}-\frac{p}{4} &  \gray 0 \\
    -\frac{1-p}{2} & \gray 0 & \gray 0 & \frac{1}{2}- \frac{2-p}{4} 
   \end{array}
\right).
\end{align}
Note that $\frac{1}{2}- \frac{2-p}{4}= \frac{p}{4}$ holds.
By swapping rows and columns number 2 and 4, we obtain the block-diagonal matrix 
\begin{align}
\left( 
\def\arraystretch{1.3}
\begin{array}{cccc}
    \frac{p}{4} & -\frac{1-p}{2} & \gray 0 & \gray 0  \\
  - \frac{1-p}{2}   & \frac{p}{4}  &\gray 0   & \gray 0  \\
    \gray 0  &  \gray 0 & \frac{1}{2}-\frac{p}{4} & \gray 0 \\
   \gray 0 &  \gray 0   &  \gray 0 &\frac{1}{2}-\frac{p}{4}
   \end{array}
\right)
\end{align}
which has the same eigenvalues as $\mathbbm{1}_\mathrm{A} \otimes \Tr_\mathrm{A}[\rho_\mathrm{W}(p)] \hspace{.2em}-\rho_\mathrm{W}(p)$ because the determinant is an alternating multilinear form.
Up to a sign on the off-diagonal, this is the same $2\times2$-block as in Eq.~\eqref{eq:PPT_Werner}, i.e., its eigenvalues are again given by $p/4 \mp (1-p)/2$. 
Therefore, the corresponding noise threshold is $p^\mathrm{Red}_\mathrm{crit}=2/3$ as well~\cite{CeAdGi99, Horodecki99}. 

\newpage

\subsection{\label{sec:2.2}\protect Multipartite entanglement}In the bipartite setting, one only distinguishes between states which are separable or entangled. 
In the multipartite case, the notion of entanglement is much richer as one can define separability with respect to one or more specific partitions.
After introducing these notions in Sec.~\ref{sec:2.2.1}, we discuss the abilities and limitations that bipartite entanglement criteria face when they are applied in the multipartite case in Sec.~\ref{sec:2.2.2}.
Finally, in Sec.~\ref{sec:entanglement_witnesses}, we introduce entanglement witnesses which provide an experimentally accessible alternative to entanglement criteria based on positive maps.

\subsubsection{Partial separability}
\label{sec:2.2.1}

A \emph{partition} of the set of all parties, $I= \{ 1,\ldots, n \}$, is a set of disjoint subsets $I_1, \ldots, I_k \subset I$ for which $I = \bigcupdot _{i=1}^k I_i$ holds.
A quantum state is called \emph{separable with respect to this partition} if its density operator is of the form
\begin{align}
 \rho &= \sum_{j=1}^N p_j \rho_j^{I_1} \otimes \ldots \otimes \rho_j^{I_k},
\end{align}
where $\{ p_j\}$ is a probability distribution and each $\rho_j^{I_i}$ is some quantum state on the systems specified by $I_i$~\cite{Horodecki09}. 
The most refined partition, $I_1=\{1\}, \ldots, I_n=\{n\}$, corresponds to a fully separable state as we have already defined in Eq.~\eqref{eq:separable}.

Consider $k$ natural numbers $n>n_1 \ge \ldots \ge  n_k > 0$ which sum up to $n$.
We call a state \emph{$(n_1,\ldots,n_k)$-separable} if it is a convex combination of states which are separable with respect to some partition $\{I_1,\ldots, I_k\}$ with $\vert I_i \vert = n_i$ for all $1\le i \le k$. \footnote{The specific partition may be different for each state in the convex combination. Only the sizes of the subsets are fixed.}
States which are $(n-1,1)$-separable are also called \emph{semiseparable}~\cite{Horodecki09}.

More generally, a state is called $k$-separable if it is a convex combination of states which are separable with respect to any partition of $I$ into $k$ subsets $I_1,\ldots, I_k$~\cite{WaGrEi16}.
Since $k$-separability implies $(k-1)$-separability, the strongest form of entanglement is when a state does not even allow for a  biseparable ($k=2$) decomposition. Such states are called \emph{genuinely multipartite entangled} (GME).

\subsubsection{Positive maps in the multipartite setting}
 \label{sec:2.2.2}
 
In general, it is not possible to investigate multipartite entanglement by means of bipartite entanglement criteria with respect to all possible partitions of the parties. This is best illustrated by the following example~\cite{GueTo09}.
 
 Let $\Phi^+= \ket{\Phi^+}\bra{\Phi^+}$ and $\pi^0=\ket{0}\bra{0}$ denote the projectors onto a qubit Bell pair and a computational basis state, respectively.
 The three qubit state
 \begin{align}\label{eq:3qubits}
  \rho = \frac{1}{3} \left(  \Phi^+ _\mathrm{AB} \otimes \pi^0_\mathrm{C} +\Phi^+ _\mathrm{AC} \otimes \pi^0_\mathrm{B} +\Phi^+ _\mathrm{BC} \otimes\pi^0_\mathrm{A}
  \right)
 \end{align}
is  obviously a convex combination of biseparable states, i.e., it is $(2,1)$-separable. 
However, it is readily verified that $\ket{001}+\ket{010}- \sqrt{2}\ket{100}$ is an eigenvector of
\begin{align}
 \rho\TA &=  \frac{1}{6} \left( \begin{array}{cccc:cccc}
 3&\gray 0&\gray 0&1 &  \gray 0&\gray 0&\gray 0&\gray 0 \\
 \gray 0&\gray 0&\gray 0&\gray 0 &  1&\gray 0&\gray 0&\gray 0 \\
 \gray 0&\gray 0&\gray 0&\gray 0 &  1&\gray 0&\gray 0&\gray 0 \\
 1&\gray 0&\gray 0&1 &  \gray 0&\gray 0&\gray 0&\gray 0 \\
             \hdashline
\gray  0&1&1&\gray 0 &\gray  0&\gray 0&\gray 0&\gray 0 \\
 \gray 0&\gray 0&\gray 0&\gray 0 & \gray 0&1&\gray 0&\gray 0 \\
 \gray 0&\gray 0&\gray 0&\gray 0 & \gray 0&\gray 0&1&\gray 0 \\
 \gray 0&\gray 0&\gray 0&\gray 0 & \gray 0&\gray 0&\gray 0&\gray 0
            \end{array} \right)
\end{align}
 to the eigenvalue $-{1}/{\sqrt{18}}$.
Since the state is symmetric under exchange of the parties,  $-{1}/{\sqrt{18}}$ is also an eigenvalue of $\rho^{\mathrm{T_B}}$ and $\rho^{\mathrm{T_C}}$. 
That is, $\rho$ is biseparable although it is NPT with respect to every nontrivial partition. 

This example shows that a straightforward application of the Peres-Horodecki criterion cannot be used to determine whether a given state is GME or not (it can only rule out full separability). 
Because of this, recently a more sophisticated criterion based on positive maps has been  developed~\cite{CHLM17}. 
There, the idea is to consider maps which are positive when applied to biseparable states but can map to a non-positive operator if applied to a GME state.

\subsubsection{Entanglement witnesses}
\label{sec:entanglement_witnesses}

For the detection of (genuine multipartite) entanglement, so-called \emph{entanglement witnesses} provide an alternative to criteria based on positive maps~\cite{Horodecki96, GueTo09, Terhal00}. 
If a state $\rho$ is entangled, it is always possible (since the set of fully separable states is closed and convex) to find a Hermitian operator $W$ which fulfills the following two conditions:
\begin{enumerate}[label=(\roman*)]
 \item  $\Tr[W\rho] < 0$. 
 \item  $\Tr[W\sigma] \ge 0$ for all fully separable states $\sigma$. 
\end{enumerate}
That is, one can experimentally verify that a given state is entangled by measuring the observable $W$. Even if the exact form of $\rho$ is not known, a negative expectation value would verify entanglement.
For that reason an operator $W$ fulfilling (i) and (ii) was given the name \emph{entanglement witness for $\rho$}.
If one changes ``fully separable'' into ``biseparable'' in condition (ii), an experimental verification of $\Tr[W\rho]<0$ would imply that $\rho$ is GME.
Note, however, that no single entanglement witness can certify (genuine multipartite) entanglement for all entangled states simultaneously.

Once again, consider the example of the  Werner state. 
It can be shown that the Hermitian operator $W:= \mathbbm{1}/2 - \ket{\Phi^+}\bra{\Phi^+}$ is an entanglement witness~\cite{GueTo09}.
To find the values of $p$ for which the expectation value of $W$ is negative, we insert Eq.~\eqref{eq:Werner_state} and obtain
\begin{align}
 \Tr[W\rho_\mathrm{W}(p)] &= \Tr\left [\left( \frac{\mathbbm{1}}{2}- \ket{\Phi^+}\bra{\Phi^+} \right)\left( (1-p) \ket{\Phi^+}\bra{\Phi^+} + p\frac{\mathbbm 1}{4} \right ) \right] \\
  &= \left(\frac{1-p}{2} - (1-p) -\frac{p}{4} \right) \Tr\left[\ket{\Phi^+}\bra{\Phi^+}\right ] + \frac{p}{2} \Tr\left[ \frac{\mathbbm 1}{4} \right ] 
&= \frac{3p-2}{4} ,
\end{align}
where we have used the normalization of the Bell state and the completely mixed state.
Therefore, also entanglement witnesses can yield the tight noise threshold  $p^\mathrm{Wit}_\mathrm{crit}=2/3$.   
For more examples of entanglement witnesses for given entangled (or GME) state see e.g., Refs.~\cite{ToGue05, GueTo09, JuMoGue11, FVMH19}.

\subsection{\label{sec:2.3}\protect Pure, genuinely multipartite entangled quantum states}
Let us assume that the parties have perfect control over their own quantum systems and that they  can exchange classical information among each other. This is the so-called \emph{distant laboratory paradigm}~\cite{Horodecki09}.
The protocols that can be performed with a nonzero probability within this paradigm are commonly referred to as stochastic local operations and classical communication (SLOCC).
While the Bell pair plays the role of a universal unit of bipartite entanglement as every bipartite state can be produced by means of SLOCC from a sufficient amount of Bell pairs~\cite{Horodecki09},
the multipartite situation is more complicated. 
For three qubits, there are exactly two inequivalent classes of genuine tripartite entanglement~\cite{DVC2000}. They are represented by the Greenberger–Horne–Zeilinger state~\cite{GHZ89} 
and the so-called $W$-state,
\begin{align} \label{def:GHZ_W}
 \ket{\mathrm{GHZ}} = \frac{1}{\sqrt{2}} \left( \ket{000}  + \ket{111}  \right )
 \hspace{2em}
 \text{and}
 \hspace{2em}
 \ket{W} = \frac{1}{\sqrt{3}}\left( \ket{100}  + \ket{010}  + \ket{001}  \right ).
\end{align}
It was recently shown that for four or more qubits every set of GME states from which every other state can be reached by means of SLOCC must have full measure, i.e., almost all states must be contained in such a set~\cite{dVSpKr13, GouKrWa17, SWGK18}. In particular, there are infinitely many inequivalent SLOCC-entanglement classes.
This makes it difficult to achieve a complete and operationally meaningful classification of multipartite entanglement.

Here, we will therefore concentrate on a special class of multipartite entangled states called $m$-uniform states which we introduce in Sec.~\ref{sec:m_uniform}.
In Sec.~\ref{sec:AME}, we separately treat the extremal case of $m=\lfloor n/2 \rfloor$-uniform states which are also known as absolutely-maximally-entangled states.

\subsubsection{\textit{m}-uniform states}
\label{sec:m_uniform} 
Fix an integer $m\le n$.
An $n$-partite pure quantum state $\ket{\psi}$ is called \emph{$m$-uniform} if for every subset $J\subset I=\{1,\ldots,n\}$ with $\vert J \vert = m$ elements, the reduced $m$-partite state 
\begin{align}\label{eq:marginal}
 \rho^{(J)} = \Tr_{J^\mathrm{C}} \left[ \ket{\psi}\bra{\psi}  \right]
\end{align}
is maximally mixed, i.e., $\rho^{(J)} = \mathbbm{1}/D^m$. 
Hereby, the quantum systems which are labeled by  $J^\mathrm{C}:=I \backslash J$, i.e., the complement of $J$, have been traced out. 
As pure states $\rho =\ket{\psi}\bra{\psi}$ have a minimal von Neumann entropy of $S[\rho]=0$ and maximally mixed states have a maximal von Neumann entropy of $S[\mathbbm{1}/D^m] = m \log_2(D)$, 
$m$-uniform states constitute extremal cases  of the entropic inequality~\eqref{eq:entropic_inequality}.
In particular, $m$-uniform states are entangled if $m\ge 1$. In fact, they are even not semiseparable. 
Note that every $m$-uniform state is also $(m-1)$-uniform~\cite{Scott04}.

While the $W$ state is not even 1-uniform as tracing out two parties yields the reduced state $\Tr_{2,3} \left[ \ket{W}\bra{W}  \right] = \frac{2}{3}\ket{0}\bra{0} + \frac{1}{3}\ket{1} \bra{1}$, all generalized Greenberger–Horne–Zeilinger states
\begin{align} \label{def:GHZnD}
 \ket{\mathrm{GHZ}_{D}^n} = \frac{1}{\sqrt{D}}\sum_{j=0}^{D-1} \ket{j}^{\otimes n}
\end{align}
are $1$-uniform  for all $D,n\ge 2$ (but not $m$-uniform for $m>1$).
Note that, for qubits, any state that is symmetric under exchange of parties is $m$-uniform for at most $m=1$~\cite{ArCerf13}. 
We are not aware about an analogous statement about higher-dimensional states.

\subsubsection{Absolutely-maximally-entangled states}
\label{sec:AME}

As it can be shown using the Schmidt decomposition, any pure $n$-partite quantum state can only be an $m$-uniform state if $m\le n/2$~\cite{ArCerf13}.
The limit case $m=\lfloor n/2\rfloor$ has received particular attention.
In Ref.~\cite{HCLRL12}, the term 
\emph{absolutely-maximally-entangled} (AME) state has been introduced for states which are
$\lfloor n/2\rfloor$-uniform. 
Note that AME implies GME~\cite{WaGrEi16} but not vice versa as the example of the $W$-state shows.
Also note that AME states can be used as a resource for multipartite quantum teleportation schemes and quantum secret sharing~\cite{HCLRL12}.

Later, in Sec.~\ref{sec:3.4}, we will present a family of AME states for $n=4$ parties and all odd dimensions $D\ge3$.
It is an open problem for which combinations of $n\ge2$ and $D\ge3$ AME states do exist. 
For qubits, i.e., $D=2$, however, the classification has been completed recently:
AME states do not exist for 4 parties~\cite{GouWal10}, 7 parties~\cite{HuGueSie17} or more.
However, they do exist for 2, 3, 5 and 6 parties~\cite{HCLRL12, ArCerf13}.
The four Bell states 
\begin{align} \label{def:Bell_states}
\ket{\Phi^\pm}= \frac{1}{\sqrt{2}} (\ket{00}\pm\ket{11})
\hspace{1em} \text{and} \hspace{1em}
\ket{\Psi^\pm}= \frac{1}{\sqrt{2}} (\ket{01}\pm\ket{10})
 \end{align}
are examples of bipartite AME states.
As we have mentioned already, the state $\ket{\mathrm{GHZ}_2^3}$ is a 3-partite, 1-uniform state, thus an AME state.
The two logical states of the five qubit code~\cite{LMPZ96},
\begin{align}
\ket{\bar 0^5} := \frac{1}{\sqrt{8}} \big(
&-\ket{00000} +\ket{01111} -\ket{10011}  +\ket{11100} \\ \nonumber
&+\ket{00110} +\ket{01001} +\ket{10101}  +\ket{11010} \big)
\\ 
\text{ and } 
 \hspace{1em} \ket{\bar 1^5} := \frac{1}{\sqrt{8}} \big(
&-\ket{11111} +\ket{10000} +\ket{01100} -\ket{00011} \\ \nonumber
&+\ket{11001} +\ket{10110} -\ket{01010} -\ket{00101} \big),
\end{align} 
and the states $\frac{1}{\sqrt{2}} \left( \ket{0}\otimes\ket{\bar 0^5} \pm \ket{1}\otimes \ket{\bar 1^5} \right)  $ and $ \frac{1}{\sqrt{2}} \left( \ket{0}\otimes\ket{\bar 1^5} \pm \ket{1}\otimes \ket{\bar 0^5} \right)$
are known examples of AME states for $n=5$ and $n=6$ parties, respectively~\cite{ArCerf13}. 

\section{\label{sec:3}\protect Qudit stabilizer formalism} 
 
The essential idea of the \emph{stabilizer formalism}~\cite{GottesmanPhD} is to organize the exponentially fast growing Hilbert space of a multipartite quantum system using algebraic methods.
This is done by labeling the basis states of a single quantum system by the elements of an algebraic ring~\cite{Bosch}.\footnote{A \emph{ring} $R$ is a set closed under addition and multiplication, both of which are commutative. Furthermore, any ring contains a zero element and a one element. If every nonzero element of $R$ has a multiplicative inverse, $R$ is called a \emph{field}.}
The addition of the ring is employed to define \emph{generalized Pauli operators}. 
More specifically, if $a,b$ are two elements of the ring, the action $X(b)\ket{a}= \ket{a+b}$ defines a unitary operator $X(b)$.
All operators arising in this way commute with each other because the addition in the ring is commutative. 
Therefore, there is a unitary operator $F$, called the \emph{quantum Fourier transform}, which simultaneously
diagonalizes all the Pauli $X$ operators~\cite{ConradMinPol}.
The diagonalized operators play the role of generalized Pauli $Z$ operators. 

In this thesis, we restrict ourselves to the choice of the ring of integers modulo $D$ as this is the simplest case which includes all possible qudit dimensions.
This choice leads to a specific generalization of the Pauli group and Clifford group to qudits which we cover in Sec.~\ref{sec:3.1}.
In Sec.~\ref{sec:3.2}, we introduce stabilizer states and explain how they can  efficiently be described within the stabilizer formalism.
The important subclass of qudit graph states is introduced in Sec.~\ref{sec:3.3}.
Finally, in Sec.~\ref{sec:3.4} we apply the stabilizer formalism to construct tetrapartite odd-dimensional qudit graph states which are absolutely-maximally-entangled.

\subsection{Pauli group and Clifford group for qudits}
\label{sec:3.1}

A \emph{qudit} is a quantum system with a Hilbert space $\mathcal{H}$ of finite dimension $D\ge2$. 
One can choose an orthonormal basis of $\mathcal{H}$ and label it using $\ZDZ=\{0,1,\ldots, D-1\}$, the ring of integers modulo $D$, i.e.,
\begin{align}
 \mathcal{H} = \Span_\mathbb{C} \{\, \ket{k} \  \vert \ k\in \ZDZ   \}.
\end{align}
This basis is referred to as \emph{computational basis}.
Any pure state of an $n$-qudit system can be written as
\begin{align}
 \ket{\psi} &= \sum_{\mathbf{j}\in (\ZDZ)^n} z_\mathbf{j} \ \ket{\mathbf{j}},
\end{align} 
where the {probability amplitudes} $z_\mathbf{j}\in \mathbb{C}$ are normalized to $\sum_{\mathbf{j} } \vert z_\mathbf{j} \vert ^2 = 1$, and the multi-qudit computational basis states $\ket{\mathbf{j}}$ are labeled by vectors $\mathbf{j}=(j_1, \ldots, j_n)$ in the free module $(\ZDZ)^n$.\footnote{\emph{Modules} over a ring are defined analogously to vector spaces over a field. In general, a module over a ring does not necessarily have a basis. A module which has a basis is called \emph{free module}.}
If $\ket{\psi}$ is measured in the computational basis, the result will be a random vector $\mathbf{j}$ with probability $\vert z_\mathbf{j} \vert ^2 \in [0,1]$.

For quantum information processing purposes, it is crucial to manipulate such quantum states by means of unitary operations.
In this subsection, we will describe two important groups of unitary qudit operations; the qudit Pauli group and the qudit Clifford group. 
We start with their abstract definitions in Sec.~\ref{sec:3.1.1} and discuss a possible physical implementation in Sec.~\ref{sec:3.1.2}.
 
\subsubsection{Abstract definition of the Pauli group and Clifford group for qudits}
\label{sec:3.1.1}
Let $\omega_D:=\e^{2\pi i / D}$ be a primitive $D^\mathrm{th}$ complex root of unity. 
The operators 
\begin{align}\label{def:ZDZ_Pauli}
 X_D := \sum_{k\in \ZDZ} \ket{k+1}\bra{k}
 \hspace{8mm}  \text{ and } \hspace{8mm}
 Z_D := \sum_{k\in \ZDZ} \omega_D^k \ket{k}\bra{k} 
\end{align}
are called qudit Pauli $X$ and $Z$ operator, respectively~\cite{Gottesman99}.  
The product of two Pauli operators is again a Pauli operator. For $n$ qudits there are (up to a global phase) $D^{2n}$ different Pauli operators, each of which can be written as
\begin{align}\label{eq:ZDZ_pauli_standard_form}
X_D^\mathbf{r}Z_D^\mathbf{s}   
&:= \bigotimes_{i=1}^n X_D^{r_i}Z_D^{s_i} 
\hspace{5mm}=\sum_{\mathbf{k}\in (\ZDZ)^n} \omega_D^\mathbf{k\cdot s} \ket{\mathbf{k+r}}\bra{\mathbf{k}} 
\end{align}
for unique vectors $\mathbf{r,s} \in (\ZDZ)^n$, where $\mathbf{k\cdot s}= \sum_{i=1}^n k_is_i$ is the standard bilinear form, and $\mathbf{k+r}=(k_1+r_1, \ldots, k_n+r_n)$ is  the vector addition in $(\ZDZ)^n$.
From the fact that two Pauli operators commute up to a phase, $\omega_DX_DZ_D=Z_DX_D$, the \emph{Pauli group law},
\begin{align} \label{eq:pauli_commutation}
    (X_D^\mathbf{r}Z_D^\mathbf{s})(X_D^\mathbf{r'}Z_D^\mathbf{s'}) &= \omega_D^{ \mathbf{r'\cdot s-r\cdot s'}} (X_D^\mathbf{r'}Z_D^\mathbf{s'})(X_D^\mathbf{r}Z_D^\mathbf{s}),
\end{align} 
is readily verified. 
Besides $X_D$ and $Z_D$, it is important to include the phase $\sqrt{\omega_D}=\omega_{2D} $ as a generator into the definition of the \emph{single qudit Pauli group},
\begin{align}\label{def:ZDZ_Pauli_group}
 \mathcal{P}_D^1 := \left\langle \sqrt{\omega_D}\mathbbm{1}_D , X_D, Z_D \right \rangle.
\end{align}
This ensures that for all $r,s\in \ZDZ $, there is a Pauli operator which is proportional to $X_D^rZ_D^s$ and has $1$ as an eigenvalue~\cite[Sec. 1.2.1]{Bachelor}. 
This will turn out to be the crucial feature in the stabilizer state-stabilizer group correspondence which we will treat in Sec.~\ref{sec:3.2}.
The $n$-qudit Pauli group $\mathcal{P}_D^n$ is defined to contain all tensor products of single qudit Pauli operators. 
Thus, any operator $P\in \mathcal{P}_D^n$ can be written as $P=\sqrt{\omega_D}^q X_D^\mathbf{r}Z_D^\mathbf{s}$ for unique $q\in \mathbb{Z}/2D\mathbb{Z}$ and $\mathbf{r}, \mathbf{s} \in (\ZDZ)^n$.

As in the case of qubits~\cite{QEC}, the $n$-qudit  \emph{Clifford group} is defined as the normalizer of the Pauli group,
\begin{align}\label{def:Clifford}
\cliff _D^n := \left\{  U \in \mathcal{U}_D^n \ \big\vert \ \forall P\in \mathcal{P}_D^n:  UPU^\dagger \in \mathcal{P}_D^n
\right\},
\end{align}
where $\mathcal{U}_D^n$ is the group of unitary ${D^n\times D^n} $-matrices.
The elements of $\cliff_D^n$ are called qudit Clifford operators or qudit Clifford gates.
An exemplary class of single-qudit Clifford gates is that of 
\emph{multiplication-with-$\ell$ gates}
\begin{align}
 M_D(\ell) := \sum_{k\in \ZDZ} \ket{k\ell}\bra{k},
\end{align}
where $\ell\in \ZDZ$ has to be invertible. Note that a ring is a field iff every $\ell\neq 0$ is invertible.
A direct computation shows that the inverse of the operator $M_D(\ell)$ is given by $M_D(\ell)^\dagger = M_D(\ell^{-1})$.
Another important Clifford gate is the  \emph{Fourier gate},
\begin{align}
 F_D:= \frac{1}{\sqrt D}\sum_{j,k\in \ZDZ} \omega_D^{jk}\ket{j}\bra{k},
\end{align}
which satisfies $F_DX_DF_D^\dagger = Z_D$ and $F_DZ_DF_D^\dagger =X_D^{-1}$ as well as $F_D^4=\mathbbm{1}_D$~\cite{Gottesman99}. The Fourier gate is the qudit generalization of the Hadamard gate $F_2=H=(X_2+Z_2)/\sqrt{2}$.
There are also multi-qudit Clifford gates, for example
\begin{align}\label{def:CX_CZ}
       \mathrm{C}X_D^r := \sum_{k\in \ZDZ} \ket{k}\bra{k} \otimes X_D^{rk}    
       \hspace{2em} \text{and} \hspace{2em}
  \mathrm{C}Z_D^s := \sum_{k\in \ZDZ} \ket{k}\bra{k} \otimes Z_D^{sk},
\end{align}
where $r,s\in\ZDZ$~\cite{Gottesman99}.

\subsubsection{Possible implementations of unitary qudit operations}
\label{sec:3.1.2}

A physical system associated to a quantum spin $s\in\{\frac{1}{2},1,\frac{3}{2},2,\ldots \}$
has a $D=2s+1$ dimensional Hilbert space. 
If $m_z$, the secondary spin quantum number~\cite{Wigner31}, is used to define the computational basis as 
$\ket{0} := \ket{m_s=s}$, $\ket{1}:= \ket{m_s=s-1}$, $\ldots$, $\ket{D-1}:=\ket{m_z=-s}$, one can implement the Pauli $Z$ gates by turning on the spin Hamiltonian 
$S_z = \hbar \times \diag(s,s-1,\ldots, -(s-1), -s)$ for a time $t=\pi/D$, i.e., 
\begin{align}
Z_D \propto \exp\left[\frac{i \pi }{D\hbar}\times S_z \right].
\end{align}
For Pauli $X$ gates, as similar statement is only true for qubits, i.e., for $D>2$ it holds  $X_D \not \propto \exp\left[-\frac{i \pi }{D\hbar}\times S_x \right]$, where  $S_x$ is the spin operator along the {$x$-direction}.
However, one can still implement the Pauli $X$ gate as well as all qudit Clifford gates, if one has full unitary control over the quantum spin system.

\subsection{Stabilizer state-stabilizer group correspondence   } \label{sec:3.2}
The stabilizer formalism provides an efficient way to describe a certain class of pure quantum states.
Instead of writing down $D^{n}$ complex probability amplitudes into a huge state vector, one can characterize a so-called \emph{stabilizer state} as the unique joint eigenstate to the eigenvalue $1$ of a set of Pauli operators called its \emph{stabilizers}. 
The easiest example of a stabilizer state is the computational basis state $\ket{0}$.
It is the unique  eigenstate to the eigenvalue $1$ of a single operator, namely $Z_D$ ($Z_D^\dagger$ is also possible).  

The fact that the product of two stabilizers is again a stabilizer implies that all stabilizers of a given $n$-qudit stabilizer state $\ket{\psi}$ form a group
\begin{align}
 \mathcal{S}_{\ket{\psi}} := \left \{ S\in \mathcal{P}_D^n\,  \big\vert \, S\ket{\psi}=\ket{\psi}   \right\},
\end{align}
called the \emph{stabilizer group} of $\ket{\psi}$. 
Using this notion it is possible to characterize all stabilizer states:

\theorem \label{thrm:stab} \cite{Gottesman99, Gheorghiu14} 
\emph{For a subgroup $\mathcal{S}\subset\mathcal{P}_D^n$ of the $n$-qudit Pauli group, the following statements are equivalent:}
\begin{enumerate}[label=(\roman*)]
 \item\emph{There is a unique $n$-qudit stabilizer state $\ket{\psi}$ that has  $\mathcal{S}$ as its stabilizer group, i.e., $\mathcal{S}=\mathcal{S}_{\ket{\psi}}$.}
 \item\emph{The group $\mathcal{S}$ is an Abelian group of cardinality $\vert \mathcal{S}\vert=D^n$ which does not contain a nontrivial multiple of the identity, i.e., $z\mathbbm{1}\in \mathcal{S}\Rightarrow z=1$. }
\end{enumerate}
\begin{proof}
 (i)$\Rightarrow $(ii):
 We have to show that every stabilizer group $\mathcal{S}=\mathcal{S}_{\ket{\psi}}$ fulfills {property (ii)}.
 Indeed, every $z\in\mathbb{C}$ with $z\mathbbm{1}\in \mathcal{S}$ has to be equal to 1 because $z\ket{\psi} = \ket{\psi}$ holds and  $\ket{\psi}$ is nonzero.
 To show that $\mathcal{S}$ is Abelian, let $S,S'\in \mathcal{S}$. 
 Like all Pauli operators, $S$ and $S'$ commute up to a phase, i.e., $SS'=zS'S$ for some $z\in\mathbb{C}$.
 Since $\mathcal{S}$ is closed under inversion and multiplication, it contains $S'^\dagger S S' S^\dagger = S'^\dagger z S'SS^\dagger =\mathbbm{1}z$, thus $z=1$, i.e., $SS'=S'S$. 
 We now prove that $\mathcal{S}$ contains exactly $D^n$ elements by employing the theory of group actions, see Chapter 5.1 of Ref.~\cite{Bosch} for an introduction into this theory.  
Consider the set of states $\mathcal{X}:=\{P\ket{\psi} \,\vert\, P \in \mathcal{P}_D^n \} $ which is obtained by applying all Pauli operators to the stabilizer state $\ket{\psi}$.
 This gives rise to a group action
 \begin{align}
  \mathcal{P}_D^n \times \mathcal{X} \longrightarrow \mathcal{X}, 
  \hspace{2em}
  (P,\ket{\phi}) \longmapsto P\ket{\phi}.
 \end{align}
By construction, the stabilizer group of $\ket{\psi}$ with respect to this group action coincides with the stabilizer group $\mathcal{S}$ with which we started in the beginning. 
Furthermore, each orbit is of size $\vert \mathcal{X}\vert$ (because the group action is transitive).
Hence, by the \emph{Bahnformel}~\cite[5.1/ Bem. 5]{Bosch}, the length of the orbit of $\ket{\psi}$ is equal to the index of $\mathcal{S}$ in $\mathcal{P}_D^n$, i.e., $\vert \mathcal{X}\vert = \vert \mathcal{P}_D^n/ \mathcal{S}\vert$, or equivalently
$  \vert \mathcal{S} \vert = \vert \mathcal{P}_D^n \vert / \vert \mathcal{X} \vert$.  
As the Pauli group contains $\vert \mathcal{P}_D^n \vert = 2D\times D^{2n}$ elements, it suffices to show that $\mathcal{X}$ contains exactly $2D\times D^n$ states.
 Indeed, after expanding the stabilizer state as $\ket{\psi} = \sum_{\mathbf{j}\in (\ZDZ)^n}z_\mathbf{j}\ket{\mathbf{j}}$,  
 we can rewrite the set of states as
 \begin{align}
 \mathcal{X}  =
 \left\{\, \omega_{2D}^t  \sum_{\mathbf{j}\in (\ZDZ)^n} z_\mathbf{j} \ket{\mathbf{j}+\mathbf{k}} \, \bigg\vert \, t\in\ZZ/ 2D \ZZ, \mathbf{k}\in (\ZDZ)^n \right\}
 \end{align}
which thus contains $\vert \mathcal{X} \vert = 2D\times D^n$ elements in total.
  
    (ii)$\Rightarrow$(i):
Now, conversely,  assume that $\mathcal{S}\subset\mathcal{P}_D^n$ is an Abelian group of cardinality $D^n$ not containing any nontrivial multiple of the identity. 
We have to show the existence and uniqueness of a joint eigenstate $\ket{\psi}$ to the eigenvalue $1$ for  all $S\in \mathcal{S}$. 
This we do by recycling the proof of Thrm. 1 in Ref.~\cite{Gheorghiu14}: 
The operator $\Pi:=  \frac{1}{D^n} \sum_{S\in\mathcal{S}}S$ fulfills
\begin{align}
 \Pi^2  
 = \frac{1}{D^{2n}} \sum_{S,S'\in\mathcal{S}} SS' \hspace{1em}
 =  \frac{1}{D^{2n}} \sum_{S''\in\mathcal{S}} S'' \sum_{S\in\mathcal{S}} 1 \hspace{1em}
= \frac{1}{D^n} \sum_{S''\in\mathcal{S}} S'' \hspace{1em}= \Pi,
\end{align}
where we have used the substitution $S'\mapsto S'':= SS'$ which is bijective since $\mathcal{S}$ is a group. 
Furthermore, $\Pi$ is Hermitian, i.e., $\Pi=\Pi^\dagger$, because every $S\in\mathcal{S}$ has a unique inverse.
Thus, the operator $\Pi$ is an orthogonal projector. 
We claim that $\Pi$ is the projector onto the joint eigenspace to the eigenvalue 1 of all $S\in\mathcal{S}$.
Indeed, let $\ket{\psi}$ be a state in this joint eigenspace, i.e., $S\ket{\psi}=\ket{\psi}$. By construction, we obtain ${\Pi\ket{\psi}= \frac{1}{D^n} \sum_{S\in \mathcal{S}}\ket{\psi}=\ket{\psi}}$. Conversely, let $\ket{\phi}$ be a state in the space onto which $\Pi$ projects, i.e., $\ket{\phi}= \Pi\ket{\phi}$. Multiplying the latter equation with an arbitrary $S\in\mathcal{S}$ from the left yields
\begin{align}
 S\ket{\phi} 
 = S\Pi\ket{\phi}
 \hspace{1em}
 = \frac{1}{D^n} \sum_{S'\in \mathcal{S}} SS' \ket{\phi} 
 \hspace{1em}
 = \frac{1}{D^n} \sum_{S''\in \mathcal{S}} S'' \ket{\phi}  
 \hspace{1em}
 = \Pi\ket{\phi} 
 = \ket{\phi},
\end{align}
where again we substituted $S'':=SS'$. 
Now that we know that $\Pi$ is the projector onto the joint eigenspace to the eigenvalue 1 of all operators $S\in \mathcal{S}$, we can compute its dimension which is equal to the trace of $\Pi$,
\begin{align}
 \Tr[\Pi]
 =\frac{1}{D^n} \sum_{S\in\mathcal{S}}  \Tr[S] =\frac{1}{D^n} \left(\Tr[\mathbbm{1}_{D^n}] + \sum_{S\neq \mathbbm{1}} \Tr[S]  \right) 
 =1,
\end{align}
where we have used that every Pauli-operator which is not a multiple of the identity 
has trace zero.
Therefore, there exists exactly one joint eigenstate $\ket{\psi}$ to the eigenvalue $1$ for all $S\in\mathcal{S}$ which shows that $\ket{\psi}$ is an $n$-qudit stabilizer state with stabilizer group $\mathcal{S}_{\ket{\psi}}= \mathcal{S}$, as claimed.
\end{proof}

The theorem reveals a useful property of stabilizer groups.
By the classification of finite Abelian groups~{\cite[2.9/Kor. 9]{Bosch}}, condition {(ii)}  implies  the existence of numbers $D_1,\ldots,D_N$ such that  $\mathcal{S}$ is isomorphic to the group  $\Pi_{i=1}^N \ZZ/D_i\ZZ$, where $\Pi_{i=1}^ND_i=D^n$. 
Therefore, one can find operators $S_1,\ldots,S_N\in \mathcal{S}$ generating the group $\mathcal{S}$, where $D_i$ is the order of $S_i$. 
Such operators are referred to as \emph{stabilizer generators}. 
Note that in Ref.~\cite{Gottesman99}, the content of Theorem~\ref{thrm:stab} is only discussed in the case where $\mathcal{S}\cong \prod_{i=1}^n \ZDZ$.
In general, however,  various choices of $N$ and $D_1,\ldots, D_N$ are possible.
An example which shows this difference is given by the two ququart states $\ket{0}$ and $\frac{1}{\sqrt{2}}(\ket{0}+\ket{2})$ whose stabilizer group is given by $\langle Z_4 \rangle\simeq \ZZ/4\ZZ$ and  $\langle X_4^2, Z_4^2\rangle \simeq \ZZ/2\ZZ \times \ZZ/2\ZZ$, respectively.
That is, if one would only allow for the case $D_1=\ldots=D_N=D$, one would loose the possibility to describe many states (such as $\frac{1}{\sqrt{2}}(\ket{0}+\ket{2})$) within the stabilizer formalism.
However, we also include the more general case where $D_i\neq D$ is possible.
 
Since every $D_i$ is a divisor of $D$ (because of $S_i^D=\mathbbm{1}$),
the stabilizer group $\mathcal{S} = \Pi_{i=1}^N \ZZ/D_i\ZZ$ carries the additional algebraic structure of a $\ZDZ$-module.
To better understand this structure, denote the exponent vectors of the stabilizer generators by $\mathbf{r}_i,\mathbf{s}_i\in(\ZDZ)^n$, i.e., $S_i \propto X^{\mathbf{r}_i}Z^{\mathbf{s}_i}$ for all $i\in\{1,\ldots, N\}$.
This yields an embedding of $\mathcal{S}$ 
into the free module $(\ZDZ)^n \times (\ZDZ)^n$ 
via the $\ZDZ$-linear injection
\begin{align}\label{eq:Z/DZ_Stab_embedding}
 \mathcal{S}  \longrightarrow (\ZDZ)^n \times (\ZDZ)^n, 
 \hspace{2em}
 S_i \longmapsto  (\mathbf{r}_i,\mathbf{s}_i).
 \end{align}
Note that, the vectors $(\mathbf{r}_i,\mathbf{s}_i)$ only form a basis of the image of $\mathcal{S}$ if $D_1=\ldots=D_N=D$. Otherwise, this submodule does not have a basis, i.e., it is not a free module over $\ZDZ$.
Throughout this thesis, we will use the parametrization in Eq.~\eqref{eq:Z/DZ_Stab_embedding} to explicitly work with stabilizer groups.
 
The next lemma characterizes how the stabilizer generators of a given stabilizer state change after the  application of a Clifford operator.
\lemma\label{lem:cliff_stab}
\emph{ If $\ket{\psi}$ is a stabilizer state with stabilizer generators $S_1,\ldots,S_N\in \mathcal{P}_D^n$ and $U\in\cliff_D^n$ is a Clifford gate, 
then $\ket{\psi'}:=U\ket{\psi}$ is also a stabilizer state and its stabilizer group is generated by
$S_i' := U S_i U^\dagger$, where $i\in\{1,\ldots, N\}$. }  
\begin{proof}
For each $1\le i\le N$, the operator $S_i'=US_iU^\dagger$ is a Pauli operator
by the definition of the Clifford group, 
and because $S_i$ is assumed to be an element of the $n$-qudit Pauli group. 
By setting $\mathcal{S}=\langle S_1,\ldots, S_N\rangle $ 
and $\mathcal{S'}=\langle S_1',\ldots,S_N' \rangle $,
we obtain an isomorphism
\begin{align}
\mathcal{S} \overset{\simeq}{\longrightarrow} \mathcal{S'}, 
\hspace{2em}
P\longmapsto UPU^\dagger.
\end{align}
In particular, both Pauli subgroups have the same number of elements,
i.e., $\vert \mathcal{S}'\vert = \vert \mathcal{S}\vert = D^n$.
Thus, Theorem~\ref{thrm:stab} yields that $\mathcal{S}'$ is the stabilizer group of a unique stabilizer state.  As claimed, this unique state is $\ket{\psi'}$ because of
$S_i'\ket{\psi'}= US_iU^\dagger U\ket{\psi} = U S_i\ket{\psi} = U\ket{\psi}= \ket{\psi'}$, i.e., the state $\ket{\psi'}$ indeed is a common eigenstate of $S_1',\ldots S_N'$ to the eigenvalue 1. 
\end{proof}

An immediate consequence of Lemma~\ref{lem:cliff_stab} is that
it is only possible to find a Clifford operator 
that maps a stabilizer state $\ket{\psi}$ to a different stabilizer state $\ket{\psi'}$
if their stabilizer groups are isomorphic, i.e., if $\ket{\psi'}=U\ket{\psi}$ for an Clifford gate $U$, then $\mathcal{S}_{\ket{\psi}} \cong \mathcal{S}_{\ket{\psi'}}$.
For example, the aforementioned ququart states $ \ket{0}$ and $ \frac{1}{\sqrt 2}(\ket{0}+\ket{2})$ cannot be mapped onto each other using Clifford gates only.
Physically, this implies that for higher-dimensional qudits, the set of stabilizer states is richer in the sense that not every stabilizer state can be reached by applying a sequence of Clifford gates to a single initial state such as  $\ket{0}^{\otimes n}$. 
If one wants to produce arbitrary stabilizer states using Clifford gates only, one must be able to initialize the qudits into more than one initial state.
Alternatively, one could start with a single initial state and change the isomorphic class of the stabilizer group either via application of non-Clifford gates or via suited projective Pauli measurements~\cite{NielsenChuang}.

\subsection{Qudit graph states}
\label{sec:3.3}

\emph{Graph states} are specific stabilizer states with a pictorial description related to graphs~\cite{HeinEisBri04,  SchlWer01, GrKlRoe02, HDERVdNB06, BahBei06, LooiGriff11}.
We consider graphs whose edges are weighted by elements in the ring $\ZDZ$.
Formally, such a graph is given by a finite set of vertices $V$ and a set of weighted edges $E$.
Each vertex corresponds to one of $n$ qudits; so we use the notation $V=\{1,2,\ldots, n\}$ (and sometimes  $V=\{\mathrm{A,B,C,\ldots}\}$).
The edges $e\in E$ are denoted by $e=(\{i,j\},\gamma_{i,j})$, where $\gamma_{i,j}\in \ZDZ$ is the weight of the edge between party $i$ and $j$. 
The whole information about a graph $(V,E)$ is summarized into its \emph{adjacency matrix}
$\Gamma=(\gamma_{i,j})_{i,j\in V} 
$ where the entry $\gamma_{i,j}$ is the weight of the edge $(\{i,j\},\gamma_{i,j})\in E$. 
If there is no such edge for two given qudits $i,j \in V$, the corresponding weight is $\gamma_{i,j}=0$. 
As we do not consider graphs with loops, we additionally require that the diagonal elements $\gamma_{i,i}$ of the adjacency matrix are equal to zero. 
See Fig.~\ref{fig:ABCD_Graph} 
\begin{figure}
\centering
\begin{minipage}{0.85\textwidth}
\begin{framed} 
\centering
\begin{minipage}{.5\textwidth}
\centering
\begin{tikzpicture}
 \draw (0,0) node[text=black]{A}; 
 \draw (3,0) node[text=black]{D}; 
 \draw (3,-3) node[text=black]{C}; 
 \draw (0,-3) node[text=black]{B}; 
 \draw[-, line width=.2em] (0.3,0) -- (2.7,0); 
 \draw[-, line width=.2em] (0,-0.3) -- (0,-2.7); 
 \draw[-, line width=.2em] (3,-2.7) -- (3,-0.3); 
 \draw[-, line width=.2em] (2.7,-3) -- (0.3,-3); 
 \draw[line width=.1em] (0,0) circle (0.3);
 \draw[line width=.1em] (3,0) circle (0.3);
 \draw[line width=.1em] (0,-3) circle (0.3);
 \draw[line width=.1em] (3,-3) circle (0.3); 
 \draw (1.5, 0.3) node[text=black]{$-1$}; 
 \draw (-0.5, -1.5) node[text=black]{$1$}; 
 \draw (1.5, -3.3) node[text=black]{$1$}; 
 \draw (3.3, -1.5) node[text=black]{$1$}; 
 \end{tikzpicture}
\end{minipage}
\begin{minipage}{.4\textwidth}
\centering
 \begin{align}\nonumber   
\Gamma^4_D := \left[\begin{array}{rrrr}0&1&0&-1 \\ 1&0&1&0\\ 0&1&0&1 \\ -1&0&1&0 \end{array} \right]
\end{align}
\end{minipage}
\caption{Example of a tetrapartite family of qudit graph states.
Pictorial depiction (left)  and adjacency matrix  (right) of a $\ZDZ$-weighted directed graph $(V,E)$ with vertex set $V=\{\mathrm{A,B,C,D}\}$ and edge set $E=\{(\{\mathrm{A,B}\},1),(\{\mathrm{B,C}\},1),(\{\mathrm{C,D}\},1),(\{\mathrm{A,D}\},-1)\}$. }
 \label{fig:ABCD_Graph}
\end{framed}
\end{minipage}
 \end{figure}
 for an example of a tetrapartite graph  and its adjacency matrix.

Given such a graph, we define its corresponding graph state as the state obtained from $n$ copies of the plus state 
\begin{align}\label{def:plus}
 \ket{+_D} = \frac{1}{\sqrt D}\sum_{j\in \ZDZ} \ket{j}
\end{align}
by applying a $\gamma_{i,j}$-fold controlled-phase gate with control qudit $i$ and target qudit $j$. 
That is, the corresponding graph state is given by $\ket{\Gamma } := U_\Gamma \ket{+_D}^{\otimes n}$, where
\begin{align}\label{def:U_Gamma}
U_\Gamma :=  \prod\limits_{i=1}^n\prod\limits_{j=i+1}^n  \mathrm{C}_{i,j}Z_D^{\gamma_{i,j}} .
\end{align}
Using Definition~\eqref{def:CX_CZ} and \eqref{def:plus}, we find the alternative, useful  expression
\begin{align}\label{eq:ketGamma}
 \ket{\Gamma} = \frac{1}{\sqrt{D^n}} \sum_{\mathbf{r}\in (\ZDZ)^n} \omega_D^{\sum\limits_{i=1}^n\sum\limits_{j=i+1}^n \gamma_{i,j} r_i r_j} \ket {\mathbf{r}}.
\end{align}
 
 As we now show, for every adjacency matrix $\Gamma\in \ZDZ^{n\times n}$, the corresponding graph state $\ket{\Gamma}$ is a stabilizer state~\cite{Helwig03}.
The $n$-qudit state $\ket{+_D}^{\otimes n}$ is a stabilizer state with stabilizer group $\mathcal{S}_{\ket{+}^{\otimes n}} = \langle X_D^{(1)}, \ldots, X_D^{(n)}  \rangle$, where $P^{(k)}\in\mathcal{P}^n_D$ denotes a single qudit Pauli operator $P\in\mathcal{P}^1_D$ acting on qudit $k$.
Since the state $\ket{\Gamma}$ is obtained by applying the $n$-qudit Clifford gate 
$U_\Gamma$
to the state $\ket{+_D}^{\otimes n}$, 
Lemma~\ref{lem:cliff_stab} yields that $\ket{\Gamma}$ is a stabilizer state with stabilizer generators formally given by
\begin{align}\label{eq:graph_stab_formal}
S_ k 
:= U_\Gamma X_D^{(k)} U_\Gamma^\dagger ,
\end{align}
where $k\in V$.   
The key to make this expression more explicit is the relation
\begin{align}\label{l:CZXCZ_ZDZ_1}
 CZ_D^\gamma &(\mathbbm 1_D \otimes X_D) (CZ_D^\gamma ) ^\dagger   = Z_D^\gamma \otimes X_D,
\end{align} 
where $\gamma \in \ZDZ$~\cite{Gheorghiu14}.
Thereby, it does not matter which qudit is the target as the controlled-$Z$ gate is symmetric.
Note that $C_{i,j}Z_D^{\gamma_{i,j}}X_D^{(k)} (C_{i,j}Z_D^{\gamma_{i,j}})^\dagger=X_D^{(k)}$ if neither $i$ nor $j$ are equal to $k$ because the operators have different support qudits so that the controlled-$Z$ gates cancel each other. 
That is, only neighbors of $k$ contribute to the product in Eq.~\eqref{eq:graph_stab_formal}.
Since every neighbor appears exactly once in the definition of $U_\Gamma$ in Eq.~\eqref{def:U_Gamma}, the stabilizer generators of $\ket{\Gamma}$ follow as 
\begin{align}  \label{eq:graph_stab}
S_k 
=X_D^{(k)} \prod_{j=1}^n (Z_D^{(j)})^{\gamma_{j,k} }
\end{align} 
where $k\in V$.

\subsection{Absolutely-maximally-entangled states on four qudits}
\label{sec:3.4} 

In this subsection, we further investigate the tetrapartite qudit graph state $\ket{\Gamma_D^4}$ which is defined by the adjacency matrix in Fig.~\ref{fig:ABCD_Graph}. By Eq.~\eqref{eq:ketGamma}, it can be written as
\begin{align} \label{eq:AME4D}
 \ket{\Gamma_D^4} = \frac{1}{D^2} \sum_{k_1,\ldots,k_4 \in \ZDZ} \omega_D^{k_1k_2 +k_2k_3+k_3k_4-k_1k_4} \ket{k_1,k_2,k_3,k_4}.
\end{align}
We will show that $\ket{\Gamma^4_D}$ is an AME state whenever $D$ is odd. Note that this result is already known~\cite{Helwig03}.
As a preparation, in Sec.~\ref{sec:3.4.1} we establish a lemma with which one can prove that the marginals of a given state are maximally mixed. 
In Sec.~\ref{sec:3.4.2}, we apply this lemma to $\ket{\Gamma^4_D}$ in the odd-dimensional case.
Finally, in Sec.~\ref{sec:3.4.3}, we show how this procedure fails in the even-dimensional case.

\subsubsection{Lemma for the verification of $m$-uniformness}
\label{sec:3.4.1}

The following lemma relates the qudit stabilizer formalism to maximally mixed states.
It will be key to show that $\ket{\Gamma_D^4}$ is an AME state in the odd-dimensional case.
\lemma\label{lem:mixed}
\emph{ For an $m$-qudit state $\rho$, the following statements are equivalent:}
\begin{itemize} 
\item[(i)] \emph{The state is maximally mixed, i.e., $\rho=\mathbbm{1}/D^m$. }  
 \item[(ii)] \emph{The state is stabilized by $X_D$ and $Z_D$ on every qudit.}
 \item[(iii)] \emph{The state is stabilized by all Pauli operators $P\in \mathcal{P}_D^m$, i.e., $P\rho P^\dagger=\rho$.}
\end{itemize} 
 
\begin{proof} 
(i)$\Rightarrow$(ii): 
Since the maximally mixed state commutes with every other operator, it holds $X_D^{(i)}\rho X_D^{\dagger (i)} = \rho $ and $Z_D^{(i)}\rho Z_D^{\dagger (i)} = \rho $ for all  $1\le i \le m$.

(ii)$\Rightarrow$(iii): Follows directly from $\mathcal{P}_D^m=\langle \omega_{2D}, X_D^{(i)}, Z_D^{(i)} \ \vert 1\le i \le m\rangle$.

(iii)$\Rightarrow$(i):
To prove that every state $\rho$ which is stabilized by all Pauli operators is necessarily  maximally mixed, we expand the state as
\begin{align}
 \rho = \sum_{\mathbf{r,s}\in (\ZDZ)^m} z_\mathbf{r,s} \ket{\mathbf{r}}\bra{\mathbf{s}}.
\end{align}
We start by showing that $\rho$ is diagonal. Let $\mathbf{r,s}\in (\ZDZ)^m$ with $\mathbf{r}\neq\mathbf{s}$.
We have to show $z_\mathbf{r,s}=0$. 
Because of $\mathbf{r}\neq\mathbf{s}$, these two vectors have to differ in at least one entry $i\in\{1,\ldots,k\}$, i.e., $r_i\neq s_i$. 
By our assumption, $P\rho P^\dagger = \rho$ for all $P \in \mathcal{P}_k$, we have 
\begin{align} 
z_\mathbf{r,s}=\bra{\mathbf{r}}\rho \ket{\mathbf{s}} =\bra{\mathbf{r}} (Z_{D}^{(i)})^\dagger \, \rho\, Z_{D}^{(i)}\ket{\mathbf{s}}=\bra{\mathbf{r}}\omega_D^{-r_i} \rho \omega_D^{s_i} \ket{\mathbf{s}}  = \omega_D^{s_i-r_i} \bra{\mathbf{r}}\rho \ket{\mathbf{s}} =\omega_D^{s_i-r_i}z_\mathbf{r,s}
\end{align} 
where $P^{(i)}$ denotes a single qudit operator $P$ acting on qudit $i$.   
This equation is equivalent to 
\begin{align}
(1-\omega_D^{s_i-r_i})z_\mathbf{r,s}=0.
\end{align} 
The first factor, $
(1-\omega_D^{s_i-r_i})$, is nonzero because $r_i\neq s_i$ implies $\omega_D^{s_i-r_i} \neq 1$.  
Thus $z_\mathbf{r,s}$ must be zero, i.e., the state $\rho$ is diagonal. 
Abbreviating the diagonal entries as $z_\mathbf{r}:= z_\mathbf{r,r}$, we can denote any two of them as $z_\mathbf{r}, z_\mathbf{s} $ for some $\mathbf{r},\mathbf{s} \in (\ZDZ)^m$. 
By assumption, we have  
\begin{align}
 z_\mathbf{r}= \bra{\mathbf{r}} \rho \ket{\mathbf{r}}
=\bra{\mathbf{r}} (X_D^{\mathbf{s}-\mathbf{r}})^\dagger \, \rho \, X_D^{\mathbf{s}-\mathbf{r}}  \ket{\mathbf{r}}
=\bra{\mathbf{s}} \rho \ket{\mathbf{s}} = z_\mathbf{s},
\end{align} 
i.e., all diagonal elements coincide and the normalization condition $\Tr{[\rho]}=1$ finishes the proof. 
\end{proof}
 
This lemma can be used to construct stabilizer states which are $m$-uniform in the following way.
If $S=S_1\otimes \ldots \otimes S_n$ stabilizes an $n$-qudit state $\ket{\psi}$ and $I$ is a subset of $\{1,\ldots,n\}$, then 
\begin{align}
 S^{(I)}:=\Tr_{I^\mathrm{C}} [S] = \bigotimes_{i\in I} S_i
\end{align}
stabilizes the marginal state $\rho^{I}= \Tr_{I^\mathrm{C}}[\ket{\psi} \bra{\psi}]$. 
That is, every $n$-qudit stabilizer state  $\ket{\psi}$ with the property 
$\{ S^{(I)} \vert S \in \mathcal{S}_{\ket{\psi}} \} = \mathcal{P}_D^m$ for all subsets $I\subset \{1,\ldots,n\}$ with exactly $m$ elements is an $m$-uniform state by Lemma~\ref{lem:mixed}.

\subsubsection{Proof in odd dimensions}
\label{sec:3.4.2}

Now, we can show that, for every odd qudit dimension $D$, the tetrapartite  graph state $\ket{\Gamma^4_D}$ from Eq.~\eqref{eq:AME4D} is an example of a $2$-uniform state, 
thus,  an AME state.
Recall that the associated graph is given by the set of vertices $V=\{\mathrm{A,B,C,D}\}$ and the set of edges $E=\{(\{\mathrm{A,B}\},1),(\{\mathrm{B,C}\},1),(\{\mathrm{C,D}\},1),(\{\mathrm{A,D}\},-1)\}$. 
Hence, by Eq.~\eqref{eq:graph_stab}, the stabilizers of $\ket{\Gamma^4_D}$ are given by
\begin{align}\label{eq:stab_family}
 S_{\mathrm{A}} &= X_D \otimes Z_D \otimes \mathbbm 1 \otimes Z_D^{-1},
&& S_{\mathrm{B}} =  Z_D \otimes X_D \otimes Z_D \otimes \mathbbm 1, \\
\label{eq:stab_family2}
 S_{\mathrm{C}} &= \mathbbm{1}\otimes Z_D \otimes X_D \otimes Z_D
 &\text{ and } \hspace{2em}
& S_{\mathrm{D}} = Z_D^{-1} \otimes \mathbbm 1 \otimes  Z_D \otimes  X_D.
\end{align}
It is more convenient to consider the corresponding 
vectors $(\mathbf{r}_{k},\mathbf{s}_{k}) \in (\ZDZ)^4\times (\ZDZ)^4$ defining these stabilizers via $S_{k}= X_D^{\mathbf{r}_{k}}Z_D^{\mathbf{s}_{k}}$
for all $k\in V$. 
These are given by 
\begin{align} \label{eq:AME4D_exponent_vectors}
\begin{array}{rccrccr}
 (\mathbf{r}_\mathrm{A},\mathbf{s}_\mathrm{A})&=&\boldsymbol{(} (1,0,0,0),&(\hspace{.78em} 0,&1,&0,&-1)\boldsymbol{)}, \\  
 (\mathbf{r}_\mathrm{B},\mathbf{s}_\mathrm{B})&=&\boldsymbol{(} (0,1,0,0),&(\hspace{.78em} 1,&0,&1,& 0)\boldsymbol{)}, \\
 (\mathbf{r}_\mathrm{C},\mathbf{s}_\mathrm{C})&=&\boldsymbol{(} (0,0,1,0),&(\hspace{.78em}  0,&1,&0,& 1)\boldsymbol{)}, \\ 
 \text{ and } \hspace{1em}
 (\mathbf{r}_\mathrm{D},\mathbf{s}_\mathrm{D})&=&\boldsymbol{(} (0,0,0,1),&(-1,&0,&1,&0) \boldsymbol{)}.
\end{array}  
\end{align} 
Products of stabilizer generators correspond to linear combinations of such vectors with coefficients in $\ZDZ$. 
Likewise, marginals of $\rho = \ket{\Gamma_D^4}\bra{\Gamma_D^4}$ are stabilized by Pauli operators defined by vectors where the columns which correspond to the traced-out systems have been removed.
For instance, the bipartite reduced state $\rho^{\{\mathrm{A,B}\}}$ has stabilizers with exponent vectors
$\mathbf v_1:= \boldsymbol{(} (1,0),(0,1) \boldsymbol{)},
\mathbf v_2:= \boldsymbol{(} (0,1),(1,0) \boldsymbol{)},
\mathbf v_3:= \boldsymbol{(} (0,0),(0,1) \boldsymbol{)},$ 
 and 
$\mathbf v_4:= \boldsymbol{(} (0,0),(-1,0) \boldsymbol{)}$.
The linear combinations $\mathbf v_1-\mathbf v_3$, $\mathbf v_2+\mathbf v_4$, $-\mathbf v_4$, and $\mathbf v_3$ are the standard basis vectors.
Thus, $\rho^{\{\mathrm{A,B}\}}$ is stabilized by $X_D\otimes \mathbbm{1}$,  $ \mathbbm{1} \otimes X_D$,  $Z_D\otimes \mathbbm{1}$, and  $\mathbbm{1} \otimes Z_D$ and Lemma~\ref{lem:mixed} yields $\rho^{\{\mathrm{A,B}\}}=\mathbbm1/D^2$. 
By analogous arguments, one can show $\mathbbm1/D^2=\rho^{\{\mathrm{A,C}\}}=\rho^{\{\mathrm{A,D}\}}=\rho^{\{\mathrm{B,C}\}}=\rho^{\{\mathrm{B,D}\}}=\rho^{\{\mathrm{C,D}\}}$, i.e., $\ket{\Gamma^4_D} $ is indeed a 2-uniform state, thus AME.

\subsubsection{Obstruction in even dimensions}
\label{sec:3.4.3} 

Here, we show that we cannot apply Lemma~\ref{lem:mixed} in the even-dimensional case to construct 
a tetrapartite ring-graph state which is also AME. Let $\ket{\Gamma}$ denote a potential candidate where the weights $a,b,c,d \in \ZDZ$ have not been fixed yet. We will show that there is no choice  of $a,b,c$ and $d$ such that $\ket{\Gamma}$ is AME.
Analogous to Eqs.~\eqref{eq:stab_family} and \eqref{eq:stab_family2}, $\ket{\Gamma}$ has stabilizer generators
\begin{align}
 S_{\mathrm{A}} &= X_D \otimes Z_D^a \otimes \mathbbm 1 \otimes Z_D^d,
&& S_{\mathrm{B}} =  Z_D^a \otimes X_D \otimes Z_D^b \otimes \mathbbm 1, \\
 S_{\mathrm{C}} &= \mathbbm{1}\otimes Z_D ^b \otimes X_D \otimes Z_D^c
 &\text{ and } \hspace{2em}
& S_{\mathrm{D}} = Z_D^d \otimes \mathbbm 1 \otimes  Z_D^c \otimes  X_D.
\end{align} 
This time, the exponent vectors are given by 
\begin{align}
\begin{array}{rcccccl}
 (\mathbf{r}_\mathrm{A},\mathbf{s}_\mathrm{A})&=&\boldsymbol{(} (1,0,0,0),&(0,&a,&0,& d)\boldsymbol{)}, \\  
 (\mathbf{r}_\mathrm{B},\mathbf{s}_\mathrm{B})&=&\boldsymbol{(} (0,1,0,0),&(a,&0,&b,& 0)\boldsymbol{)}, \\
 (\mathbf{r}_\mathrm{C},\mathbf{s}_\mathrm{C})&=&\boldsymbol{(} (0,0,1,0),&(0,&b,&0,& c)\boldsymbol{)}, \\
 \text{ and }
 (\mathbf{r}_\mathrm{D},\mathbf{s}_\mathrm{D})&=&\boldsymbol{(} (0,0,0,1),&(d,&0,&c,&0) \boldsymbol{)}.
\end{array}  
\end{align}
After tracing out party C and D, one obtains reduced stabilizer exponents 
$\boldsymbol{(} (1,0),(0,a) \boldsymbol{)}$,
$\boldsymbol{(} (0,1),(a,0) \boldsymbol{)}$,
$\boldsymbol{(} (0,0),(b,0) \boldsymbol{)}$, and
$\boldsymbol{(} (0,0),(0,d) \boldsymbol{)}$.
Thus, $\rho^\mathrm{\{A,B\}}$ is stabilized by  $X_D^\mathrm{A}$, $X_D^\mathrm{B}$, $Z_D^\mathrm{A}$, and $Z_D^\mathrm{B}$, iff it is possible to turn the matrix 
\begin{align}
\left[
 \begin{array}{cccc}
  1 &0&0&a \\
  0&1&a&0\\
  0&0&0&b \\
  0&0&d&0
 \end{array}\right] \in (\ZDZ)^{4\times 4}
\end{align}
into the unit matrix using the Gaussian algorithm.
This, in turn, is possible iff $b$ and $d$ are invertible in $\ZDZ$.
The same argument with D instead of B shows that also $a$ and $c$ have to be invertible in $\ZDZ$.
Since $D$ is even, $a,b,c,$ and $d$ have to be odd integers (modulo $D$).
Finally, to ensure that also $\rho^\mathrm{\{A,C\}}$ is stabilized by $X_D$ and $Z_D$ on both qudits, the matrix which has to be turned into the unit matrix using the Gaussian algorithm is given by 
\begin{align}
\left[
 \begin{array}{cccc}
  1 &0&0&0 \\
  0&0&a&b\\
  0&1&0&0 \\
  0&0&d&c
 \end{array}\right] 
 \longrightarrow
\left[
 \begin{array}{cccc}
  1 &0&0&0 \\
  0&1&0&0  \\
  0&0&a&b  \\
  0&0&d&c 
 \end{array}\right] ,
\end{align}
where we have swapped row 2  and 3.
This imposes that the $2\times2 $-determinant $ac-bd$ is invertible in $\ZDZ$.
However, $a,b,c,$ and $d$ are odd. Thus $ac-bd$ is even and cannot be invertible in $\ZDZ$.
This shows why other methods are needed for the construction of tetrapartite, ring-graph  AME states in even dimensions.  

Let us comment on when even-dimensional, tetrapartite AME states do exist in general.
While it is known that there is no such state in the case of qubits~\cite{GouWal10},
they do exist for all $D=2^n$ and $n\ge2$. They can be explicitly constructed from Theorem 14 of Ref.~\cite{GrBeRoe04} using a correspondence established in Ref.~\cite{RGRA18}. 
Alternatively, one can directly construct them using a procedure analogous to that in Sec.~\ref{sec:3.4.2} where the ring $\ZZ/2^n \ZZ$ is replaced by the finite field $\FF_{2^n}$.
Via tensor products, it is straightforward to combine these even-dimensional AME states with the odd-dimensional AME states of Sec.~\ref{sec:3.4.2} to also obtain tetrapartite AME states for all $D$ which are divisible by 4.

\section[Noise thresholds for qudit graph states with global white noise]{\label{sec:4}\protect Noise thresholds for qudit graph states}
In this Section, we will consider graph states $\ket{\Gamma}$ which are replaced by a completely mixed state, globally on all parties, with probability $p$ which is referred to as white noise.
We will denote such states by
\begin{align} \label{eq:Gamma_glob}
 \rhoGlobGammaP  := (1-p) \ket{\Gamma} \bra{\Gamma} + p \frac{\mathbbm{1}}{D^n},
\end{align}
and apply bipartite entanglement criteria to find critical noise values $p_\mathrm{glob}$ such that $p<p_\mathrm{glob}$ implies that $\rhoGlobGammaP$ is entangled. 
Thereby, we will only consider a bipartition of size $(n-1,1)$ for which the distinguished party (Alice) is incident to at least one edge because we have numerical evidence that this leads to the best thresholds.

In Sec.~\ref{sec:4.1}, we derive expressions for the von Neumann entropies $S\left[\rhoGlobGammaP\right]$ and $ S\left[\Tr_\mathrm{A}[\rhoGlobGammaP]\right]$ which directly depend on $p, D$ and $n$. 
This makes it computationally feasible to determine a  critical noise threshold $p_\mathrm{glob}^\mathrm{Entr}$ for the entropy criterion.
Afterwards, we apply the Peres-Horodecki criterion in Sec.~\ref{sec:4.2} and the reduction criterion in Sec.~\ref{sec:4.3}. In both cases, we will establish the noise threshold $p_\mathrm{glob}^\mathrm{PPT}(D,n)=p_\mathrm{glob}^\mathrm{Red}(D,n)={1-\frac{1}{D^{n-1}+1}}$ by explicitly computing an eigenvalue of  
$\rhoGlobGammaP\TA$ and $\mathbbm 1_\mathrm{A}\otimes \Tr_\mathrm{A}[\rhoGlobGammaP]-\rhoGlobGammaP$, respectively, which is negative for all $p<{1-\frac{1}{D^{n-1}+1}}$.
Finally, in Sec.~\ref{sec:4.4}, we will briefly comment on the range of applicability of the here-established noise thresholds.

\subsection{Entropy criterion for qudit graph states}
\label{sec:4.1}

In order to apply the entropy criterion,
\begin{align}\label{eq:entr_crit_gamma_glob}
 S\left[\rhoGlobGammaP\right] <  S\left[\Tr_\mathrm{A}[\rhoGlobGammaP]\right] 
 \hspace{2em}
 \Longrightarrow  
 \hspace{2em}
\rhoGlobGammaP \text{ is entangled},
\end{align}
to noisy graph states, we need to compute the complete spectra of eigenvalues for both $\rhoGlobGammaP$ and $\Tr_\mathrm{A}[\rhoGlobGammaP]$.
For the unreduced state, no assumption on $\Gamma$ is needed as we have
\begin{align}
  \rhoGlobGammaP  \ket{\Gamma}= (1-p) \ket{\Gamma} \braket{\Gamma| \Gamma}+ p \frac{\mathbbm{1}}{D^n} \ket{\Gamma} = \left(   
  1-p +\frac{p}{D^n}
  \right) \ket{\Gamma}
\end{align}
and for every state $\ket{\psi}$  which is orthogonal to  $\ket{\Gamma}$, we obtain
\begin{align}\label{eq:red_total_perp}
  \rhoGlobGammaP \ket{\psi}= (1-p) \ket{\Gamma} \braket{\Gamma| \psi}+ p \frac{\mathbbm{1}}{D^n} \ket{\psi} = \frac{p}{D^n}
   \ket{\psi}.
\end{align}
Since the whole Hilbert state of all parties decomposes into the one-dimensional span of $\ket{\Gamma}$ and its $D^n-1$-dimensional orthogonal complement, the eigenvalue $1-p +\frac{p}{D^n}$ and $\frac{p}{D^n}$ has degeneracy $1$ and $D^n-1$, respectively.

Obviously, we cannot expect to detect entanglement in $\rhoGlobGammaP$ if the adjacency matrix $\Gamma\in(\ZDZ)^{n \times n}$ is trivial. 
For technical reasons, here we will only consider the case where at least one entry in $\Gamma$ is invertible, w.l.o.g. $\gamma_{1,2}$. 
We will comment on the general case of arbitrary $\Gamma$ in Sec.~\ref{sec:4.2.1}.
Note that in prime dimension this technicality is trivial.
Let us first obtain an expression for the reduced state where party 1 is discarded:
\begin{align}\label{eq:Gamma_glob_reduced}
 \Tr_\mathrm{A}[\rhoGlobGammaP] &= \sum_{k\in\ZDZ} \bra{k}_\mathrm{A}  \left( (1-p) \ket{\Gamma} \bra{\Gamma} + p \frac{\mathbbm{1}}{D^n} \right) \ket{k}_\mathrm{A} 
 \\ 
 &=(1-p) \left(\sum_{k\in \ZDZ} \ket{\Gamma_k} \bra{\Gamma_k} \right) +  p\frac{\mathbbm1}{D^{n-1}}, 
\end{align}
Thereby, we have substituted $\ket{\Gamma_k} :=\bra{k}_\mathrm{A} \ket{\Gamma}$. 
From Eq.~\eqref{eq:ketGamma} follows that for $k,l\in\ZDZ$ the inner product of two such (unnormalized) vectors is given by 
\begin{align}
 \braket{\Gamma_l | \Gamma_k} &= 
 \frac{1}{D^n} \sum_{\substack{r_2,\ldots,r_n \in \ZDZ \\ s_2,\ldots,s_n \in \ZDZ }} 
 \omega_D^{\sum\limits_{j=2}^n \gamma_{1,j} r_j (k-l) }
 \omega_D^{\sum\limits_{i=2}^n \sum\limits_{j=i+1}^n \gamma_{i,j} (r_i r_j- s_i s_j ) } \braket{s_2,\ldots, s_n | r_2, \ldots , r_n} 
 \\
  &= 
 \frac{1}{D^n} \sum_{ r_2,\ldots,r_n \in \ZDZ} 
 \omega_D^{\sum\limits_{j=2}^n (k \gamma_{1,j} -l\gamma_{1,j} )r_j }  = \frac{1}{D^n} D^{n-1}\delta_{(k\gamma_{1,2},\ldots, k\gamma_{1,n}), (l\gamma_{1,2},\ldots, l\gamma_{1,n})}.
\end{align}
Note that we have used the relation $\sum_{r\in (\ZDZ)^{n-1}} \omega_D^\mathbf{(a-b)\cdot r} = D^n \delta_\mathbf{a,b}$, i.e., complex roots sum up to zero.
Since we assume that $\gamma_{1,2}$ is invertible, we obtain $\delta_{k\gamma_{1,2}, l\gamma_{1,2}} = \delta_{k,l}$ and can establish 
\begin{align}\label{eq:inner_Gamma_k}
 \braket{\Gamma_l | \Gamma_k} &=   \frac{\delta_{k,l} }{D}.
\end{align}
From this, we obtain for each $k\in \ZDZ$ an eigenequation of the form
\begin{align}
 \Tr_\mathrm{A}[\rhoGlobGammaP] \ket{\Gamma_k}
 &=(1-p) \sum_{l\in \ZDZ} \ket{\Gamma_l} \braket{\Gamma_l| \Gamma_k} +  p\frac{\mathbbm1}{D^{n-1}}\ket{\Gamma_k} 
 = \left( \frac{1-p}{D} +\frac{p}{D^{n-1}} \right) \ket{\Gamma_k}.
\end{align}
Similarly to Eq.~\eqref{eq:red_total_perp}, we obtain for every state $\ket{\psi} $ in the orthogonal complement of	
$\Span_\mathbb{C}\left \{ \ket{\Gamma_k} \ \big \vert \ k \in \ZDZ   \right  \}$ in the Hilbert space of all parties but Alice an eigenequation of the form
\begin{align}
 \Tr_\mathrm{A}[\rhoGlobGammaP] \ket{\psi}
 &=(1-p) \sum_{l\in \ZDZ} \ket{\Gamma_l} \braket{\Gamma_l| \psi} +  p\frac{\mathbbm1}{D^{n-1}}\ket{\psi} = \frac{p}{D^{n-1}}\ket{\psi}.
\end{align}
Again, by counting the dimensions, we obtain that $\frac{1-p}{D} +\frac{p}{D^{n-1}} $ is $D$-fold degenerate  and $\frac{p}{D^{n-1}}$ is $(D^{n-1}-D)$-fold degenerate.
From their spectra of eigenvalues, we conclude the von Neumann entropies
\begin{align}
 S\left[\rhoGlobGammaP \right] = -\left(1-p + \frac{p}{D^n} \right) \log_2\left(1-p + \frac{p}{D^n} \right)
 -(D^n-1)\left(\frac{p}{D^n}\right) \log_2\left(\frac{p}{D^n}\right) 
 \\
 \text{and} 
 \hspace{0.5em}
  S\left[\Tr_\mathrm{A}[\rhoGlobGammaP]\right] = -D \left(\frac{1-p}{D}+\frac{p}{D^{n-1}} \right) \log_2 \left(\frac{1-p}{D}+\frac{p}{D^{n-1}} \right)
 \\
 -(D^{n-1}-D)\left( \frac{p}{D^{n-1}} \right)\log_2 \left( \frac{p}{D^{n-1}} \right).
\end{align}
Note that this result is the generalization of the Werner state, our initial example in Sec.~\ref{sec:2}, to graph states on $n\ge2$ qudits in  dimension $D\ge2$.
We defer a numerical evaluation of the noise threshold resulting from the entropy criterion as in Eq.~\eqref{eq:entr_crit_gamma_glob} to Secs.~\ref{sec:5.5.3} and~\ref{sec:6.3}.

\subsection{Peres–Horodecki criterion for qudit graph states}
\label{sec:4.2}

Before we consider arbitrary qudit graph states, it is instructive to first discuss the generalized $n$-qudit Werner state
 \begin{align}
  \GHZp := (1-p)\ket{\mathrm{GHZ}_D^n}\bra{\mathrm{GHZ}_D^n} + p\frac{\mathbbm{1}}{D^n}.
 \end{align}
Although $\ket{\mathrm{GHZ}_D^n}$ itself is not a qudit graph state, it can be transformed into a graph state by a local application of the quantum Fourier transform $F$ on all qudits but one. 
Such a transformation does not change the entanglement properties of the state.
The following lemma is an application of the Peres-Horodecki criterion to $\GHZp$.
\lemma \label{lem:PPT_GHZ} \cite{DuCiTa99, GHH10}
\emph{The operator $\GHZp\TA$ has $\lambda(p)= p/D^n-(1-p)/D$ as an eigenvalue. In particular, $\GHZp $ is entangled for all $p< {1-\frac{1}{D^{n-1}+1}}$.}
\begin{proof}
 Using the expression
 \begin{align}
\ket{\mathrm{GHZ}_D^n}\bra{\mathrm{GHZ}_D^n}\TA & = \frac{1}{D}\sum_{i,j=0}^n  \ket{ji\ldots i} \bra{ij\ldots j} 
 \end{align}
 we can easily show that  $\ket{010\ldots0} - \ket{100\ldots0} $ is an eigenvector of $\GHZp\TA$ to the eigenvalue $\lambda(p)= p/D^n-(1-p)/D$. Indeed,
 \begin{align}
  &\left(  (1-p)\ket{\mathrm{GHZ}_D^n}\bra{\mathrm{GHZ}_D^n} \TA + p\frac{\mathbbm{1}}{D^n} \right) \left(\ket{010\ldots0} - \ket{100\ldots0} \right) \\
  &= \frac{1-p}{D}\left(\ket{100\ldots0} - \ket{010\ldots0} \right) + \frac{p}{D^n} \left(\ket{010\ldots0} - \ket{100\ldots0} \right)\\
  &= \left(\frac{p}{D^n} -\frac{1-p}{D} \right)\left(\ket{010\ldots0} - \ket{100\ldots0} \right).
 \end{align}
\end{proof}

It is possible to generalize this noise threshold to all graph states $\ket{\Gamma}$ whose adjacency matrix has at least one invertible ($\Leftrightarrow$ nonzero for prime $D$) edge, w.l.o.g. $\gamma_{1,2}$ is invertible in $\ZDZ$. This result is captured in the following theorem which we prove in Appendix~\ref{app:proof_PPT}.

\theorem\label{thrm:PPT_graph} 
\emph{Let $\Gamma\in (\ZDZ)^{n\times n}$ be the adjacency matrix of a graph state such that $\gamma_{1,2}$ is invertible. Then, $\lambda(p)= p/D^n-(1-p)/D$ is an eigenvalue of the operator $\rhoGlobGammaP\TA$.  In particular, $\rhoGlobGammaP $ is entangled for all $p< p_\mathrm{glob}^\mathrm{PPT}(D,n):= {1-\frac{1}{D^{n-1}+1}}$.} \\

The proof of Theorem~\ref{thrm:PPT_graph} relied on the technicality that $\gamma_{1,2}$ is invertible in $\ZDZ$.
Interestingly, if we drop this condition, the critical noise threshold becomes better such that we can draw the following conclusion.

\corollary\label{cor:PPT_graph} 
\emph{Let $\Gamma\in (\ZDZ)^{n\times n}$ be a non-trivial adjacency matrix, i.e., $\Gamma \neq 0$. 
The state $\rhoGlobGammaP $ is entangled for all $p< p_\mathrm{glob}^\mathrm{PPT}(D,n)={1-\frac{1}{D^{n-1}+1}}$.} 
\begin{proof}
In order to apply Theorem~\ref{thrm:PPT_graph}, we will reduce to the case of an adjacency matrix $\Gamma'$ with at least one invertible entry. 
For this, we will have to regard each qudit of dimensions $D$ as $g$ qudits of dimension $d:=D/g$, where we choose $g:=\gcd(\gamma_{i,j})$ as the greatest common divisor of all entries of the original matrix $\Gamma\in(\ZDZ)^{n\times n}$. This yields a new adjacency matrix $\Gamma'\in(\ZdZ)^{n\times n}$ for lower-dimensional qudits via $\gamma_{i,j}':=\gamma_{i,j}/g$.
Recall from Eq.~\eqref{eq:ketGamma}, that  graph states can be written as
\begin{align}
 \ket{\Gamma} &= \frac{1}{\sqrt{D^n}} \sum_{\mathbf{r}\in (\ZDZ)^n} \omega_D^{\sum\limits_{i=1}^n\sum\limits_{j=i+1}^n \gamma_{i,j} r_i r_j} \ket {\mathbf{r}}.
\end{align}
Using the $d$-ary decomposition $r_k= \sum_{l=0}^{g-1}d^lR_{k,l}$ (such that all $0\le r_k \le D-1$ and $0\le R_{k,l}\le d-1$ are integers), and the fact that $\omega_D^{\gamma_{i,j}}= \omega_D^{g\gamma'_{i,j}}=\omega_d^{\gamma'_{i,j}}$, we can rewrite the graph state as
\begin{align}
 \ket{\Gamma} = \frac{1}{\sqrt{D^n}} \sum_{R\in (\ZdZ)^{n\times g}}\omega_d^{\sum\limits_{i=1}^n\sum\limits_{j=i+1}^n \gamma'_{i,j} R_{i,0} R_{j,0} } \ket {R}
\end{align}
since $\omega_d^d=1$ implies $\omega_d^{r_k}= \omega_d^{ \sum_{l=0}^{g-1}d^lR_{k,l} } = \omega_d^{R_{k,0}}$.
Note that (for a fixed $k$) the coefficient in front of $\ket{R_{k,l}}$ is the same for all $l=1,\ldots,g-1$. 
We introduce the notation $r_k':= R_{k,0}$ and $R'_{k,l}:=R_{k,l}$ for $l=1,\ldots,g$, to express the matrix $R$ as $[\mathbf{r'}\vert R']$. 
In this way, we can rewrite the summation over $R\in (\ZdZ)^{n\times g}$ as
\begin{align}
 \ket{\Gamma} &= \left( \frac{1}{\sqrt{d^n}} 
 \sum_{\mathbf{r}\in(\ZdZ)^n}
 \omega_d^{\sum\limits_{i=1}^n\sum\limits_{j=i+1}^n \gamma'_{i,j} r'_i r'_j } \ket{\mathbf{r'}} 
 \right) \otimes \left(
 \frac{1}{\sqrt{d}^{n(g-1)}} 
 \sum_{R'\in(\ZdZ)^{n\times(g-1)}} \ket{R'} \right) \\
 &=
 \ket{\Gamma'} \otimes \ket{+_d}^{\otimes n(g-1)}.
\end{align}
In this way, we can regard the graph state $\ket{\Gamma}$ on $n$ qudits of dimension $D$ as a graph state on $ng$ qudits of dimension $d=D/g$ with at least one invertible edge (say $\gamma_{1,2}$).
Therefore, the operator
$(1-p)\ket{\Gamma}\bra{\Gamma}\TA + p\mathbbm{1}/D^n$ has an eigenvalue
$p/d^{ng}-(1-p)/d$ which is negative for all $p<p_\mathrm{glob}^\mathrm{PPT}(d,ng)=1-1/(d^{ng-1}+1)$ and $p_\mathrm{glob}^\mathrm{PPT}(D,n)=1-1/(D^{n-1}+1)= 1-1/(d^{ng-g}) \le 1-1/(d^{ng-1}+1)$ finishes the proof.
\end{proof}

\subsubsection{Invertible vs non-invertible edges}
\label{sec:4.2.1}

To better understand the technicality of $\gamma_{1,2}$ being invertible or not consider the easiest example of two ququarts, i.e., $D=4$ and $n=2$. Further consider the adjacency matrices 
\begin{align}
 \Gamma^2_4(1) = \left[\begin{array}{cc}0&1\\ 1&0\end{array} \right]
 \hspace{2em}
 \text{and}
 \hspace{2em} 
 \Gamma^2_4(2) = \left[\begin{array}{cc}0&2\\ 2&0\end{array} \right].
\end{align}
The important difference is that 1 is invertible in $\ZZ/4\ZZ$ but 2 is not.
By Theorem~\ref{thrm:PPT_graph}, $\rho_{\mathrm{glob},\Gamma^2_4(1)}(p)$ is NPT, thus entangled, for all $p<1-\frac{1}{4^{2-1}+1} = \frac{4}{5}$ and from the proof of Corollary~\ref{cor:PPT_graph} it is clear that $\rho_{\mathrm{glob},\Gamma^2_4(1)}(p)$ is NPT  for all $p<1-\frac{1}{2^{4-1}+1} = \frac{8}{9}>\frac{4}{5}$. That is, $ \ket{\Gamma^2_4(2)} $ is more robust against global white noise than $ \ket{\Gamma^2_4(1)}$. 
Next, let us bring the graph states into a form from which we can read off their entanglement properties.
Since $\omega_4 =i$ and $\omega_4^2=-1$,  Eq.~\eqref{eq:ketGamma} yields
\begin{align}
 \ket{\Gamma^2_4(1)} = \frac{1}{4} \sum_{j,k=0}^3 i^{jk} \ket{j,k}
 \hspace{2em}
 \text{and}
 \hspace{2em} 
 \ket{\Gamma^2_4(2)} = \frac{1}{4} \sum_{j,k=0}^3 (-1)^{jk} \ket{j,k}.
\end{align}
From this, it is straightforward to compute
\begin{align}\label{eq:gamma24_1}
F_4^\dagger \otimes \mathbbm 1_4 \ket{\Gamma^2_4(1)} =\frac{1}{2} \left( \ket{0}\otimes\ket{0}+\ket{1}\otimes\ket{1}+\ket{2}\otimes\ket{2}+\ket{3}\otimes\ket{3} \right)
\end{align}
as well as 
\begin{align}\label{eq:gamma24_2} 
(F_2\otimes F_2) \otimes \mathbbm 1_4 \ket{\Gamma^2_4(2)} =\frac{1}{\sqrt{2}} \left( \ket{0}\otimes \frac{1}{\sqrt{2}}(\ket{0}+\ket{2})+\ket{1}\otimes \frac{1}{\sqrt{2}}(\ket{1}+\ket{3}) 
\right).
\end{align}
If we use the binary identification $0\leftrightarrow00$, $1\leftrightarrow01$, $2\leftrightarrow10$, and $3\leftrightarrow11$, we see from Eq.~\eqref{eq:gamma24_1} that $ \ket{\Gamma^2_4(1)}$ is locally unitary equivalent to the state
\begin{align}
  &\frac{1}{2} \left( \ket{00}_\mathrm{A_1A_2}  \ket{00}_\mathrm{B_1B_2} +\ket{01}_\mathrm{A_1A_2}  \ket{01}_\mathrm{B_1B_2} +\ket{10}_\mathrm{A_1A_2}  \ket{10}_\mathrm{B_1B_2} +\ket{11}_\mathrm{A_1A_2}  \ket{11}_\mathrm{B_1B_2}  \right) \\
  =& \frac{1}{2}\left( \ket{00}_\mathrm{A_1 B_1}  + \ket{11}_\mathrm{A_1 B_1}  \right) \left(\ket{00}_\mathrm{A_2B_2}+\ket{11}_\mathrm{A_2B_2} \right)
  = \ket{\Phi^+}  _\mathrm{A_1 B_1} \ket{\Phi^+}  _\mathrm{A_2 B_2}.
\end{align}
That is, $\ket{\Gamma^2_4(1)} $ carries the same amount of entanglement as two qubit Bell pairs, while $\ket{\Gamma^2_4(2)}$ only is worth one qubit Bell pair as one can read off from Eq.~\eqref{eq:gamma24_2}.
To conclude our discussion, $\ket{\Gamma^2_4(1)}$ is more entangled than $\ket{\Gamma^2_4(2)}$ but its entanglement is less robust against global white noise.

\subsection{Reduction criterion for qudit graph states}
\label{sec:4.3}

Here, we will reestablish the noise threshold of Theorem~\ref{thrm:PPT_graph} with a much easier proof based on the reduction criterion,
\begin{align}
 \rhoGlobGammaP\text{ is separable}
 \hspace{2em}
 \Longrightarrow  
 \hspace{2em}
\mathbbm{1}_D \otimes \Tr_\mathrm{A}[\rhoGlobGammaP] \ - \rhoGlobGammaP  \  \ge\ 0 .
\end{align}
Although the reduction criterion is in general not better that the Peres-Horodecki criterion~\cite{Horodecki09}, the resulting noise thresholds  for qudit graph states coincide.

In order to find a negative eigenvalue, we use Eqs.~\eqref{eq:Gamma_glob} and \eqref{eq:Gamma_glob_reduced} and rewrite the operator of interest  as $\mathbbm{1}_D \otimes \Tr_\mathrm{A}[\rhoGlobGammaP] \ - \rhoGlobGammaP $
\begin{align} 
= &(1-p) \left(\left(\mathbbm{1}_D \otimes \sum_{l\in \ZDZ} \ket{\Gamma_l} \bra{\Gamma_l}  \right)  - \ket{\Gamma}\bra{\Gamma} \right) +  p \left( \frac{1}{D^{n-1}} -  \frac{1}{D^n} \right) \mathbbm{1}_{D^n} ,
\end{align}
where $\ket{\Gamma_l} =\bra{l}_\mathrm{A} \ket{\Gamma}$. 
By expanding the graph state as 
\begin{align}
 \ket{\Gamma} = \sum_{k\in \ZDZ}  \ket{k}_\mathrm{A} \otimes\ket{\Gamma_k},
\end{align}
we obtain the eigenequation
\begin{align}
 & \big( \mathbbm{1}_D \otimes \Tr_\mathrm{A}[\rhoGlobGammaP]\ -\rhoGlobGammaP \big) \ket{\Gamma} \\
 &= (1-p)\left(\left(\sum_{k\in \ZDZ} \ket{k}_\mathrm{A} \otimes \sum_{l\in \ZDZ} \ket{\Gamma_l} \braket{\Gamma_l| \Gamma_k}  \right)  - \ket{\Gamma} \right) +p\left( \frac{1}{D^{n-1}} -  \frac{1}{D^n} \right) \ket{\Gamma}\\
 &= \left( (1-p)\left(\frac{1}{D}-1\right) + p\left(\frac{1}{D^{n-1}}-\frac{1}{D^n} \right)\right) \ket{\Gamma},
\end{align}
where we have used Eq.~\eqref{eq:inner_Gamma_k}.
That is,  $\mathbbm{1}_D \otimes \Tr_\mathrm{A}[\rhoGlobGammaP] \ - \rhoGlobGammaP $ has the eigenvalue,
\begin{align}
  \left( (1-p)\left(\frac{1}{D}-1\right) + p\left(\frac{1}{D^{n-1}}-\frac{1}{D^n} \right)\right) =  \left( 1-\frac{1}{D}\right) \left(p \left( 1+\frac{1}{D^{n-1}} \right)-1\right)  
\end{align}
which is negative for all $p$ smaller than
\begin{align}
 p_{\mathrm{crit,glob}}^\mathrm{Red}(D,n) := \frac{1}{1+ \frac{1}{D^{n-1}}} = 1- \frac{1}{D^{n-1}+1}.
\end{align}
This is exactly the bound from Theorem~\ref{thrm:PPT_graph}, however, it takes very little effort to establish it using the reduction criterion.

\subsection{Range of applicability of the established noise thresholds}
\label{sec:4.4}

The noise thresholds established in Sec.~\ref{sec:4} apply to all entangled qudit graph states with arbitrary qudit dimension $D$.
In Lemma 7 of Ref.~\cite{BahBei06} it is shown that every stabilizer state is local-unitary equivalent to a graph state if $D$ is prime. Therefore, our noise thresholds even hold for all stabilizer states (at least if $D$ is prime). 
A counterexample of a non-stabilizer state with a worse noise threshold is the $W$-state defined in Eq.~\eqref{def:GHZ_W}. Numerically, we find that $(1-p)\ket{W} \bra{W} + p\mathbbm{1}/8$ is NPT iff $p< 0.7904$ while the three-qubit GHZ state tolerates up to $p_\mathrm{glob}^\mathrm{PPT}(2,3)=0.8$ global white noise.
This implies that the noise threshold established in Corollary~\ref{cor:PPT_graph} cannot be extended to general entangled states. The result only holds for stabilizer states.

We have restricted ourselves to bipartitions with a single isolated party (Alice) and did not consider other bipartitions because we have numerical evidence that this leads to the best noise thresholds.
Since we only applied bipartite entanglement criteria, 
we are not able make statements about multipartite entanglement aspects of the noisy states using this approach, recall the example in Sec.~\ref{sec:2.2.2}.
Furthermore, we could only apply these criteria to the simplest noise model of a global depolarizing channel.
In order to capture also multipartite entanglement aspects, even for a more complicated noise model, we will introduce a helpful technical tool in the next section.

\section{\label{sec:5}\protect Sector lengths}

The sector lengths of a multipartite quantum state are properties which can sometimes capture certain aspects of multipartite entanglement~\cite{ACHB04, deVicHub11, KloeHub15, WydGueh19, EltSiew19}. 
In general, the resulting entanglement criteria are not strong enough to verify genuine multipartite entanglement. This weakness has been attributed to the otherwise convenient fact that sector lengths are invariant under local unitary operations~\cite{KloeHub15} which will enable us to establish noise thresholds to rule out $(n_1,\ldots, n_k)$-separability (recall Sec.~\ref{sec:2.2.1}) for two types of noise: global white noise and local white noise.
However, we will be able to rule out semiseparability in some cases which elude other approaches.

We begin with the definition of sector lengths and some of their elementary properties in Sec.~\ref{sec:5.1}. 
In Sec.~\ref{sec:5.2}, we explain how entanglement criteria can be derived from sector lengths.
In Sec.~\ref{sec:5.3}, we discuss how sector lengths of stabilizer states are influenced by global and local white noise, respectively. 
Since knowing the sector lengths of the corresponding pure stabilizer state will turn out to be crucial, we have devoted Sec.~\ref{sec:5.4} for their calculation for important families of states.
In Sec.~\ref{sec:5.5} we compare the resulting noise thresholds with other thresholds from the literature.

\subsection{Definition and basic properties}
\label{sec:5.1}

Recall from the proof of Theorem~\ref{thrm:stab} from Sec.~\ref{sec:3.2} that the projector onto every stabilizer state $\ket{\psi}$ can be rewritten as 
\begin{align}\label{eq:Pj_decomp_stab}
 \ket{\psi}\bra{\psi} &= \frac{1}{D^n} \sum_{S \in \mathcal{S}_{\ket{\psi}}} S 
 \hspace{2em}=\frac{1}{D^n}\sum_{j=0}^n  P_j.
\end{align} 
Hereby, we have combined the stabilizers which (non-trivially) act on the same number $j$ of qudits into
\begin{align}\label{eq:Pj_stab}
 P_j := \sum _{\substack{ S\in\mathcal{S}_{\ket{\psi}} \\ S \text{ acts on (exactly) }j \text{ qudits} } }
 \hspace{-2.0em}
 S.
\end{align} 
As the Pauli operators $X_D^\mathbf{r}Z_D^\mathbf{s}$, where $\mathbf{r,s}\in(\ZDZ)^n$, constitute an orthonormal basis of the complex $D^n\times D^n$-matrices, with respect to the Hilbert-Schmidt inner product, i.e., 
\begin{align}
 \frac{1}{D^n} \Tr\left[ (X_D^\mathbf{r'}Z_D^\mathbf{s'})^\dagger X_D^\mathbf{r}Z_D^\mathbf{s} \right] =  \delta_\mathbf{r,r'}\delta_\mathbf{s,s'},
\end{align}
every arbitrary state $\rho$ can be decomposed as
\begin{align}\label{eq:Pj_decomp_general}
 \rho &= \frac{1}{D^n}\sum_{\mathbf{r,s}\in (\ZDZ)^n} w_\mathbf{r,s} X_D^\mathbf{r}Z_D^\mathbf{s}
 \hspace{1.5em} = \frac{1}{D^n}\sum_{j=0}^n P_j 
\end{align}
for some coefficients $w_\mathbf{r,s}\in \mathbb{C}$~\cite{VolWol03}.
Hereby, we have again combined the terms which act on the same number $j$ of qudits into
\begin{align}\label{eq:Pj_general}
 P_j &:= \sum_{\substack{\mathbf{r,s}\in(\ZDZ)^n\\ \swt(\mathbf{r,s})=j}}
 w_\mathbf{r,s} X_D^\mathbf{r} Z_D^\mathbf{s},
\end{align}
where the \emph{symplectic weight} of $(\mathbf{r,s})$ is formally defined as 
$\swt(\mathbf{r,s}):= \left \vert \{i\vert r_i\neq0 \vee s_i\neq 0\} \right \vert$.
In the case where $\Psi=\ket{\psi}\bra{\psi}$ is the projector onto a stabilizer state,
Eq.~\eqref{eq:Pj_decomp_general} and Eq.~\eqref{eq:Pj_general} simplify to Eq.~\eqref{eq:Pj_decomp_stab} and Eq.~\eqref{eq:Pj_stab}, respectively.

Based on such decompositions, the \emph{sector lengths} $\ell^n_j[\rho]$ of an $n$-partite quantum state $\rho$ are defined as the Hilbert-Schmidt norm of the operators $P_j$ 
~\cite{ACHB04, WydGueh19}, i.e.,
\begin{align}\label{def:sector_length}
 \ell^n_j[\rho] := \norm{P_j}^2 := \frac{1}{D^n} \Tr\left[ P_j^\dagger P_j \right]
 \hspace{1em}
 = \sum_{\substack{\mathbf{r,s}\in(\ZDZ)^n\\ \swt(\mathbf{r,s})=j}}
 \vert w_\mathbf{r,s} \vert ^2.
\end{align}
Note that the normalization condition $\Tr[\rho]=1$ implies $P_0 =\mathbbm{1}_{D^n}$, thus $\ell^n_0[\rho] = 1$.
For pure stabilizer states $\Psi=\ket{\psi}\bra{\psi}$, $\vert w_\mathbf{r,s} \vert ^n$ is either 0 or 1, cf. Eq.~\eqref{eq:Pj_stab}. Thus, the sector lengths are simply given by
\begin{align} \label{eq:sec_len_counting}
 \ell^n_j[\Psi] 
 =  \left \vert \{S\in\mathcal{S}_{\ket{\psi}} \ \vert \ S \text{ acts on }j\text{ qudits} \} \right \vert.
\end{align} 

Under tensor products, sector lengths behave as follows~\cite{WydGueh19}.
Consider an $n$-qudit state $\rho = \rho'\otimes \rho''$ which is the tensor product of an $m$- and an $(n-m)$-qudit state, say 
\begin{align}
 \rho'\otimes \rho''= \left(\frac{1}{D^m} \sum_{k=0}^m P'_k \right)\otimes\left(\frac{1}{D^{n-m}} \sum_{l=0}^{n-m} P''_l \right) = \frac{1}{D^n} \sum_{k=0}^m \sum_{l=0}^{n-m}  P'_k \otimes P''_l
  = \frac{1}{D^n}\sum_{j=0}^n P_j,
\end{align}
where $P'_k$ and $P''_l$ are defined in analogy to Eq.~\eqref{eq:Pj_general}.
The operators $P_j$ in the decomposition of $\rho = \rho'\otimes \rho''$ follow as
\begin{align}\label{eq:Pj_product}
 P_j = \sum_{i=0}^j P'_i \otimes P''_{j-i},
\end{align}
where we set $P'_i=0$ if $i>m$. 
As the terms in Eq.~\eqref{eq:Pj_product} are mutually orthogonal (with respect to the Hilbert-Schmidt inner product),
the sector lengths of $\rho'\otimes \rho''$ and  of $\rho',\rho''$ are related by
\begin{align}\label{eq:sec_tensor}
 \ell^n_j[\rho' \otimes \rho''] =  \sum_{i=0}^j \ell^m_i[\rho']\, \ell^{n-m}_{j-i}[\rho''],
\end{align}
where we set $\ell^m_i[\rho]=0$ if $i>m$ for consistency with Eq.~\eqref{def:sector_length}.

\subsection{Tailoring entanglement criteria from sector lengths}
\label{sec:5.2}
 
The key to construct entanglement criteria for an $n$-qudit state $\rho$ is its \emph{purity} which is defined as $\Tr[\rho^2]$ and can take values between ${1}/{D^n}$ and $1$.
It is equal to $1$ iff $\rho$ is a pure state. Conversely, it is equal to $1/D^n$ iff $\rho=\mathbbm{1}/D^n$ is the completely mixed state~\cite{NielsenChuang}.
Since the operators $P_j$ in the decomposition of $\rho$  
are mutually orthogonal, we can relate the purity of a quantum state to its sector lengths,
\begin{align}\label{eq:purity}
 \Tr[\rho^2]= \frac{1}{D^n} \sum_{j=0}^n\ell^n_j[\rho] .
\end{align}
For single qudit states, the condition $\Tr[\rho^2]\le 1$ is equivalent to  
$\ell^1_1[\rho]\le D-1$, 
with equality iff $\rho $ is pure.
More generally, omitting all but one of the (nonnegative) terms in Eq.~\eqref{eq:purity} yields inequalities of the form
\begin{align}\label{eq:sectorlength_bound}
 \ell^n_j[\rho] \le D^n -1
\end{align}
for all $1\le j \le n$.  
In less generic cases, tighter inequalities are known~\cite{WydGueh19}.
Using the behavior of sector lengths under tensor products, Eq.~\eqref{eq:sec_tensor}, one can recursively construct upper bounds on the sector lengths of states that are $(n_1,\ldots,n_k)$-separable~\cite{deVicHub11}.
As the triangle inequality of the Hilbert Schmidt norm implies 
\begin{align}\label{eq:ell_convex}
 \ell^n _j[p\rho + (1-p) \tilde \rho] \le p\ell^n _j[\rho] +(1-p)\ell^n _j[\tilde{\rho}],
\end{align}
for all convex combinations of $n$-partite states $\rho$ and $\tilde\rho$, it suffices to construct such upper bounds for product states of corresponding separability. 
This provides entanglement criteria for states which exceed these bounds. 
To demonstrate the procedure~\cite{deVicHub11}, we will now recursively derive such bounds for up to $n=4$ parties as we will separately  study tetrapartite quantum networks in Sec.~\ref{sec:5.5.3}.  

For two qudits, the only nontrivial separability type is $(1,1)$. Such a fully separable state $\rho=\rho_\mathrm{A}\otimes\rho_\mathrm{B}$ fulfills
 \begin{align} \label{eq:ell2}
 \def\arraystretch{1.2}
\begin{array}{rlll}
 \ell^2_1[\rho_\mathrm{A} \otimes\rho_\mathrm{B}] &= \ell^1_1 [\rho_\mathrm{A}]+\ell^1_1 [\rho_\mathrm{B}] &\le 2(D-1) &=: b_1^{(1,1)} \\
\hspace{1em} \text{ and }
\ell^2_2[\rho_\mathrm{A} \otimes\rho_\mathrm{B}] &= \ell^1_1 [\rho_\mathrm{A}]\ell^1_1 [ \rho_\mathrm{B}] &\le (D-1)^2 &=: b_2^{(1,1)},
\end{array}
\end{align}
as a repeated use of Ineq.~\eqref{eq:sectorlength_bound} with $n=j=1$ shows.
Therefore, a bipartite state $\rho_\mathrm{AB}$ with $\ell^2_j[ \rho_\mathrm{AB}] > b_j^{(1,1)}$ for $j=1$ or $j=2$, is necessarily entangled.
%

%
For three qudits, there are two nontrivial separability types: full separability $(1,1,1)$ and biseparability $(2,1)$. 
Similar to Eq.~\eqref{eq:ell2}, we obtain the bounds 
 \begin{align} \label{eq:ell3fully}
 \def\arraystretch{1.2}
\begin{array}{rlll}
 \ell^3_1[\rho 
 ] &=  \ell^1_1 [\rho_\mathrm{A}]+\ell^1_1 [\rho_\mathrm{B}]+\ell^1_1[\rho_\mathrm{C}] 
 &\le 3(D-1) &=: b_1^{(1,1,1)}, \\
 \ell^3_2[\rho 
 ] &= \ell^1_1 [\rho_\mathrm{A}]\ell^1_1 [\rho_\mathrm{B}] + \ell^1_1 [\rho_\mathrm{A}]\ell^1_1 [\rho_\mathrm{C}]+ \ell^1_1 [\rho_\mathrm{B}]\ell^1_1 [\rho_\mathrm{C}] 
 & \le 3(D-1)^2 & =: b_2^{(1,1,1)}, \\
\text{and } \hspace{0.1em}
\ell^3_3[\rho 
] &= 
\ell^1_1 [\rho_\mathrm{A}] \ell^1_1 [\rho_\mathrm{B}]  \ell^1_1 [\rho_\mathrm{C}] 
&\le (D-1)^3  &=: b_3^{(1,1,1)}
\end{array}
\end{align}
for fully separable states $\rho 
= \rho_\mathrm{A}\otimes\rho_\mathrm{B}\otimes\rho_\mathrm{C}$. 
Thus, if a sector length of a tripartite state exceeds one of the above bounds, it cannot be fully separable.
Such a state, however, could be biseparable still.
To derive a bound which could rule out this possibility, consider w.l.o.g. a biseparable product state of the form $\rho=\rho_\mathrm{A}\otimes\rho_\mathrm{BC}$. Its sector lengths fulfill 
 \begin{align} \label{eq:ell3bi}
 \def\arraystretch{1.2}
\begin{array}{rlll}
 \ell^3_1[\rho_\mathrm{A}\otimes\rho_\mathrm{BC}] &= \ell^1_1 [\rho_\mathrm{A}]+\ell^2_1 [\rho_\mathrm{BC}] &\le D^2 + D-2  &=: b_1^{(2,1)}   
 , \\
 \ell^3_2[\rho_\mathrm{A}\otimes\rho_\mathrm{BC}] &= \ell^2_2 [\rho_\mathrm{BC}]+ \ell^1_1 [\rho_\mathrm{A}]\ell^2_1[\rho_\mathrm{BC}]
 &\le D^3 - D  &=: b_2^{(2,1)}, \\ 
\text{ and } 
\hspace{1em}
\ell^3_3[\rho_\mathrm{A}\otimes\rho_\mathrm{BC}] &= \ell^1_1 [\rho_\mathrm{A}] \ell^2_2 [\rho_\mathrm{BC}] &\le D^3-D^2-D+1 &=: b_3^{(2,1)}.   
\end{array}
\end{align}
Here, we used Ineq.~\eqref{eq:sectorlength_bound} with $n,j\in\{1,2\}$.  
Again, if a sector length of a tripartite state exceeds one of the bounds in Eq.~\eqref{eq:ell3bi}, the state cannot be biseparable and is thus genuinely tripartite entangled.
%
%

For four qudits, there are five separability types as displayed in Fig.~\ref{fig:4_separablity}.
\begin{figure}
\centering
\begin{minipage}{0.85\textwidth}
\begin{framed} 
\centering
\vspace{1.5em}

\begin{tikzpicture} 
 \draw (0,0) node[text=black]{(1,1,1,1)}; 
 \draw (0,-1.5) node[text=black]{(2,1,1)}; 
 \draw (-1.5,-3) node[text=black]{(2,2)}; 
 \draw (1.5,-3) node[text=black]{(3,1)}; 
 \draw (0,-4.5) node[text=black]{(4)};
 \draw[->, line width=.1em] (0, -0.3) -- (0, -1.2); 
 \draw[->, line width=.1em] (0.3,-1.9) -- (1.2,-2.6); 
 \draw[->, line width=.1em] (-0.3,-1.9) -- (-1.2,-2.6); 
 \draw[->, line width=.1em] (1.2,-3.4)  -- (0.3, -4.2); 
 \draw[->, line width=.1em] (-1.2,-3.4) -- (-0.3,- 4.2); 

 
\end{tikzpicture}

\caption{The five partitions of the number four. Each partition corresponds to a separability type of a tetrapartite quantum system. 
The  arrows indicate relations between the separability types, e.g., $(1,1,1,1)$-separability implies $(2,1,1)$-separability.
}
 \label{fig:4_separablity}
\end{framed}
\end{minipage}
 \end{figure} 
Using the trivial bound $b_j^{(n)}=D^n-1$ and the bounds previously established, we obtain in analogy to Eq.~\eqref{eq:sec_tensor} the nontrivial tetrapartite bounds 
\begin{align}\label{eq:b14}
\def\arraystretch{1.2}
&
\begin{array}{lll} 
b_1^{(3,1)}    	&:= b_1^{(3)}+ b_1^{(1)}	 &=\hspace{0em} D^3+D-2, \\ 
b_1^{(2,2)}	&:= b_1^{(2)}+ b_1^{(2)} 	 &=2(D^2-1), \\
b_1^{(2,1,1)}	&:= b_1^{(2)}+ b_1^{(1,1)} 	 &=\hspace{0em} D^2 + 2D - 3, \\ 
b_1^{(1,1,1,1)} 	&:= b_1^{(1,1)} + b_1^{(1,1)}	 &=4(D-1),
\end{array}
\\ &
\label{eq:b24}
\def\arraystretch{1.2}
\begin{array}{lll} 
b_2^{(3,1)}    	&:= b_2^{(3)}+ b_1^{(3)}b_1^{(1)}	 	 &=\hspace{0em} D^4-D^3, \\ 
b_2^{(2,2)}	&:= b_2^{(2)}+ b_1^{(2)}b_1^{(2)}+b_2^{(2)} 	 &=\hspace{0em} D^4-1 ,\\ 
b_2^{(2,1,1)}	&:= b_2^{(2)}+ b_1^{(2)}b_1^{(1,1)}+b_2^{(1,1)}  &= 2(D^3 -2D +1) ,\\  
b_2^{(1,1,1,1)}	&:= b_2^{(1,1)}+ b_1^{(1,1)} b_1^{(1,1)} + b_2^{(1,1)}	 	 &=6(D^2-2D+1),
\end{array}
\\ &
\label{eq:b34}
\def\arraystretch{1.2}
\begin{array}{lll} 
b_3^{(3,1)}    	&:= b_3^{(3)}+ b_2^{(3)}b_1^{(1)}	 	&=\hspace{0em} D^4-D , \\ 
b_3^{(2,2)}	&:= b_2^{(2)}b_1^{(2)}+ b_1^{(2)}b_2^{(2)}	&= 2(D^4-2D^2+1), \\ 
b_3^{(2,1,1)}	&:= b_2^{(2)}b_1^{(1,1)}+b_1^{(2)}b_2^{(1,1)}  &=\hspace{0em} D^4-2D^2+1, \\   
b_3^{(1,1,1,1)}	&:= b_2^{(1,1)} b_1^{(1,1)} + b_1^{(1,1)}b_2^{(1,1)}  	 &=4(D^3-3D^2+3D-1),
\end{array}
\\ &
\label{eq:b44}
\def\arraystretch{1.2}
\begin{array}{lll} 
b_4^{(3,1)}    	&:= b_3^{(3)}b_1^{(1)}	 	&=D ^4-D^3-D+1 , \\ 
b_4^{(2,2)}	&:= b_2^{(2)}b_2^{(2)} 		&=D^4 -2D^2+1, \\ 
b_4^{(2,1,1)}	&:= b_2^{(2)}b_2^{(1,1)}	&=D^4-2D^3+2D-1 ,  \text{ and} \\ 
b_4^{(1,1,1,1)}	&:= b_2^{(1,1)}b_2^{(1,1)} 		&= D^4-4D^3+6D^2-D+1. 
\end{array}
\end{align}

In general, there are exactly $p(n)$ separability types for an $n$-qudit system, where $p$ is the number-theoretical partition function. 
Since not even an explicit expression of $p(n)$ is known, we do not hope to find a closed expression of all such bounds. However, the following lemma provides the bounds for full separability in the general $n$-qudit case.
\lemma \label{lem:fully_sep_bound}
\emph{The sector lengths of every fully separable state $\rho$ fulfill}
\begin{align}
 \ell^n_j [\rho] \le b_j^{(1,\ldots,1)} := \nchoosej (D-1)^j. 
\end{align}
\begin{proof}
We prove this lemma by induction. The base case was treated already. Now, assume
$
 \ell^{n-1}_j [\rho] \le
\genfrac{(}{)}{0pt}{1}{n-1}{j} 
(D-1)^j
$ 
for all fully separable $n$-qudit states $\rho$ and all $0\le j \le n-1$. 
Let $\rho'$ be some additional single qudit state.
From Eq.~\eqref{eq:sec_tensor}, we obtain (for $j\neq n$)
\begin{align}
 \ell^{n}_j [\rho'\otimes \rho] =&  \ell^{1}_0[\rho']\ell^{n-1}_j[\rho]+ \ell^{1}_1[\rho']\ell^{n-1}_{j-1}[\rho'] \\
 \le &
\genfrac{(}{)}{0pt}{1}{1}{0}      (D-1)^0  
\genfrac{(}{)}{0pt}{1}{n-1}{j} (D-1)^j  
+
\genfrac{(}{)}{0pt}{1}{1}{1}      (D-1)^1  
\genfrac{(}{)}{0pt}{1}{n-1}{j-1}  (D-1)^{j-1}  \\
=& \left(  
\genfrac{(}{)}{0pt}{1}{n-1}{j-0}  +
\genfrac{(}{)}{0pt}{1}{n-1}{j-1} \right ) (D-1)^j
= \nchoosej (D-1)^j.
 \end{align}
And for $j=n$, we have
\begin{align}
 \ell^{n}_n [\rho'\otimes \rho] =& \ell^1_1[\rho'] \ell^{n-1}_{n-1} [\rho] \le 
\genfrac{(}{)}{0pt}{1}{1}{1}     (D-1)^1 \genfrac{(}{)}{0pt}{1}{n-1}{n-1} (D-1)^{n-1} 
= 
\genfrac{(}{)}{0pt}{1}{n}{n}  (D-1)^n.
\end{align} 
\end{proof}

However, most of these bounds are trivial as they are fulfilled by all states: In the qubit case, for example, it is conjectured that for all $j$ there exists an $n_j$, such that for all $n\ge n_j$, $\ell^n_j[\rho]\le \nchoosej$ holds for all $n$-qubit states~\cite{WydGueh19}.
The conjecture is proven for all $j\le 3$ with $n_1=1$, $n_2=3$, and $n_3=5$~\cite{WydGueh19}.
However, as we will show in Appendix~\ref{app:bound_full_body_sec}, at least the full-body sector length $\ell^n_n[\rho]$ exceeds $b_n^{(1,\ldots,1)}=(D-1)^n$ for a very large class of pure stabilizer states $\rho=\ket{\Gamma}\bra{\Gamma}$.
As we show next, this leads to nontrivial noise thresholds based on sector lengths.

\subsection{Sector lengths of noisy stabilizer states} 
\label{sec:5.3}

In this section, we establish how sector lengths of a stabilizer state $\ket{\psi}$ change in the presence of noise. This will lead to certain noise thresholds similar to those in Sec.~\ref{sec:4}.
As in Eq.\eqref{eq:Gamma_glob}, we use the abbreviation
 \begin{align} \label{eq:psi_glob}
 \rhoGlobPsiP  := (1-p) \ket{\psi} \bra{\psi} + p \frac{\mathbbm{1}}{D^n}
\end{align} 
for the state that has passed a global noise channel 
\begin{align}
 \mathcal{E}^{(p)}_{\mathrm{depol};D^n}: \rho \longmapsto (1-p) \rho +p \frac{\mathbbm{1}}{D^n}.
\end{align}
Furthermore, we will be able to draw conclusions about states of the form
\begin{align} \label{eq:psi_loc}
 \rhoLocPsiP  := ( \mathcal{E}^{(p)}_{\mathrm{depol};D})^{\otimes n} \left[ \ket{\psi} \bra{\psi}\right].
\end{align} 
For spatially separated qudits, this is the physically more relevant noise model as it corresponds to $n$ independent depolarization processes.
We will refer to these noise models as global and local white noise, respectively.
The following proposition shows how precisely they diminish sector lengths.
 
\proposition\label{prop:noisy_sector_lengths}
\emph{Let $\ket{\psi}$ be an $n$-qudit stabilizer state and write $\Psi=\ket{\psi}\bra{\psi}$.
For every $j\in\{1,\ldots,n\}$ and every $p\in [0,1]$ it holds:} 
\begin{align} \label{eq:sectorlength_glob}
 \ell^n_j \left[\rhoGlobPsiP    \right] &= (1-p)^2 \ell^n_j   [\Psi ]  \\ \label{eq:sectorlength_loc}
 \ell^n_n \left[\rhoLocPsiP   \right] &= (1-p)^{2n} \ell^n_n   [\Psi ]  
\end{align}
\begin{proof} 
First note that the depolarizing channel which adds global white noise can be regarded as an error channel where a discrete Pauli error $X_D^\mathbf{r}Z_D^\mathbf{s}$ occurs with probability 
\begin{align}
p^\mathrm{glob}_{\mathbf{r,s}} :=   
\begin{cases} 
1-p + \frac{p}{D^{2n}} & \text{ if $\mathbf{r=s}=(0,\ldots,0)$}\\
      \frac{p}{D^{2n}} & \text{ otherwise,} 
\end{cases}
\end{align}
i.e., $\mathcal{E}^{(p)}_{\mathrm{depol};D^n} [\rho] = \sum_{\mathbf{r,s}\in (\ZDZ)^n } p_\mathbf{r,s}^\mathrm{glob} (X_D^\mathbf{r}Z_D^\mathbf{s}) \rho (X_D^\mathbf{r}Z_D^\mathbf{s}) ^\dagger$; see Ref.~\cite{MHKB18} for a proof of this. From this, it follows that also  $(\mathcal{E}^{(p)}_{\mathrm{depol};D})^{\otimes n}$ can be regarded as a Pauli error channel with error probabilities 
\begin{align}
 \tilde p^\mathrm{loc}_\mathbf{r,s} :=  \left(1-\tilde p + \frac{\tilde p}{D^{2}} \right)^{n-\swt(\mathbf{r,s})} \left(\frac{\tilde p}{D^{2}} \right)^{\swt(\mathbf{r,s})},
\end{align}
i.e., $ ( \mathcal{E}^{(1-\tilde p)}_{\mathrm{depol};D})^{\otimes n}[\rho] = \sum_{\mathbf{r,s}\in (\ZDZ)^n } \tilde p_\mathbf{r,s}^\mathrm{loc} (X_D^\mathbf{r}Z_D^\mathbf{s}) \rho (X_D^\mathbf{r}Z_D^\mathbf{s}) ^\dagger$,
as this is the probability that the number of qudits on which nontrivial errors occur is exactly $\swt(\mathbf{r,s})$.

Since it will be convenient to express the stabilizer operators in terms of their exponent vectors we introduce the following notation.
As a $\ZDZ$-module, the  stabilizer group of $\ket{\psi}$ is isomorphic to 
\begin{align}
M_{\ket{\psi}}:=\{ (\mathbf{k,l})\in (\ZDZ)^n\times (\ZDZ)^n \ \vert \ \exists S \in \mathcal{S}_{\ket{\psi}}: S \propto X_D^\mathbf{k}Z_D^\mathbf{l} \}.
\end{align}
Note that the corresponding phase of a  stabilizer operator is uniquely determined by $(\mathbf{k,l})\in M_{\ket{\psi}}$, i.e., there is a function $f: M_{\ket{\psi}} \rightarrow  \ZDZ$ such that the isomorphism from $M_{\ket{\psi}} $ to $\mathcal{S}_{\ket{\psi}}$ is given by $(\mathbf{k,l}) \mapsto \omega_D^{f(\mathbf{k,l})}X_D^\mathbf{k}Z_D^\mathbf{l}$. 
The module $M_{\ket{\psi}}$ is the disjoint union of 
\begin{align}
 M_{j} := \{(\mathbf{k,l})\in M_{\ket{\psi}} \ \vert \ \swt(\mathbf{k,l})=j \} 
\end{align}
for $j \in \{0,\ldots,n\}$.
With this notation we can rewrite the projector onto $\ket{\psi}$ as 
\begin{align}
 \Psi = \frac{1}{D^n} \sum_{S\in \mathcal{S}_{\ket{\psi}}} S
 =\frac{1}{D^n} \sum_{j=0}^n \sum_{(\mathbf{k,l})\in M_{j}} \omega_D^{f(\mathbf{k,l})} X_D^\mathbf{k} Z_D^\mathbf{l}
\end{align}
and the application of a Pauli error channel with an error distribution $(p_\mathbf{r,s})_{\mathbf{r,s}\in(\ZDZ)^n}$ as 
\begin{align} 
\rho := \sum_{\mathbf{r,s}\in\ZDZ} p_\mathbf{r,s} (X_D^\mathbf{r}Z_D^\mathbf{s}) \Psi (X_D^\mathbf{r}Z_D^\mathbf{s})^\dagger  
 = \frac{1}{D^n} \sum_{j=0}^n  P_j
\end{align}
with 
\begin{align}
P_j = \sum_{(\mathbf{k,l})\in M_{ j}} \omega_D^{f(\mathbf{k,l})} X_D^\mathbf{k} Z_D^\mathbf{l}  \sum_{\mathbf{r,s}\in(\ZDZ)^n} p_\mathbf{r,s} \omega_D^{\mathbf{k}\cdot \mathbf{s}-\mathbf{r}\cdot\mathbf{l}},
\end{align}
where we have used Eq.~\eqref{eq:pauli_commutation} to cancel $ (X_D^\mathbf{r}Z_D^\mathbf{s})$ with $ (X_D^\mathbf{r}Z_D^\mathbf{s}) ^\dagger$ 
at the expense of the phase factors
$\omega_D^{\mathbf{k}\cdot \mathbf{s}-\mathbf{r}\cdot\mathbf{l}}$.
By Eq.~\eqref{def:sector_length}, the sector lengths of $\rho$ follow as 
\begin{align}\label{eq:noisy_sector_lengths}
 \ell^n_j[\rho] = \norm{P_j}^2 =  \sum_{(\mathbf{k,l})\in M_{ j}} \Bigg \vert  \sum_{\mathbf{r,s}\in(\ZDZ)^n} p_\mathbf{r,s} \omega_D^{\mathbf{k}\cdot \mathbf{s}-\mathbf{r}\cdot\mathbf{l}} \Bigg \vert^2.
\end{align}
We will proceed by simplifying this expression in the case of $p^\mathrm{glob}_{\mathbf{r,s}}$ and $\tilde p^\mathrm{loc}_\mathbf{r,s}$, respectively.

First, consider global white noise. Let $(\mathbf{k,l})\in M_j$ for some $j\ge1$. We are interested in 
\begin{align}\label{eq:glob_scaling_factor}
  \sum_{\mathbf{r,s}\in(\ZDZ)^n} p^\mathrm{glob}_\mathbf{r,s} \omega_D^{\mathbf{k}\cdot \mathbf{s}-\mathbf{r}\cdot\mathbf{l}} 
  = 
  p_0 + p_\mathrm{e} \sum_{\substack{\mathbf{r,s}\in \ZDZ \\ (\mathbf{r,s})\neq(0,0)}} \omega_D^{\mathbf{k}\cdot \mathbf{s}-\mathbf{r}\cdot\mathbf{l}},
\end{align}
where we have used the abbreviation $p_0:=1-p + \frac{p}{D^{2n}}$ and $p_\mathrm{e} := \frac{p}{D^{2n}}$.
By splitting the second sum into three parts (first: $\mathbf{r}=0$, second: $\mathbf{s}=0$, third: $\mathbf{r},\mathbf{s}\neq0$) we obtain
\begin{align}\label{eq:phase_glob_1}
  \sum_{\substack{\mathbf{r,s}\in \ZDZ \\ (\mathbf{r,s})\neq(0,0)}} \omega_D^{\mathbf{k}\cdot \mathbf{s}-\mathbf{r}\cdot\mathbf{l}}
&=
  \sum_{\mathbf{s}\neq0} \omega_D^{  \mathbf{k}\cdot \mathbf{s}}
  +
  \sum_{\mathbf{r}\neq0} \omega_D^{ -\mathbf{r}\cdot \mathbf{l}}
  +
  \sum_{\mathbf{s}\neq0} \omega_D^{  \mathbf{k}\cdot \mathbf{s}}
  \sum_{\mathbf{r}\neq0} \omega_D^{ -\mathbf{r}\cdot \mathbf{l}} 
  \\ \label{eq:phase_glob_2}
  &= (\delta_{\mathbf{k},0}D^n-1) + (\delta_{\mathbf{l},0}D^n-1) +(\delta_{\mathbf{k},0}D^n-1)(\delta_{\mathbf{l},0}D^n-1) 
  \\ \label{eq:phase_glob_3}
  &=\delta_{\mathbf{k},0}\delta_{\mathbf{l},0}D^{2n} -1 =\delta_{j,0}D^{2n} -1 = -1.
\end{align}
From line~\eqref{eq:phase_glob_1} to \eqref{eq:phase_glob_2}, we have used the fact that (nontrivial) complex roots of unity sum up to zero in the form of  $\sum_{\mathbf{s}\in (\ZDZ)^n} \omega_D^{\mathbf{k}\cdot\mathbf{s}}= \delta_{\mathbf{k},0}D^n$, and likewise for $-\mathbf{r}$ and $ \mathbf{l}$.
Since this result does not depend on $(\mathbf{k,l})\in M_j$, Eqs.~\eqref{eq:noisy_sector_lengths} and \eqref{eq:glob_scaling_factor} simplify to 
\begin{align}
 \ell^n_j[\rhoGlobPsiP] =   \vert p_0 -p_\mathrm{e}\vert ^2 \sum_{(\mathbf{k,l})\in M_{ j}} 1 = (1-p)^2 \ell^n_j[\Psi],
\end{align}
as claimed in Eq.~\eqref{eq:sectorlength_glob}.

Finally, consider local white noise. Let $(\mathbf{k,l})\in M_n$. We are interested in 
\begin{align}\label{eq:loc_scaling_factor}
  \sum_{\mathbf{r,s}\in(\ZDZ)^n} \tilde p^\mathrm{loc}_\mathbf{r,s} \omega_D^{\mathbf{k}\cdot \mathbf{s}-\mathbf{r}\cdot\mathbf{l}} 
  = \sum_{m=0}^n \tilde p_0^{n-m} \tilde p_\mathrm{e}^m 
  \sum_{\substack{\mathbf{r,s}\in \ZDZ \\ \swt(\mathbf{r,s})=m}} \omega_D^{\mathbf{k}\cdot \mathbf{s}-\mathbf{r}\cdot\mathbf{l}},
\end{align}
where we have again used an abbreviation $\tilde p_0:=1-\tilde p + \frac{\tilde p}{D^{2}}$ and $\tilde p_\mathrm{e} := \frac{\tilde p}{D^{2}}$.
For each term in the second sum, we can permute the nonzero pairs of $(r_i,s_i)$ to the first $m$ positions.
As there are $\nchoosem$ choices for this, the sum can be rewritten as 

\begin{align}
  \sum_{\substack{\mathbf{r,s}\in \ZDZ \\ \swt(\mathbf{r,s})=m}} \omega_D^{\mathbf{k}\cdot \mathbf{s}-\mathbf{r}\cdot\mathbf{l}}
  &=
  \nchoosem \sum_{(r'_1,s'_1)\neq (0,0)} \omega_D^{k_1's_1'-r_1'l_1'} \ldots \sum_{(r'_m,s'_m)\neq (0,0)} \omega_D^{k_m's_m' - r_m'l_m'}  
  \\ 
  &= \nchoosem(-1)^m,
\end{align}
where we have used $\sum_{(r'_i,s'_i)\neq (0,0)} \omega_D^{k_i's_i'-r_i'l_i'} = \delta_{k_i',0}\delta_{l_i',0} D^2-1 =-1$. 
Again, this result does not depend on $(\mathbf{k,l})\in M_n$. 
Using the binomial theorem, Eqs.~\eqref{eq:noisy_sector_lengths} and \eqref{eq:loc_scaling_factor} simplify to 
\begin{align}
 \ell^n_n[\rhoLocPsiP] =   \vert \tilde p_0 -\tilde p_\mathrm{e}\vert ^{2n} \sum_{(\mathbf{k,l})\in M_{n}} 1 = (1-\tilde p)^{2n} \ell^n_n[\Psi],
\end{align}
as claimed in Eq.~\eqref{eq:sectorlength_loc}.
\end{proof}

From this proposition, we are now able to express critical noise thresholds in terms of sector lengths as follows.

\corollary\label{cor:crit_noise_sector_lengths}
\emph{Let $\ket{\psi}$ be an $n$-qudit stabilizer state and $\Psi=\ket{\psi}\bra{\psi}$.
Furthermore, let $b_j^{(n_1,\ldots,n_k)}$ be the bound on sector lengths of a specific partition $(n_1,\ldots,n_k)$ of $n$ for  $j\in\{1,\ldots,n\}$. 
If $p<p_{\mathrm{glob},j}^{\mathrm{Sec}{(n_1,\ldots,n_k)}}[\Psi]$ (or $\tilde p< p_{\mathrm{loc}}^{\mathrm{Sec}{(n_1,\ldots,n_k)}}[\Psi]$), where}
\begin{align}\label{eq:sector_threshold_glob}
 p_{\mathrm{glob},j}^{\mathrm{Sec}{(n_1,\ldots,n_k)}}[\Psi] &= 1- \sqrt    {b_j^{(n_1,\ldots,n_k)}/\ell^n_j\left[\Psi \right]} 
\hspace{2em} \text{ and }   
\\
\label{eq:sector_threshold_loc}
p_{\mathrm{loc}}^{\mathrm{Sec}{(n_1,\ldots,n_k)}}[\Psi]& = 1- \sqrt[2n]{b_n^{(n_1,\ldots,n_k)}/\ell^n_n\left[\Psi \right]},
\end{align}
\emph{then $\ell^n_j \left[ \rhoGlobPsiP  \right] > b_j^{(n_1,\ldots,n_k)}$ 
 (or $\ell^n_n \left[\rho_{\mathrm{loc},\psi}(\tilde p) \right] > b_n^{(n_1,\ldots,n_k)}$). That is,  $\rhoGlobPsiP$ (or $\rho_{\mathrm{loc},\psi}(\tilde p) $) is not $(n_1,\ldots, n_k)$-separable.}
\begin{proof} 
Equating $b_j^{(n_1,\ldots,n_k)}$ to Eq.~\eqref{eq:sectorlength_glob}, and $b_n^{(n_1,\ldots,n_k)}$ to Eq.~\eqref{eq:sectorlength_loc}, respectively, and solving for $p$ yields the corresponding noise thresholds. The solutions given in Eqs.~\eqref{eq:sector_threshold_glob} and \eqref{eq:sector_threshold_loc} are the only physically relevant solution (i.e., between 0 and 1) as $ \ell^n_j \left[\rhoGlobPsiP \right]$ and $ \ell^n_n \left[\rhoLocPsiP \right]$ are strictly monotonically decreasing in $p$.
\end{proof}

If we combine this result with Theorem~\ref{thrm:PPT_graph}, it becomes clear that the global-white-noise thresholds based on sector lengths are not tight: It will turn out that best sector-length entanglement criterion is obtained for $j=n$ which is still outperformed by the Peres-Horodecki noise threshold $p_\mathrm{glob}^\mathrm{PPT}(D,n) ={1-\frac{1}{D^{n-1}+1}}$. Indeed, since each sector length is trivially bounded by $D^n-1$ [recall Eq.~\eqref{eq:sectorlength_bound}], no sector-length threshold can exceed $1-\sqrt{{(D-1)^n}/{(D^n-1)}}$.
To see that the Peres-Horodecki criterion is superior, it suffices to verify the inequality
\begin{align}  1-\sqrt{\frac{(D-1)^n}{D^n-1}} 
< {1-\frac{1}{D^{n-1}+1}} 
\end{align}
which is equivalent to 
\begin{align}
(D-1)^n(D^{n-1}+1)^2 > D^n-1 .
\end{align}
This, in turn, holds true because the relations $(D-1)^n\ge1$ (for  $D\ge2$) and $D^{2n-2}\ge D^n$ (for $D,n\ge 2$) imply the chain of inequalities
\begin{align}
(D-1)^n(D^{n-1}+1)^2 \ge  D^{2n-2} +2D^{n-1} + 1 > D^{2n-2} - 1  \ge D^n-1.
\end{align}

Nevertheless, the sector-length criterion is still better than the entropy criterion. For example, the noise thresholds for the Werner state defined in Eq.~\eqref{eq:Werner_state} are given by $p^\mathrm{Entr}_\mathrm{glob}\approx 0.2524$, $p^\mathrm{Sec}_{\mathrm{glob},2}[\Phi^+] =1-1/\sqrt{3}\approx 0.4226$ and $ p_\mathrm{glob}^\mathrm{PPT}(2,2)=2/3 \approx 0.6667$.
Furthermore, sector-length entanglement criteria are not limited to ruling out full separability; in contrast to the Peres-Horodecki criterion as in Theorem~\ref{thrm:PPT_graph}.

\subsection{Explicit derivation of sector lengths}
\label{sec:5.4}

In order to make use of the noise thresholds established in Corollary~\ref{cor:crit_noise_sector_lengths}, 
we have to know two things: Sector lengths $\ell^n_j[\Psi]$ of the pure stabilizer state $\Psi=\ket{\psi}\bra{\psi}$ and corresponding bounds $b_j^{(n_1,\ldots,n_k)}$ which are exceeded by $\ell^n_j[\Psi]$. While we have already explained in Sec.~\ref{sec:5.2} how $b_j^{(n_1,\ldots,n_k)}$ can be derived, explicit derivations of sector lengths are still missing. 
To fill this gap, we will now establish sector lengths of important families of qudit stabilizer states by combinatorial investigations; recall from Eq.~\eqref{eq:sec_len_counting}, 
\begin{align}\tag{\ref{eq:sec_len_counting}}
 \ell^n_j[\Psi] 
 =  \left \vert  \{S\in\mathcal{S}_{\ket{\psi}} \ \vert \ S \text{ acts on }j\text{ qudits} \} \right  \vert, 
\end{align}  
that the sector lengths of a stabilizer state are given by the number of stabilizer operators with a fixed number $j$ of support qudits.
Using Eq.~\eqref{eq:sec_len_counting}, we can show that the
sector lengths of the Greenberger-Horne-Zeilinger state as defined in Eq.~\eqref{def:GHZnD} are as follows:

\proposition\label{prop:GHZ_sector_lengths}~\cite{EltSiew19}
\emph{The sector length distribution of the GHZ state is given by } 
\begin{align} \label{eq:GHZ_sectorlength}
 \ell^n_j \left[  \ket{ \mathrm{GHZ}^n_D }\bra{\mathrm{GHZ}^n_D } \right] &= \delta_{j,n}(D-1)D^{n-1} + 
\genfrac{(}{)}{0pt}{0}{n}{j} \frac{(D-1)^j+(-1)^j(D-1)}{D}.
\end{align}

We defer the lengthy proof to Appendix~\ref{app:proof_sec_of_GHZ}.  
Here, we just calculate the sector lengths of the tetrapartite AME state $\ket{\Gamma^4_D}$ from Eq.~\eqref{eq:AME4D}.

\proposition \label{prop:AME4D_sector_lengths}
\emph{Let $D$ be odd. The sector length distribution of the tetrapartite AME state $\Psi^4_D=\ket{\Gamma^4_D}\bra{\Gamma^4_D}$ is given by
  $\ell^4_0[\Psi^4_D ]=1$, $\ell^4_1[\Psi^4_D ]=\ell^4_2[\Psi^4_D ]=0$, $\ell^4_3[\Psi^4_D ]=4(D^2-1)$, and $\ell^4_4[ \Psi^4_D]= (D^2-1)(D^2-3)$.}
\begin{proof}
Recall from Sec.~\ref{sec:3.4} that the stabilizer group of $\ket{\Gamma^4_D}$ is isomorphic to the submodule $M \subset (\ZDZ)^4 \times (\ZDZ)^4$ which is generated by the four exponent vectors in Eq.~\eqref{eq:AME4D_exponent_vectors}. By forming all possible linear combinations, we obtain the parametrization 
\begin{align}
  M = \left\{  \textbf{(}(a,b,c,d)  ,(b-d,a+c,b+d,-a+c)\textbf{)} \  \big \vert \ a,b,c,d\in \ZDZ
  \right\}.
\end{align}
As we are interested in the number of stabilizers of $\ket{\Gamma^4_D}$ acting on exactly $j$ qudits, we decompose $M$ into the disjoint union of $M_j:= \{\mathbf{(r,s)}\in M\ \vert \ \swt(\mathbf{r,s})=j \}$ 
 for $j\in\{0,1,2,3,4\}$. The sector lengths follow as $\ell^4_j[\Psi^4_D] = \left \vert  M_j \right \vert$. 
We will now consecutively count the vectors in $M_j$. For this we will use the more convenient notation 
\begin{align} \label{eq:notation_AME4D_exponent}
 \left(\begin{array}{c|c|c|c}
a &b&c&d  \\
b-d & a+c & b+d& -a+c
 \end{array} \right) 
 :=\textbf{(}(a,b,c,d)  ,(b-d,a+c,b+d,-a+c)\textbf{)}
\end{align} 
from which the symplectic weight can be easily seen by the number of nonzero columns.

As always, there is exactly one vector in $M_0$, namely the one with $a=b=c=d=0$. Hence, $\ell^4_0[\Psi^4_D] =0$.
For $j=1$, exactly one column in Eq.~\eqref{eq:notation_AME4D_exponent} must be nonzero; w.l.o.g. $b=c=d=0$ to achieve this in the upper part.
The resulting vector is of the form 
\begin{align} 
 \left(\begin{array}{c|c|c|c}
a &0&0&0  \\
0 & a & 0 & -a
 \end{array} \right) .
\end{align} 
If $a=0$ and $a\neq0$, it has symplectic weight $0$ and $3$, respectively. Thus, $M_1=\emptyset$, i.e., $\ell^4_1[\Psi^4_D ]=0$.

Similarly, for $j=2$, exactly two columns must be nonzero. 
Because of the upper row in Eq.~\eqref{eq:notation_AME4D_exponent}, this implies that at least two of the four number $a,b,c$ and $d$ are equal to zero. Although there are $\genfrac{(}{)}{0pt}{1}{4}{2}=6$ possible cases, it suffices to consider the two of them: $c,d=0$ ($b,c=0$, $a,d=0$ and $a,b=0$ are analogous) and $b,d=0$ ($a,c=0$ is analogous). The first (second) case is that where the stabilizer generators of two (non-)neighboring  vertices of $\Gamma^4_D$ are ``turned off''. The corresponding vectors are of the form
\begin{align} \label{eq:AME4D_case2}
 \left(\begin{array}{c|c|c|c}
a &b&0&0  \\
b & a & b & a
 \end{array} \right) 
 \hspace{2em}
 \text{and}
 \hspace{2em}
  \left(\begin{array}{c|c|c|c}
a &0&c&0  \\
0 & a+c & 0 & -a+c
 \end{array} \right).
\end{align} 
There is no choice for $a$ and $b$ such that the left vector in Eq.~\eqref{eq:AME4D_case2} has symplectic weight $2$, i.e., no such vector lies in $M_2$.
Similarly, if the right vector in  Eq.~\eqref{eq:AME4D_case2} would have symplectic weight $2$, $a$ and $c$ must fulfill $a+c=-a+c=0$, thus $2c=0$. But since we have assumed that $D$ is odd, no $c\in \ZDZ$ fulfills $2c=0$; i.e., no vector with $a,c\neq0$ lies in $M_2$.
Therefore, also $M_2=\emptyset$, i.e., $\ell^4_2[\Psi^4_D ]=0$. 

For $j=3$, exactly three of the of the columns must be nonzero. Thus, at least one of the four numbers  $a,b,c$ and $d$ is equal to zero. 
Consider for example $d=0$. In this case the rightmost column in Eq.~\eqref{eq:notation_AME4D_exponent} is supposed to be zero, i.e., $-a+c=0$.  Thus, $a=c$ and the corresponding vector is of the form 
\begin{align}  
 \left(\begin{array}{c|c|c|c}
a &b& a&0  \\
b & 2a & b& 0
 \end{array} \right)  .
\end{align} 
Since $2a\neq 0$ is equivalent to $a\neq 0$ ($D$ is odd), each combinations of  $a,b\in \ZDZ$ for which not both $a$ and $b$ are zero yields a vector in $M_3$, i.e., there are $D^2-1$ vectors in $M_3$ with $d=0$.
Analogously, the three other cases ($c=0$, $b=0$, and $a=0$) yield $D^2-1$ vectors each.
Therefore, $\ell^4_3[\Psi^4_D ]= 4(D^2-1)$.

From normalization, the full-body sector length follows as $\ell^4_4[\Psi^4_D ]=D^4- 4(D^2-1)-1$.
The polynomial is equal to $(D^2-1)(D^2-3)$ as one can easily verify by checking that both polynomials share the same roots $\pm1$ and $\pm\sqrt{3}$. This finishes the proof.
\end{proof}

\begin{figure}
\centering
\begin{minipage}{0.85\textwidth}
\begin{framed} 
\centering
\begin{minipage}{.99\textwidth}
    \includegraphics[height = 14.5em]{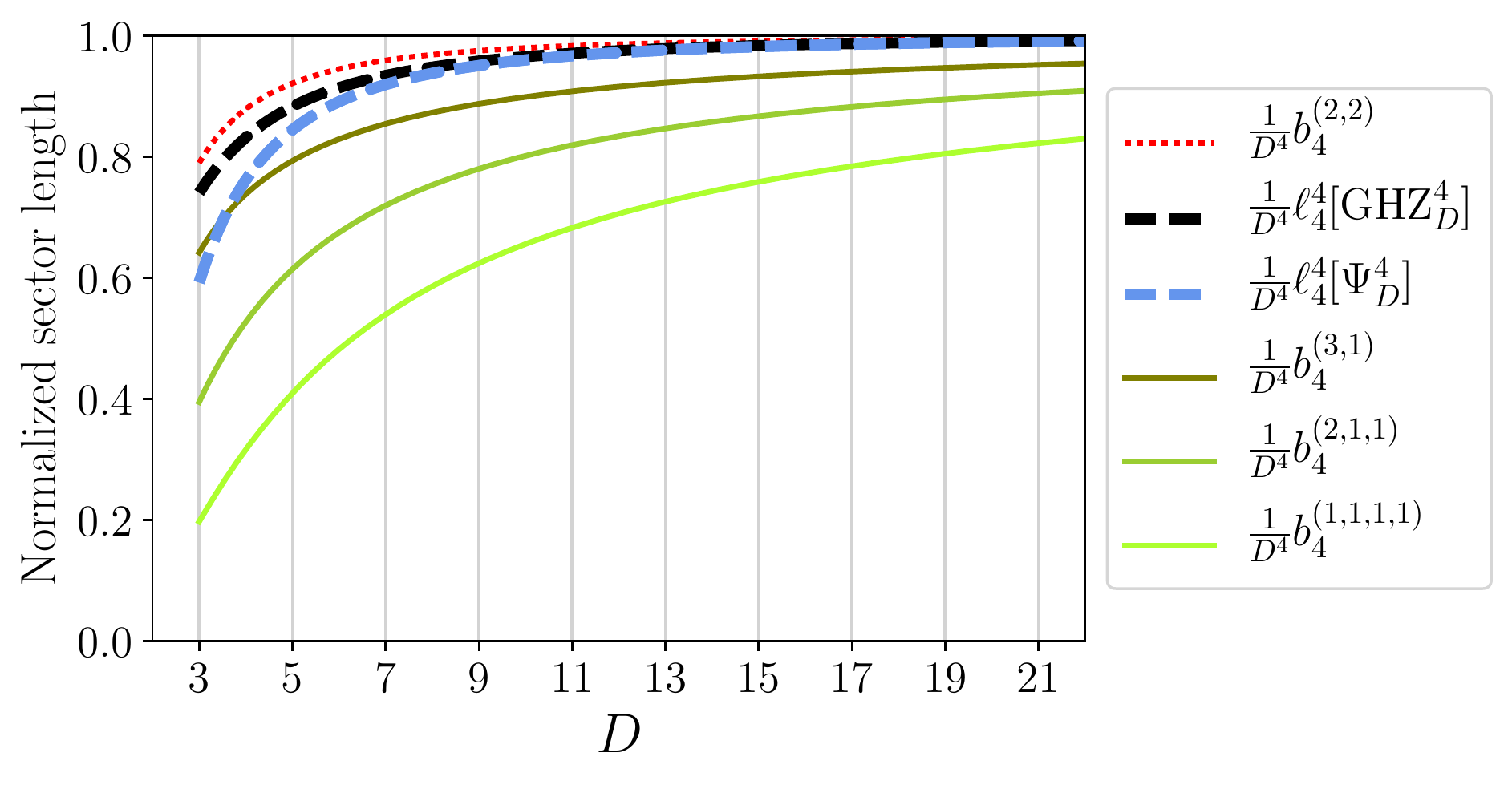} 
\end{minipage}
 
\caption{Comparison of the full-body sector lengths of tetrapartite GHZ (black) and AME (blue) states to the corresponding separability bounds as a function of the qudit dimension. 
To provide a better overview, we depict all these quantities with a normalization factor of $1/D^4$.  
}
 \label{fig:sector_bounds}
\end{framed}
\end{minipage}
 \end{figure}

Since the sector length distributions of the tetrapartite AME and GHZ state are dominated by their corresponding full-body sector length,
$\ell^4_4[\Psi_D^4] = D^4 -4D^2 +3$ and $\ell^4_4[\mathrm{GHZ}_D^4] = D^4 -4D^2 +3(2D-1)$, respectively,
the best noise thresholds are obtained by choosing $j=4$ in Corollary~\ref{cor:crit_noise_sector_lengths}.
The corresponding separability bounds are $b_4^{(1,1,1,1)}$, $b_4^{(2,1,1)}$, $b_4^{(3,1)}$, and $b_4^{(2,2)}$ as defined in Eq.~\eqref{eq:b44}. 
In Fig.~\ref{fig:sector_bounds},
we display these sector lengths (dashed lines) and separability bounds (dotted and solid lines) with an overall scaling factor of $1/D^4$ as a function of $D$.
Vertical lines indicate odd qudit dimensions.
Every case where a sector length exceeds a bound yields a nontrivial noise threshold.  
Because of $\ell^4_4[\mathrm{GHZ}_D^4] > \ell^4_4[\Psi_D^4]$, these thresholds will be higher for GHZ states.
The green lines $b_4^{(1,1,1,1)}$, $b_4^{(2,1,1)}$, and $b_4^{(3,1)}$ are surpassed by the dashed lines $\ell^4_4[\Psi_D^4]  $ and $\ell^4_4[\mathrm{GHZ}_D^4]$. In particular, this allows us to rule out semiseparability (except for the qutrit AME state, where  $\ell^4_4[\Psi_3^4]<  b_4^{(3,1)} $).
However, the red line $b_4^{(2,2)}$ is strictly larger than the dashed lines.
This is because for four qudits, the state with the largest full-body sector length is a  tensor product of two Bell states, $\ell^4_4[\Phi_D\otimes\Phi_D]=(D^2-1)^2= b_4^{(2,2)}$~\cite{EltSiew19}.
Because of this limitation, sector length criteria are not strong enough to establish a noise threshold to establish genuine multipartite entanglement.

\subsection{Noise robustness of small quantum networks for qudits}
\label{sec:5.5}

In this section, we compare the noise thresholds based on sector lengths with other noise thresholds from the literature which we briefly review in Secs.~\ref{sec:5.5.1} and \ref{sec:5.5.2}.
In Sec.~\ref{sec:5.5.3}, we carry out the comparison.

\subsubsection{Witnessing genuine multipartite entanglement}
\label{sec:5.5.1}

For every pure GME state $\ket{\psi}$ 
the Hermitian operator
\begin{align} \label{eq:generic_witness}
 W := \alpha \mathbbm{1} - \ket{\psi}\bra{\psi}
\end{align}
is a GME witness~\cite{BEKGWGHBLS04}, where 
\begin{align}
 \alpha = \max_{\ket{\phi} \in B} \vert \braket{\phi|\psi}\vert ^2,
\end{align}
denotes the maximal overlap of $\ket{\psi}$ with the set of biseparable states $B$.
Equating 
\begin{align}
 \Tr \left[W \left( (1-p)\ket{\psi}\bra{\psi} + p \frac{\mathbbm{1}}{D^n} \right) \right] =  p\left(1-\frac{1}{D^n}\right) - (1-\alpha) 
\end{align}
and zero yields the critical noise threshold
\begin{align} \label{eq:threshold_WitGME}
 p^\mathrm{Wit:GME}_\mathrm{glob} = 
 \frac{D^n}{D^n-1} (1-\alpha).
\end{align}
To make use of this noise threshold, one has to know $\alpha$. For our purposes, the following result will suffice:

\lemma\label{lem:maximal_overlap}
\emph{Let $\ket{\psi}$ be an AME state. The maximal overlap of $\ket{\psi}$ with the biseparable states is given by $\alpha=1/D$.}
\begin{proof}
 First note that $\alpha$ can be rewritten as the maximal squared Schmidt coefficient over all nontrivial bipartitions~\cite{GueTo09}. 
 For every bipartition, the Schmidt decomposition of $\ket{\psi}$ is given by 
\begin{align}
 \ket{\psi} = \sum_{i=1}^{D^k} \lambda_i \ket{v_i}_J \otimes \ket{w_i}_{J^\mathrm{C}},
\end{align}
 where the subset $J\subset I=\{1,\ldots,n\}$ corresponds to the bipartition and $\{\ket{v_i}\}$ and  $\{\ket{w_i}\}$ are sets of orthonormal vectors in the Hilbert space of the parties in $J$ and $J^\mathrm{C}$, respectively. 
 We assume w.l.o.g. $\left\vert J \right \vert > \left \vert J^\mathrm{C} \right \vert=:k$.
 The real numbers $\lambda_i$ are referred to as Schmidt coefficients.
  Since $\ket{\psi}$ is AME, tracing out the larger set of parties $J$ yields a completely mixed state,
 \begin{align}
  \frac{\mathbbm{1}}{D^k} &= \Tr_J\left[ \ket{\psi}\bra{\psi}  \right]
  = \sum_{j=1}^{D^{n-k}} \bra{a_j}_J  \left (  \sum_{i=1}^{D^k} \lambda_i^2 \ket{v_i}\bra{v_i}_J \otimes \ket{w_i} \bra{w_i}_{J^\mathrm{C}} \right ) \ket{a_j}_J  \\
  &= \sum_{i=1}^{D^k} \lambda_i^2 \ket{w_i} \bra{w_i}_{J^\mathrm{C}} \sum_{j=1}^{D^{n-k}} \big\vert  \braket{a_j|v_i} \big \vert ^2
  = \sum_{i=1}^{D^k} \lambda_i^2 \ket{w_i} \bra{w_i}_{J^\mathrm{C}}  
  ,
 \end{align}
where $\{\ket{a_j}_J\}$  is a basis for the Hilbert space associated to $J$.
Since $\{\ket{w_i}\}$ is an orthonormal set, we obtain $\lambda_i=\frac{1}{\sqrt{D^k}}$ for all $i$. Maximizing over $1\le k\le n/2$ (the size of the bipartition) yields $\alpha =\lambda_\mathrm{max}^2 = 1/D$. This finishes the proof.
\end{proof}

\subsubsection{Noise thresholds for Greenberger-Horne-Zeilinger states}
\label{sec:5.5.2}

Because of the simple form of $\ket{\mathrm{GHZ}^n_D}$, there exists a plethora of noise thresholds for this state in the literature. 
The first one we review here is due to Huber \emph{et al.}~\cite{HMGH10}.
There, the critical global white noise threshold below which one can rule out a certain separability type is given by
\begin{align} \label{eq:Huber_threshold}
 p^\mathrm{Huber}_\mathrm{glob} = 1- \frac{\gamma}{\gamma+D^{n-1}}
\end{align}
where $\gamma$ is the corresponding number of possible partitions, see Ref.~\cite{GHH10} for more details. 
We will only use Eq.~\eqref{eq:Huber_threshold} with $\gamma=1$ for full separability, $\gamma=n$ for semiseparability, and $\gamma=2^{n-1}-1$ for biseparability~\cite{PrivateHuber}.
Note that Eq.~\eqref{eq:Huber_threshold} with $\gamma =1$ coincides with the Peres-Horodecki (or reduction) noise threshold for qudit graph states which we have derived in Sec.~\ref{sec:4}.

There is also a GME-criterion based on positive maps~\cite{CHLM17} which leads to the global white noise threshold
\begin{align}
 p^\mathrm{PosMap:GME}_\mathrm{glob} = 1- \frac{1+(D-2)(2^{n-1}-1)}{1+(D-2)(2^{n-1}-1)+(D-2)D^{n-1}}.
\end{align}

In the case of local white noise, the only noise threshold we found in the literature is based on the Peres-Horodecki criterion~\cite{LiuFan09},
\begin{align}\label{eq:threshold_GHZ_local_PPT}
p_\mathrm{loc}^\mathrm{PPT} = 1-\frac{\sqrt[n]{4}+\sqrt[n]{2}\sqrt{4+\sqrt[n]{2D}}}{\sqrt[n]{4}+\sqrt[n]{2}\sqrt{4+\sqrt[n]{2D}} + 2D}.
\end{align}
It can only be used to rule out full separability.


\subsubsection{Comparison of noise thresholds for tetrapartite qudit states}
\label{sec:5.5.3}

\begin{figure}
\centering
\begin{minipage}{0.85\textwidth}
\begin{framed} 
\centering
\begin{minipage}{\textwidth}
   \includegraphics[height = 15em]{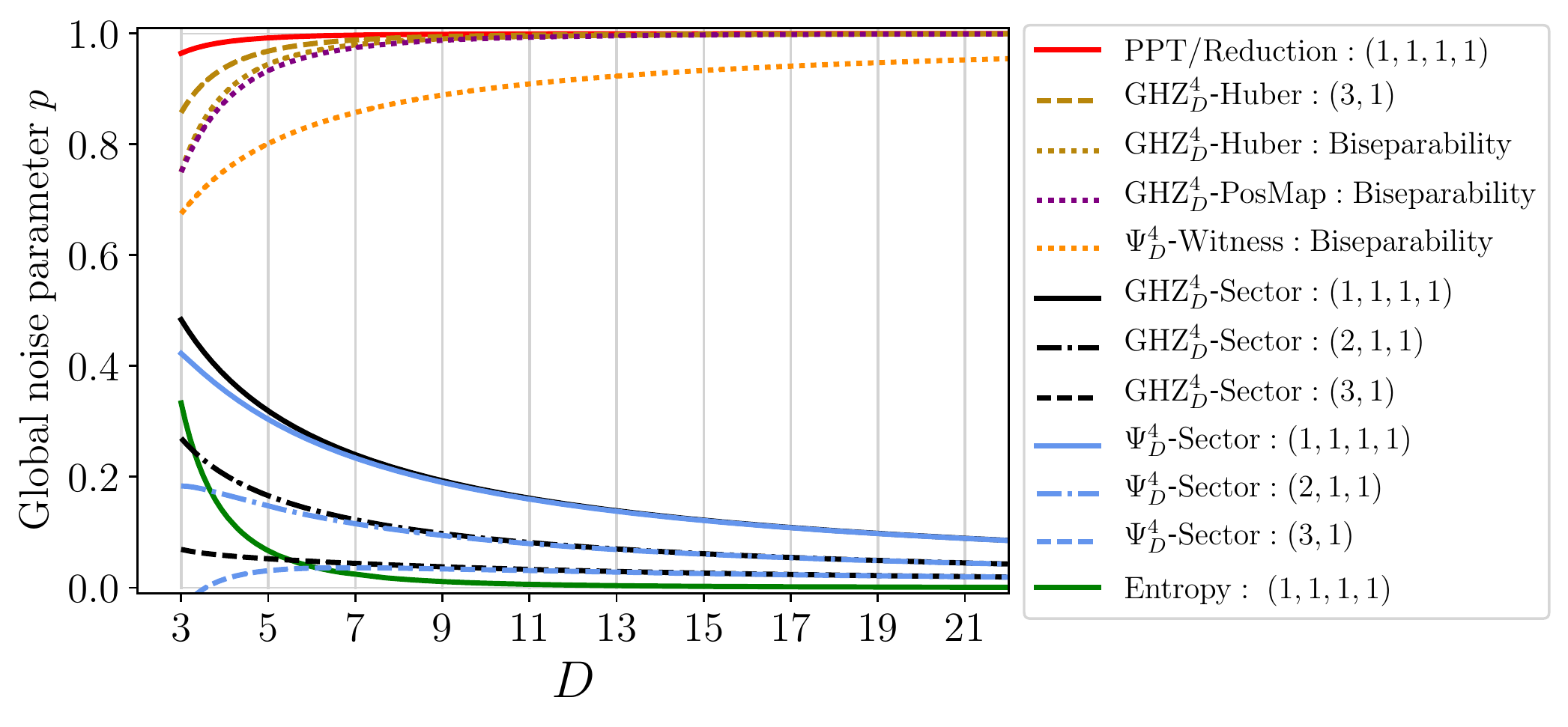} \\
   \includegraphics[height = 15em]{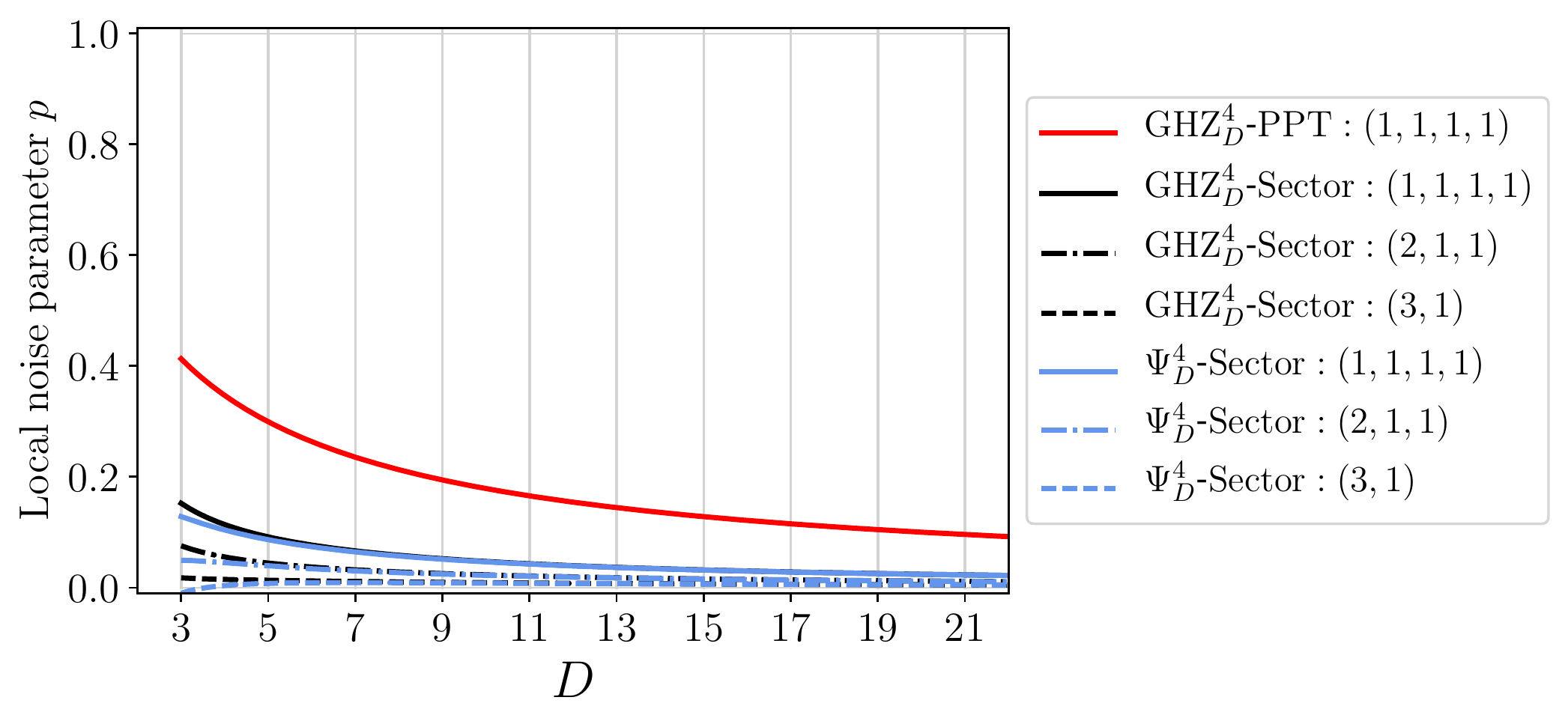} 
\end{minipage}
 
\caption{Noise thresholds for global (top) and local (bottom) white noise on tetrapartite GHZ and AME states as a function of the qudit dimension. For noise parameters below a given curve, the corresponding separability type is ruled out. For example, below a solid (dotted) line, the corresponding state is entangled (GME).}
 \label{fig:tetrapartite_bounds}
\end{framed}
\end{minipage}
 \end{figure}

In Fig.~\ref{fig:tetrapartite_bounds},
we have plotted all previously-discussed noise thresholds for the  tetrapartite GHZ state and the AME state $\Psi^4_D= \ket{\Gamma_D^4}\bra{\Gamma_D^4}$ as defined in Eqs.~\eqref{def:GHZnD} and~\eqref{eq:AME4D} as a function of $D$. 
If the noise parameter $p$ is below a solid, dashed and dotted curve, one can rule out full separability, semiseparability and biseparability, respectively. 

First, consider the upper part of the figure which depicts global white noise thresholds. 
One can see that the PPT (and reduction) noise threshold from Sec.~\ref{sec:4.2} (and Sec.~\ref{sec:4.3}) as well as the literature thresholds from Secs.~\ref{sec:5.5.1} and~\ref{sec:5.5.2} are good in the sense that they converge to one in the limit of large $D$. That is, these criteria can certify genuine multipartite entanglement for even  almost completely mixed states in the limit of large dimensions.
Our thresholds based on sector lengths, on the other hand, converge to zero. Only the entropy criterion from Sec.~\ref{sec:4.1} is worse than the sector-length thresholds.
The reason why the latter converge to zero is because, both the full-body sector lengths $\ell^4_4[\mathrm{GHZ}^4_D] $ and $\ell^4_4[\Psi^4_D]$ as well as the bounds $b_4^{(1,1,1,1)}$, $b_4^{(2,1,1)}$, and $b_4^{(3,1)}$ scale as $D^4$. To yield good noise thresholds, however, they would need to be large compared to these bounds.
Also note that the sector-length noise thresholds are slightly better for the GHZ than for the AME state because of $\ell^4_4[\mathrm{GHZ}^4_D]>\ell^4_4[\Psi^4_D]$. For $D\ge7$, however, this effect is barely visible anymore. 

Now, consider  the lower part of Fig.~\ref{fig:tetrapartite_bounds} which shows local white noise thresholds. 
In this case, the only threshold we found in the literature is based on the Peres-Horodecki criterion, recall Eq.~\eqref{eq:threshold_GHZ_local_PPT}.
Although also this curve  converges to zero here, it is still better than the corresponding sector-length threshold for the GHZ state. Also note that here the sector length thresholds are smaller than in the upper plot. This is always the case because one has to take a larger root in Eq.~\eqref{eq:sector_threshold_loc} than in Eq.~\eqref{eq:sector_threshold_glob}.
An intuitive argument for why global white noise thresholds are generally higher than their local counterparts is the following.
Regarded as generalized Pauli error channels, all (nontrivial) errors occur with equal probability for global noise while large-weight Pauli errors are suppressed for local white noise, recall the proof of Proposition~\ref{prop:noisy_sector_lengths}.
The errors which are stabilizers of a given state do not deteriorate it. 
As we have seen at the example of GHZ states (Prop.~\ref{prop:GHZ_sector_lengths}) and the tetrapartite AME state (Prop.~\ref{prop:AME4D_sector_lengths}), most stabilizer operators have a large symplectic weight.
That is, for local noise it is less likely that stabilizer errors occur.

In conclusion, the methods developed in Sec.~\ref{sec:5} yield noise thresholds for tetrapartite GHZ and AME states.
Since for the case of global white noise GME-thresholds are known, they only provide new insights in the case of local white noise: For both GHZ and AME states, we obtain for the first time thresholds to rule out semiseparability; for the AME state also the full-separability threshold is the only threshold we are aware of.
Since our new noise thresholds based on sector lengths scale very bad in the dimension of the qudits, we have devoted the final section of this thesis for a detailed investigation of the qubit case.

\section{\label{sec:6}\protect Small quantum quantum networks for qubits}
 
In the special case of qubits, graph states have received the most attention and are thus best understood.
In particular, a meaningful classification of qubit graph states has been carried out which we review in Sec.~\ref{sec:6.1}.
In Sec.~\ref{sec:6.2}, we introduce a graph-theoretical puzzle whose solution yields the full-body sector length and solve it for four important families of graph states.
In Sec.~\ref{sec:6.3} we numerically investigate noise thresholds established in the previous sections and compare them to other thresholds from the literature.


\subsection{Classification of few-qubit graph states}
\label{sec:6.1}
  
In order to keep the number of inequivalent classes of graph states small and as lucid as possible, one puts states with the same entanglement properties into a single class:
In the distant laboratory paradigm, the parties can apply unitary single qubit operations without changing the entanglement properties of their state. Thus, graph states which are local-unitary (LU) equivalent or the same up to relabeling of the qubits are grouped together.
Since the graph state of a graph with two disconnected components is the same as the tensor product of two graph states corresponding to these components, it suffices to establish a classification for connected graphs.

In Sec.~\ref{sec:6.1.1}, we present the technique of local complementation which is used to illustrate the change of graphs under the action of certain local unitaries. 
In Sec.~\ref{sec:6.1.2}, we introduce four families of $n$-qubit graph states which are sufficient to classify graph states up to $n=5$ qubits.
In Sec.~\ref{sec:6.1.3}, we present the classification for $n\le 8$ and in Sec.~\ref{sec:6.1.4}, we comment on which sector lengths are possible for graph states.

\subsubsection{Local complementation} \label{sec:6.1.1}

Here, we review an important class of local operations which transform certain LU-equivalent graph states into one another~\cite{HeinEisBri04, VdNDDM04}.
Let $k\in V$ be a vertex of a graph with adjacency matrix $\Gamma\in \FF_2^{n\times n}$ and consider the unitary operator
\begin{align} \label{def:complementation_gate}
 U_k := \sqrt{-iX^{(k)}} \prod_{j=1}^n \left (\sqrt{i Z^{(j)}} \right )^{\gamma_{j,k}},
\end{align}
where 
\begin{align}
 \sqrt{iX} =  (\sqrt{-iX})^\dagger = \frac{1}{\sqrt{2}}\left(\begin{array}{cc} 1 & -i \\ -i & 1 \end{array}\right)
 \hspace{1em} \text{and}
 \hspace{1.2em}
 \sqrt{iZ} = \left(\begin{array}{cc}  \e^{i\pi/4} & 0 \\ 0 & \e^{-i\pi/4} \end{array}\right)
\end{align}
are single-qubit Clifford gates~\cite{AndBri06}.
Applying this operation to $\ket{\Gamma}$ yields again a graph state,
\begin{align}
 U_k \ket{\Gamma} = \ket{L_k(\Gamma)},
\end{align}
where the entries of the new adjacency matrix are given by $L_k(\Gamma)_{i,j} = \gamma_{i,j} + \gamma_{i,k}\gamma_{j,k}$. 
Graphically, the new graph arises via \emph{local complementation about $k$} from the original one. In graph-theoretical terms, this means that the subgraph induced by the neighborhood of $k$ is inverted.
In Fig.~\ref{fig:local_complementation},
\begin{figure}
\centering
\begin{minipage}{0.85\textwidth}
\begin{framed} 
\centering  
\begin{tabular}{cccc} 
 \begin{tikzpicture}   
\draw[-, line width=.1em] (0.2,0  ) -- (0.8,0  );       
\draw[-, line width=.1em] (0.2,1  ) -- (0.8,1  );       
\draw[-, line width=.1em] (0,  0.2) -- (0,  0.8);       
\draw[-, line width=.1em] (1,  0.2) -- (1,  0.8);        
\fill[red,opacity=0.4] (0,1) circle (0.2); 
\draw[line width=.1em] (0,0) circle (0.2);   
\draw[line width=.1em] (1,0) circle (0.2);    
\draw[line width=.1em] (0,1) circle (0.2);   
\draw[line width=.1em] (1,1) circle (0.2);     
 \draw (0,1) node[text=black]{\scriptsize 1};  
 \draw (0,0) node[text=black]{\scriptsize 2};  
 \draw (1,0) node[text=black]{\scriptsize 3};  
 \draw (1,1) node[text=black]{\scriptsize 4};   
\draw (2,0.5) node[text= red]{$\overset{U_1}{\longrightarrow}$};
 \end{tikzpicture}    
 &
 \begin{tikzpicture}    
\draw[-, line width=.1em] (0.2,0  ) -- (0.8,0  );       
\draw[-, line width=.1em] (0.2,1  ) -- (0.8,1  );       
\draw[-, line width=.1em] (0,  0.2) -- (0,  0.8);       
\draw[-, line width=.1em] (1,  0.2) -- (1,  0.8);       
\draw[-, line width=.1em] (0.1414,0.1414) -- (0.8586,0.8586);      
\fill[green,opacity=0.4] (0,0) circle (0.2); 
\draw[line width=.1em] (0,0) circle (0.2);   
\draw[line width=.1em] (1,0) circle (0.2);    
\draw[line width=.1em] (0,1) circle (0.2);   
\draw[line width=.1em] (1,1) circle (0.2);     
 \draw (0,1) node[text=black]{\scriptsize 1};  
 \draw (0,0) node[text=black]{\scriptsize 2};  
 \draw (1,0) node[text=black]{\scriptsize 3};  
 \draw (1,1) node[text=black]{\scriptsize 4};   
\draw (2,0.5) node[text= green]{$\overset{U_2}{\longrightarrow}$};
 \end{tikzpicture}    
 &
 \begin{tikzpicture}    
\draw[-, line width=.1em] (0.2,0  ) -- (0.8,0  );        
\draw[-, line width=.1em] (0,  0.2) -- (0,  0.8);        
\draw[-, line width=.1em] (0.1414,0.1414) -- (0.8586,0.8586);       
\draw[-, line width=.1em] (0.1414,0.8586) -- (0.8586,0.1414);      
\fill[blue,opacity=0.4] (1,0) circle (0.2); 
\draw[line width=.1em] (0,0) circle (0.2);   
\draw[line width=.1em] (1,0) circle (0.2);    
\draw[line width=.1em] (0,1) circle (0.2);   
\draw[line width=.1em] (1,1) circle (0.2);     
 \draw (0,1) node[text=black]{\scriptsize 1};  
 \draw (0,0) node[text=black]{\scriptsize 2};  
 \draw (1,0) node[text=black]{\scriptsize 3};  
 \draw (1,1) node[text=black]{\scriptsize 4};   
\draw (2,0.5) node[text= blue]{$\overset{U_3}{\longrightarrow}$};
 \end{tikzpicture}    
 &
 \begin{tikzpicture}   
 \draw (0,1) node[text=black]{\scriptsize 1};  
 \draw (0,0) node[text=black]{\scriptsize 2};  
 \draw (1,0) node[text=black]{\scriptsize 3};  
 \draw (1,1) node[text=black]{\scriptsize 4};  
\draw[-, line width=.1em] (0.2,0  ) -- (0.8,0  );         
\draw[-, line width=.1em] (0.1414,0.1414) -- (0.8586,0.8586);       
\draw[-, line width=.1em] (0.1414,0.8586) -- (0.8586,0.1414);             
\draw[line width=.1em] (0,0) circle (0.2);   
\draw[line width=.1em] (1,0) circle (0.2);    
\draw[line width=.1em] (0,1) circle (0.2);   
\draw[line width=.1em] (1,1) circle (0.2);    
 \end{tikzpicture}   
 \end{tabular}
 
\caption{Exemplary sequence of local complementations on a tetrapartite graph. At every step, local Clifford gates according to Eq.~\eqref{def:complementation_gate} are applied to the current graph state $\ket{\Gamma}$. In particular, all depicted graphs correspond to LU-equivalent graph states.}
 \label{fig:local_complementation}
\end{framed}
\end{minipage}
 \end{figure} 
we depict a series of local complementations which shows that the graph states 
$(\mathrm{C}_{1,2}Z)(\mathrm{C}_{2,3}Z)(\mathrm{C}_{3,4}Z)(\mathrm{C}_{4,1}Z)\ket{+}^{\otimes 4}$ and 
$(\mathrm{C}_{1,3}Z)(\mathrm{C}_{3,2}Z)(\mathrm{C}_{2,4}Z)\ket{+}^{\otimes 4}$ are local-Clifford (LC) equivalent.
It was shown that for each pair of LC-equivalent graph states, there exists a sequence of local complementations which transforms them into each other~\cite{VdNDDM04}.
For up to eight qubits, one can prove that every pair of LU-equivalent graph states is already LC-equivalent~\cite{HeinEisBri04, CaLTMoPo09, CaLTMoPo09B}.
This motivated to conjecture that this is also true for arbitrary graph states~\cite{LU_LC_conjecture, GroVdN08}.
However, explicit counterexamples with $n=27$ and $n=28$ qubits have been constructed which prove that this so-called \emph{LU-LC conjecture} is wrong~\cite{JCWY10, TsiGueh17}.
In particular, this implies that there exist LU-equivalent graph states for which the corresponding graphs differ by more than a sequence of local complementations.

\subsubsection{Star, dandelion, line, and ring graphs} \label{sec:6.1.2}

Consider the graphs in Fig.~\ref{fig:graph_star}.
\begin{figure}
\centering
\begin{minipage}{0.85\textwidth}
\begin{framed} 
\centering  
\begin{tabular}{ccccc} 
\begin{tikzpicture} 
\draw[white, line width=.1em] (0,1) circle (0.2);   
\draw[-, line width=.1em](0,-0.2) -- (0,-0.8);    
\draw[line width=.1em] (0,0) circle (0.2);   
\draw[line width=.1em] (0,-1) circle (0.2);  

\draw (0,0) node[text=black]{\scriptsize 1}; 
\draw (0,-1) node[text=black]{\scriptsize 2}; 
 \end{tikzpicture}  
 \hspace{2em}
 &
\begin{tikzpicture} 
\draw[-, line width=.1em](0,0.2) -- (0,0.8); 
\draw[-, line width=.1em] (0,-0.2) -- (0,-0.8);    
\draw[line width=.1em] (0,0) circle (0.2);   
\draw[line width=.1em] (0,-1) circle (0.2);  
\draw[line width=.1em] (0,1) circle (0.2);  

\draw (0,0) node[text=black]{\scriptsize 1}; 
\draw (0,-1) node[text=black]{\scriptsize 2}; 
\draw (0,1) node[text=black]{\scriptsize 3}; 
 \end{tikzpicture}  
 \hspace{2em}
 &
 \begin{tikzpicture} 
 \draw[white, line width=.1em] (0,1) circle (0.2);   
\draw[-, line width=.1em] (0,-0.2) -- (0,-0.8);  
\draw[-, line width=.1em] (0.1732,0.1) -- (0.6928, 0.4);  
\draw[-, line width=.1em] (-0.1732,0.1) -- (-0.6928, 0.4);   
\draw[line width=.1em] (0,0) circle (0.2);  
\draw[line width=.1em] (0,-1) circle (0.2);  
\draw[line width=.1em] (0.866,0.5) circle (0.2);     
\draw[line width=.1em] (-0.866,0.5) circle (0.2);     

\draw (0,0) node[text=black]{\scriptsize 1}; 
\draw (0,-1) node[text=black]{\scriptsize 2}; 
\draw (0.866,.5) node[text=black]{\scriptsize 3}; 
\draw (-0.866,.5) node[text=black]{\scriptsize 4}; 
 \end{tikzpicture}  
 \hspace{1em}
 &
\begin{tikzpicture} 
\draw[-, line width=.1em] (0.2,0) -- (0.8,0); 
\draw[-, line width=.1em] (-0.2,0) -- (-0.8,0);  
\draw[-, line width=.1em](0,0.2) -- (0,0.8); 
\draw[-, line width=.1em] (0,-0.2) -- (0,-0.8);    
\draw[line width=.1em] (0,0) circle (0.2);  
\draw[line width=.1em] (1,0) circle (0.2);  
\draw[line width=.1em] (-1,0) circle (0.2);     
\draw[line width=.1em] (0,-1) circle (0.2);  
\draw[line width=.1em] (0,1) circle (0.2);   

\draw (0,0) node[text=black]{\scriptsize 1}; 
\draw (0,-1) node[text=black]{\scriptsize 2};
\draw (1,0) node[text=black]{\scriptsize 3};
\draw (0,1) node[text=black]{\scriptsize 4};
\draw (-1,0) node[text=black]{\scriptsize 5};
\end{tikzpicture}  
 \hspace{1em}
 &
\begin{tikzpicture}
\draw[white, line width=.1em] (0,1) circle (0.2);  
\draw[-, line width=.1em] (0,-0.2) -- (0,-0.8);   
\draw[-, line width=.1em] (0.1902,-0.0618) -- (0.7608,-0.2472);   
\draw[-, line width=.1em] (-0.1902,-0.0618) -- (-0.7608,-0.2472);   
\draw[-, line width=.1em] ( 0.1176,0.1618) -- ( 0.4702, 0.6472);   
\draw[-, line width=.1em] (-0.1176,0.1618) -- (-0.4702, 0.6472);   
 
\draw[line width=.1em] (0,0) circle (0.2);  
\draw[line width=.1em] (0,-1) circle (0.2);  
\draw[line width=.1em] ( 0.9511,-0.3090) circle (0.2);      
\draw[line width=.1em] (-0.9511,-0.3090) circle (0.2);    
\draw[line width=.1em] ( 0.5878,0.8090) circle (0.2);      
\draw[line width=.1em] (-0.5878,0.8090) circle (0.2);      

\draw (0,0) node[text=black]{\scriptsize 1}; 
\draw (0,-1) node[text=black]{\scriptsize 2}; 
\draw ( 0.9511,-0.3090) node[text=black]{\scriptsize 3}; 
\draw ( 0.5878,0.8090) node[text=black]{\scriptsize 4}; 
\draw (-0.5878,0.8090) node[text=black]{\scriptsize 5}; 
\draw (-0.9511,-0.3090) node[text=black]{\scriptsize 6}; 
 \end{tikzpicture}  
 \end{tabular}
 
\caption{Star graphs with two, three, four, five, and six vertices. We call vertex $1$ \emph{central vertex}. The other vertices are leaves.}
 \label{fig:graph_star}
\end{framed}
\end{minipage}
 \end{figure}
 They are referred to as \emph{star graphs} because they have one distinguished vertex  which is connected via an edge to all of the other vertices which have no direct connections among each other, i.e., they are are leaves.
 If one applies a Hadamard gate to all leaf qubits of a star-graph state $\ket{\Gamma^n_\mathrm{star}}$, one obtains the $n$-qubit GHZ state $\ket{\mathrm{GHZ}^n_2}$. 
Via local complementation about the central vertex, one can show that the GHZ state is also LU-equivalent to a graph state with a fully connected graph~\cite{HeinEisBri04, VdNDDM04}.  

A slight variation of the star-graph  is obtained when one of the leaves gets connected to two additional vertices, in which case we call them \emph{dandelion graphs} 
, see Fig.~\ref{fig:graph_dandelion}.
\begin{figure}
\centering
\begin{minipage}{0.85\textwidth}
\begin{framed} 
\centering  
\begin{tabular}{cccc}
\begin{tikzpicture}
 
 
\draw[-, line width=.1em](0,0.2) -- (0,0.8);

\draw[line width=.1em] (0,0) circle (0.2);  
\draw[line width=.1em] (0,1) circle (0.2);  

\draw[-, line width=.1em] (0,-0.2) -- (0,-1.8);   
\draw[line width=.1em] (0,-2) circle (0.2);  
\draw[-, line width=.1em] (0.1732,-2.1) -- (0.6928, -2.4);  
\draw[-, line width=.1em] (-0.1732,-2.1) -- (-0.6928, -2.4);   
\draw[line width=.1em] ( 0.866,-2.5) circle (0.2);     
\draw[line width=.1em] (-0.866,-2.5) circle (0.2);   

\draw (0,-2) node[text=black]{\scriptsize 1}; 
\draw (-0.866,-2.5) node[text=black]{\scriptsize 2}; 
\draw ( 0.866,-2.5) node[text=black]{\scriptsize 3}; 
\draw (0,0) node[text=black]{\scriptsize 4}; 
\draw (0,1) node[text=black]{\scriptsize 5}; 
 \end{tikzpicture}  
 \hspace{2em}
 &
 \begin{tikzpicture} 

 \draw[white, line width=.1em] (0,1) circle (0.2);  
 
\draw[-, line width=.1em] (0.1732,0.1) -- (0.6928, 0.4);  
\draw[-, line width=.1em] (-0.1732,0.1) -- (-0.6928, 0.4);  
 
\draw[line width=.1em] (0,0) circle (0.2);   
\draw[line width=.1em] (0.866,0.5) circle (0.2);     
\draw[line width=.1em] (-0.866,0.5) circle (0.2);

\draw[-, line width=.1em] (0,-0.2) -- (0,-1.8);   
\draw[line width=.1em] (0,-2) circle (0.2);  
\draw[-, line width=.1em] (0.1732,-2.1) -- (0.6928, -2.4);  
\draw[-, line width=.1em] (-0.1732,-2.1) -- (-0.6928, -2.4);   
\draw[line width=.1em] (0.866,-2.5) circle (0.2);     
\draw[line width=.1em] (-0.866,-2.5) circle (0.2);  

\draw (0,-2) node[text=black]{\scriptsize 1}; 
\draw (-0.866,-2.5) node[text=black]{\scriptsize 2}; 
\draw ( 0.866,-2.5) node[text=black]{\scriptsize 3}; 
\draw (0,0) node[text=black]{\scriptsize 4}; 
\draw ( 0.866,.5) node[text=black]{\scriptsize 5}; 
\draw (-0.866,.5) node[text=black]{\scriptsize 6};
 \end{tikzpicture}  
 \hspace{1em}
 &
\begin{tikzpicture}

\draw[-, line width=.1em] (0.2,0) -- (0.8,0); 
\draw[-, line width=.1em] (-0.2,0) -- (-0.8,0);  
\draw[-, line width=.1em](0,0.2) -- (0,0.8);

\draw[line width=.1em] (0,0) circle (0.2);  
\draw[line width=.1em] (1,0) circle (0.2);  
\draw[line width=.1em] (-1,0) circle (0.2);      
\draw[line width=.1em] (0,1) circle (0.2);  

\draw[-, line width=.1em] (0,-0.2) -- (0,-1.8);   
\draw[line width=.1em] (0,-2) circle (0.2);  
\draw[-, line width=.1em] (0.1732,-2.1) -- (0.6928, -2.4);  
\draw[-, line width=.1em] (-0.1732,-2.1) -- (-0.6928, -2.4);   
\draw[line width=.1em] (0.866,-2.5) circle (0.2);     
\draw[line width=.1em] (-0.866,-2.5) circle (0.2);  

\draw (0,-2) node[text=black]{\scriptsize 1}; 
\draw (-0.866,-2.5) node[text=black]{\scriptsize 2}; 
\draw ( 0.866,-2.5) node[text=black]{\scriptsize 3}; 
\draw (0,0) node[text=black]{\scriptsize 4};  
\draw (1,0) node[text=black]{\scriptsize 5};  
\draw (0,1) node[text=black]{\scriptsize 6};  
\draw (-1,0) node[text=black]{\scriptsize 7}; 
 \end{tikzpicture}  
 \hspace{1em}
 &
\begin{tikzpicture}  \draw[white, line width=.1em] (0,1) circle (0.2);   
 

\draw[-, line width=.1em] (0.1902,-0.0618) -- (0.7608,-0.2472);   
\draw[-, line width=.1em] (-0.1902,-0.0618) -- (-0.7608,-0.2472);   
\draw[-, line width=.1em] ( 0.1176,0.1618) -- ( 0.4702, 0.6472);   
\draw[-, line width=.1em] (-0.1176,0.1618) -- (-0.4702, 0.6472);    
\draw[line width=.1em] (0,0) circle (0.2);   
\draw[line width=.1em] ( 0.9511,-0.3090) circle (0.2);      
\draw[line width=.1em] (-0.9511,-0.3090) circle (0.2);    
\draw[line width=.1em] ( 0.5878,0.8090) circle (0.2);      
\draw[line width=.1em] (-0.5878,0.8090) circle (0.2);    
\draw[-, line width=.1em] (0,-0.2) -- (0,-1.8);   
\draw[line width=.1em] (0,-2) circle (0.2);  
\draw[-, line width=.1em] (0.1732,-2.1) -- (0.6928, -2.4);  
\draw[-, line width=.1em] (-0.1732,-2.1) -- (-0.6928, -2.4);   
\draw[line width=.1em] (0.866,-2.5) circle (0.2);     
\draw[line width=.1em] (-0.866,-2.5) circle (0.2);  
 
 \draw (0,-2) node[text=black]{\scriptsize 1}; 
 \draw (-0.866,-2.5) node[text=black]{\scriptsize 2}; 
 \draw ( 0.866,-2.5) node[text=black]{\scriptsize 3}; 
 \draw (0,0) node[text=black]{\scriptsize 4}; 
 \draw ( 0.9511,-0.3090) node[text=black]{\scriptsize 5};
 \draw ( 0.5878, 0.8090) node[text=black]{\scriptsize 6};
 \draw (-0.5878, 0.8090) node[text=black]{\scriptsize 7};
 \draw (-0.9511,-0.3090) node[text=black]{\scriptsize 8};
  \end{tikzpicture}  
 \end{tabular}
 
\caption{Dandelion graphs with five, six, seven, and eight vertices. We call vertex 1 \emph{lower central vertex}, vertices 2 and 3  \emph{root vertices}, vertex 4 \emph{upper central vertex}, and all other vertices \emph{seed vertices} as they resemble the seeds of a dandelion seed head.}
 \label{fig:graph_dandelion}
\end{framed}
\end{minipage}
 \end{figure}
 One can define dandelion graphs with $n\ge 5$ vertices. 
The third family of graph states we consider here corresponds to \emph{line graphs}, see Fig.~\ref{fig:graph_line}. 
\begin{figure}
\centering
\begin{minipage}{0.85\textwidth}
\begin{framed} 
\centering  
\begin{tabular}{cccc}
 \begin{tikzpicture}   
\draw[-, line width=.1em] (0.2,0) -- (0.8,0);       
\draw[line width=.1em] (0,0) circle (0.2);   
\draw[line width=.1em] (1,0) circle (0.2);    

\draw ( 0, 0) node[text=black]{\scriptsize 1};
\draw ( 1, 0) node[text=black]{\scriptsize 2};
 \end{tikzpicture}  
 \hspace{2em}
 &
\begin{tikzpicture}  
\draw[-, line width=.1em] (0.2,0) -- (0.8,0);      
\draw[-, line width=.1em] (1.2,0) -- (1.8,0);      
\draw[line width=.1em] (0,0) circle (0.2);   
\draw[line width=.1em] (1,0) circle (0.2);    
\draw[line width=.1em] (2,0) circle (0.2);    

\draw ( 0, 0) node[text=black]{\scriptsize 1};
\draw ( 1, 0) node[text=black]{\scriptsize 2};
\draw ( 2, 0) node[text=black]{\scriptsize 3};
 \end{tikzpicture}  
 \hspace{2em}
 &
\begin{tikzpicture}  
\draw[-, line width=.1em] (0.2,0) -- (0.8,0);      
\draw[-, line width=.1em] (1.2,0) -- (1.8,0);  
\draw[-, line width=.1em] (2.2,0) -- (2.8,0);             
\draw[line width=.1em] (0,0) circle (0.2);   
\draw[line width=.1em] (1,0) circle (0.2);    
\draw[line width=.1em] (2,0) circle (0.2);    
\draw[line width=.1em] (3,0) circle (0.2);    

\draw ( 0, 0) node[text=black]{\scriptsize 1};
\draw ( 1, 0) node[text=black]{\scriptsize 2};
\draw ( 2, 0) node[text=black]{\scriptsize 3};
\draw ( 3, 0) node[text=black]{\scriptsize 4};
 \end{tikzpicture}  
 \end{tabular}
 
\caption{Line graphs with two, three,  and four vertices.}
 \label{fig:graph_line}
\end{framed}
\end{minipage}
 \end{figure}
 If a controlled-$Z$ gate is applied to the two leaf-qubits of a line graph state $\ket{\Gamma^n_\mathrm{line}}$, one obtains a \emph{ring graph} state $\ket{\Gamma^n_\mathrm{ring}}$ as displayed in Fig.~\ref{fig:graph_ring}.
\begin{figure}
\centering
\begin{minipage}{0.85\textwidth}
\begin{framed} 
\centering  
\begin{tabular}{cccc}
 \begin{tikzpicture} 
\draw[white, line width=.1em] (0,-1) circle (0.2);  \draw[white, line width=.1em] (0,1) circle (0.2);  
 
\draw[-, line width=.1em] (0.8268,0.1) -- (-0.3268,0.766 );     
\draw[-, line width=.1em] (0.8268,-0.1) -- (-0.3268,-0.766 );     
\draw[-, line width=.1em] (-0.5,0.666) -- (-0.5,-0.666);     
\draw[line width=.1em] (1,0) circle (0.2);  
\draw[line width=.1em] (-0.5, 0.866) circle (0.2);     
\draw[line width=.1em] (-0.5,-0.866) circle (0.2);    

\draw ( 1, 0) node[text=black]{\scriptsize 1};
\draw (-0.5, 0.866) node[text=black]{\scriptsize 2};
\draw (-0.5,-0.866) node[text=black]{\scriptsize 3};
\end{tikzpicture}  
 \hspace{1em}
 &
\begin{tikzpicture} 
\draw[-, line width=.1em] (0.8586, 0.1414) -- (0.1414,0.8586);  
\draw[-, line width=.1em] (0.8586, -0.1414) -- (0.1414,-0.8586);  
\draw[-, line width=.1em] (-0.8586, 0.1414) -- (-0.1414,0.8586);  
\draw[-, line width=.1em] (-0.8586, -0.1414) -- (-0.1414,-0.8586);    
\draw[line width=.1em] (1,0) circle (0.2);  
\draw[line width=.1em] (-1,0) circle (0.2);     
\draw[line width=.1em] (0,-1) circle (0.2);  
\draw[line width=.1em] (0,1) circle (0.2);  

\draw ( 1, 0) node[text=black]{\scriptsize 1};
\draw ( 0, 1) node[text=black]{\scriptsize 2};
\draw (-1, 0) node[text=black]{\scriptsize 3};
\draw ( 0,-1) node[text=black]{\scriptsize 4};
 \end{tikzpicture}  
 \hspace{1em}
 &
\begin{tikzpicture}
\draw[white, line width=.1em] (0,-1) circle (0.2);  \draw[white, line width=.1em] (0,1) circle (0.2);  
\draw[-, line width=.1em] (0.8824, 0.1618) --  (0.4266, 0.7893); 
\draw[-, line width=.1em] (0.8824,-0.1618) --  (0.4266,-0.7893);  
\draw[-, line width=.1em] (0.1188, 0.8893) -- (-0.6188, 0.6496);  
\draw[-, line width=.1em] (0.1188,-0.8893) -- (-0.6188,-0.6496);  
\draw[-, line width=.1em] (-0.8090, 0.3878)-- (-0.8090,-0.3878); 
\draw[line width=.1em] (1,0) circle (0.2);   
\draw[line width=.1em] (0.3090, 0.9511) circle (0.2);      
\draw[line width=.1em] (0.3090, -0.9511) circle (0.2);    
\draw[line width=.1em] (-0.8090, 0.5878) circle (0.2);      
\draw[line width=.1em] (-0.8090, -0.5878) circle (0.2);       

\draw ( 1, 0) node[text=black]{\scriptsize 1};
\draw ( 0.3090, 0.9511) node[text=black]{\scriptsize 2};
\draw (-0.8090, 0.5878) node[text=black]{\scriptsize 3};
\draw (-0.8090,-0.5878) node[text=black]{\scriptsize 4};
\draw ( 0.3090,-0.9511) node[text=black]{\scriptsize 5};
 \end{tikzpicture}  
 \hspace{1em}
 &
\begin{tikzpicture}
\draw[white, line width=.1em] (0,-1) circle (0.2);  \draw[white, line width=.1em] (0,1) circle (0.2);  
\draw[-, line width=.1em] (-0.3,0.866) -- (0.3,0.866);  
\draw[-, line width=.1em] (-0.3,-0.866) -- (0.3,-0.866);  
\draw[-, line width=.1em] (0.9, 0.1732) -- (0.6,0.6928);  
\draw[-, line width=.1em] (0.9, -0.1732) -- (0.6,-0.6928);  
\draw[-, line width=.1em] (-0.9, 0.1732)  -- (-0.6,0.6928);  
\draw[-, line width=.1em] (-0.9, -0.1732) --(-0.6,-0.6928);  
 
\draw[line width=.1em] (1,0) circle (0.2);    
\draw[line width=.1em] (-1,0) circle (0.2);     
\draw[line width=.1em] (-0.5, 0.866) circle (0.2);     
\draw[line width=.1em] (-0.5,-0.866) circle (0.2);    
\draw[line width=.1em] (0.5, 0.866) circle (0.2);     
\draw[line width=.1em] (0.5,-0.866) circle (0.2);   

\draw ( 1  , 0    ) node[text=black]{\scriptsize 1};
\draw ( 0.5, 0.866) node[text=black]{\scriptsize 2};
\draw (-0.5, 0.866) node[text=black]{\scriptsize 3};
\draw (-1  , 0    ) node[text=black]{\scriptsize 4};
\draw (-0.5,-0.866) node[text=black]{\scriptsize 5};
\draw ( 0.5,-0.866) node[text=black]{\scriptsize 6};
 \end{tikzpicture}  
 \end{tabular}
 
\caption{Ring graphs with  three, four, five, and six vertices.}
 \label{fig:graph_ring}
\end{framed}
\end{minipage}
 \end{figure}
 One can define ring graphs with $n\ge 3$ vertices.

\subsubsection{Classification of graph states on up to eight qubits} \label{sec:6.1.3}

Here, we present the classification of graph states for an increasing number of qubits $n$ up to $n=8$.
For a single qubit, the only graph state is $\ket{+}$ which corresponds to the trivial graph with a single vertex. 
For $n=2$, there is also only one connected graph which can be regarded either as star graph or line graph and it corresponds to the graph state
\begin{align}
 \ket{\Gamma^2_\mathrm{star}}=
 \ket{\Gamma^2_\mathrm{line}} = \frac{1}{2}\left(\begin{array}{r}1\\1\\1\\-1 \end{array} \right)
\end{align}
which is LU-equivalent to the Bell states defined in Eq.~\eqref{def:Bell_states}.
In Ref.~\cite{HeinEisBri04}, the classification starts at $n=2$ and $\ket{\Gamma^2_\mathrm{star}}$ has been labeled graph state No.~1.

For three qubits, the still-coinciding star and line graph has been labeled graph state No.~2.
However, there is also the ring graph which looks different at a first glance. But for three parties, the ring graph is fully connected, i.e., local complementation of a star (or line) graph about the central vertex shows that the states
$ \ket{\Gamma^3_\mathrm{star}}=  \ket{\Gamma^3_\mathrm{line}} $ and $\ket{\Gamma^3_\mathrm{ring}}$ are LU-equivalent.  
Since these are all connected graphs with three vertices, there is also only one equivalence class of graph states on $n=3$ qubits.

For $n=4$ qubits, it turns out that there are exactly two equivalence classes which can be represented by the star graph (No.~3) and the line graph (No.~4). The sequence of local complementations in Fig.~\ref{fig:local_complementation} shows that $\ket{\Gamma^4_\mathrm{ring}}$ also belongs to class No.~4.

For $n=5$ qubits, there are exactly four equivalence classes with representatives $\ket{\Gamma^5_\mathrm{star}}$ (No.~5),  $\ket{\Gamma^5_\mathrm{dandelion}}$ (No.~6),  $\ket{\Gamma^5_\mathrm{line}}$ (No.~7), and  $\ket{\Gamma^5_\mathrm{ring}}$ (No.~8). 
That is, the four families  introduced in Sec.~\ref{sec:6.1.2} are sufficient for the classification on up to five qubits.

For $n=6$ qubits, there are already eleven inequivalent classes of connected graph states, 
%
%
see Fig.~4  in Ref.~\cite{HeinEisBri04} 
for their graphical representation. For even more qubits, the number of inequivalent classes is growing very fast, see Table~\ref{tab:number_of_graphs}
 \begin{table}
\centering
\begin{minipage}{0.85\textwidth}
\begin{framed} 
\centering
\begin{tabular}{c|ccccccc} \hline\hline
$n$      & 2 & 3 & 4 & 5 & 6  & 7  & 8    \\ \hline 
number of graphs       & 1 & 1 & 2 & 4 & 11 & 26 & 101 \\ \hline\hline
\end{tabular}
\caption{Number of connected graphs which correspond to inequivalent $n$-qubit graph states. More information about this sequence is available in \emph{The On-Line Encyclopedia of Integer Sequences~\textsuperscript{\textregistered}}~\href{https://oeis.org/A090899}{\texttt{[OEIS:A090899]}}. }
 \label{tab:number_of_graphs}
\end{framed}
\end{minipage}
 \end{table}
  for the first few values.
  %
  %
  There are exactly 146 inequivalent classes of connected graph states on up to eight qubits, see Figs.~4 and 5 in Ref.~\cite{HeinEisBri04} and Fig.~2 in Ref.~\cite{CaLTMoPo09} for the corresponding graphs.

\subsubsection{Sector length distributions of graph states} \label{sec:6.1.4}

With the classification of qubit graph states at hand, we can answer the question which sector length distributions can occur for few-qubit graph states.
Note that these are independent of the representative of the equivalence class as sector lengths are LU-invariant. Also note that the sector length distribution of a graph state with a disconnected graph can be obtained my means of Eq.~\eqref{eq:sec_tensor}. 
Here, we make use of Eq.~\eqref{eq:sec_len_counting} which, in the case of graph states, simplifies to
\begin{align} \label{eq:sec_len_qubits}
\ell ^n_j [\Psi_\Gamma^n] = \left \vert \left\{  \mathbf{r} \in \FF_2^n \ \big \vert \ \swt(\mathbf{r},\Gamma \mathbf{r})=j \right\} \right \vert .
\end{align}
We have written a C++ routine which, for a given input graph $\Gamma$, computes $\ell ^n_j [\Psi_\Gamma^n]$ by counting the number of bitstrings $\mathbf{r}\in \FF_2^n$ for which $\swt(\mathbf{r}, \Gamma \mathbf{r})$. 
It is computationally feasible to run the routine for $n\le 28$ qubits. 
Table~\ref{tab:sector_upto6} 
 \begin{table}
\centering
\begin{minipage}{0.85\textwidth}
\begin{framed} 
\centering
   \begin{tabular}{|l|ccccccc|l|}
   \hline
  Name  & $\ell^n_0$& $\ell^n_1$ & $\ell^n_2$ & $\ell^n_3$ & $\ell^n_4$ & $\ell^n_5$& $\ell^n_6$ & Family
    \\
   \hline \hline
No. 1     &  1 &  0 &  3 &&&&                  	&star/line \\ \hline
No. 2     &  1 &  0 &  3 &  4 &&& 		&star/line\\ \hline
No. 3     &  1 &  0 &  6 &  0 &  9 && 		&star\\
No. 4     &  1 &  0 &  2 &  8 &  5 && 		&line/ring\\ \hline
No. 5     &  1 &  0 &  10 &  0 &  5 &  16&  	&star\\
No. 6     &  1 &  0 &  4 &  6 &  11 &  10&  	&dandelion\\
No. 7     &  1 &  0 &  2 &  8 &  13 &  8 & 	&line\\
No. 8     &  1 &  0 &  0 &  10 &  15 &  6&  	&ring\\ \hline
No. 9     &  1 &  0 &  15 &  0 &  15 &  0 &  33 &star\\
No. 10    &  1 &  0 &  7 &  8 &  7 &  24 &  17  &\\
No. 11    &  1 &  0 &  6 &  0 &  33 &  0 &  24  & dandelion\\
No. 12    &  1 &  0 &  4 &  8 &  13 &  24 &  14 &\\
No. 13    &  1 &  0 &  3 &  8 &  15 &  24 &  13 &\\
No. 14    &  1 &  0 &  2 &  8 &  17 &  24 &  12 & line\\
No. 15    &  1 &  0 &  3 &  8 &  15 &  24 &  13 &\\
No. 16    &  1 &  0 &  3 &  0 &  39 &  0 &  21  &\\
No. 17    &  1 &  0 &  1 &  8 &  19 &  24 &  11 &\\
No. 18    &  1 &  0 &  0 &  8 &  21 &  24 &  10 & ring\\
No. 19    &  1 &  0 &  0 &  0 &  45 &  0 &  18 	&\\
   \hline
   \end{tabular} 
  \caption{Reproduced from Ref.~\cite{CaLTMoPo09B}. Sector length distributions for graph states up to six qubits.   
  The graphs for which no family is mentioned do not belong to one of the four families of Sec.~\ref{sec:6.1.2}.
  See Tables~{\ref{tab:sector_7}--\ref{tab:sector_8c}} in Appendix~\ref{app:QubitTables} for the sector length distributions for graph states on seven and eight qubits.}
 \label{tab:sector_upto6}
\end{framed}
\end{minipage}
 \end{table}
 shows the sector length distributions obtained in this way for all graph states on up to six qubits.
 Our method is consistent with previous results~\cite{CaLTMoPo09B}.
 For seven and eight qubits, we provide similar tables in Appendix~\ref{app:QubitTables}.
 Note that it can happen that two LU-inequivalent states have the same sector length distribution, e.g., state No.~13 and No.~15.

From the tables, one finds, in consistency with the recently proven fact that the GHZ state has the largest full-body sector length~\cite{TDLP16, EltSiew19}, that the star graph state has the largest full-body sector length among the graph states on $n\le 8$ qubits.
It is given by
\begin{align} \label{eq:lnn_star}
 \ell^n_n[\Psi^n_\mathrm{star}] = \begin{cases} 2^{n-1} & \text{ if $n$ is odd} \\ 2^{n-1}+1 & \text{ if $n$ is even .} \end{cases}
\end{align}
We can also see that for $5\le n \le 8$, the second largest full-body sector length is obtained by graph state No.~11, No.~22, and No.~48, all of which belong to the family of dandelion graph states.
In fact, among all possible generalizations of these three graphs, we focus on dandelion graphs because we have numerical evidence that they have the second largest full-body sector length; even for $n>8$.
In Sec.~\ref{sec:6.2.2}, we will show that it is given by 
\begin{align} \label{eq:lnn_dandelion}
 \ell^n_n[\Psi^n_\mathrm{dandelion}] = \begin{cases} 5\times2^{n-4} & \text{ if $n$ is odd} \\ 5\times2^{n-4}+ 4 & \text{ if $n$ is even .} \end{cases}
\end{align} 
Note that line graph states have a rather small full-body sector length.
However, we find that for $4\le n \le 8$ ring graph states have the smallest full-body sector length, 
which, for general $n$, is given by:
 \begin{align} \label{eq:lnn_ring}
  \ell^n_n[\Psi^n_\mathrm{ring}] &= 1 + \sum_{k=1}^{\lfloor n/3 \rfloor} 
{\genfrac{(}{)}{0pt}{0}{n-2k-1}{k-1}} \frac{n}{k} 
 \end{align}
as we will show in Sec.~\ref{sec:6.2.4}.
We conjecture that there is no $n$-qubit graph state $\ket{\Gamma}$ with a connected graph for which $\ell^n_n[\Psi_\Gamma]$ is smaller than $\ell^n_n[\Psi^n_\mathrm{ring}]$.
In any case, we can conclude using Theorem~\ref{thrm:PPT_graph} and Proposition~\ref{prop:noisy_sector_lengths} that the state 
$\rho = \frac{1}{2}(\Psi^n_\mathrm{ring}+ \frac{\mathbbm{1}}{2^n})$ is a GME mixed state with a full-body sector length as low as $\ell^n_n[\rho]= \frac{1}{4}\ell^n_n[\Psi^n_\mathrm{ring}]$.
Similarly follows 
\begin{align}
 \inf_{\rho \text{ GME}} \ell^n_n[\rho] <  \frac{1}{4} +\frac{1}{4} \sum_{k=1}^{\lfloor n/3 \rfloor} 
{\genfrac{(}{)}{0pt}{}{n-2k-1}{k-1}} \frac{n}{k},
\end{align}
as an upper bound on the lowest possible full-body sector length that an $n$-qudit state $\rho$ can have while still being genuinely multipartite entangled.

\subsection{A graph-theoretical puzzle related to the full-body sector length}
\label{sec:6.2}

Here, we will derive the full-body sector lengths for the graph states families introduced in Sec.~\ref{sec:6.1.2}.
It turns out that this is possible by means of the following graph-theoretical puzzle.

\paragraph{Puzzle.}
\emph{Given a graph, in how many ways can one color its vertices such that every white vertex has an odd number of black neighbors?}

\proposition\label{prop:puzzle}
\emph{For a graph with adjacency matrix $\Gamma \in \FF_2^{n\times n}$, the solution of the puzzle above is given by~$\ell^n_n[\Psi_\Gamma]$.}
\begin{proof}
 Recall from Eq.~\eqref{eq:sec_len_qubits} that $\ell^n_n[\Psi_\Gamma]$ is equal to the number of bitstrings $\mathbf{r} \in \FF_2^n$ for which $\swt(\mathbf{r},\Gamma \mathbf{r})=n$. That is, a given bitstring $\mathbf{r}$ contributes to the full-body sector length of $\Psi^n_\Gamma$ iff for all $i\in\{ 1, \ldots, n\} $ it holds $r_i=1$ or $\sum_{j=1}^n \gamma_{i,j}r_j =1$ (or both). Now, color the vertices $i \in V$ for which $r_i=1$ black and all other vertices white. This bitstring will contribute iff $\sum_{j=1}^n \gamma_{i,j} r_j =1$ for every white vertex $i\in V$, i.e., if every white vertex has an odd (the sum is evaluated in $\FF_2$) number of black ($r_j=1$) neighbors ($\gamma_{i,j}=1$). 
 This finishes the proof.
\end{proof}
 
\begin{figure}
\centering
\begin{minipage}{0.85\textwidth}
\begin{framed} 
\centering 

\begin{tabular}{cc}
\begin{tikzpicture}
\fill[black] (0,0) circle (0.2); 
\fill[black] (1,0) circle (0.2); 
\fill[black] (-1,0) circle (0.2); 
 \draw[-, line width=.1em] (0.2,0) -- (0.8,0); 
 \draw[-, line width=.1em] (-0.2,0) -- (-0.8,0);  
 \draw[line width=.1em] (0,0) circle (0.2);  
 \draw[line width=.1em] (1,0) circle (0.2);  
 \draw[line width=.1em] (-1,0) circle (0.2);

\draw (-1,0) node[text=white]{\scriptsize 1}; 
\draw ( 0,0) node[text=white]{\scriptsize 2}; 
\draw ( 1,0) node[text=white]{\scriptsize 3}; 
 \end{tikzpicture}  
 & 
 $X_2^{(1,1,1)}Z_2^{(1,0,1)}$ 
 \vspace{.8em}
 
 \\\begin{tikzpicture}
\fill[black] (0,0) circle (0.2);  
\fill[black] (-1,0) circle (0.2); 
 \draw[-, line width=.1em] (0.2,0) -- (0.8,0); 
 \draw[-, line width=.1em] (-0.2,0) -- (-0.8,0);  
 \draw[line width=.1em] (0,0) circle (0.2);  
 \draw[line width=.1em] (1,0) circle (0.2);  
 \draw[line width=.1em] (-1,0) circle (0.2);     

\draw (-1,0) node[text=white]{\scriptsize 1}; 
\draw ( 0,0) node[text=white]{\scriptsize 2}; 
\draw ( 1,0) node[text=black]{\scriptsize 3};  \end{tikzpicture}  
 & 
 $X_2^{(1,1,0)}Z_2^{(1,1,1)}$ 
 \vspace{.8em}
 
 \\
\begin{tikzpicture}
\fill[black] (0,0) circle (0.2); 
\fill[black] (1,0) circle (0.2);  
 \draw[-, line width=.1em] (0.2,0) -- (0.8,0); 
 \draw[-, line width=.1em] (-0.2,0) -- (-0.8,0);  
 \draw[line width=.1em] (0,0) circle (0.2);  
 \draw[line width=.1em] (1,0) circle (0.2);  
 \draw[line width=.1em] (-1,0) circle (0.2);     

\draw (-1,0) node[text=black]{\scriptsize 1}; 
\draw ( 0,0) node[text=white]{\scriptsize 2}; 
\draw ( 1,0) node[text=white]{\scriptsize 3};  \end{tikzpicture}  
 &  
 $X_2^{(0,1,1)}Z_2^{(1,1,1)}$
 \vspace{.8em}
 
 \\ 
\begin{tikzpicture}
\fill[black] (0,0) circle (0.2);  
 \draw[-, line width=.1em] (0.2,0) -- (0.8,0); 
 \draw[-, line width=.1em] (-0.2,0) -- (-0.8,0);  
 \draw[line width=.1em] (0,0) circle (0.2);  
 \draw[line width=.1em] (1,0) circle (0.2);  
 \draw[line width=.1em] (-1,0) circle (0.2);     

\draw (-1,0) node[text=black]{\scriptsize 1}; 
\draw ( 0,0) node[text=white]{\scriptsize 2}; 
\draw ( 1,0) node[text=black]{\scriptsize 3};  \end{tikzpicture}  
 & 
 $X_2^{(0,1,0)}Z_2^{(1,0,1)}$  
\end{tabular}
 
\caption{All four colorings of a line graph with three vertices (left) according to the rule \emph{every white vertex has an odd number of black neighbors}. They are in one-to-one correspondence to the full-weight stabilizer operators (right) of the graph state, i.e., $\ell^n_n[\Psi^3_\mathrm{line}]=4$.}
 \label{fig:puzzle_line_3}
\end{framed}
\end{minipage}
 \end{figure}
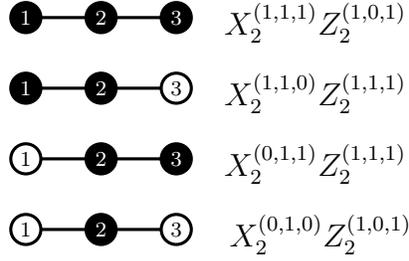

In Fig.~\ref{fig:puzzle_line_3},
we display the solution of the puzzle for a line graph with three vertices. 
Each allowed coloring of the vertices corresponds to the full weight stabilizer operator $\prod_{i=1}^n S_i^{r_i}= X^\mathbf{r}Z^\mathbf{\Gamma r}$, where we write $r_i=0$ ($r_i=1$) if vertex $i$ is white (black). 
For all graphs it is possible to color all vertices black (uppermost coloring in Fig.~\ref{fig:puzzle_line_3}) as this corresponds to the stabilizer operator $S=\prod_{i=1}^nS_i=X^{(1,\ldots,1)}Z^{\Gamma(1,\ldots,1)}$ for which all stabilizer generators are ``activated'', i.e., $r_i=1$ for all $i\in\{1,\ldots,n\}$.
If one of the stabilizer generators $S_i$ is ``deactivated'' ($r_i=0$), vertex $i$ has to posses activated neighbors which induce a $Z$ operator on qubit $i$ (for which $S$ does not has an $X$ operator anymore). If, however, the number activated neighbors was even, the  $Z$ operators induced on qubit $i$ would cancel. That is why, for example, the coloring ``black-white-black'' does not occur in Fig.~\ref{fig:puzzle_line_3}. 

In the next four subsections, we solve the graph-theoretical puzzle for the four families of graph states introduced in Sec.~\ref{sec:6.1.2}. While the full-body sector length of star graph states can be found, e.g., in  Ref.~\cite{TDLP16, WydGueh19, EltSiew19}, the results for dandelion, line and ring graph states were to our knowledge not known before.

 \subsubsection{Solution of the puzzle for star graphs}
 \label{sec:6.2.1}

\begin{figure}
\centering
\begin{minipage}{0.85\textwidth}
\begin{framed} 
\centering  

\begin{tikzpicture}
\draw[white, line width=.1em] (0,1) circle (0.2);  
\draw[-, line width=.1em] (0,-0.2) -- (0,-0.8);   
\draw[-, line width=.1em] (0.1902,-0.0618) -- (0.7608,-0.2472);   
\draw[-, line width=.1em] (-0.1902,-0.0618) -- (-0.7608,-0.2472);   
\draw[-, line width=.1em] ( 0.1176,0.1618) -- ( 0.4702, 0.6472);   
\draw[-, line width=.1em] (-0.1176,0.1618) -- (-0.4702, 0.6472);   
  
\fill[black] (0,-1) circle (0.2);  
\fill[black] ( 0.9511,-0.3090) circle (0.2);      
\fill[black] (-0.9511,-0.3090) circle (0.2);    
\fill[black] ( 0.5878,0.8090) circle (0.2);      
\fill[black] (-0.5878,0.8090) circle (0.2);      

\draw[line width=.1em] (0,0) circle (0.2);  
\draw[line width=.1em] (0,-1) circle (0.2);  
\draw[line width=.1em] ( 0.9511,-0.3090) circle (0.2);      
\draw[line width=.1em] (-0.9511,-0.3090) circle (0.2);    
\draw[line width=.1em] ( 0.5878,0.8090) circle (0.2);      
\draw[line width=.1em] (-0.5878,0.8090) circle (0.2);

\draw (0,0) node[text=black]{\scriptsize 1}; 
\draw (0,-1) node[text=white]{\scriptsize 2}; 
\draw ( 0.9511,-0.3090) node[text=white]{\scriptsize 3}; 
\draw ( 0.5878,0.8090) node[text=white]{\scriptsize 4}; 
\draw (-0.5878,0.8090) node[text=white]{\scriptsize 5}; 
\draw (-0.9511,-0.3090) node[text=white]{\scriptsize 6}; 
 \end{tikzpicture}   
 
\caption{The additional coloring of a star graph where the central vertex is white;
here depicted for $n=6$. Because of the rule \emph{every white vertex has an odd number of black neighbors}, this coloring is only possible if $n$ is even.}
 \label{fig:puzzleGHZ6_extra_stab}
\end{framed}
\end{minipage}
 \end{figure}
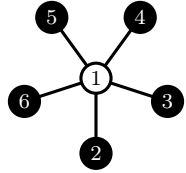

Here, we verify using our graph-theoretical puzzle that the full-body sector length of a star graph state on $n$ qubits is given by Eq.~\eqref{eq:lnn_star}.
First note that, as long as the central vertex is black, any coloring of the leaf vertices is in accordance to the rule of the puzzle since their only neighbor is the (black) central vertex. This gives $2^{n-1}$ colorings. 
If $n$ is even, there is one additional coloring, see Fig.~\ref{fig:puzzleGHZ6_extra_stab}.
Since $n$ is even,  the number of the neighbors of the central vertex is odd.
Thus, if all  leaf vertices are black, the rule is fulfilled. 
No further colorings according to the rule of the puzzle are possible.
This proves Eq.~\eqref{eq:lnn_star}.


 \subsubsection{Solution of the puzzle for dandelion graphs}
 \label{sec:6.2.2}
 
 
 In order to explain why the solution of the puzzle for dandelion graphs is indeed given by Eq.~\eqref{eq:lnn_dandelion}, we use the naming for its vertices which is given in the caption of Fig.~\ref{fig:graph_dandelion}:
  Let us now discuss the possible colorings which are in accordance to the rule of the puzzle.
 We distinguish the two cases where the lower central vertex is black and white, respectively.
 If it is black, every coloring (of which there are four) of the root vertices is valid. 
 Similar to the case of the star graph state, all $2^{n-4}$ colorings of the seed vertices are valid as long as the upper central vertex is black. 
 If, on the other hand, the upper central vertex is white, all seed vertices must be colored black. This gives one additional coloring; but only if $n-3$ is odd as also the number of marked neighbors of the upper central vertex must be odd. 
 This gives $4\times \left(2^{n-4} + \left(\frac{1+(-1)^{n}}{2}\right) \right) $ valid colorings for the first case.
 In the second case, the lower central vertex is white. 
 Then, all of its three neighbors must be black. Again, there are $2^{n-4}$ possible choices for the seed vertices. Since the upper central vertex is necessarily black now, there are no further colorings.
 We thus obtain the total number of colorings according to Eq.~\eqref{eq:lnn_dandelion}.

 \subsubsection{Solution of the puzzle for line graphs} 
 \label{sec:6.2.3} 
 
 For line graphs, each vertex has at most two neighbors and the rule of the puzzle can be simplified as follows:
 \emph{Every white vertex has exactly one black neighbor.}
 Using Proposition~\ref{prop:puzzle}, we are able to derive the following recursive relation:
 
 \corollary~\label{cor:sequence_line}
 \emph{The full-body sector length of a line graph state is given by ${\ell^1_1[\Psi^1_\mathrm{star}]=1}$, $\ell^2_2[\Psi^2_\mathrm{star}]=3$, $\ell^3_3[\Psi^3_\mathrm{star}]=4$, and $\ell^{n}_{n}[\Psi^{n}_\mathrm{star}]= \ell^{n-1}_{n-1}[\Psi^{n-1}_\mathrm{star}]+ \ell^{n-3}_{n-3}[\Psi^{n-3}_\mathrm{star}]$  for all $n \ge 4$.}
 \\
  More information about this sequence, e.g., its generating function,  is available in \emph{The On-Line Encyclopedia of Integer Sequences~\textsuperscript{\textregistered}}~\href{https://oeis.org/A179070}{\texttt{[OEIS:A179070]}}.
 \begin{proof}
 According to the rule of the puzzle, it is only allowed to color a leaf vertex white if its neighbor is black. Furthermore,  white vertices in the bulk are allowed iff also (exactly) one of its neighbors is white. Thus, white vertices can only come in pair of two, see Fig.~\ref{fig:example_line8} 
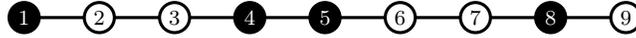
\begin{figure}
\centering
\begin{minipage}{0.85\textwidth}
\begin{framed} 
\centering  
\begin{tikzpicture}
\fill[black] (1,0) circle (0.2); 
\fill[black] (4,0) circle (0.2);  
\fill[black] (5,0) circle (0.2);  
\fill[black] (8,0) circle (0.2);   
\draw[-, line width=.1em] (1.2,0) -- (1.8,0);  
\draw[-, line width=.1em] (2.2,0) -- (2.8,0);  
\draw[-, line width=.1em] (3.2,0) -- (3.8,0);  
\draw[-, line width=.1em] (4.2,0) -- (4.8,0);  
\draw[-, line width=.1em] (5.2,0) -- (5.8,0);  
\draw[-, line width=.1em] (6.2,0) -- (6.8,0);  
\draw[-, line width=.1em] (7.2,0) -- (7.8,0);  
\draw[-, line width=.1em] (8.2,0) -- (8.8,0);  
\draw[line width=.1em] (1,0) circle (0.2);  
\draw[line width=.1em] (2,0) circle (0.2);  
\draw[line width=.1em] (3,0) circle (0.2);  
\draw[line width=.1em] (4,0) circle (0.2);  
\draw[line width=.1em] (5,0) circle (0.2);  
\draw[line width=.1em] (6,0) circle (0.2);  
\draw[line width=.1em] (7,0) circle (0.2);  
\draw[line width=.1em] (8,0) circle (0.2);  
\draw[line width=.1em] (9,0) circle (0.2);  

\draw (1,0) node[text=white]{\scriptsize 1}; 
\draw (2,0) node[text=black]{\scriptsize 2}; 
\draw (3,0) node[text=black]{\scriptsize 3}; 
\draw (4,0) node[text=white]{\scriptsize 4}; 
\draw (5,0) node[text=white]{\scriptsize 5}; 
\draw (6,0) node[text=black]{\scriptsize 6}; 
\draw (7,0) node[text=black]{\scriptsize 7}; 
\draw (8,0) node[text=white]{\scriptsize 8}; 
\draw (9,0) node[text=black]{\scriptsize 9};  
 \end{tikzpicture}    
\caption{Example of coloring vertices according to the rule \emph{every white vertex has exactly one  black neighbor} on a line graph with $n=9$ vertices. In the bulk, white vertices can only come in pairs of exactly two which are separated by at least one (here two) black vertex. 
}
 \label{fig:example_line8}
\end{framed}
\end{minipage}
 \end{figure}
 for an example.
 To solve the problem, we introduce the notation 
 $\ell^n_n[\Psi^n_\mathrm{line}]=c^{(n)}_2+c^{(n)}_1+c^{(n)}_0$, where $c^{(n)}_i$ is the
 number colorings of an $n$-vertex line graph where exactly $i$ of the two leaf vertices are black, e.g., $c^{(3)}_2 = 1$, $c^{(3)}_1 = 2$, and $c^{(3)}_0 = 1$ as one can infer from Fig.~\ref{fig:puzzle_line_3}.
 In Table~\ref{tab:line_choices} 
 \begin{table}
\centering
\begin{minipage}{0.85\textwidth}
\begin{framed} 
\centering
\begin{tabular}{|c|cccccccccc|} \hline\hline
$n$         & 2 & 3 & 4 & 5 & 6 & 7 & 8 & 9 & 10 & 11   \\ \hline 
$\ell^n_n[\Psi^n_\mathrm{line}]$ 
	    & 3 & 4 & 5 & 8 & 12 & 17 &	25 & 37 & 54 & 79\\ \hline
$c^{(n)}_2$ & 1 & 1 & 2 & 3 &  4 &  6 & 9  & 13	& 19 & 28	\\  
$c^{(n)}_1$ & 2 & 2 & 2 & 4 &  6 &  8 &	12 & 18	& 26 & 38	\\  
$c^{(n)}_0$ & 0 & 1 & 1 & 1 &  2 &  3 & 4  &  6 &  9 & 13
	    \\ \hline\hline
\end{tabular}
\caption{Number of allowed colorings $c_i^{(n)}$ of a line graph with $n$ vertices such that $i$ leaf vertices are black. We find three relations: $c_1^{(n)}=2c_2^{(n-1)}$ for $n\ge3$, $c^{(n)}_2=c^{(n-1)}_2+c^{(n-3)}_2$ for $n\ge 5$, and $c^{(n)}_0=c^{(n-2)}_2$ for $n\ge4$.}
 \label{tab:line_choices}
\end{framed}
\end{minipage}
 \end{table}
 we show the first few values of $c^{(n)}_i$. 
 There are a few relations among the values in this table. 
 First, note that for every allowed coloring on $n-1$ vertices where the two leaves are black, one can attach an $n^\mathrm{th}$ vertex which is white. As one can attach this additional vertex on either side, one obtains $c_1^{(n)}=2c_2^{(n-1)}$ for $n\ge3$.  
This relation is best illustrated in Fig.~\ref{fig:puzzle_line_4_5}.
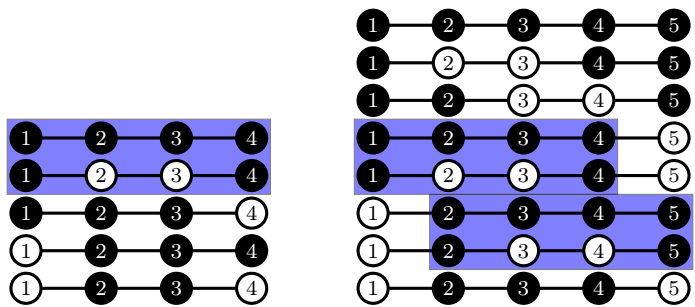
\begin{figure}
\centering
\begin{minipage}{0.85\textwidth}
\begin{framed} 
\centering  
\begin{tikzpicture}
\draw [fill=blue, opacity = .5] (.75,0.25) rectangle (4.25,-0.75);

\fill[white] (2,-.5) circle (0.2); 
\fill[white] (3,-.5) circle (0.2);

\fill[black] (1,0) circle (0.2); 
\fill[black] (2,0) circle (0.2); 
\fill[black] (3,0) circle (0.2); 
\fill[black] (4,0) circle (0.2);  
 \draw[-, line width=.1em] (1.2,0) -- (1.8,0); 
 \draw[-, line width=.1em] (2.2,0) -- (2.8,0);  
 \draw[-, line width=.1em] (3.2,0) -- (3.8,0); 
 \draw[line width=.1em] (1,0) circle (0.2);  
 \draw[line width=.1em] (2,0) circle (0.2);  
 \draw[line width=.1em] (3,0) circle (0.2);     
 \draw[line width=.1em] (4,0) circle (0.2);   
 
\fill[black] (1,-.5) circle (0.2); 
\fill[black] (4,-.5) circle (0.2);  
 \draw[-, line width=.1em] (1.2,-.5) -- (1.8,-.5); 
 \draw[-, line width=.1em] (2.2,-.5) -- (2.8,-.5);  
 \draw[-, line width=.1em] (3.2,-.5) -- (3.8,-.5); 
 \draw[line width=.1em] (1,-.5) circle (0.2);  
 \draw[line width=.1em] (2,-.5) circle (0.2);  
 \draw[line width=.1em] (3,-.5) circle (0.2);     
 \draw[line width=.1em] (4,-.5) circle (0.2);   
 
\fill[black] (1,-1) circle (0.2); 
\fill[black] (2,-1) circle (0.2); 
\fill[black] (3,-1) circle (0.2); 
 \draw[-, line width=.1em] (1.2,-1) -- (1.8,-1); 
 \draw[-, line width=.1em] (2.2,-1) -- (2.8,-1);  
 \draw[-, line width=.1em] (3.2,-1) -- (3.8,-1); 
 \draw[line width=.1em] (1,-1) circle (0.2);  
 \draw[line width=.1em] (2,-1) circle (0.2);  
 \draw[line width=.1em] (3,-1) circle (0.2);     
 \draw[line width=.1em] (4,-1) circle (0.2);  
  
\fill[black] (2,-1.5) circle (0.2); 
\fill[black] (3,-1.5) circle (0.2); 
\fill[black] (4,-1.5) circle (0.2);  
 \draw[-, line width=.1em] (1.2,-1.5) -- (1.8,-1.5); 
 \draw[-, line width=.1em] (2.2,-1.5) -- (2.8,-1.5);  
 \draw[-, line width=.1em] (3.2,-1.5) -- (3.8,-1.5); 
 \draw[line width=.1em] (1,-1.5) circle (0.2);  
 \draw[line width=.1em] (2,-1.5) circle (0.2);  
 \draw[line width=.1em] (3,-1.5) circle (0.2);     
 \draw[line width=.1em] (4,-1.5) circle (0.2);  
  
\fill[black] (2,-2) circle (0.2); 
\fill[black] (3,-2) circle (0.2); 
 \draw[-, line width=.1em] (1.2,-2) -- (1.8,-2); 
 \draw[-, line width=.1em] (2.2,-2) -- (2.8,-2);  
 \draw[-, line width=.1em] (3.2,-2) -- (3.8,-2); 
 \draw[line width=.1em] (1,-2) circle (0.2);  
 \draw[line width=.1em] (2,-2) circle (0.2);  
 \draw[line width=.1em] (3,-2) circle (0.2);     
 \draw[line width=.1em] (4,-2) circle (0.2);

\draw (1,0) node[text=white]{\scriptsize 1}; 
\draw (2,0) node[text=white]{\scriptsize 2}; 
\draw (3,0) node[text=white]{\scriptsize 3}; 
\draw (4,0) node[text=white]{\scriptsize 4}; 

\draw (1,-.5) node[text=white]{\scriptsize 1}; 
\draw (2,-.5) node[text=black]{\scriptsize 2}; 
\draw (3,-.5) node[text=black]{\scriptsize 3}; 
\draw (4,-.5) node[text=white]{\scriptsize 4}; 

\draw (1,-1) node[text=white]{\scriptsize 1}; 
\draw (2,-1) node[text=white]{\scriptsize 2}; 
\draw (3,-1) node[text=white]{\scriptsize 3}; 
\draw (4,-1) node[text=black]{\scriptsize 4}; 

\draw (1,-1.5) node[text=black]{\scriptsize 1}; 
\draw (2,-1.5) node[text=white]{\scriptsize 2}; 
\draw (3,-1.5) node[text=white]{\scriptsize 3}; 
\draw (4,-1.5) node[text=white]{\scriptsize 4}; 

\draw (1,-2) node[text=black]{\scriptsize 1}; 
\draw (2,-2) node[text=white]{\scriptsize 2}; 
\draw (3,-2) node[text=white]{\scriptsize 3}; 
\draw (4,-2) node[text=black]{\scriptsize 4};

 \end{tikzpicture}  
 \hspace{2em}
\begin{tikzpicture}
\draw [fill=blue, opacity = .5] (.75,-1.25) rectangle (4.25,-2.25); 
\draw [fill=blue, opacity = .5] (1.75,-2.25) rectangle (5.25,-3.25); 

\fill[white] (2,-2) circle (0.2); 
\fill[white] (3,-2) circle (0.2); 
\fill[white] (3,-3) circle (0.2); 
\fill[white] (4,-3) circle (0.2); 
 
\fill[black] (1,0) circle (0.2); 
\fill[black] (2,0) circle (0.2); 
\fill[black] (3,0) circle (0.2); 
\fill[black] (4,0) circle (0.2); 
\fill[black] (5,0) circle (0.2);  
 \draw[-, line width=.1em] (1.2,0) -- (1.8,0); 
 \draw[-, line width=.1em] (2.2,0) -- (2.8,0);  
 \draw[-, line width=.1em] (3.2,0) -- (3.8,0); 
 \draw[-, line width=.1em] (4.2,0) -- (4.8,0); 
 \draw[line width=.1em] (1,0) circle (0.2);  
 \draw[line width=.1em] (2,0) circle (0.2);  
 \draw[line width=.1em] (3,0) circle (0.2);     
 \draw[line width=.1em] (4,0) circle (0.2);    
 \draw[line width=.1em] (5,0) circle (0.2);   
 
\fill[black] (1,-.5) circle (0.2); 
\fill[black] (4,-.5) circle (0.2); 
\fill[black] (5,-.5) circle (0.2);  
 \draw[-, line width=.1em] (1.2,-.5) -- (1.8,-.5); 
 \draw[-, line width=.1em] (2.2,-.5) -- (2.8,-.5);  
 \draw[-, line width=.1em] (3.2,-.5) -- (3.8,-.5); 
 \draw[-, line width=.1em] (4.2,-.5) -- (4.8,-.5); 
 \draw[line width=.1em] (1,-.5) circle (0.2);  
 \draw[line width=.1em] (2,-.5) circle (0.2);  
 \draw[line width=.1em] (3,-.5) circle (0.2);     
 \draw[line width=.1em] (4,-.5) circle (0.2);    
 \draw[line width=.1em] (5,-.5) circle (0.2);   
 
\fill[black] (1,-1) circle (0.2); 
\fill[black] (2,-1) circle (0.2); 
\fill[black] (5,-1) circle (0.2);  
 \draw[-, line width=.1em] (1.2,-1) -- (1.8,-1); 
 \draw[-, line width=.1em] (2.2,-1) -- (2.8,-1);  
 \draw[-, line width=.1em] (3.2,-1) -- (3.8,-1); 
 \draw[-, line width=.1em] (4.2,-1) -- (4.8,-1); 
 \draw[line width=.1em] (1,-1) circle (0.2);  
 \draw[line width=.1em] (2,-1) circle (0.2);  
 \draw[line width=.1em] (3,-1) circle (0.2);     
 \draw[line width=.1em] (4,-1) circle (0.2);    
 \draw[line width=.1em] (5,-1) circle (0.2);   
\fill[black] (1,-1.5) circle (0.2); 
\fill[black] (2,-1.5) circle (0.2); 
\fill[black] (3,-1.5) circle (0.2); 
\fill[black] (4,-1.5) circle (0.2); 

 \draw[-, line width=.1em] (1.2,-1.5) -- (1.8,-1.5); 
 \draw[-, line width=.1em] (2.2,-1.5) -- (2.8,-1.5);  
 \draw[-, line width=.1em] (3.2,-1.5) -- (3.8,-1.5); 
 \draw[-, line width=.1em] (4.2,-1.5) -- (4.8,-1.5); 
 \draw[line width=.1em] (1,-1.5) circle (0.2);  
 \draw[line width=.1em] (2,-1.5) circle (0.2);  
 \draw[line width=.1em] (3,-1.5) circle (0.2);     
 \draw[line width=.1em] (4,-1.5) circle (0.2);    
 \draw[line width=.1em] (5,-1.5) circle (0.2);   
\fill[black] (1,-2) circle (0.2); 
\fill[black] (4,-2) circle (0.2); 
 \draw[-, line width=.1em] (1.2,-2) -- (1.8,-2); 
 \draw[-, line width=.1em] (2.2,-2) -- (2.8,-2);  
 \draw[-, line width=.1em] (3.2,-2) -- (3.8,-2); 
 \draw[-, line width=.1em] (4.2,-2) -- (4.8,-2); 
 \draw[line width=.1em] (1,-2) circle (0.2);  
 \draw[line width=.1em] (2,-2) circle (0.2);  
 \draw[line width=.1em] (3,-2) circle (0.2);     
 \draw[line width=.1em] (4,-2) circle (0.2);    
 \draw[line width=.1em] (5,-2) circle (0.2);  
 
\fill[black] (2,-2.5) circle (0.2); 
\fill[black] (3,-2.5) circle (0.2); 
\fill[black] (4,-2.5) circle (0.2); 
\fill[black] (5,-2.5) circle (0.2);  
 \draw[-, line width=.1em] (1.2,-2.5) -- (1.8,-2.5); 
 \draw[-, line width=.1em] (2.2,-2.5) -- (2.8,-2.5);  
 \draw[-, line width=.1em] (3.2,-2.5) -- (3.8,-2.5); 
 \draw[-, line width=.1em] (4.2,-2.5) -- (4.8,-2.5); 
 \draw[line width=.1em] (1,-2.5) circle (0.2);  
 \draw[line width=.1em] (2,-2.5) circle (0.2);  
 \draw[line width=.1em] (3,-2.5) circle (0.2);     
 \draw[line width=.1em] (4,-2.5) circle (0.2);    
 \draw[line width=.1em] (5,-2.5) circle (0.2);   
\fill[black] (2,-3) circle (0.2); 
\fill[black] (5,-3) circle (0.2); 

 \draw[-, line width=.1em] (1.2,-3) -- (1.8,-3); 
 \draw[-, line width=.1em] (2.2,-3) -- (2.8,-3);  
 \draw[-, line width=.1em] (3.2,-3) -- (3.8,-3); 
 \draw[-, line width=.1em] (4.2,-3) -- (4.8,-3); 
 \draw[line width=.1em] (1,-3) circle (0.2);  
 \draw[line width=.1em] (2,-3) circle (0.2);  
 \draw[line width=.1em] (3,-3) circle (0.2);     
 \draw[line width=.1em] (4,-3) circle (0.2);    
 \draw[line width=.1em] (5,-3) circle (0.2);   
\fill[black] (2,-3.5) circle (0.2); 
\fill[black] (3,-3.5) circle (0.2); 
\fill[black] (4,-3.5) circle (0.2); 
 \draw[-, line width=.1em] (1.2,-3.5) -- (1.8,-3.5); 
 \draw[-, line width=.1em] (2.2,-3.5) -- (2.8,-3.5);  
 \draw[-, line width=.1em] (3.2,-3.5) -- (3.8,-3.5); 
 \draw[-, line width=.1em] (4.2,-3.5) -- (4.8,-3.5); 
 \draw[line width=.1em] (1,-3.5) circle (0.2);  
 \draw[line width=.1em] (2,-3.5) circle (0.2);  
 \draw[line width=.1em] (3,-3.5) circle (0.2);     
 \draw[line width=.1em] (4,-3.5) circle (0.2);    
 \draw[line width=.1em] (5,-3.5) circle (0.2);

\draw (1,0) node[text=white]{\scriptsize 1}; 
\draw (2,0) node[text=white]{\scriptsize 2}; 
\draw (3,0) node[text=white]{\scriptsize 3}; 
\draw (4,0) node[text=white]{\scriptsize 4};  
\draw (5,0) node[text=white]{\scriptsize 5}; 

\draw (1,-.5) node[text=white]{\scriptsize 1}; 
\draw (2,-.5) node[text=black]{\scriptsize 2}; 
\draw (3,-.5) node[text=black]{\scriptsize 3}; 
\draw (4,-.5) node[text=white]{\scriptsize 4};  
\draw (5,-.5) node[text=white]{\scriptsize 5}; 

\draw (1,-1) node[text=white]{\scriptsize 1}; 
\draw (2,-1) node[text=white]{\scriptsize 2}; 
\draw (3,-1) node[text=black]{\scriptsize 3}; 
\draw (4,-1) node[text=black]{\scriptsize 4};  
\draw (5,-1) node[text=white]{\scriptsize 5}; 

\draw (1,-1.5) node[text=white]{\scriptsize 1}; 
\draw (2,-1.5) node[text=white]{\scriptsize 2}; 
\draw (3,-1.5) node[text=white]{\scriptsize 3}; 
\draw (4,-1.5) node[text=white]{\scriptsize 4};  
\draw (5,-1.5) node[text=black]{\scriptsize 5}; 

\draw (1,-2) node[text=white]{\scriptsize 1}; 
\draw (2,-2) node[text=black]{\scriptsize 2}; 
\draw (3,-2) node[text=black]{\scriptsize 3}; 
\draw (4,-2) node[text=white]{\scriptsize 4};  
\draw (5,-2) node[text=black]{\scriptsize 5}; 

\draw (1,-2.5) node[text=black]{\scriptsize 1}; 
\draw (2,-2.5) node[text=white]{\scriptsize 2}; 
\draw (3,-2.5) node[text=white]{\scriptsize 3}; 
\draw (4,-2.5) node[text=white]{\scriptsize 4};  
\draw (5,-2.5) node[text=white]{\scriptsize 5}; 

\draw (1,-3) node[text=black]{\scriptsize 1}; 
\draw (2,-3) node[text=white]{\scriptsize 2}; 
\draw (3,-3) node[text=black]{\scriptsize 3}; 
\draw (4,-3) node[text=black]{\scriptsize 4};  
\draw (5,-3) node[text=white]{\scriptsize 5}; 

\draw (1,-3.5) node[text=black]{\scriptsize 1}; 
\draw (2,-3.5) node[text=white]{\scriptsize 2}; 
\draw (3,-3.5) node[text=white]{\scriptsize 3}; 
\draw (4,-3.5) node[text=white]{\scriptsize 4};  
\draw (5,-3.5) node[text=black]{\scriptsize 5};  
 \end{tikzpicture}   
 
\caption{Illustration of the relation $c^{(n)}_1=2c^{(n-1)}_2$ at the example $n=5$.
Depicted are all colorings of a line graph with four (left) and five (right) vertices according to the rule \emph{every white vertex has exactly one black neighbor}. 
We have highlighted in blue $c^{(4)}_2=2$ choices where both leaves of a four-vertex line graph are black. 
}
 \label{fig:puzzle_line_4_5}
\end{framed}
\end{minipage}
 \end{figure}
 We also find the recursive formula $c^{(n)}_2=c^{(n-1)}_2+c^{(n-3)}_2$ for $n\ge 5$ which we illustrate in   Fig.~\ref{fig:puzzle_line_markedleaves}.
 \begin{figure}
\centering
\begin{minipage}{0.85\textwidth}
\begin{framed} 
\centering 

\begin{tabular}{|c|l|} 

\hline
$n$ & Colorings   \\ \hline \hline
$2$&
\begin{tikzpicture}
\fill[black] (1,0) circle (0.2); 
\fill[black] (2,0) circle (0.2);  
\draw[-, line width=.1em] (1.2,0) -- (1.8,0);  
\draw[line width=.1em] (1,0) circle (0.2);  
\draw[line width=.1em] (2,0) circle (0.2);   

\draw (1,0) node[text=white]{\scriptsize 1}; 
\draw (2,0) node[text=white]{\scriptsize 2};
\end{tikzpicture}
\\ \hline
$3$ &
\begin{tikzpicture}
\fill[black] (1,0) circle (0.2); 
\fill[black] (2,0) circle (0.2);  
\fill[black] (3,0) circle (0.2);  
\draw[-, line width=.1em] (1.2,0) -- (1.8,0);  
\draw[-, line width=.1em] (2.2,0) -- (2.8,0);  
\draw[line width=.1em] (1,0) circle (0.2);  
\draw[line width=.1em] (2,0) circle (0.2);   
\draw[line width=.1em] (3,0) circle (0.2);  

\draw (1,0) node[text=white]{\scriptsize 1}; 
\draw (2,0) node[text=white]{\scriptsize 2}; 
\draw (3,0) node[text=white]{\scriptsize 3}; 
\end{tikzpicture}
\\\hline

$4$ &
\begin{tikzpicture}
\fill[black] (1,0) circle (0.2); 
\fill[black] (2,0) circle (0.2);  
\fill[black] (3,0) circle (0.2); 
\fill[black] (4,0) circle (0.2);  
\draw[-, line width=.1em] (1.2,0) -- (1.8,0);  
\draw[-, line width=.1em] (2.2,0) -- (2.8,0);  
\draw[-, line width=.1em] (3.2,0) -- (3.8,0);  
\draw[line width=.1em] (1,0) circle (0.2);  
\draw[line width=.1em] (2,0) circle (0.2);  
\draw[line width=.1em] (3,0) circle (0.2);  
\draw[line width=.1em] (4,0) circle (0.2);  
\fill[black] (1,-.5) circle (0.2); 
\fill[black] (4,-.5) circle (0.2);  
\draw[-, line width=.1em] (1.2,-.5) -- (1.8,-.5);  
\draw[-, line width=.1em] (2.2,-.5) -- (2.8,-.5);  
\draw[-, line width=.1em] (3.2,-.5) -- (3.8,-.5);  
\draw[line width=.1em] (1,-.5) circle (0.2);  
\draw[line width=.1em] (2,-.5) circle (0.2);  
\draw[line width=.1em] (3,-.5) circle (0.2);  
\draw[line width=.1em] (4,-.5) circle (0.2);

\draw (1,0) node[text=white]{\scriptsize 1}; 
\draw (2,0) node[text=white]{\scriptsize 2}; 
\draw (3,0) node[text=white]{\scriptsize 3}; 
\draw (4,0) node[text=white]{\scriptsize 4}; 

\draw (1,-.5) node[text=white]{\scriptsize 1}; 
\draw (2,-.5) node[text=black]{\scriptsize 2}; 
\draw (3,-.5) node[text=black]{\scriptsize 3}; 
\draw (4,-.5) node[text=white]{\scriptsize 4}; 

\end{tikzpicture} 
\\\hline
$5$ & 
\begin{tikzpicture}
\draw [fill=blue, opacity = .5] (0.75,0.25) rectangle (4.25,-0.75);
\draw [fill=red, opacity = .5] (0.75,-0.75) rectangle (2.25,-1.25);

\fill[white] (2,-0.5) circle (0.2); 
\fill[white] (3,-0.5) circle (0.2);

\fill[black] (1,0) circle (0.2); 
\fill[black] (2,0) circle (0.2);  
\fill[black] (3,0) circle (0.2); 
\fill[black] (4,0) circle (0.2);  
\fill[black] (5,0) circle (0.2);  
\draw[-, line width=.1em] (1.2,0) -- (1.8,0);  
\draw[-, line width=.1em] (2.2,0) -- (2.8,0);  
\draw[-, line width=.1em] (3.2,0) -- (3.8,0);  
\draw[-, line width=.1em] (4.2,0) -- (4.8,0);  
\draw[line width=.1em] (1,0) circle (0.2);  
\draw[line width=.1em] (2,0) circle (0.2);  
\draw[line width=.1em] (3,0) circle (0.2);  
\draw[line width=.1em] (4,0) circle (0.2);  
\draw[line width=.1em] (5,0) circle (0.2);  
\fill[black] (1,-.5) circle (0.2); 
\fill[black] (4,-.5) circle (0.2);  
\fill[black] (5,-.5) circle (0.2);  
\draw[-, line width=.1em] (1.2,-.5) -- (1.8,-.5);  
\draw[-, line width=.1em] (2.2,-.5) -- (2.8,-.5);  
\draw[-, line width=.1em] (3.2,-.5) -- (3.8,-.5);  
\draw[-, line width=.1em] (4.2,-.5) -- (4.8,-.5);  
\draw[line width=.1em] (1,-.5) circle (0.2);  
\draw[line width=.1em] (2,-.5) circle (0.2);  
\draw[line width=.1em] (3,-.5) circle (0.2);  
\draw[line width=.1em] (4,-.5) circle (0.2);  
\draw[line width=.1em] (5,-.5) circle (0.2);  
\fill[black] (1,-1) circle (0.2); 
\fill[black] (2,-1) circle (0.2);  
\fill[black] (5,-1) circle (0.2);  
\draw[-, line width=.1em] (1.2,-1) -- (1.8,-1);  
\draw[-, line width=.1em] (2.2,-1) -- (2.8,-1);  
\draw[-, line width=.1em] (3.2,-1) -- (3.8,-1);  
\draw[-, line width=.1em] (4.2,-1) -- (4.8,-1);  
\draw[line width=.1em] (1,-1) circle (0.2);  
\draw[line width=.1em] (2,-1) circle (0.2);  
\draw[line width=.1em] (3,-1) circle (0.2);  
\draw[line width=.1em] (4,-1) circle (0.2);  
\draw[line width=.1em] (5,-1) circle (0.2);  

\draw (1,0) node[text=white]{\scriptsize 1}; 
\draw (2,0) node[text=white]{\scriptsize 2}; 
\draw (3,0) node[text=white]{\scriptsize 3}; 
\draw (4,0) node[text=white]{\scriptsize 4};  
\draw (5,0) node[text=white]{\scriptsize 5}; 

\draw (1,-.5) node[text=white]{\scriptsize 1}; 
\draw (2,-.5) node[text=black]{\scriptsize 2}; 
\draw (3,-.5) node[text=black]{\scriptsize 3}; 
\draw (4,-.5) node[text=white]{\scriptsize 4};  
\draw (5,-.5) node[text=white]{\scriptsize 5}; 

\draw (1,-1) node[text=white]{\scriptsize 1}; 
\draw (2,-1) node[text=white]{\scriptsize 2}; 
\draw (3,-1) node[text=black]{\scriptsize 3}; 
\draw (4,-1) node[text=black]{\scriptsize 4};  
\draw (5,-1) node[text=white]{\scriptsize 5}; 
\end{tikzpicture} 
\\\hline

$6$ &
\begin{tikzpicture}
\draw [fill=blue, opacity = .5] (0.75,0.25) rectangle  (5.25,-1.25);
\draw [fill=red, opacity = .5] (0.75,-1.25) rectangle (3.25,-1.75);

\fill[white] (2,-0.5) circle (0.2);  
\fill[white] (3,-0.5) circle (0.2); 
\fill[white] (3,-1) circle (0.2);  
\fill[white] (4,-1) circle (0.2); 

\fill[black] (1,0) circle (0.2); 
\fill[black] (2,0) circle (0.2);  
\fill[black] (3,0) circle (0.2); 
\fill[black] (4,0) circle (0.2);  
\fill[black] (5,0) circle (0.2);  
\fill[black] (6,0) circle (0.2);  
\draw[-, line width=.1em] (1.2,0) -- (1.8,0);  
\draw[-, line width=.1em] (2.2,0) -- (2.8,0);  
\draw[-, line width=.1em] (3.2,0) -- (3.8,0);  
\draw[-, line width=.1em] (4.2,0) -- (4.8,0);  
\draw[-, line width=.1em] (5.2,0) -- (5.8,0);  
\draw[line width=.1em] (1,0) circle (0.2);  
\draw[line width=.1em] (2,0) circle (0.2);  
\draw[line width=.1em] (3,0) circle (0.2);  
\draw[line width=.1em] (4,0) circle (0.2);  
\draw[line width=.1em] (5,0) circle (0.2);  
\draw[line width=.1em] (6,0) circle (0.2);  
\fill[black] (1,-.5) circle (0.2); 
\fill[black] (4,-.5) circle (0.2);  
\fill[black] (5,-.5) circle (0.2);  
\fill[black] (6,-.5) circle (0.2);  
\draw[-, line width=.1em] (1.2,-.5) -- (1.8,-.5);  
\draw[-, line width=.1em] (2.2,-.5) -- (2.8,-.5);  
\draw[-, line width=.1em] (3.2,-.5) -- (3.8,-.5);  
\draw[-, line width=.1em] (4.2,-.5) -- (4.8,-.5);  
\draw[-, line width=.1em] (5.2,-.5) -- (5.8,-.5);  
\draw[line width=.1em] (1,-.5) circle (0.2);  
\draw[line width=.1em] (2,-.5) circle (0.2);  
\draw[line width=.1em] (3,-.5) circle (0.2);  
\draw[line width=.1em] (4,-.5) circle (0.2);  
\draw[line width=.1em] (5,-.5) circle (0.2);  
\draw[line width=.1em] (6,-.5) circle (0.2);  
\fill[black] (1,-1) circle (0.2); 
\fill[black] (2,-1) circle (0.2);  
\fill[black] (5,-1) circle (0.2);  
\fill[black] (6,-1) circle (0.2);  
\draw[-, line width=.1em] (1.2,-1) -- (1.8,-1);  
\draw[-, line width=.1em] (2.2,-1) -- (2.8,-1);  
\draw[-, line width=.1em] (3.2,-1) -- (3.8,-1);  
\draw[-, line width=.1em] (4.2,-1) -- (4.8,-1);  
\draw[-, line width=.1em] (5.2,-1) -- (5.8,-1);  
\draw[line width=.1em] (1,-1) circle (0.2);  
\draw[line width=.1em] (2,-1) circle (0.2);  
\draw[line width=.1em] (3,-1) circle (0.2);  
\draw[line width=.1em] (4,-1) circle (0.2);  
\draw[line width=.1em] (5,-1) circle (0.2);  
\draw[line width=.1em] (6,-1) circle (0.2);  
\fill[black] (1,-1.5) circle (0.2); 
\fill[black] (2,-1.5) circle (0.2);  
\fill[black] (3,-1.5) circle (0.2); 
\fill[black] (6,-1.5) circle (0.2);  
\draw[-, line width=.1em] (1.2,-1.5) -- (1.8,-1.5);  
\draw[-, line width=.1em] (2.2,-1.5) -- (2.8,-1.5);  
\draw[-, line width=.1em] (3.2,-1.5) -- (3.8,-1.5);  
\draw[-, line width=.1em] (4.2,-1.5) -- (4.8,-1.5);  
\draw[-, line width=.1em] (5.2,-1.5) -- (5.8,-1.5);  
\draw[line width=.1em] (1,-1.5) circle (0.2);  
\draw[line width=.1em] (2,-1.5) circle (0.2);  
\draw[line width=.1em] (3,-1.5) circle (0.2);  
\draw[line width=.1em] (4,-1.5) circle (0.2);  
\draw[line width=.1em] (5,-1.5) circle (0.2);  
\draw[line width=.1em] (6,-1.5) circle (0.2); 

\draw (1,0) node[text=white]{\scriptsize 1}; 
\draw (2,0) node[text=white]{\scriptsize 2}; 
\draw (3,0) node[text=white]{\scriptsize 3}; 
\draw (4,0) node[text=white]{\scriptsize 4};  
\draw (5,0) node[text=white]{\scriptsize 5}; 
\draw (6,0) node[text=white]{\scriptsize 6}; 

\draw (1,-.5) node[text=white]{\scriptsize 1}; 
\draw (2,-.5) node[text=black]{\scriptsize 2}; 
\draw (3,-.5) node[text=black]{\scriptsize 3}; 
\draw (4,-.5) node[text=white]{\scriptsize 4};  
\draw (5,-.5) node[text=white]{\scriptsize 5};  
\draw (6,-.5) node[text=white]{\scriptsize 6}; 

\draw (1,-1) node[text=white]{\scriptsize 1}; 
\draw (2,-1) node[text=white]{\scriptsize 2}; 
\draw (3,-1) node[text=black]{\scriptsize 3}; 
\draw (4,-1) node[text=black]{\scriptsize 4};  
\draw (5,-1) node[text=white]{\scriptsize 5};  
\draw (6,-1) node[text=white]{\scriptsize 6};  

\draw (1,-1.5) node[text=white]{\scriptsize 1}; 
\draw (2,-1.5) node[text=white]{\scriptsize 2}; 
\draw (3,-1.5) node[text=white]{\scriptsize 3}; 
\draw (4,-1.5) node[text=black]{\scriptsize 4};  
\draw (5,-1.5) node[text=black]{\scriptsize 5};  
\draw (6,-1.5) node[text=white]{\scriptsize 6}; 
\end{tikzpicture} 
\\\hline
 
$7$&
\begin{tikzpicture}
\draw [fill=blue, opacity = .5] (0.75,0.25)  rectangle (6.75,-1.75);
\draw [fill=red, opacity = .5] (0.75,-1.75) rectangle (4.25,-2.75);

\fill[white] (2,-0.5) circle (0.2);  
\fill[white] (3,-0.5) circle (0.2); 
\fill[white] (3,-1) circle (0.2);  
\fill[white] (4,-1) circle (0.2); 
\fill[white] (4,-1.5) circle (0.2);  
\fill[white] (5,-1.5) circle (0.2); 

\fill[white] (2,-2.5) circle (0.2);  
\fill[white] (3,-2.5) circle (0.2);

\fill[black] (1,0) circle (0.2); 
\fill[black] (2,0) circle (0.2);  
\fill[black] (3,0) circle (0.2); 
\fill[black] (4,0) circle (0.2);  
\fill[black] (5,0) circle (0.2);  
\fill[black] (6,0) circle (0.2);  
\fill[black] (7,0) circle (0.2);  
\draw[-, line width=.1em] (1.2,0) -- (1.8,0);  
\draw[-, line width=.1em] (2.2,0) -- (2.8,0);  
\draw[-, line width=.1em] (3.2,0) -- (3.8,0);  
\draw[-, line width=.1em] (4.2,0) -- (4.8,0);  
\draw[-, line width=.1em] (5.2,0) -- (5.8,0);  
\draw[-, line width=.1em] (6.2,0) -- (6.8,0);  
\draw[line width=.1em] (1,0) circle (0.2);  
\draw[line width=.1em] (2,0) circle (0.2);  
\draw[line width=.1em] (3,0) circle (0.2);  
\draw[line width=.1em] (4,0) circle (0.2);  
\draw[line width=.1em] (5,0) circle (0.2);  
\draw[line width=.1em] (6,0) circle (0.2);  
\draw[line width=.1em] (7,0) circle (0.2);  
\fill[black] (1,-.5) circle (0.2); 
\fill[black] (4,-.5) circle (0.2);  
\fill[black] (5,-.5) circle (0.2);  
\fill[black] (6,-.5) circle (0.2);  
\fill[black] (7,-.5) circle (0.2);  
\draw[-, line width=.1em] (1.2,-.5) -- (1.8,-.5);  
\draw[-, line width=.1em] (2.2,-.5) -- (2.8,-.5);  
\draw[-, line width=.1em] (3.2,-.5) -- (3.8,-.5);  
\draw[-, line width=.1em] (4.2,-.5) -- (4.8,-.5);  
\draw[-, line width=.1em] (5.2,-.5) -- (5.8,-.5);  
\draw[-, line width=.1em] (6.2,-.5) -- (6.8,-.5);  
\draw[line width=.1em] (1,-.5) circle (0.2);  
\draw[line width=.1em] (2,-.5) circle (0.2);  
\draw[line width=.1em] (3,-.5) circle (0.2);  
\draw[line width=.1em] (4,-.5) circle (0.2);  
\draw[line width=.1em] (5,-.5) circle (0.2);  
\draw[line width=.1em] (6,-.5) circle (0.2);  
\draw[line width=.1em] (7,-.5) circle (0.2);  
\fill[black] (1,-1) circle (0.2); 
\fill[black] (2,-1) circle (0.2);  
\fill[black] (5,-1) circle (0.2);  
\fill[black] (6,-1) circle (0.2);  
\fill[black] (7,-1) circle (0.2);  
\draw[-, line width=.1em] (1.2,-1) -- (1.8,-1);  
\draw[-, line width=.1em] (2.2,-1) -- (2.8,-1);  
\draw[-, line width=.1em] (3.2,-1) -- (3.8,-1);  
\draw[-, line width=.1em] (4.2,-1) -- (4.8,-1);  
\draw[-, line width=.1em] (5.2,-1) -- (5.8,-1);  
\draw[-, line width=.1em] (6.2,-1) -- (6.8,-1);  
\draw[line width=.1em] (1,-1) circle (0.2);  
\draw[line width=.1em] (2,-1) circle (0.2);  
\draw[line width=.1em] (3,-1) circle (0.2);  
\draw[line width=.1em] (4,-1) circle (0.2);  
\draw[line width=.1em] (5,-1) circle (0.2);  
\draw[line width=.1em] (6,-1) circle (0.2);  
\draw[line width=.1em] (7,-1) circle (0.2);  
\fill[black] (1,-1.5) circle (0.2); 
\fill[black] (2,-1.5) circle (0.2);  
\fill[black] (3,-1.5) circle (0.2); 
\fill[black] (6,-1.5) circle (0.2);  
\fill[black] (7,-1.5) circle (0.2);  
\draw[-, line width=.1em] (1.2,-1.5) -- (1.8,-1.5);  
\draw[-, line width=.1em] (2.2,-1.5) -- (2.8,-1.5);  
\draw[-, line width=.1em] (3.2,-1.5) -- (3.8,-1.5);  
\draw[-, line width=.1em] (4.2,-1.5) -- (4.8,-1.5);  
\draw[-, line width=.1em] (5.2,-1.5) -- (5.8,-1.5);  
\draw[-, line width=.1em] (6.2,-1.5) -- (6.8,-1.5);  
\draw[line width=.1em] (1,-1.5) circle (0.2);  
\draw[line width=.1em] (2,-1.5) circle (0.2);  
\draw[line width=.1em] (3,-1.5) circle (0.2);  
\draw[line width=.1em] (4,-1.5) circle (0.2);  
\draw[line width=.1em] (5,-1.5) circle (0.2);  
\draw[line width=.1em] (6,-1.5) circle (0.2);  
\draw[line width=.1em] (7,-1.5) circle (0.2);  
\fill[black] (1,-2) circle (0.2); 
\fill[black] (2,-2) circle (0.2);  
\fill[black] (3,-2) circle (0.2); 
\fill[black] (4,-2) circle (0.2);  
\fill[black] (7,-2) circle (0.2);  
\draw[-, line width=.1em] (1.2,-2) -- (1.8,-2);  
\draw[-, line width=.1em] (2.2,-2) -- (2.8,-2);  
\draw[-, line width=.1em] (3.2,-2) -- (3.8,-2);  
\draw[-, line width=.1em] (4.2,-2) -- (4.8,-2);  
\draw[-, line width=.1em] (5.2,-2) -- (5.8,-2);  
\draw[-, line width=.1em] (6.2,-2) -- (6.8,-2);  
\draw[line width=.1em] (1,-2) circle (0.2);  
\draw[line width=.1em] (2,-2) circle (0.2);  
\draw[line width=.1em] (3,-2) circle (0.2);  
\draw[line width=.1em] (4,-2) circle (0.2);  
\draw[line width=.1em] (5,-2) circle (0.2);  
\draw[line width=.1em] (6,-2) circle (0.2);  
\draw[line width=.1em] (7,-2) circle (0.2);  
\fill[black] (1,-2.5) circle (0.2); 
\fill[black] (4,-2.5) circle (0.2);  
\fill[black] (7,-2.5) circle (0.2);  
\draw[-, line width=.1em] (1.2,-2.5) -- (1.8,-2.5);  
\draw[-, line width=.1em] (2.2,-2.5) -- (2.8,-2.5);  
\draw[-, line width=.1em] (3.2,-2.5) -- (3.8,-2.5);  
\draw[-, line width=.1em] (4.2,-2.5) -- (4.8,-2.5);  
\draw[-, line width=.1em] (5.2,-2.5) -- (5.8,-2.5);  
\draw[-, line width=.1em] (6.2,-2.5) -- (6.8,-2.5);  
\draw[line width=.1em] (1,-2.5) circle (0.2);  
\draw[line width=.1em] (2,-2.5) circle (0.2);  
\draw[line width=.1em] (3,-2.5) circle (0.2);  
\draw[line width=.1em] (4,-2.5) circle (0.2);  
\draw[line width=.1em] (5,-2.5) circle (0.2);  
\draw[line width=.1em] (6,-2.5) circle (0.2);  
\draw[line width=.1em] (7,-2.5) circle (0.2);

\draw (1,0) node[text=white]{\scriptsize 1}; 
\draw (2,0) node[text=white]{\scriptsize 2}; 
\draw (3,0) node[text=white]{\scriptsize 3}; 
\draw (4,0) node[text=white]{\scriptsize 4};  
\draw (5,0) node[text=white]{\scriptsize 5}; 
\draw (6,0) node[text=white]{\scriptsize 6};  
\draw (7,0) node[text=white]{\scriptsize 7}; 

\draw (1,-.5) node[text=white]{\scriptsize 1}; 
\draw (2,-.5) node[text=black]{\scriptsize 2}; 
\draw (3,-.5) node[text=black]{\scriptsize 3}; 
\draw (4,-.5) node[text=white]{\scriptsize 4};  
\draw (5,-.5) node[text=white]{\scriptsize 5};  
\draw (6,-.5) node[text=white]{\scriptsize 6};  
\draw (7,-.5) node[text=white]{\scriptsize 7}; 

\draw (1,-1) node[text=white]{\scriptsize 1}; 
\draw (2,-1) node[text=white]{\scriptsize 2}; 
\draw (3,-1) node[text=black]{\scriptsize 3}; 
\draw (4,-1) node[text=black]{\scriptsize 4};  
\draw (5,-1) node[text=white]{\scriptsize 5};  
\draw (6,-1) node[text=white]{\scriptsize 6};  
\draw (7,-1) node[text=white]{\scriptsize 7}; 

\draw (1,-1.5) node[text=white]{\scriptsize 1}; 
\draw (2,-1.5) node[text=white]{\scriptsize 2}; 
\draw (3,-1.5) node[text=white]{\scriptsize 3}; 
\draw (4,-1.5) node[text=black]{\scriptsize 4};  
\draw (5,-1.5) node[text=black]{\scriptsize 5};  
\draw (6,-1.5) node[text=white]{\scriptsize 6}; 
\draw (7,-1.5) node[text=white]{\scriptsize 7};

\draw (1,-2) node[text=white]{\scriptsize 1}; 
\draw (2,-2) node[text=white]{\scriptsize 2}; 
\draw (3,-2) node[text=white]{\scriptsize 3}; 
\draw (4,-2) node[text=white]{\scriptsize 4};  
\draw (5,-2) node[text=black]{\scriptsize 5};  
\draw (6,-2) node[text=black]{\scriptsize 6};  
\draw (7,-2) node[text=white]{\scriptsize 7};  

\draw (1,-2.5) node[text=white]{\scriptsize 1}; 
\draw (2,-2.5) node[text=black]{\scriptsize 2}; 
\draw (3,-2.5) node[text=black]{\scriptsize 3}; 
\draw (4,-2.5) node[text=white]{\scriptsize 4};  
\draw (5,-2.5) node[text=black]{\scriptsize 5};  
\draw (6,-2.5) node[text=black]{\scriptsize 6};  
\draw (7,-2.5) node[text=white]{\scriptsize 7};  
\end{tikzpicture}  
 \\\hline 
$8$&
\begin{tikzpicture}
\draw [fill=blue, opacity = .5] (0.75,0.25)  rectangle (7.25,-2.75);
\draw [fill=red, opacity = .5] (0.75,-2.75) rectangle (5.25,-4.25);

\fill[white] (2,-0.5) circle (0.2);  
\fill[white] (3,-0.5) circle (0.2); 
\fill[white] (3,-1) circle (0.2);  
\fill[white] (4,-1) circle (0.2); 
\fill[white] (4,-1.5) circle (0.2);  
\fill[white] (5,-1.5) circle (0.2); 
\fill[white] (5,-2) circle (0.2); 
\fill[white] (6,-2) circle (0.2); 

\fill[white] (2,-2.5) circle (0.2);  
\fill[white] (3,-2.5) circle (0.2); 
\fill[white] (5,-2.5) circle (0.2);  
\fill[white] (6,-2.5) circle (0.2);

\fill[white] (2,-3.5) circle (0.2);  
\fill[white] (3,-3.5) circle (0.2); 
\fill[white] (3,-4) circle (0.2);  
\fill[white] (4,-4) circle (0.2);

\fill[black] (1,0) circle (0.2); 
\fill[black] (2,0) circle (0.2);  
\fill[black] (3,0) circle (0.2); 
\fill[black] (4,0) circle (0.2);  
\fill[black] (5,0) circle (0.2);  
\fill[black] (6,0) circle (0.2);  
\fill[black] (7,0) circle (0.2);  
\draw[-, line width=.1em] (1.2,0) -- (1.8,0);  
\draw[-, line width=.1em] (2.2,0) -- (2.8,0);  
\draw[-, line width=.1em] (3.2,0) -- (3.8,0);  
\draw[-, line width=.1em] (4.2,0) -- (4.8,0);  
\draw[-, line width=.1em] (5.2,0) -- (5.8,0);  
\draw[-, line width=.1em] (6.2,0) -- (6.8,0);  
\draw[line width=.1em] (1,0) circle (0.2);  
\draw[line width=.1em] (2,0) circle (0.2);  
\draw[line width=.1em] (3,0) circle (0.2);  
\draw[line width=.1em] (4,0) circle (0.2);  
\draw[line width=.1em] (5,0) circle (0.2);  
\draw[line width=.1em] (6,0) circle (0.2);  
\draw[line width=.1em] (7,0) circle (0.2);  
\fill[black] (1,-.5) circle (0.2); 
\fill[black] (4,-.5) circle (0.2);  
\fill[black] (5,-.5) circle (0.2);  
\fill[black] (6,-.5) circle (0.2);  
\fill[black] (7,-.5) circle (0.2);  
\draw[-, line width=.1em] (1.2,-.5) -- (1.8,-.5);  
\draw[-, line width=.1em] (2.2,-.5) -- (2.8,-.5);  
\draw[-, line width=.1em] (3.2,-.5) -- (3.8,-.5);  
\draw[-, line width=.1em] (4.2,-.5) -- (4.8,-.5);  
\draw[-, line width=.1em] (5.2,-.5) -- (5.8,-.5);  
\draw[-, line width=.1em] (6.2,-.5) -- (6.8,-.5);  
\draw[line width=.1em] (1,-.5) circle (0.2);  
\draw[line width=.1em] (2,-.5) circle (0.2);  
\draw[line width=.1em] (3,-.5) circle (0.2);  
\draw[line width=.1em] (4,-.5) circle (0.2);  
\draw[line width=.1em] (5,-.5) circle (0.2);  
\draw[line width=.1em] (6,-.5) circle (0.2);  
\draw[line width=.1em] (7,-.5) circle (0.2);  
\fill[black] (1,-1) circle (0.2); 
\fill[black] (2,-1) circle (0.2);  
\fill[black] (5,-1) circle (0.2);  
\fill[black] (6,-1) circle (0.2);  
\fill[black] (7,-1) circle (0.2);  
\draw[-, line width=.1em] (1.2,-1) -- (1.8,-1);  
\draw[-, line width=.1em] (2.2,-1) -- (2.8,-1);  
\draw[-, line width=.1em] (3.2,-1) -- (3.8,-1);  
\draw[-, line width=.1em] (4.2,-1) -- (4.8,-1);  
\draw[-, line width=.1em] (5.2,-1) -- (5.8,-1);  
\draw[-, line width=.1em] (6.2,-1) -- (6.8,-1);  
\draw[line width=.1em] (1,-1) circle (0.2);  
\draw[line width=.1em] (2,-1) circle (0.2);  
\draw[line width=.1em] (3,-1) circle (0.2);  
\draw[line width=.1em] (4,-1) circle (0.2);  
\draw[line width=.1em] (5,-1) circle (0.2);  
\draw[line width=.1em] (6,-1) circle (0.2);  
\draw[line width=.1em] (7,-1) circle (0.2);  
\fill[black] (1,-1.5) circle (0.2); 
\fill[black] (2,-1.5) circle (0.2);  
\fill[black] (3,-1.5) circle (0.2); 
\fill[black] (6,-1.5) circle (0.2);  
\fill[black] (7,-1.5) circle (0.2);  
\draw[-, line width=.1em] (1.2,-1.5) -- (1.8,-1.5);  
\draw[-, line width=.1em] (2.2,-1.5) -- (2.8,-1.5);  
\draw[-, line width=.1em] (3.2,-1.5) -- (3.8,-1.5);  
\draw[-, line width=.1em] (4.2,-1.5) -- (4.8,-1.5);  
\draw[-, line width=.1em] (5.2,-1.5) -- (5.8,-1.5);  
\draw[-, line width=.1em] (6.2,-1.5) -- (6.8,-1.5);  
\draw[line width=.1em] (1,-1.5) circle (0.2);  
\draw[line width=.1em] (2,-1.5) circle (0.2);  
\draw[line width=.1em] (3,-1.5) circle (0.2);  
\draw[line width=.1em] (4,-1.5) circle (0.2);  
\draw[line width=.1em] (5,-1.5) circle (0.2);  
\draw[line width=.1em] (6,-1.5) circle (0.2);  
\draw[line width=.1em] (7,-1.5) circle (0.2);  
\fill[black] (1,-2) circle (0.2); 
\fill[black] (2,-2) circle (0.2);  
\fill[black] (3,-2) circle (0.2); 
\fill[black] (4,-2) circle (0.2);  
\fill[black] (7,-2) circle (0.2);  
\draw[-, line width=.1em] (1.2,-2) -- (1.8,-2);  
\draw[-, line width=.1em] (2.2,-2) -- (2.8,-2);  
\draw[-, line width=.1em] (3.2,-2) -- (3.8,-2);  
\draw[-, line width=.1em] (4.2,-2) -- (4.8,-2);  
\draw[-, line width=.1em] (5.2,-2) -- (5.8,-2);  
\draw[-, line width=.1em] (6.2,-2) -- (6.8,-2);  
\draw[line width=.1em] (1,-2) circle (0.2);  
\draw[line width=.1em] (2,-2) circle (0.2);  
\draw[line width=.1em] (3,-2) circle (0.2);  
\draw[line width=.1em] (4,-2) circle (0.2);  
\draw[line width=.1em] (5,-2) circle (0.2);  
\draw[line width=.1em] (6,-2) circle (0.2);  
\draw[line width=.1em] (7,-2) circle (0.2);  
\fill[black] (1,-2.5) circle (0.2); 
\fill[black] (4,-2.5) circle (0.2);  
\fill[black] (7,-2.5) circle (0.2);  
\draw[-, line width=.1em] (1.2,-2.5) -- (1.8,-2.5);  
\draw[-, line width=.1em] (2.2,-2.5) -- (2.8,-2.5);  
\draw[-, line width=.1em] (3.2,-2.5) -- (3.8,-2.5);  
\draw[-, line width=.1em] (4.2,-2.5) -- (4.8,-2.5);  
\draw[-, line width=.1em] (5.2,-2.5) -- (5.8,-2.5);  
\draw[-, line width=.1em] (6.2,-2.5) -- (6.8,-2.5);  
\draw[line width=.1em] (1,-2.5) circle (0.2);  
\draw[line width=.1em] (2,-2.5) circle (0.2);  
\draw[line width=.1em] (3,-2.5) circle (0.2);  
\draw[line width=.1em] (4,-2.5) circle (0.2);  
\draw[line width=.1em] (5,-2.5) circle (0.2);  
\draw[line width=.1em] (6,-2.5) circle (0.2);  
\draw[line width=.1em] (7,-2.5) circle (0.2); 
\fill[black] (1,-3) circle (0.2); 
\fill[black] (2,-3) circle (0.2);  
\fill[black] (3,-3) circle (0.2); 
\fill[black] (4,-3) circle (0.2);  
\fill[black] (5,-3) circle (0.2);  
\draw[-, line width=.1em] (1.2,-3) -- (1.8,-3);  
\draw[-, line width=.1em] (2.2,-3) -- (2.8,-3);  
\draw[-, line width=.1em] (3.2,-3) -- (3.8,-3);  
\draw[-, line width=.1em] (4.2,-3) -- (4.8,-3);  
\draw[-, line width=.1em] (5.2,-3) -- (5.8,-3);  
\draw[-, line width=.1em] (6.2,-3) -- (6.8,-3);  
\draw[line width=.1em] (1,-3) circle (0.2);  
\draw[line width=.1em] (2,-3) circle (0.2);  
\draw[line width=.1em] (3,-3) circle (0.2);  
\draw[line width=.1em] (4,-3) circle (0.2);  
\draw[line width=.1em] (5,-3) circle (0.2);  
\draw[line width=.1em] (6,-3) circle (0.2);  
\draw[line width=.1em] (7,-3) circle (0.2);  

\fill[black] (1,-3.5) circle (0.2); 
\fill[black] (4,-3.5) circle (0.2);  
\fill[black] (5,-3.5) circle (0.2);  
\draw[-, line width=.1em] (1.2,-3.5) -- (1.8,-3.5);  
\draw[-, line width=.1em] (2.2,-3.5) -- (2.8,-3.5);  
\draw[-, line width=.1em] (3.2,-3.5) -- (3.8,-3.5);  
\draw[-, line width=.1em] (4.2,-3.5) -- (4.8,-3.5);  
\draw[-, line width=.1em] (5.2,-3.5) -- (5.8,-3.5);  
\draw[-, line width=.1em] (6.2,-3.5) -- (6.8,-3.5);  
\draw[line width=.1em] (1,-3.5) circle (0.2);  
\draw[line width=.1em] (2,-3.5) circle (0.2);  
\draw[line width=.1em] (3,-3.5) circle (0.2);  
\draw[line width=.1em] (4,-3.5) circle (0.2);  
\draw[line width=.1em] (5,-3.5) circle (0.2);  
\draw[line width=.1em] (6,-3.5) circle (0.2);  
\draw[line width=.1em] (7,-3.5) circle (0.2); 

\fill[black] (1,-4) circle (0.2); 
\fill[black] (2,-4) circle (0.2);  
\fill[black] (5,-4) circle (0.2);  
\draw[-, line width=.1em] (1.2,-4) -- (1.8,-4);  
\draw[-, line width=.1em] (2.2,-4) -- (2.8,-4);  
\draw[-, line width=.1em] (3.2,-4) -- (3.8,-4);  
\draw[-, line width=.1em] (4.2,-4) -- (4.8,-4);  
\draw[-, line width=.1em] (5.2,-4) -- (5.8,-4);  
\draw[-, line width=.1em] (6.2,-4) -- (6.8,-4);  
\draw[line width=.1em] (1,-4) circle (0.2);  
\draw[line width=.1em] (2,-4) circle (0.2);  
\draw[line width=.1em] (3,-4) circle (0.2);  
\draw[line width=.1em] (4,-4) circle (0.2);  
\draw[line width=.1em] (5,-4) circle (0.2);  
\draw[line width=.1em] (6,-4) circle (0.2);  
\draw[line width=.1em] (7,-4) circle (0.2);

\draw (1,0) node[text=white]{\scriptsize 1}; 
\draw (2,0) node[text=white]{\scriptsize 2}; 
\draw (3,0) node[text=white]{\scriptsize 3}; 
\draw (4,0) node[text=white]{\scriptsize 4};  
\draw (5,0) node[text=white]{\scriptsize 5}; 
\draw (6,0) node[text=white]{\scriptsize 6};  
\draw (7,0) node[text=white]{\scriptsize 7}; 

\draw (1,-.5) node[text=white]{\scriptsize 1}; 
\draw (2,-.5) node[text=black]{\scriptsize 2}; 
\draw (3,-.5) node[text=black]{\scriptsize 3}; 
\draw (4,-.5) node[text=white]{\scriptsize 4};  
\draw (5,-.5) node[text=white]{\scriptsize 5};  
\draw (6,-.5) node[text=white]{\scriptsize 6};  
\draw (7,-.5) node[text=white]{\scriptsize 7}; 

\draw (1,-1) node[text=white]{\scriptsize 1}; 
\draw (2,-1) node[text=white]{\scriptsize 2}; 
\draw (3,-1) node[text=black]{\scriptsize 3}; 
\draw (4,-1) node[text=black]{\scriptsize 4};  
\draw (5,-1) node[text=white]{\scriptsize 5};  
\draw (6,-1) node[text=white]{\scriptsize 6};  
\draw (7,-1) node[text=white]{\scriptsize 7}; 

\draw (1,-1.5) node[text=white]{\scriptsize 1}; 
\draw (2,-1.5) node[text=white]{\scriptsize 2}; 
\draw (3,-1.5) node[text=white]{\scriptsize 3}; 
\draw (4,-1.5) node[text=black]{\scriptsize 4};  
\draw (5,-1.5) node[text=black]{\scriptsize 5};  
\draw (6,-1.5) node[text=white]{\scriptsize 6}; 
\draw (7,-1.5) node[text=white]{\scriptsize 7};

\draw (1,-2) node[text=white]{\scriptsize 1}; 
\draw (2,-2) node[text=white]{\scriptsize 2}; 
\draw (3,-2) node[text=white]{\scriptsize 3}; 
\draw (4,-2) node[text=white]{\scriptsize 4};  
\draw (5,-2) node[text=black]{\scriptsize 5};  
\draw (6,-2) node[text=black]{\scriptsize 6};  
\draw (7,-2) node[text=white]{\scriptsize 7};  

\draw (1,-2.5) node[text=white]{\scriptsize 1}; 
\draw (2,-2.5) node[text=black]{\scriptsize 2}; 
\draw (3,-2.5) node[text=black]{\scriptsize 3}; 
\draw (4,-2.5) node[text=white]{\scriptsize 4};  
\draw (5,-2.5) node[text=black]{\scriptsize 5};  
\draw (6,-2.5) node[text=black]{\scriptsize 6};  
\draw (7,-2.5) node[text=white]{\scriptsize 7};  

\draw (1,-3) node[text=white]{\scriptsize 1}; 
\draw (2,-3) node[text=white]{\scriptsize 2}; 
\draw (3,-3) node[text=white]{\scriptsize 3}; 
\draw (4,-3) node[text=white]{\scriptsize 4};  
\draw (5,-3) node[text=white]{\scriptsize 5};  
\draw (6,-3) node[text=black]{\scriptsize 6};  
\draw (7,-3) node[text=black]{\scriptsize 7};  

\draw (1,-3.5) node[text=white]{\scriptsize 1}; 
\draw (2,-3.5) node[text=black]{\scriptsize 2}; 
\draw (3,-3.5) node[text=black]{\scriptsize 3}; 
\draw (4,-3.5) node[text=white]{\scriptsize 4};  
\draw (5,-3.5) node[text=white]{\scriptsize 5};  
\draw (6,-3.5) node[text=black]{\scriptsize 6};  
\draw (7,-3.5) node[text=black]{\scriptsize 7};  

\draw (1,-4) node[text=white]{\scriptsize 1}; 
\draw (2,-4) node[text=white]{\scriptsize 2}; 
\draw (3,-4) node[text=black]{\scriptsize 3}; 
\draw (4,-4) node[text=black]{\scriptsize 4};  
\draw (5,-4) node[text=white]{\scriptsize 5};  
\draw (6,-4) node[text=black]{\scriptsize 6};  
\draw (7,-4) node[text=black]{\scriptsize 7};

\fill[black] (8,0) circle (0.2);  
\fill[black] (8,-.5) circle (0.2);  
\fill[black] (8,-1) circle (0.2);  
\fill[black] (8,-1.5) circle (0.2);  
\fill[black] (8,-2) circle (0.2);  
\fill[black] (8,-2.5) circle (0.2);  
\fill[black] (8,-3) circle (0.2);  
\fill[black] (8,-3.5) circle (0.2);  
\fill[black] (8,-4) circle (0.2);  
\draw[-, line width=.1em] (7.2,0) -- (7.8,0);  
\draw[-, line width=.1em] (7.2,-.5 ) -- (7.8,-.5);  
\draw[-, line width=.1em] (7.2,-1  ) -- (7.8,-1);  
\draw[-, line width=.1em] (7.2,-1.5) -- (7.8,-1.5);  
\draw[-, line width=.1em] (7.2,-2  ) -- (7.8,-2);  
\draw[-, line width=.1em] (7.2,-2.5) -- (7.8,-2.5);  
\draw[-, line width=.1em] (7.2,-3  ) -- (7.8,-3);  
\draw[-, line width=.1em] (7.2,-3.5) -- (7.8,-3.5);  
\draw[-, line width=.1em] (7.2,-4  ) -- (7.8,-4);  
\draw (8,-0) node[text=white]{\scriptsize 8}; 
\draw (8,-.5) node[text=white]{\scriptsize 8}; 
\draw (8,-1) node[text=white]{\scriptsize 8}; 
\draw (8,-1.5) node[text=white]{\scriptsize 8}; 
\draw (8,-2) node[text=white]{\scriptsize 8}; 
\draw (8,-2.5) node[text=white]{\scriptsize 8}; 
\draw (8,-3) node[text=white]{\scriptsize 8}; 
\draw (8,-3.5) node[text=white]{\scriptsize 8}; 
\draw (8,-4) node[text=white]{\scriptsize 8}; 

\end{tikzpicture}  
 \\\hline
\end{tabular}

\caption{Illustration of the relation $c^{(n)}_2=c^{(n-1)}_2+c^{(n-3)}_2$ at the example of $n\in\{5,6,7,8\}$. The $c^{(n-1)}_2$ colorings of the previous line graph are highlighted in blue; similarly for $c^{(n-3)}_2 $ and red.
} \label{fig:puzzle_line_markedleaves}
\end{framed}
\end{minipage}
 \end{figure}
 The reason for this relation is the following.  
 Consider the second vertex next to the leaf on the very right (vertex $n-1$).
 If this vertex is black, there are $c^{(n-1)}_2$ allowed colorings for the remaining vertices since we can ``remove'' the rightmost vertex and obtain the same combinatorial problem on a smaller graph.
 If, on the other hand, vertex $n-1$ is white, also the neighbor to its left (vertex $n-2$) must be white.
 Similar to the first case, we can remove the three rightmost vertices and obtain a 1:1-correspondence to the $c^{(n-3)}_2$ colorings of the smaller graph with two black leaves.
This shows the claimed recursive formula.
Finally, the allowed colorings for $n$ where both leaves are white are in 1:1-correspondence with the allowed colorings for $n-2$ where the leaves are black; just remove the two white leaves of the larger graph. Thus, $c^{(n)}_0=c^{(n-2)}_2$ for $n\ge4$.
In combination, we obtain 
\begin{align} \label{eq:puzzle_line_1}
 \ell^n_n[\Psi^n_\mathrm{line}] &= c_2^{(n)}+ c_1^{(n)}+ c_0^{(n)} =  c_2^{(n)}+ 2c_2^{(n-1)} + c_2^{(n-2)}
 \\ \label{eq:puzzle_line_2}
 &= c_2^{(n-1)} + c_2^{(n-3)} + 2 \left(c_2^{(n-2)} + c_2^{(n-4)}\right) + c_2^{(n-3)} + c_2^{(n-5)} 
 \\ \label{eq:puzzle_line_3}
 &=  c_2^{(n-1)} + c_2^{(n-3)}  + c_1^{(n-1)} + c_1^{(n-3)}  +c_0^{(n-1)} + c_0^{(n-3)} 
 \\ \label{eq:puzzle_line_4}  &=  \ell^{n-1}_{n-1}[\Psi^{n-1}_\mathrm{line}] +
 \ell^{n-3}_{n-3}[\Psi^{n-3}_\mathrm{line}],
\end{align}
where we have used $c^{(m)}_2=c^{(m-1)}_2+c^{(m-3)}_2$ with $m\in\{n-2,n-1,n\}$ to get from line~\eqref{eq:puzzle_line_1} to~\eqref{eq:puzzle_line_2}.
This finishes the proof.
 \end{proof}

 \subsubsection{Solution of the puzzle for ring graphs} 
 \label{sec:6.2.4} 
 
 Here, we will prove that the full-body sector length of the ring graph state is given by 
  \begin{align}   \tag{\ref{eq:lnn_ring}}
  \ell^n_n[\Psi^n_\mathrm{ring}] &= 1 + \sum_{k=1}^{\lfloor n/3 \rfloor} 
{\genfrac{(}{)}{0pt}{0}{n-2k-1}{k-1}} \frac{n}{k}  .
 \end{align}
 The first term ``1'' in Eq.~\eqref{eq:lnn_ring} corresponds to the coloring where all vertices are black.
 For the ring graph, the rule of the puzzle implies that white vertices can only come in pairs of exactly two, i.e., two such pairs are separated by at least one black vertex.
 The number $k$ over which the sum in Eq.~\eqref{eq:lnn_ring} runs should be regarded as the number of white pairs on a ring graph. Note  that  up to $k=\lfloor n/3 \rfloor$ white pairs fit on the ring. Hence, the validity of Eq.~\eqref{eq:lnn_ring} directly follows from the following lemma which we prove in Appendix~\ref{app:proof_choices_ring} using the theory of group actions.

 \lemma\label{lem:choices_ring}
 \emph{Let $N(n,k)$ be the number of choices to place $k$ pairs of white vertices on an $n$-vertex ring graph such that any two such pairs are separated by at least one black vertex. Then, it holds $N(n,k)={\genfrac{(}{)}{0pt}{1}{n-2k-1}{k-1}} \frac{n}{k}$.}\\

According to \emph{The On-Line Encyclopedia of Integer Sequences~\textsuperscript{\textregistered}}~\href{https://oeis.org/A001609}{\texttt{[OEIS:A001609]}}, 
the sequence fulfills
 $\ell^n_n[\Psi_n^\mathrm{ring}]= \ell^{n-1}_{n-1}[\Psi^{n-1}_\mathrm{ring}] +
 \ell^{n-3}_{n-3}[\Psi^{n-3}_\mathrm{ring}]$ for all $n\ge6$ 
 which is the same recursion relation as for the line graph.
 However, the starting values 
 $\ell^3_3[\Psi_3^\mathrm{ring}]= 4$,
 $\ell^4_4[\Psi_4^\mathrm{ring}]= 5$, and
 $\ell^5_5[\Psi_5^\mathrm{ring}]= 6$ are smaller for the ring graph. 
 This shows  $\ell^n_n[\Psi_n^\mathrm{ring}] <  \ell^n_n[\Psi_n^\mathrm{line}]$ for all $n\ge 4$.
 We conjecture that ring graph states have the smallest full-body sector length among all graph states with a connected graph.

 \newpage  
\subsection{Investigation of noise thresholds}
\label{sec:6.3}

For qubit graph states with a connected graph, the maximal overlap with the biseparable states is given by~$\alpha=1/2$~\cite{GueTo09}. Hence, the global white noise threshold from Eq.~\eqref{eq:threshold_WitGME} simplifies to
\begin{align} \label{eq:threshold_WitGME_qubit}
p^\mathrm{Wit:GME}_\mathrm{glob}
=\frac{2^n}{2^{n+1}-2}=\frac{1}{2}+\frac{1}{2^{n+1}-2}.
 \end{align}
Furthermore, there is an explicit entanglement distillation protocol~\cite{HeDuerBri05} which shows that every qubit graph state can tolerate local white noise up to at least
\begin{align} \label{eq:loc_thresh_dist}
 p^\mathrm{Dist}_\mathrm{loc} = 1- {2^{-{2}/({m_i+m_j+2})}},
\end{align}
where $m_i$ and $m_j$ are the maximal degrees (number of neighbors) of two neighboring vertices $i$ and $j$.

Here, we compare these known thresholds to our new thresholds from Corollary~\ref{cor:crit_noise_sector_lengths}. 
Thereby, we will compare the full-body sector length of the pure graph states, recall Sec.~\ref{sec:6.2}, to the bounds
 $b_n^{(1,1,\ldots,1)} = 1$, 
 $b_n^{(2,1,\ldots,1)} = 3$, 
 $b_n^{(3,1,\ldots,1)} = 4$,
 $b_n^{(4,1,\ldots,1)} = 9$, and
\begin{align}
 b_n^{(n-1,1)} = 2^{n-2} + 
 \frac{1 +(-1)^{n-1}}{2} 
\end{align}
which follow from Eq.~\eqref{eq:sec_tensor} and the fact that there is no $n$-qubit state whose full-body sector length exceeds $2^{n-1}+ \left( \frac{1 +(-1)^n}{2} \right )$~\cite{TDLP16, EltSiew19}.
We chose these bounds as they allow us to gain new insights.
We separately discuss graph states with star, dandelion, and line graphs in Sec.~\ref{sec:6.3.1}, \ref{sec:6.3.2}, and~\ref{sec:6.3.3}, respectively, and spare a discussion of ring graph states as it would be almost the same as for line graph states.
Finally, in Sec.~\ref{sec:6.3.4}, we comment on how our methods can be generalized to arbitrary graph states.

\subsubsection{Noise thresholds for qubit GHZ states}
\label{sec:6.3.1}

\begin{figure}
\centering
\begin{minipage}{0.85\textwidth}
\begin{framed} 
\centering
\begin{minipage}{\textwidth}
   \includegraphics[height = 15.3em]{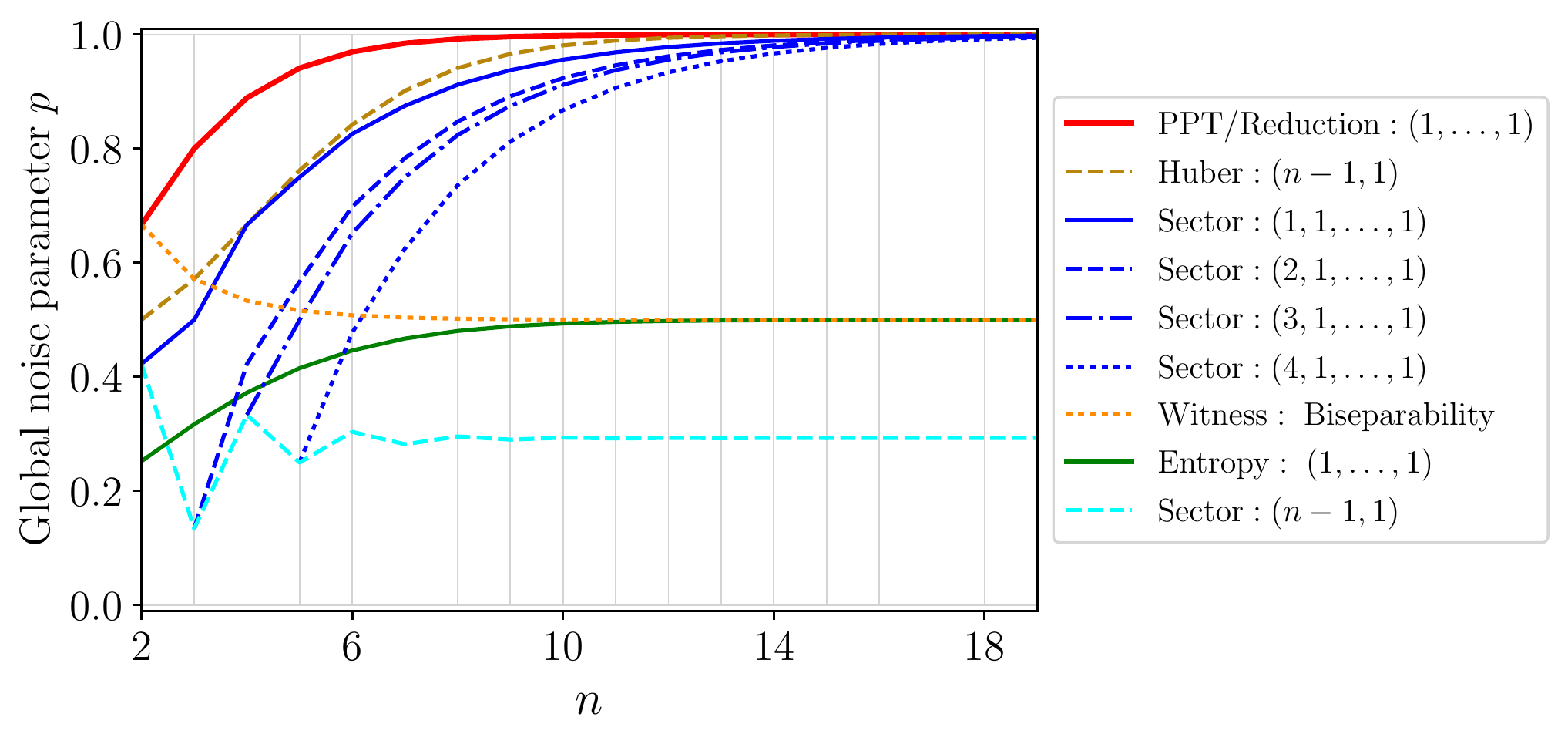} \\
   \includegraphics[height = 15.3em]{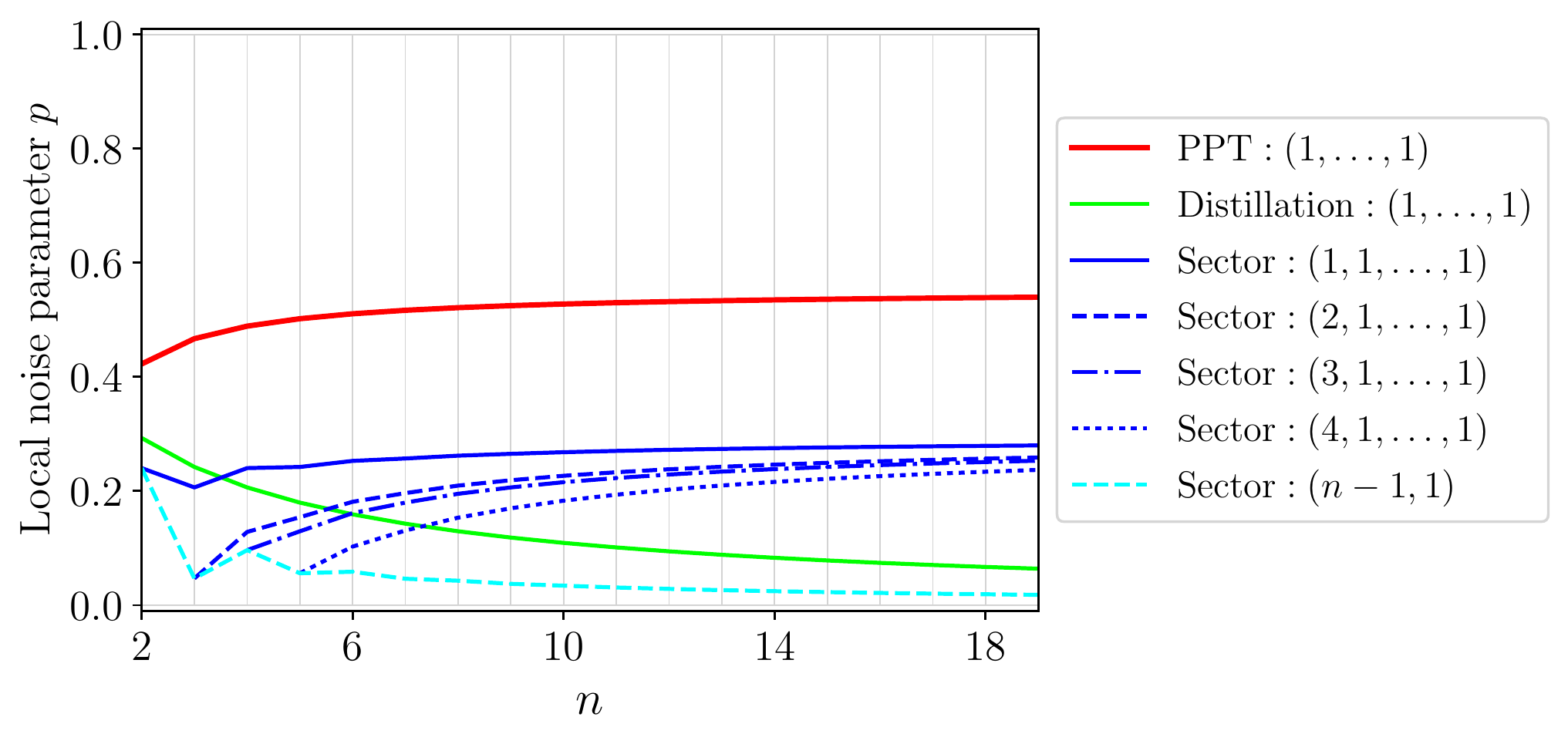} 
\end{minipage}
 
\caption{Noise thresholds for global (top) and local (bottom) white noise on $n$-qubit GHZ states. For noise parameters below a given curve, the corresponding separability type is ruled out. For example, below a solid curve, the corresponding state is not fully separable, thus entangled.}
\label{fig:qubit_GHZ_thresholds}
\end{framed}
\end{minipage}
 \end{figure}

Greenberger-Horne-Zeilinger states (LU-equivalent to star-graph states) belong to the best studied $n$-qubit states as they are simply given by an equal superposition of only two computational basis states. 
For the case of global white noise, we depict its noise thresholds in the upper part of Fig.~\ref{fig:qubit_GHZ_thresholds}.
First, consider the full-separability thresholds (solid curves). If the noise is below such a curve, $\mathrm{GHZ}^n(p):=(1-p)\ket{\mathrm{GHZ}^n_2}\bra{\mathrm{GHZ}^n_2} + p \mathbbm{1}/2^n$ is entangled.
The highest of these thresholds (red curve), $p_\mathrm{glob}^\mathrm{PPT} = 1-\frac{1}{2^{n-1}+1}$, is known for 20 years~\cite{DuCiTa99}. 
In the limit of many qubits, this threshold, as well as the full-separability threshold based on sector lengths (blue curve), converge to $1$. In contrast, the threshold based on the entropy criterion (green curve) only converges to $1/2$. This shows that our new noise threshold outperforms the entropy criterion.

Next, consider the witness threshold $p^\mathrm{Wit:GME}_\mathrm{glob}$ (dotted yellow curve) from  Eq.~\eqref{eq:threshold_WitGME_qubit} that can rule out biseparability. 
In Ref.~\cite{GueSee10} it is shown that this threshold  is tight in the sense that $\mathrm{GHZ}^n(p)$ is GME iff $p<p^\mathrm{Wit:GME}_\mathrm{glob}$.
Note that this GME-threshold can also be obtained with the approach of Ref.~\cite{HMGH10}~\footnote{There is a typo in Ref.~\cite{HMGH10}: The 
stated GME noise threshold $1-\frac{3}{2^{n-1}+3}$ contradicts the fact that threshold $\frac{1}{2}+\frac{1}{2^{n+1}-2}$ is tight~\cite{GueSee10}. Also note that in Ref.~\cite{KGWB12} a biseparable decomposition for $\mathrm{GHZ}^4(p)$ is constructed, where $p\approx 1- 0.466 = 0.534 < 0.7272 =  1-\frac{3}{2^{4-1}+3}$. 
We were informed that the correct version of the noise threshold in Ref.~\cite{HMGH10} is obtained if one replaces $3$ by $2^{n-1}-1$~\cite{PrivateHuber}.} 
 as well as with an approach based on positive maps~\cite{CHLM17}.
The results by  Huber \emph{et al.} also provide a threshold from which semiseparability can be ruled out, recall Eq.~\eqref{eq:Huber_threshold}. Here, this threshold (dashed brown curve) simplifies to
\begin{align}
 p^{\mathrm{Huber}:(n-1,1)}_\mathrm{glob} = 1 - \frac{n}{n+2^{n-1}}.
\end{align}
This curve exceeds all (blue) curves based on sector lengths; in particular, the semiseparability threshold (dashed bright blue curve) which converges to $1-1/\sqrt{2}\approx 0.29$.
Thus, our approach does not provide any new insights about the GHZ state in the case of global white noise.

In the case of local white noise, however, our approach yields a nontrivial semiseparability threshold for all $n$-qubit GHZ states as one can see in the lower part of Fig.~\ref{fig:qubit_GHZ_thresholds} (dashed bright blue curve). To our knowledge such a semiseparability threshold was not known before.
To rule out full separability (solid curves), the best known threshold (red curve) 
 is based on the Peres-Horodecki criterion. Here (for qubits), Eq.~\eqref{eq:threshold_GHZ_local_PPT} simplifies to
\begin{align}
 p_\mathrm{loc}^\mathrm{PPT} = 1- \frac{1}{\sqrt{2^{2-2/n}+1}} 
\end{align}
which was first derived in Ref.~\cite{SiKemp02}.
In the limit of many qubits, this threshold converges to $1-1/\sqrt{5}\approx 0.55$.
The sector length thresholds for ruling out ${(j,1,\ldots,1)}$-separability for a constant $j$, converge to $1-1/\sqrt{2}\approx 0.29$.
This is better than the full-separability threshold $p^\mathrm{Dist}_\mathrm{loc} = 1- 2^{-{2}/{(n+2)}}$ 
based on the entanglement distillation protocol of Ref.~\cite{HeDuerBri05} which converges to 0.

  \subsubsection{Noise thresholds for qubit dandelion-graph states}
\label{sec:6.3.2}

\begin{figure}
\centering
\begin{minipage}{0.85\textwidth}
\begin{framed} 
\centering
\begin{minipage}{\textwidth}
   \includegraphics[height = 15.3em]{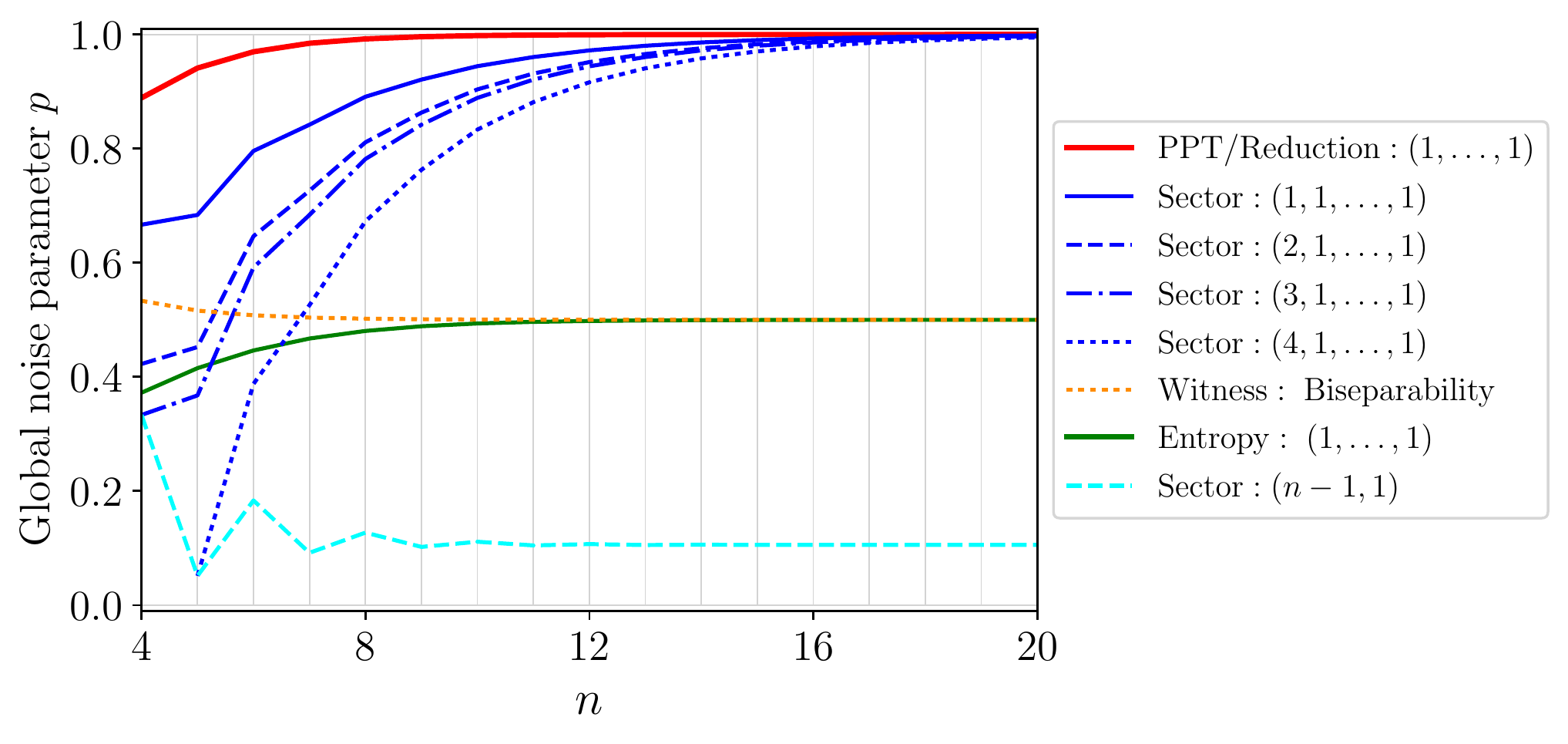} \\
   \includegraphics[height = 15.3em]{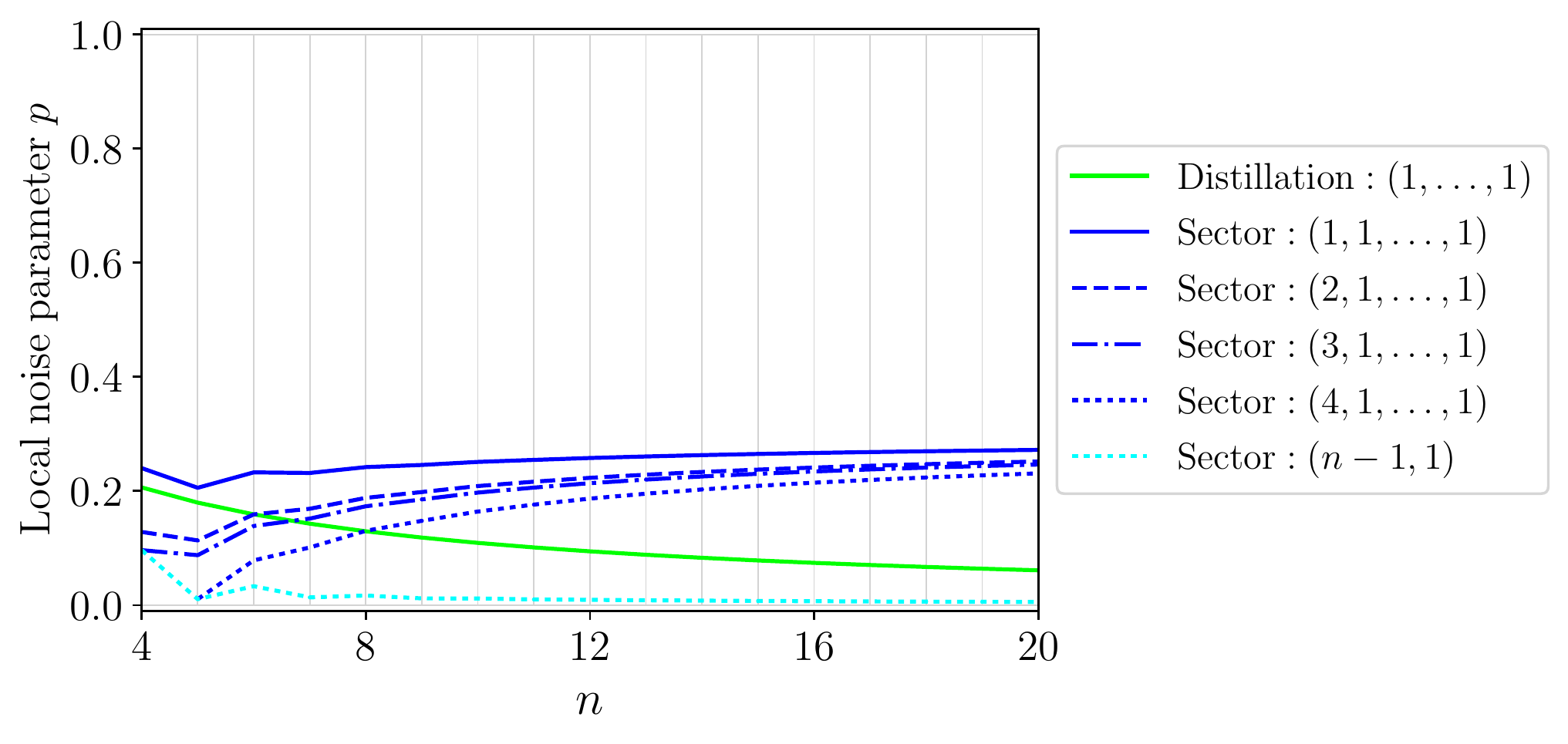} 
\end{minipage} 
\caption{Noise thresholds for global (top) and local (bottom) white noise on $n$-qubit dandelion-graph states. For noise parameters below a given curve, the corresponding separability type is ruled out. For example, below a solid curve, the corresponding state is not fully separable, thus entangled.}
\label{fig:qubit_dandelion_thresholds}
\end{framed}
\end{minipage}
 \end{figure}

As the family of dandelion graphs was defined by us, there are no entanglement thresholds particularly tailored to these states.
Still, we can compare our new thresholds to the literature thresholds for generic qubit graph states. 
In the case of global white noise, see the upper part of Fig.~\ref{fig:qubit_dandelion_thresholds},
 again the Peres-Horodecki criterion (red curve) yields the highest full-separability threshold, followed by the sector length criterion (solid blue curve), and the entropy criterion (green curve).
 As the semiseparability threshold based on sector lengths (dashed bright blue curve) is always below the biseparability threshold based on witnesses (dotted yellow curve), it does not provide any new insights. 
 However, for $n\ge 5$, sector lengths can rule out $(2,1,\ldots,1)$- and $(3,1,\ldots,1)$-separability for a larger amount of global noise than the GME-witness. 
 This comparison is not completely fair as we expect that replacing $\alpha =1/2$ in Eq.~\eqref{eq:threshold_WitGME} by the maximal overlap with the $(j,1,\ldots,1)$-separable states, where $j\in\{2,3\}$, would provide an even higher threshold to rule out $(j,1,\ldots,1)$-separability.
 However, we are not aware of an explicit expression of this overlap. 
 That is, to rule out these types of separability, the corresponding sector length thresholds are the highest which are currently known.
 
 Next, consider the case of local white noise for which we plot noise thresholds in the lower part of Fig.~\ref{fig:qubit_dandelion_thresholds}. 
 For any $n$, the sector length criterion (solid blue curve) yields a higher full-separability threshold than the distillation protocol which is here given by $p^\mathrm{Dist}_\mathrm{loc} = 1- 2^{-{2}/{(n+2)}}$ (solid green curve).
 As for GHZ-states, the sector length threshold converges to $1-1/\sqrt{2}\approx 0.29$, in the limit of many qubits, while the distillation threshold converges to 0.
 Furthermore, we obtain a nontrivial semiseparability threshold for all $n$-qubit dandelion graph states because the full-body sector length 
 $\ell^n_n[\Psi^n_\mathrm{dandelion}] = {5\times2^{n-4}+4 \left(\frac{1 +(-1)^{n-1}}{2} \right )} $ is 
 strictly larger than the corresponding semiseparability bound 
 %
 $b_n^{(n-1,1)} = 2^{n-2} + \left( \frac{1 +(-1)^{n-1}}{2} \right )$. 
 In conclusion, our sector length approach yields  new insights for dandelion graph states under local white noise.

\subsubsection{Noise thresholds for qubit line-graph states} 
\label{sec:6.3.3}

\begin{figure}
\centering
\begin{minipage}{0.85\textwidth}
\begin{framed} 
\centering
\begin{minipage}{\textwidth}
   \includegraphics[height = 15.3em]{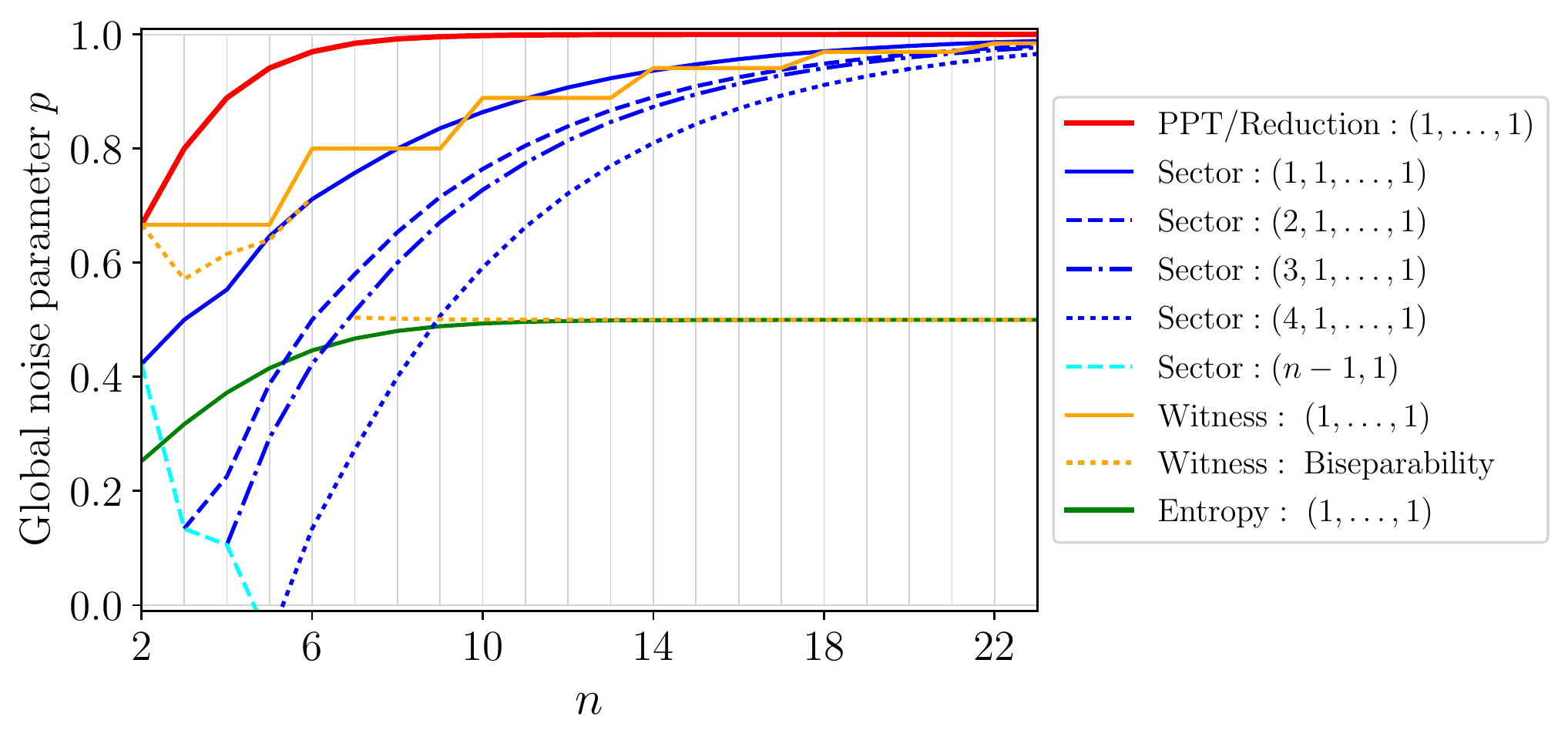} \\
   \includegraphics[height = 15.3em]{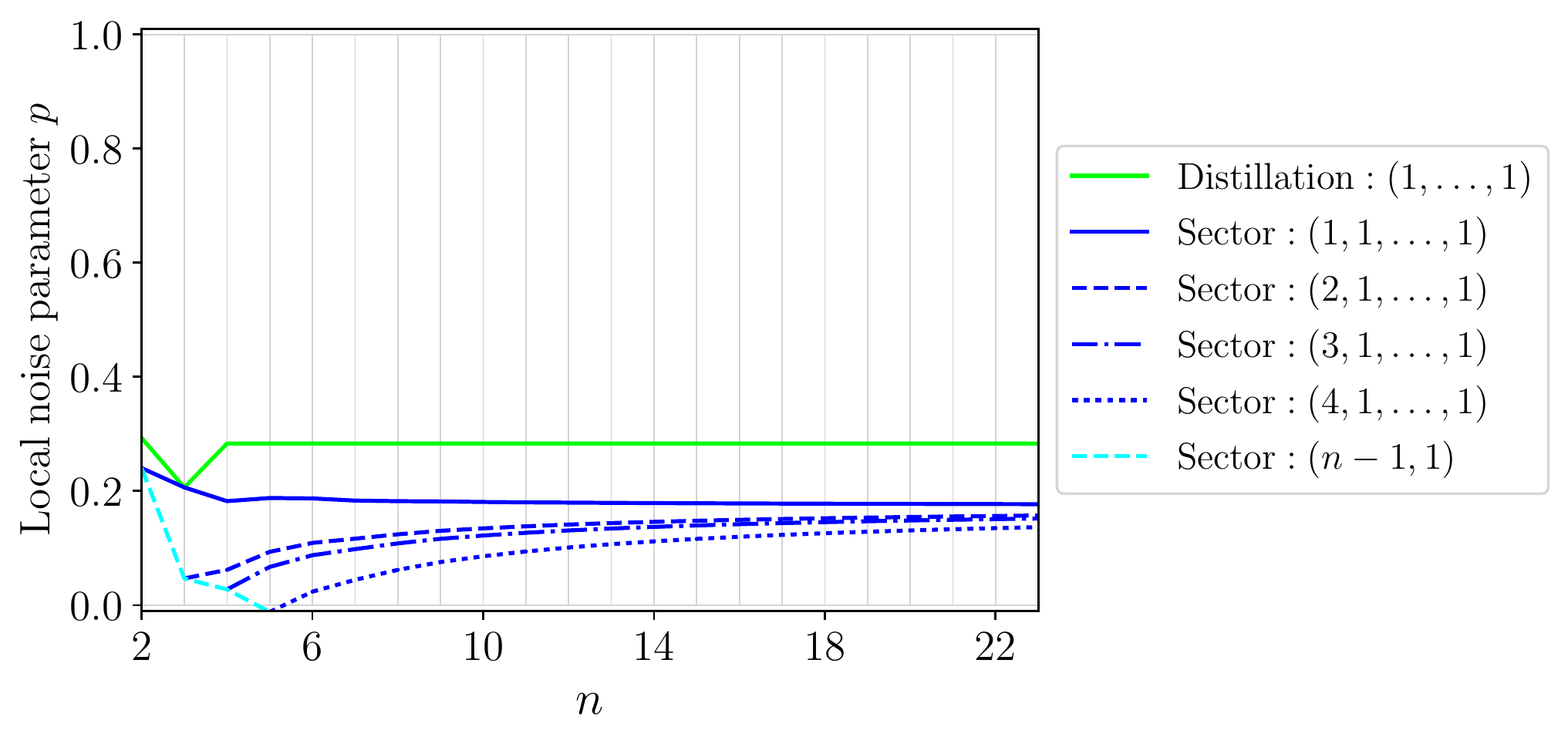} 
\end{minipage}
\caption{Noise thresholds for global (top) and local (bottom) white noise on $n$-qubit line graph states. For noise parameters below a given curve, the corresponding separability type is ruled out. For example, below a solid curve, the corresponding state is not fully separable, thus entangled.}
 \label{fig:qubit_line_thresholds}
\end{framed}
\end{minipage}
 \end{figure}

Consider the upper part of Fig.~\ref{fig:qubit_line_thresholds}, 
where we display global white noise thresholds for qubit line graph states.
For these states, there exist entanglement witnesses which rule out full separability~\cite{ToGue05}. The corresponding threshold $p^\mathrm{Wit}_\mathrm{glob} = 2^k/(2^k+1)$ is the solid yellow curve in Fig.~\ref{fig:qubit_line_thresholds}, where $k=\lfloor (n+2)/4 \rfloor$.
Also note the break in the dotted yellow curve which is due to the fact that the GME witnesses of Ref.~\cite{JuMoGue11}  outperform the generic witnesses from Eq.~\eqref{eq:generic_witness} for $n\in\{ 4,5,6\}$.

Also here, the Peres-Horodecki criterion provides the best threshold to rule out full separability (red curve). The witness threshold (solid yellow curve) and the sector length threshold (solid blue curve) perform similarly well. For large $n$, however, sector lengths are slightly superior; the smallest instance is $n=9$. 
The semiseparability threshold based on sector lengths (dashed bright blue curve) is trivial for all $n\ge 5$ because $\ell^n_n[\Psi^n_\mathrm{line}]$ is growing less quickly than the corresponding bound $b_n^{(n-1,1)}$.
As in Sec.~\ref{sec:6.3.2}, the only new insights concern $(j,1,\ldots,1)$-separability, where $j\in\{2,3,4\}$.

Now, consider local white noise thresholds for qubit line graph states which we show in the lower part of Fig.~\ref{fig:qubit_line_thresholds}.
For $n\ge4$, the distillation threshold (green curve) is independent of $n$ and constant $0.2833$, cf.~Eq.~\eqref{eq:loc_thresh_dist}~\cite{HeDuerBri05}.
This outperforms the sector length threshold (solid blue curve) for all $n$.
Using a fixed point ansatz based on the recursive formula from Corollary~\ref{cor:sequence_line}, we can show that this curve converges to
\begin{align}
 \lim _{n \rightarrow \infty} p^{\mathrm{Sec}(1,\ldots,1)}_{\mathrm{glob},n} [\Psi^n_\mathrm{line}] 
 = 1- \frac{\sqrt{\sqrt[3]{2(\sqrt{93}+9)}-\sqrt[3]{2(\sqrt{93}-9)}}}{\sqrt[3]{6}}
 \approx 0.174 .
\end{align}
Therefore, also in the case of local white noise, the only new insights concern $(j,1,\ldots,1)$-separability, where $j\in\{2,3,4\}$.

\subsubsection{On noise thresholds for general qubit graph states}
 \label{sec:6.3.4}
 
So far, we have only discussed noise thresholds for families of qubit graph states for which we are able to solve the graph-theoretical puzzle analytically.  
Given an arbitrary qubit graph state $\Psi_\Gamma=\ket{\Gamma}\bra{\Gamma}$, one can extend our methods using the following two steps. 
\begin{itemize}
 \item[(i)] Numerically compute $\ell^n_n[\Psi_\Gamma]$ or a lower bound thereof.
 \item[(ii)] The state $\Psi_\Gamma$ can tolerate at least $p^\mathrm{Sec}_\mathrm{loc}=1-\sqrt[2n]{b/\ell^n_n[\Psi_\Gamma]}$ local white noise without becoming separable ($b=1$) or semiseparable $\left(b=2^{n-2} +  \frac{1 +(-1)^{n-1}}{2}\right)$. If here a lower bound of $\ell^n_n[\Psi_\Gamma]$ is used instead of  $\ell^n_n[\Psi_\Gamma]$ itself, this noise threshold is still valid, though more coarse.
\end{itemize}

Note that Proposition~\ref{prop:puzzle} can be formally restated as 
\begin{align}
\ell^n_n[\Psi_\Gamma] = \left \vert \left \{
\mathbf{r}\in \FF_2 ^n \ \bigg \vert \ \forall i \in \{1,\ldots, n\}: (1+r_i)\left (1+\sum_{j=1}^n \gamma_{i,j}r_j\right )  =0 
\right \} \right \vert.
\end{align}
Numerically computing this cardinality is an instance of the  \emph{Boolean Multivariate Quadratic Polynomial Problem} for
which an algorithm with runtime $O(2^{0.841n})$ is known~\cite{BFSS13}.
By changing the rule of the puzzle to \emph{every vertex has an odd number of black neighbors}, one obtains the lower bound $\ell^n_n[\Psi_\Gamma] \ge \left \vert \left \{\mathbf{r}\in \FF_2 ^n \ \big \vert \ \forall i \in \{1,\ldots, n\}: \sum_{j=1}^n \gamma_{i,j}r_j =1 \right \} \right \vert$ which is efficiently computable via, e.g., Gaussian elimination. 
Such a lower bound on the full-body sector length can provide at least a coarse threshold if it is intractable to run the algorithm with runtime $O(2^{0.841n})$.

\section{\label{sec:7}\protect Conclusion}
To conclude this thesis, we recapitulate our main results in Sec.~\ref{sec:7.1} and give an outlook on future work in Sec.~\ref{sec:7.2}.

\subsection{Main results} \label{sec:7.1}
We have studied how much noise a given state can tolerate without losing its entanglement.
Using the Peres-Horodecki criterion~\cite{Peres96, Horodecki96} and the reduction criterion~\cite{CeAdGi99, Horodecki99}, we have shown that every (non-trivial) graph state on $n$ qudits of dimension $D$ does not become separable if it is replaced by the maximally mixed state $\mathbbm{1}/D^n$ with a probability smaller than $p_\mathrm{glob}^\mathrm{PPT}(D,n)=p_\mathrm{glob}^\mathrm{Red}(D,n)={1-\frac{1}{D^{n-1}+1}}$, where $D\ge2$ can be any number.  

To capture more aspects of multipartite entanglement and to establish critical noise thresholds against local white noise, we have applied the concept of sector lengths.
For an $n$-qudit stabilizer state $\Psi$, the $j$-body sector length  $\ell^n_j[\Psi]$ is given by the number of its weight-$j$ stabilizers, where $1\le j\le n$. Sector lengths can also be defined for general states $\rho$.
If $\ell^n_j[\rho]$ exceeds a certain bound $b$, one can conclude that $\rho$ is not a mixture of states with a specific multipartite separability type, e.g.,  a tripartite $\rho$ is entangled if $\ell^3_j[\rho]> b_j^{(1,1,1)}$ and genuinely tripartite entangled if $\ell^3_j[\rho]> b_j^{(2,1)}$.
More concretely, denote an $n$-qudit stabilizer state $\Psi$ that is mixed with $\mathbbm{1}/D^n$ with probability $p$ by  $\rho_\mathrm{glob}(p)$. Likewise, let $\rho_\mathrm{loc}(p)$ denote a state to which a depolarizing channel $\sigma \mapsto (1-p)\sigma + p \mathbbm{1}/D$ was applied to each qudit individually.
For noise parameters $p$ below  $p_{\mathrm{glob},j}^\mathrm{Sec}[\Psi] := 1- \sqrt{b/\ell^n_j[\Psi]}$ and 
$ p_\mathrm{loc}^\mathrm{Sec}[\Psi] := 1- \sqrt[2n]{b/\ell^n_n[\Psi]}$, we have shown that $\ell^n_j[\rho_\mathrm{glob}(p)]$ and $\ell^n_n[\rho_\mathrm{loc}(p)]$, respectively, exceeds a such a bound $b$.
 
In order to make use of these thresholds, we have computed the sector lengths of important states.
For example, the sector length distribution of a $D$-dimensional, $n$-qudit Greenberger-Horne-Zeilinger (GHZ) states~\cite{GHZ89} is given by 
$ \ell^n_j \left[ \mathrm{GHZ}^n_D  \right] = \delta_{j,n}(D-1)D^{n-1} + \nchoosej ({(D-1)^j+(-1)^j(D-1)})/{D}$. 
For the special case of qubits, we have created tables of the sector length distribution for all $146$ graph states with $n\le 8$ qubits~\cite{HeinEisBri04,CaLTMoPo09}.
The so-called full-body sector length, $\ell^n_n[\Psi]$, yields the best noise threshold.
We have related $\ell^n_n[\Psi]$ to a graph-theoretical puzzle which we solved for four important families of qubit graph states. To our knowledge, the full-body sector length of line-graph states, ring-graph states, and here-introduced dandelion-graph states were not known before.
Finally, we described a method to compute $\ell^n_n[\Psi]$ for arbitrary qubit graph states.

We have also carried out a numerical analysis of the noise thresholds derived here and compared them to preexisting thresholds. The larger the full-body sector length of the investigated state, the better the corresponding noise thresholds are. For the GHZ state which has the largest full-body sector length, however, many other criteria previously have been investigated such that our approach only yields new insights in the case of semiseparability and local white noise. 
For qubit dandelion-graph states, however, the best local-noise thresholds which now exist are based sector lengths.

\subsection{Outlook} \label{sec:7.2}
In our investigation, we have mainly focused on the full-body sector length of a given state since this provided the best thresholds among the $j$-body sector lengths.
To further improve these thresholds, it would be worthwhile to combine our results about how sector lengths are diminished under noise with the consideration linear combinations of sector lengths (and not only $\ell^n _n[\Psi]$), similar to Ref.~\cite{KloeHub15}. 
For local white noise it is then necessary extend Proposition~\ref{prop:noisy_sector_lengths} to $\ell^n_j[\rho_\mathrm{loc}(p)]$.

We have studied the sector length distributions for qubit graph states and gathered numerical evidence that the ring graph state has the smallest full-body sector length among the qubit graph states with a connected graph. Likewise, we have numerical evidence that (after the GHZ state) the dandelion-graph state has the second largest full-body sector length.
Formal proofs for this (or counterexamples) would be interesting.

Finally, it would be useful to develop a numerical method (e.g., a semidefinite programming relaxation based on the graph-theoretical puzzle) to efficiently compute a good lower bound on the full-body sector length of an arbitrary graph state. This would provide efficiently computable noise thresholds for arbitrary stabilizer states.

\section{\label{sec:8}\protect Acknowledgements}
I like  to thank my supervisors Dagmar Bru\ss, Hermann Kampermann, and Timo Holz.
A special thank goes to Timo Holz and Lucas Tendick for proofreading the manuscript. 
Furthermore, I acknowledge helpful discussions with Lennart Bittel, Jens Eisert, Jonas Haferkamp, Marcus Huber, Benno Kuckuck, Jens Siewert, and Nikolai Wyderka.

\section{\label{sec:9}\protect Clarification of the originality of my results }
Here, I clarify to what extent the results in this thesis are my own results.

\paragraph{Theorem~\ref{thrm:stab}.}
The content of this theorem was already stated in Ref.~\cite{Gottesman99}, although only in the case where $\mathcal{S}\cong \prod_{i=1}^n \ZDZ$ and without a formal proof.
For the more heavily studied case of qubits, the theorem is proven e.g., in Ref.~\cite{QEC}.
My proof, which works for arbitrary qudit dimension, consists of two parts. As already mentioned in Sec.~\ref{sec:3.2}, the essential idea of ``(ii)$\Rightarrow$(i)'' is due to Gheorghiu~\cite{Gheorghiu14} (also in this reference, the content of Thrm.~\ref{thrm:stab} is contained).
On the other hand, I formulated the proof of ``(i)$\Rightarrow$(ii)'' on my own.

\paragraph{Lemma~\ref{lem:cliff_stab}.} 
This lemma is an immediate consequence of Theorem~\ref{thrm:stab} and often used in the community in one form or another. 
Since the lemma is easily proven (no difficulty arises by considering qudits), I could not find out who used it first. 
The given proof was written by myself.

\paragraph{Lemma~\ref{lem:mixed}.}
This lemma can be regarded as a trivial application of a classical result known as \emph{Schur's lemma} \cite[p. 13]{Serre77}.
It can also be regarded as a direct consequence of the relation
 \begin{align} \label{app_dep_state}
    \frac{\mathbbm{1}}{D^n}= \frac{1}{D^{2n}}\sum_{\mathbf{r,s}\in (\ZDZ)^n} (X^\mathbf{r}Z^\mathbf{s})  \ \rho \ (X^\mathbf{r} Z^\mathbf{s})^\dagger
\end{align}
which I have formally shown in Ref.~\cite{MHKB18} for all $n$-qudit states~$\rho$.
Despite its elementary form, I did not find an adequate reference for Lemma~\ref{lem:mixed} in the literature.
Therefore, the direct proof of Lemma~\ref{lem:mixed} presented here was formulated by myself.

\paragraph{Lemma~\ref{lem:PPT_GHZ}.} 
 This result is well-known in the community. For qubits it was proven by D\"ur et al. in 1999~\cite{DuCiTa99}. For qudits it is known for at least 10 years, e.g., in Ref.~\cite{GHH10} it is mentioned without a proof. The proof presented here is straightforward and I formulated it on my own.

\paragraph{Theorem~\ref{thrm:PPT_graph}, Corollary~\ref{cor:PPT_graph}.} 
I found these results and formulated their proofs on my own. 
While the resulting noise thresholds are readily established for many special cases, I could not find a formulation of this result in the literature which is as general as here.

\paragraph{Lemma~\ref{lem:fully_sep_bound}.} 
This lemma is a straightforward application of Eq.~\eqref{eq:sec_tensor} and I would not be surprised if it is already known to the community.
Note that Thrm.~4 of Ref.~\cite{deVicHub11} also is an entanglement criterion  based on sector lengths which can be used to rule out full separability; although it is not directly related to Lemma~\ref{lem:fully_sep_bound}.

\paragraph{Proposition~\ref{prop:noisy_sector_lengths}, Corollary~\ref{cor:crit_noise_sector_lengths}.}
To the best of my knowledge, these results were not known before.

\paragraph{Proposition~\ref{prop:GHZ_sector_lengths}.} 
I figured out this result on my on. Recently, however, it was independently discovered in Ref.~\cite{EltSiew19} by Eltschka and Siewert. 
Their preprint was uploaded after I finished the formulation and the proof of Proposition~\ref{prop:GHZ_sector_lengths} (but before the submission of this
thesis).
 
\paragraph{Proposition~\ref{prop:AME4D_sector_lengths}.}
To my knowledge, the sector length distribution of the tetrapartite AME state has not been calculated before. Note that our result is consistent with the approximation $\ell^n_n[\Psi^n_D]\approx D^n\left(1-\frac{1}{D^2}\right)^n$ of the full-body sector length of an $n$-qudit AME state~\cite{EltSiew19}.

\paragraph{Lemma~\ref{lem:maximal_overlap}.}
Although proving this result is not very difficult, I did not find it in the literature.
I thank Marcus Huber for suggesting the idea of the proof.

\paragraph{Proposition~\ref{prop:puzzle}, Corollary~\ref{cor:sequence_line}.}
To my knowledge, these results were not known before.

\paragraph{Lemma~\ref{lem:choices_ring}.}
I wrote this proof on my own. It could be possible that a shorter proof based on some combinatorial argument  I am not aware of exists.


\paragraph{Lemma~\ref{lem:gamma_r_neq_0}, Lemma~\ref{lem:sum_gamma_r_neq_0}.} 
These facts directly follow from elementary group-theoretical considerations and it is very likely that they have been used before in a different context.

\paragraph{Lemma~\ref{lem:solve_sum_gamma_r_neq_0}.}
This result can be understood as a generalization of Proposition~\ref{prop:GHZ_sector_lengths} as the $\ZDZ$-module $M_Z$ generated by the vectors in Eq.~\eqref{eq:basis_GHZ_module} is an example for $\ker(\varphi)$ in the proof of Lemma~\ref{lem:solve_sum_gamma_r_neq_0}.

%

\newpage
\appendix

\section{\label{app:proof_PPT}\protect Proof of Theorem~\ref{thrm:PPT_graph}}
Here, we state the proof of Theorem~\ref{thrm:PPT_graph} from the main text. For a better readability, we repeat it now.

\paragraph{Theorem~\ref{thrm:PPT_graph}.}  
\emph{Let $\Gamma\in (\ZDZ)^{n\times n}$ be the adjacency matrix of a graph state such that $\gamma_{1,2}$ is invertible. Then, $\lambda(p)= p/D^n-(1-p)/D$ is an eigenvalue of the operator $\rhoGlobGammaP\TA$.  In particular, $\rhoGlobGammaP $ is entangled for all $p< p_\mathrm{glob}^\mathrm{PPT}(D,n):= {1-\frac{1}{D^{n-1}+1}}$.}

\begin{proof}
We will prove this statement by showing that this operator has an eigenvector $\mathbf{v}\in \mathbb{C}^{D^n}$ to the eigenvalue $\lambda(p)$ whose entries are given by
\begin{align} \label{def:PPT_graph_eigenvector}
 v_\mathbf{s} :=
 \omega_D ^{\sum\limits_{i=2}^n \sum\limits_{j=i+1}^n \gamma_{i,j}s_is_j }
 \left( \omega_D^{\sum\limits_{j=2}^n\gamma_{1,j}s_j}\delta_{s_1,0} - \delta_{s_1,1} \right) 
\end{align}
for all $\mathbf{s}=(s_1,\ldots, s_n) \in (\ZDZ)^n$.
Thereby, we will only consider the non-trivial case $p\neq 1$ and work with the operator 
\begin{align}\label{def:PPT_graph_sigma}
 \sigma := \frac{D^n}{1-p}\left( (1-p)\ket{\Gamma}\bra{\Gamma}\TA + p \frac{\mathbbm{1}}{D^n} - \frac{\mathbbm{1}}{D^n} \right) +\mathbbm{1}.
\end{align}
The claim is equivalent to the simpler eigenequation 
$\sigma \mathbf{v} = -D^{n-1}\mathbf{v}$ because by passing from $\rhoGlobGammaP\TA =(1-p)\ket{\Gamma}\bra{\Gamma}\TA + p\mathbbm{1}/D^n$ to $\sigma$, the eigenvalue $\lambda(p)$ changes into
\begin{align}
 \frac{D^n}{1-p}\left( \lambda(p) - \frac{1}{D^n} \right) +1 = \frac{D^n}{1-p}\left(- \frac{1-p}{D^n} -\frac{1-p}{D} \right) +1 =-D^{n-1}
\end{align}
while the eigenvector remains unchanged.
To obtain the explicit expression of $\sigma$, recall from Eq.~\eqref{eq:ketGamma} that the projector onto a graph state $\ket{\Gamma}$ is given by
\begin{align}
 \ket{\Gamma} \bra{\Gamma} = \frac{1}{D^n} \sum_{\mathbf{r,s}\in (\ZDZ)^n} \omega_D^{\sum\limits_{i=1}^n\sum\limits_{j=i+1}^n \gamma_{i,j} (r_i r_j- s_is_j)} \ket {\mathbf{r}} \bra{\mathbf{s}}.
\end{align}
The partial transposition with respect to Alice swaps $r_1$ and $s_1$, i.e.,
\begin{align}
 \ket{\Gamma} \bra{\Gamma}\TA = \frac{1}{D^n} \sum_{\mathbf{r,s}\in (\ZDZ)^n} \omega_D^{\sum\limits_{j=2}^n{\gamma_{1,j}(s_1r_j-r_1s_j)}+ \sum\limits_{i=2}^n\sum\limits_{j=i+1}^n \gamma_{i,j} (r_i r_j- s_is_j)} \ket {\mathbf{r}} \bra{\mathbf{s}}.
\end{align}
Inserting this expression into Eq.~\eqref{def:PPT_graph_sigma} yields that for all $\mathbf{r,s}\in (\ZDZ)^n$  the entries of the operator $\sigma$ are given by the single expression
\begin{align}
 \sigma_\mathbf{r,s} = \omega_D^{\sum\limits_{j=2}^n{\gamma_{1,j}(s_1r_j-r_1s_j)} + \sum\limits_{i=2}^n\sum\limits_{j=i+1}^n \gamma_{i,j} (r_i r_j- s_is_j)},
\end{align}
in particular they do not depend on $p$.

We now start the main work of showing $\sigma \mathbf{v} = -D^{n-1}\mathbf{v}$ for each component, i.e., for every $\mathbf{r}\in (\ZDZ)^n$, we want to prove that 
\begin{align}
 \sum_{\mathbf{s}\in (\ZDZ)^n} \sigma_\mathbf{r,s} v_\mathbf{s} 
 &= \sum_{\mathbf{s} \in (\ZDZ)^n} \omega_D^{\sum\limits_{j=2}^n{\gamma_{1,j}(s_1r_j-r_1s_j)} + \sum\limits_{i=2}^n\sum\limits_{j=i+1}^n \gamma_{i,j} r_i r_j}  
 \left( \omega_D^{\sum\limits_{j=2}^n\gamma_{1,j}s_j} \delta_{s_1,0} - \delta_{s_1,1} \right) \\ 
 &=\omega_D ^{ \sum\limits_{i=2}^n\sum\limits_{j=i+1}^n \gamma_{i,j} r_i r_j}   \sum_{\mathbf{s} \in (\ZDZ)^n} \omega_D^{\sum\limits_{j=2}^n{\gamma_{1,j}(s_1r_j-r_1s_j)} }
 \left( \omega_D^{\sum\limits_{j=2}^n\gamma_{1,j}s_j}\delta_{s_1,0} - \delta_{s_1,1} \right)
\\
\text{is}&\text{ equal to} -D^{n-1}v_\mathbf{r} = -D^{n-1} 
 \omega_D ^{\sum\limits_{i=2}^n \sum\limits_{j=i+1}^n \gamma_{i,j}r_ir_j } \left(\omega_D^{\sum\limits_{j=2}^n\gamma_{1,j}r_j} \delta_{r_1,0}- \delta_{r_1,1} \right) .
\end{align} 
Thus, it suffices to show that 
\begin{align}\label{def:PPT_graph_An}
 A_n(\mathbf{r}) &:=  \sum_{\mathbf{s} \in (\ZDZ)^n} \omega_D^{\sum\limits_{j=2}^n{\gamma_{1,j}(s_1r_j-r_1s_j)} }
 \left(\omega_D^{\sum\limits_{j=2}^n\gamma_{1,j}s_j} \delta_{s_1,0}- \delta_{s_1,1} \right) \\
 \text{and} \hspace{2em} \label{def:PPT_graph_Bn}
 B_n(\mathbf{r})& :=-D^{n-1} 
 \left(\omega_D^{\sum\limits_{j=2}^n\gamma_{1,j}r_j} \delta_{r_1,0}- \delta_{r_1,1} \right) 
\end{align}
are equal for all $n$ and all $\mathbf{r}\in(\ZDZ)^n$.
This we do with a proof by induction over the parties $n\ge2$.

For $n=2$, this theorem is only about adjacency matrices of the form 
\begin{align}
 \Gamma = \left[\begin{array}{cc} 0 & \gamma \\ \gamma & 0\end{array}\right] ,
\end{align}
where $\gamma \in \ZDZ$ is invertible. In that case, however, $\ket{\Gamma}$ is locally Clifford equivalent to $\ket{\mathrm{GHZ}^2_D} = \mathbbm{1} \otimes M(\gamma) F^\dagger \ket{\Gamma}$. 
From the proof of lemma~\ref{lem:PPT_GHZ}, we know
\begin{align}
 \left((1-p)\ket{\mathrm{GHZ}_D^2}\bra{\mathrm{GHZ}_D^2}\TA + p\frac{\mathbbm{1}}{D^2}\right) \left(\ket{0,1} - \ket{1,0} \right)= \lambda(p) \left(\ket{0,1} - \ket{1,0} \right),
\end{align}
thus, the vector
\begin{align}
\mathbbm{1}\otimes  M(\gamma )^\dagger F \left(\ket{0,1} - \ket{1,0} \right) 
&= \frac{1}{\sqrt{D}} \sum_{k,l\in \ZDZ} \omega_D^{kl}\ket{\gamma^{-1}k}_\mathrm{B}\bra{l}_\mathrm{B} \left(\ket{0,1} - \ket{1,0} \right) \\
&= \frac{1}{\sqrt{D}} \sum_{k \in \ZDZ} \left(\omega_D^{k\cdot 1} \ket{0,\gamma^{-1}k} - \omega_D^{k \cdot 0} \ket{1,\gamma^{-1}k} \right) \\
&= \frac{1}{\sqrt{D}} \sum_{s \in \ZDZ} \left(\omega_D^{\gamma s} \ket{0,s} - \ket{1,s} \right) \label{eq:eigenvector_GHZ_to_Gamma}
\end{align}
is an eigenvector of $\rhoGlobGammaP$ to the eigenvalue $\lambda(p)$. 
Note that we have substituted $k=\gamma s$ in the last step.
Identifying $s$ with $s_2$ and $\gamma$ with $\gamma_{1,2}$ is becomes clear that (up to rescaling) the vector in Eq.~\eqref{eq:eigenvector_GHZ_to_Gamma} is the same as $\mathbf{v}$ which was defined in Eq.~\eqref{def:PPT_graph_eigenvector}.
We have thus established $\sigma \mathbf{v}= -D \mathbf{v}$ which implies the base case of the induction, $A_2(r_1,r_2)=B_2(r_1,r_2)$ for all $r_1,r_2\in \ZDZ$.
 
We now move on with the inductive step: Let $n\ge 2$ and assume $ A_n(\mathbf{r})  =  B_n(\mathbf{r}) $ for all $\mathbf{r}\in (\ZDZ)^n$. Additionally let $r_{n+1}\in\ZDZ$.
There is a relation between $ A_{n+1}(\mathbf{r},r_{n+1})$ and  $ A_n(\mathbf{r})$: 
For $n+1$, definition~\eqref{def:PPT_graph_An} reads
\begin{align}
 A_{n+1}(\mathbf{r},r_{n+1}) &:=  \sum_{s_1, s_{n+1} \in \ZDZ} \sum_{s_2,\ldots, s_n \in \ZDZ} \omega_D^{\sum\limits_{j=2}^{n+1}{\gamma_{1,j}(s_1r_j-r_1s_j)} }
 \left(\omega_D^{\sum\limits_{j=2}^{n+1}\gamma_{1,j}s_j} \delta_{s_1,0}- \delta_{s_1,1} \right) \\
 &= \sum_{s_{n+1} \in \ZDZ} \Big(
 \omega_D^{\gamma_{1,n+1}(s_{n+1}-r_1s_{n+1})} x_0(\mathbf{r})
 - \omega_D^{\gamma_{1,n+1}(r_{n+1}-r_1s_{n+1})} x_1(\mathbf{r})
 \Big)
 \label{eq:PPT_graph_An+1}
\end{align}
where we have used the substitution 
\begin{align}\label{def:PPT_graph_xsi}
 x_0(\mathbf{r}) :=&  \sum_{s_2,\ldots, s_n \in \ZDZ}  \omega_D^{\sum\limits_{j=2}^{n}\gamma_{1,j}(s_j-r_1s_j)} 
 \hspace{1.5em}\text{and} \hspace{1.5em}
 x_1(\mathbf{r}) := & \sum_{s_2,\ldots, s_n \in \ZDZ}  \omega_D^{\sum\limits_{j=2}^{n}\gamma_{1,j}(r_j-r_1s_j)} 
\end{align}
which helps to separate all terms with index $j=n+1$ from the rest.
Carrying out the summation over $s_1$ in definition~\eqref{def:PPT_graph_An} yields
$A_n(\mathbf{r})= x_0(\mathbf{r}) -x_1(\mathbf{r})$. This is the relation  between $ A_{n+1}(\mathbf{r},r_{n+1})$ and  $ A_n(\mathbf{r})$.
 For $B$, we obtain
\begin{align}
 B_{n+1}(\mathbf{r}) =
& -D^n  \left( \omega_D^{\sum\limits_{j=2}^{n+1}\gamma_{1,j}r_j} \delta_{r_1,0}- \delta_{r_1,1} \right)  
 = D\omega_D^{\gamma_{1,n+1}r_{n+1} \delta_{r_1,0} }  B_n(\mathbf{r})
\end{align}
 from definition~\eqref{def:PPT_graph_Bn}. 
Thus, the inductive hypothesis $A_n(\mathbf{r}) = B_n(\mathbf{r})$ implies
\begin{align}\label{eq:PPT_graph_Bn+1}
 B_{n+1}(\mathbf{r})  =    D\omega_D^{\gamma_{1,n+1}r_{n+1} \delta_{r_1,0} }  \left(x_0(\mathbf{r}) -x_1(\mathbf{r})\right ).
\end{align}

 Now, we will show
 $x_0(\mathbf{r})=0$ if $r_1 \neq1$, and
 $x_1(\mathbf{r})=0$ if $r_1 \neq0$ as we will need these facts below.
 Indeed, if $r_1 \neq 1$, Eq.~\eqref{def:PPT_graph_xsi} simplifies to 
 \begin{align}
  x_0(\mathbf{r})= \sum_{s_2,\ldots, s_n \in \ZDZ}  \omega_D^{\sum\limits_{j=2}^{n}\gamma_{1,j}(s_j-r_1s_j)} 
  =  \sum_{s_3,\ldots, s_n \in \ZDZ}  \omega_D^{\sum\limits_{j=3}^{n}\gamma_{1,j}(s_j-r_1s_j)} \sum_{s_2\in\ZDZ} \omega_D^{\gamma_{1,2}s_2} =0,
 \end{align}
 since $\gamma_{1,2}\neq 0$ implies $\sum_{s_2} \omega_D^{\gamma_{1,2}s_2} =0$ (the powers of a complex root of unity sum up to zero as long as not all of them are equal to one).
 Likewise, if $r_1 \neq 0$, we have
 \begin{align}
  x_1(\mathbf{r}) 
&= \sum_{s_2,\ldots, s_n \in \ZDZ}  \omega_D^{\sum\limits_{j=2}^{n}\gamma_{1,j}(r_j-r_1s_j)} \\
&= \sum_{s_3,\ldots, s_n \in \ZDZ}  \omega_D^{\gamma_{1,2}r_2+ \sum\limits_{j=3}^{n}\gamma_{1,j}(r_j-r_1s_j)}\sum_{s_2\in\ZDZ} \omega_D^{-\gamma_{1,2}r_1s_2} =0.
 \end{align}
Note that this is the step where it is crucial that the graph has an invertible edge. 
(Otherwise, $-\gamma_{1,2}r_1$ would be zero for at least one $r_1\in\ZDZ$ besides $r_1=0$. In that case, we could not have used  $\sum_{s_2} \omega_D^{-\gamma_{1,2}r_1s_2} =0$.)

To show $A_{n+1}(\mathbf{r}) =
 B_{n+1}(\mathbf{r}) $ we will distinguish the three cases. 
 \emph{First case:} $r_1=0$. We have $x_0(\mathbf{r})=0$ since $r_1\neq1$.
 Thus, Eqs.~\eqref{eq:PPT_graph_An+1} and \eqref{eq:PPT_graph_Bn+1} simplify to 
 \begin{align}
 A_{n+1}(0,r_2,\ldots,r_{n+1}) 
 &=- \sum_{s_{n+1} \in \ZDZ}  \omega_D^{\gamma_{1,n+1}r_{n+1}} x_1(\mathbf{r})
  =-D\omega_D^{\gamma_{1,n+1}r_{n+1} }  x_1(\mathbf{r}) 
  \\
  &=  B_{n+1}(0,r_2,\ldots,r_{n+1}).
 \end{align}
  \emph{Second case:} $r_1=1$.
 We have $x_1(\mathbf{r})=0$ since $r_1\neq0$.
 Thus, Eqs.~\eqref{eq:PPT_graph_An+1} and \eqref{eq:PPT_graph_Bn+1} simplify to 
 \begin{align}
 A_{n+1}(1,r_2,\ldots,r_{n+1}) 
 &=  \sum_{s_{n+1} \in \ZDZ}   x_0(\mathbf{r}) = D  x_0(\mathbf{r}) = B_{n+1}(1,r_2,\ldots,r_{n+1}) 
 \end{align}
 \emph{Third case:} $r_1\in\{ 2,\ldots, n\}$.
 We have $x_0(\mathbf{r})=x_1(\mathbf{r})=0$ since $r_1$ is neither $0$ nor $1$.
 Thus, Eqs.~\eqref{eq:PPT_graph_An+1} and \eqref{eq:PPT_graph_Bn+1} simplify to  $A_{n+1}(\mathbf{r}) = 0 =
 B_{n+1}(\mathbf{r})$.

 This finishes the proof by induction, i.e., we have established that $\mathbf{v}\in\mathbb{C}^{D^n}$ as defined in Eq.~\eqref{def:PPT_graph_eigenvector} is indeed an eigenvector of  $(1-p)\ket{\Gamma}\bra{\Gamma}\TA + p\mathbbm{1}/D^n$ to the eigenvalue  $\lambda(p)= p/D^n-(1-p)/D$.
\end{proof}

\section{\label{app:bound_full_body_sec}\protect Analytical lower bounds on the full-body sector length of general qudit graph states}
 
Whenever a sector length of a given graph state $\Psi_\Gamma = \ket{\Gamma}\bra{\Gamma}$ exceeds a corresponding separability bound, Corollary~\ref{cor:crit_noise_sector_lengths} yields a certain noise threshold. 
Here, we show that the full-body sector length, $\ell^n_n[\Psi_\Gamma]$, is never smaller than  $b_n^{(1,\ldots,1)}= {(D-1)^n} $, the full-separability bound. 
We also provide coarse lower bounds on $\ell^n_n[\Psi_\Gamma]$ which  already suffice to establish a nontrivial noise threshold in many cases.

Recall from Eq.~\eqref{eq:sec_len_counting} that the full-body sector length of $\Psi_\Gamma$ is equal to the 
number of its full-weight stabilizer operators.
Using Eq.~\eqref{eq:graph_stab} we can parametrize its stabilizer group via the isomorphism
\begin{align}
  (\ZDZ)^n \overset{\simeq}{\longrightarrow} \mathcal{S}_{\ket{\Gamma}}, \hspace{2em} \mathbf{r} \longmapsto \prod_{i=1}^n S_i^{r_i} = X_D^\mathbf{r} Z_D^{\Gamma \mathbf{r}},
\end{align}
where $\Gamma \mathbf{r}\in (\ZDZ)^n$  is the vector obtained by applying the adjacency matrix $\Gamma$ 
to the vector $\mathbf{r}$.
Thus, the full-body sector length can be rewritten as 
\begin{align} \label{eq:full_body_sector_length_graph_state}
 \ell^n_j[\Psi_\Gamma] = \left \vert  \left \{  \mathbf{r} \in (\ZDZ)^n\ \big\vert \ j=\swt(\mathbf{r},\Gamma \mathbf{r}) \right \} \right \vert.
\end{align}
Since every vector $\mathbf{r}$ which has only nonzero entries (i.e., $\wt(\mathbf{r})=n \Rightarrow \swt(\mathbf{r},\Gamma \mathbf{r})=n$) contributes to this quantity, it follows at once that $\ell^n _n [\Psi_\Gamma] \ge(D-1)^n$ holds for all graphs.
Thus, to establish a nontrivial noise threshold, it suffices to find a single vector $\mathbf{r}$ which has at least one entry equal to zero, say $r_i=0$, and which fulfills $\sum_{j=1}^n \gamma_{i,j} r_j \neq 0$.
In the case where $i$ is a leaf, i.e., there is exactly one $j$ with $\gamma_{i,j}\neq 0$, the latter condition simplifies to $\gamma_{i,j}r_j \neq 0$ and the following group-theoretical fact terminates the search for the additional vector we look for.
\lemma \label{lem:gamma_r_neq_0}
\emph{Let $\gamma \in \ZDZ$. The number of elements $r\in \ZDZ$ fulfilling $\gamma r \neq 0$ is given by $D- \gcd(D,\gamma)$.~\footnote{We use the convention $\gcd(D,0)=\gcd(D,D)=D$.} } 
\begin{proof}
We are looking for the number of elements in the set 
 $L:= \{r \in \ZDZ \ \vert \ \gamma r \neq 0 \}$ which can be rewritten as 
 $L=\ZDZ \backslash \ker(\varphi)$ for the group homomorphism 
 \begin{align}
  \varphi: \ZDZ \longrightarrow \ZDZ, \hspace{1em} r\longmapsto \gamma r.
 \end{align}
By the first isomorphism theorem~\cite[1.2/  Kor.7]{Bosch}, the groups $\mathrm{im}(\varphi)$ and $(\ZDZ)/\ker(\varphi)$ are isomorphic. Since the image $\mathrm{im}(\varphi)= \langle \gamma \rangle$ contains $\ord(\gamma)= D/\gcd(D,\gamma)$ elements, we obtain
\begin{align}
 \left \vert L\right\vert= \left\vert \ZDZ \right\vert- \left\vert \ker(\varphi) \right\vert = \left\vert\ZDZ\right\vert- \frac{\left\vert \ZDZ\right\vert}{\left\vert \mathrm{im}(\varphi)\right\vert} 
 = D- \gcd(D,\gamma).
\end{align}
This finishes the proof.
\end{proof}
Note that this indeed provides the additional vector $\mathbf{r}$ since $D-\gcd(D,\gamma)>0$ holds for all nonzero $\gamma\in\ZDZ$.
For general graphs, however, the situation is  more complicated as the sum $\sum_{j=1}^n \gamma_{i,j} r_j$ consists of more than one term. 
In that case, we need the following generalization of Lemma~\ref{lem:gamma_r_neq_0}:
\lemma \label{lem:sum_gamma_r_neq_0}
\emph{Let $ \boldsymbol{\gamma}=(\gamma_1,\ldots, \gamma_m) \in (\ZDZ)^m$. Denote the number of vectors $\mathbf r$ fulfilling $\boldsymbol{\gamma}^\mathrm{T} \mathbf{r} \neq 0$ and $r_j\neq 0$ for all $j$ by  }
\begin{align}
 N_m(\boldsymbol{\gamma}) := \left\vert \left\{ \mathbf{r}\in(\ZDZ)^m
\ \bigg \vert \ \sum_{j=1}^m \gamma_j r_j\neq 0 \text{ and } r_1, \ldots, r_m \neq 0 \right\} \right\vert.
\end{align}
\emph{Then, the recurrence relation }
\begin{align}\label{eq:recurrence_counting}
 N_m(\boldsymbol{\gamma}) = D^{m-1}\left (D -\gcd(D,\boldsymbol \gamma) \right ) -  \sum_{k=1}^{m-1} \sum_{\substack{I\subset\{1,\ldots,m\}\\ \left\vert I\right \vert =k}} N_k(\boldsymbol \gamma \vert _ I)
\end{align}
\emph{is fulfilled, where $\gcd(D,\boldsymbol \gamma)$ denotes the greatest common divisor of $D, \gamma_1, \ldots,$ and $\gamma_m$, and $\boldsymbol \gamma \vert _I\in(\ZDZ)^k$ is the restricted vector which results from $\boldsymbol \gamma $ after removing all entries except for those labeled by $I$.}
\begin{proof}
 The key is to count the elements in the set 
 \begin{align}
  L:=  \left \{\mathbf{r}\in (\ZDZ)^m \ \bigg \vert \ \sum_{j=1}^m \gamma_j r_j \neq 0\right \}
 \end{align}
in two different ways.
The first possibility follows a direct generalization of Lemma~\ref{lem:gamma_r_neq_0}: Rewrite the set as $L=(\ZDZ)^m\backslash \ker(\varphi)$ for the $\ZDZ$-linear map 
\begin{align}
  \varphi: (\ZDZ)^m \longrightarrow \ZDZ, \hspace{1em}
 \mathbf{r}\longmapsto \boldsymbol{\gamma}^\mathrm{T} \mathbf{r} =  \sum_{j=1}^m \gamma_j r_j.
\end{align}
Since the image $\mathrm{im}(\varphi)= \langle \gamma_1,\ldots, \gamma_m \rangle$ contains $ D/\gcd(D,\gamma_1,\ldots,\gamma_m)$ elements, we obtain
\begin{align} \label{eq:count_L_1}
 \left \vert L  \right\vert  
 = \left\vert (\ZDZ)^m \right\vert - \frac{\left\vert (\ZDZ)^m\right\vert }{\left\vert  \mathrm{im}(\varphi) \right\vert } 
 = D^{m-1}(D- \gcd(D,\boldsymbol \gamma) ).
\end{align}
For the second possibility to count the elements in $L$, consider the decomposition  
 \begin{align}
 L  = \bigcupdot_{k=1}^m \bigcupdot_{\substack{I\subset\{1,\ldots,m\}\\ \left\vert I\right\vert  =k}}
  \left\{ \mathbf{r}\in(\ZDZ)^m
\ \bigg \vert \ \sum_{j=1}^m \gamma_j r_j \neq 0, \substack{r_j =0 \text{ for } j\not \in I\\r_j \neq0 \text{ for } j \in I} \right\}
 \end{align}
 into disjoint subsets where, for all possible subsets $I\subset \{1,\ldots, m\}$ with $k$ elements, the $k$ entries of $\mathbf{r}$ which correspond to $I$ are nonzero while the other $m-k$ entries are fixed to zero.  
The cardinality follows as
\begin{align} 
\left\vert  L \right\vert 
 = \sum_{k=1}^{m} \sum_{\substack{I\subset\{1,\ldots,m\}\\ \left\vert I\right\vert  =k}} N_k(\boldsymbol \gamma \vert _ I) 
 = N_m(\boldsymbol \gamma) +  \sum_{k=1}^{m-1} \sum_{\substack{I\subset\{1,\ldots,m\}\\ \left\vert I\right\vert  =k}} N_k(\boldsymbol \gamma \vert _ I) 
\end{align}
and solving for $N_m(\boldsymbol \gamma)$ under the use of Eq.~\eqref{eq:count_L_1} finishes the proof.
\end{proof}
In the special case where all entries of $\boldsymbol \gamma$ are invertible, we can use the recurrence relation~\eqref{eq:recurrence_counting} to derive an explicit expression for
 $N_m(\boldsymbol{\gamma})$. As we comment on in Sec.~\ref{sec:9}, it not a coincidence that this expression is very similar to Eq.~\eqref{eq:GHZ_sectorlength}. 
\lemma \label{lem:solve_sum_gamma_r_neq_0}
\emph{An explicit expression for $N_m:= N_m(\gamma_1,\ldots,\gamma_n)$ where $\gamma_1,\ldots,\gamma_n$ are invertible is given by}
\begin{align}\label{eq:explicit_counting}
 N_m = \frac{(D-1)^{m+1} + (-1)^{m+1}(D-1)}{D}.
\end{align}
\begin{proof}
We verify this expression by induction. For the base case $m=1$, Lemma~\ref{lem:gamma_r_neq_0} yields $N_1= D-1$ since $\gamma\in \ZDZ$ is invertible implies $\gamma$ does not divide $D$, thus, $\gcd(D,\gamma)=1$.
 It is straightforward to verify that also the right-hand side of Eq.~\eqref{eq:explicit_counting} is equal to $D-1$.
 
 For the induction step, let $m\ge2$ and assume that  $ N_k = \frac{(D-1)^k + (-1)^{k+1}(D-1)  }{D}$  holds for all $k <m$. Since all $\gamma_i$ are invertible, Eq.~\eqref{eq:recurrence_counting} simplifies to
 \begin{align}
   N_m  = D^{m-1}\left (D -1  \right ) -  \sum_{k=1}^{m-1} N_k  \sum_{\substack{I\subset\{1,\ldots,m\}\\ \left\vert I\right\vert  =k}}1. 
 \end{align}
Since the number of subsets $I\subset\{1,\ldots, m\}$ with exactly $k$ elements is given by the binomial coefficient $\mchoosek$, inserting the induction hypothesis yields the expression
\begin{align}
 N_m  = \frac{D-1}{D} \left(  D^m -\sum_{k=1}^{m-1} \mchoosek \left((D-1)^k+(-1)^{k+1} \right )
 \right).
\end{align}
By applying the binomial theorem $\sum_{k=0}^m a^k= (a+1)^m$ for $a=D-1$ and $a=-1$, respectively, we obtain 
\begin{align} 
 N_m & = \frac{D-1}{D} \left( D^m-(D^m-1-(D-1)^m)-(0+ 1+ (-1)^m)  \right)\\
 &= \frac{(D-1)^{m+1} + (-1)^{m+1}(D-1)}{D}
.
\end{align}
This finishes the proof. 
\end{proof}

Denote the set of neighbors of a vertex $i$ by $I(i):=\{j \ \vert \ \gamma_{i,j} \neq 0 \}$ and its degree, i,e, the number of its neighbors,  by $m_i := \left\vert  I(i)\right\vert $.
Our investigation shows that the number of vectors $\mathbf{r}\in( \ZDZ)^n$ fulfilling $\swt(\mathbf{r},\Gamma \mathbf{r})=n$ can be lower bounded as
\begin{align} \label{eq:lower_bound_lnn_general_graph}
 \ell^n_n[\Psi_\Gamma]& \ge (D-1)^n + \sum_{i=1}^{n} (D-1)^{n-1-m_i} N_{m_i}(( \gamma_{i,j})_{j\in I(i)}).
\end{align}
The term $(D-1)^n$ corresponds to the vectors $\mathbf{r}$ which have only nonzero entries.
The sum runs over all parties $i$ and counts the choices where $r_i=0$ while $\sum_{j=1}^n \gamma_{i,j}r_j \neq 0$ and $r_j\neq 0$ for all $j$ but $i$ such that $\swt(\mathbf{r},\Gamma \mathbf{r})=n$ is fulfilled.
Each term in the sum contributes $(D-1)^{n-1-m_i}$ choices for $r_j\neq0$ where $j$ is not $i$ and not a neighbor of $i$ times $N_{m_i}(( \gamma_{i,j})_{j\in I(i)})$ choices for the neighbors of $i$ (times 1 choice for $r_i=0$).
We can simplify the ratio that enters the noise thresholds of Corollary~\ref{cor:crit_noise_sector_lengths} to
 \begin{align}\label{eq:lower_bound_threshold_general_graph}
  \frac{b_n^{(1,\ldots,1)}}{\ell^n_n[\Psi_\Gamma]} &\le  \left( 1 + \sum_{i=1}^{n}   \frac{N_{m_i}(( \gamma_{i,j})_{j\in I(i)})}{(D-1)^{m+1}}  \right)^{-1} \le 1.
 \end{align}
 
We can further simplify these expressions in the case where all entries of $\Gamma$ are either invertible or zero, which is automatically fulfilled if $D$ is prime. 
In that case, Inequality~\eqref{eq:lower_bound_lnn_general_graph} can be rewritten as
\begin{align}
 \ell^n_n[\Psi_\Gamma]& \ge (D-1)^n + \sum_{m=1}^{n-1}  (D-1)^{n-1-m}  M_m N_m 
\end{align}
where $M_m$ is the number of vertices with exactly $m$ neighbors.
After inserting Eq.~\eqref{eq:explicit_counting}, Inequality~\eqref{eq:lower_bound_threshold_general_graph} simplifies to
 \begin{align}\label{eq:lower_bound_threshold_invertible_entries_graph}
  \frac{b_n^{(1,\ldots,1)}}{\ell^n_n[\Psi_\Gamma]} &\le \left( 1 + \frac{1}{D}\sum_{m=1}^{n-1}   \left(1+\frac{1}{(D-1)^{m+1}}  \right)M_m  \right)^{-1} \le 1.
 \end{align}
For this coarse bound on the noise threshold, the only thing one has to know about the graph are the numbers $M_1,\ldots,M_{n-1}$. 
However, for $D=2$ and graphs where each vertex has an even number of neighbors, the sum in Eq.~\eqref{eq:lower_bound_threshold_invertible_entries_graph} vanishes, i.e., the considerations in this section are not sufficient to obtain a nontrivial noise threshold in this case. 
However, we solve this issue in Sec.~\ref{sec:6} by a detailed investigation of the qubit case. 

\section{\label{app:proof_sec_of_GHZ}\protect Proof of Proposistion~\ref{prop:GHZ_sector_lengths}}

Here, we calculate the sector lengths of state $ \mathrm{GHZ}^n_D :=\ket{ \mathrm{GHZ}^n_D }\bra{\mathrm{GHZ}^n_D }$ which is defined in Eq.~\eqref{def:GHZnD}.  

\paragraph{Proposition~\ref{prop:GHZ_sector_lengths}.}~\cite{EltSiew19}
\emph{The sector length distribution of the GHZ state is given by } 
\begin{align} \label{eq:GHZ_sectorlength}
 \ell^n_j \left[ \mathrm{GHZ}^n_D  \right] &= \delta_{j,n}(D-1)D^{n-1} + \genfrac{(}{)}{0pt}{0}{n}{j} \frac{(D-1)^j+(-1)^j(D-1)}{D}.
\end{align}
\begin{proof} 
We have to count the number of stabilizer operators in $\mathcal{S}:=\mathcal{S}_{\ket{\mathrm{GHZ}^n_D}} $ which act on exactly $j$ qudits. Note that $\mathcal{S}$ is generated by
\begin{align}
 S_1 &= X_D^{\otimes n}, \text{ and for all } i \in \{2,\ldots,n\} \\
 S_i &=  \mathbbm{1}_{D^{i-2}} \otimes Z_D \otimes Z_D^{-1}\otimes \mathbbm{1}_{D^{n-i}},
\end{align}
i.e., $\mathcal{S}=\{\prod_{i=1}^n S_j ^{t_j} \ \vert \ t_j \in \ZDZ \}$.
It follows at once, that every stabilizer $S\in\mathcal{S}$ with an $X_D$ part (i.e., $t_1\neq 0$) acts on all qudits. Since there are $(D-1)$ choices for $t_1 \neq 0$ and $D^{n-1}$ choices for $t_2,\ldots, t_n$ there are at least  $(D-1)D^{n-1}$ stabilizers which contribute to $\ell^n_n \left[ \mathrm{GHZ}^n_D  \right]$. This is the first term in Eq.~\eqref{eq:GHZ_sectorlength}.

It remains to count the stabilizer without an $X_D$ part (i.e., $t_1=0$). 
For this we consider the free $\ZDZ$-module of $Z$-like exponents $M_Z= \Span_{\ZDZ}\{\mathbf{b}_2,\ldots, \mathbf{b}_{n}\}$ with basis vectors given by
\begin{align} \label{eq:basis_GHZ_module}
 \mathbf{b}_2 &= (1,-1,0,0, \ldots, 0)\\  \nonumber
 \mathbf{b}_3 &= (0,1,-1,0, \ldots, 0)\\  \nonumber
 &\vdots \\ \nonumber
 \mathbf{b}_n &= (0,0,\ldots, 0,1,-1).
\end{align}
The remaining stabilizers are in one-to-one correspondence with the vectors $\mathbf{x}\in M_Z$ and the number of qudits a stabilizer acts on is equal to the number of nonzero entries of $x$.
Hence, we have to count the vectors in $M_Z$ which contain exactly $j$ nonzero entries. 
We denote this number by 
\begin{align}\label{eq:knjD1}
 k^n_j(D) :=  \left \vert  \left\{ \mathbf{x} = (x_1,\ldots,x_n)\in M_Z \ \big\vert \ j=\left \vert \{i\vert x_i\neq0\}\right \vert  \right\} \right \vert 
\end{align}
and the sector lengths will follow as 
 $\ell^n_j \left[ \mathrm{GHZ}^n_D  \right]= \delta_{j,n}(D-1)D^{n-1} + k_j^n(D)$. 
 
For every $j\in \{0,\ldots, n\}$, the number of vectors of the form $(x_1,\ldots,x_j,\ldots,0,0)\in M_Z$ with $x_1,\ldots,x_j\neq 0$ is independent of $n$ because such vectors are given by some linear combination of $\mathbf{b}_2,\ldots,\mathbf{b}_j$. 
Since the coefficients of such a linear combination are in $\ZDZ$, this number is  a function of $D$ and $j$ which we denote by $f_j(D)$.
However, there are $n$ choose $j$ possible permutations of the nonzero entries which yield the other vectors in $M_Z$ with exactly $j$ nonzero entries (since permuting the entries of all $\mathbf{b}_j$ simultaneously gives rise to a new basis of $M_Z$).
Therefore,
the total number of vectors in $M_Z$ with exactly $j$ nonzero entries is of the form 
\begin{align} \label{eq:knjD2}
 k^n_j(D)  = \nchoosej f_j(D).
\end{align}

The function $f_j(D)$ can be determined \emph{recursively}: 
Clearly, the null vector is the only vector in $M_Z$ without any nonzero entries, i.e., $f_0(D) = 1$.
For $n \ge 1$, the total number of vectors in $M_Z$ (with arbitrarily many nonzero entries) can be rewritten as 
\begin{align}
 D^{n-1} = \left \vert  M_Z \right \vert  = \sum_{j=0}^n  k_j^n (D), 
\end{align}
or, equivalently 
\begin{align}\label{eq:recursion_f}
 f_n (D) = D^{n-1} - \sum _{j=0}^{n-1} \nchoosej f_j (D).
\end{align}
In particular, the functions $f_n(D)$ are in fact polynomials  $f_n\in \mathbb{Z}[D]$ of degree $f_n\le n-1$, i.e., they can be expressed as 
\begin{align} \label{eq:fnD2}
 f_n (D) = \sum_{i=0}^{n-1} a_i^{(n)}D^i,
\end{align}
where the vectors of coefficients $\mathbf{a}^{(n)} := (a_0^{(n)},\ldots, a_{n-1}^{(n)})\in \mathbb{Z}^n$ follow from Eq.~\eqref{eq:recursion_f} and are recursively given by
\begin{align} \label{eq:recursion_a}
 \mathbf{a}^{(n)} = (0,\ldots,0,1)- \sum_{j=0}^{n-1} \nchoosej \mathbf{a}^{(j)}
\end{align}
for $n\ge1 $, and $(a_0^{(0)})\in \mathbb{Z}^0$ is the trivial vector. For the individual coefficients, this gives the final recursive formula
\begin{align} \label{eq:ain1}
 a_i^{(n)} = \delta_{i,n-1} - \sum_{j=0}^{n-1} \nchoosej  a_i^{(j)}
\end{align}
for all $i\in\{0,\ldots, n-1\}$. 
Using a proof by induction, we will now verify the non-recursive formula
\begin{align}\label{eq:ain2}
 a_i^{(n)} = (-1)^n\delta_{i,0} + (-1)^{n+i+1} \genfrac{(}{)}{0pt}{1}{n}{i+1}.
\end{align}

The case $n=0$ is trivial. Let $n\ge1 $ and assume Eq.~\eqref{eq:ain2} holds for all $j\le n-1$ (and all $i$). Inserting this into Eq.~\eqref{eq:ain1} and rearranging the terms yields that also $a_i^{(n)}$ fulfills Eq.~\eqref{eq:ain2} for all $i$:
\begin{align} \label{l:induction1}
 a_i^{(n)} &= \delta_{i,n-1} - \sum_{j=0}^{n-1} \nchoosej \left(  (-1)^j\delta_{i,0} + (-1)^{j+i+1} \genfrac{(}{)}{0pt}{1}{j}{i+1} \right) \\ \label{l:induction2}
 &=  \delta_{i,n-1} + \delta_{i,0}\left ( (-1)^n \genfrac{(}{)}{0pt}{1}{n}{n}- \sum_{j=0}^n (-1)^j \nchoosej  \right)
 \\ \label{l:induction3}
 & \hspace{2em} +\left((-1)^{n+i+1} \genfrac{(}{)}{0pt}{1}{n}{n} \genfrac{(}{)}{0pt}{1}{n}{i+1}- \sum_{j=0}^n (-1)^{j+i+1} \nchoosej \genfrac{(}{)}{0pt}{1}{j}{i+1}
 \right)\\ \label{l:induction4}
 &=  \delta_{i,n-1} +(-1)^n \delta_{i,0} + (-1)^{n+i+1}\genfrac{(}{)}{0pt}{1}{n}{i+1}+ (-1)^{n+i}\delta_{i,n-1} 
 \\ \label{l:induction5}
 &= (-1)^n\delta_{i,0} + (-1)^{n+i+1} \genfrac{(}{)}{0pt}{1}{n}{i+1}
\end{align}
In the first step, we have increased the range of the sums and subtracted the extra term.
In line~\eqref{l:induction2}, the alternating sum over binomial coefficients equals $\delta_{n,0}=0$ (since $n\ge 1$). From line~\eqref{l:induction3} to \eqref{l:induction4}, we have used the fact
\begin{align} \label{eq:combinatorix}
 \sum_{\tilde k =0}^{\tilde l} (-1)^{\tilde k}
 \genfrac{(}{)}{0pt}{1}{\tilde l }{\tilde m + \tilde k }
 \genfrac{(}{)}{0pt}{1}{\tilde s + \tilde k}{\tilde n} 
 = (-1)^{\tilde l + \tilde m } 
 \genfrac{(}{)}{0pt}{1}{\tilde s - \tilde m }{\tilde n - \tilde l} ,
\end{align}
from Ref.~\cite[Eq.~(5.24)]{GKP94}. In this case ($\tilde k = j$, $\tilde l=n$, $\tilde n = i+1$, and $\tilde m = \tilde s =0$), we used Eq.~\eqref{eq:combinatorix} to evaluate the term 
\begin{align}
 - \sum_{j=0}^n (-1)^{j+i+1} \nchoosej \genfrac{(}{)}{0pt}{1}{j}{i+1}
 &= (-1)^{i} \sum_{j=0}^n (-1)^{j} \nchoosej \genfrac{(}{)}{0pt}{1}{j}{i+1}
 \\ \label{l:apply_combinatorix}
 = (-1)^{i} (-1)^n \genfrac{(}{)}{0pt}{1}{0-0}{i+1-n} 
 &= (-1) ^{n+i} \delta_{i,n-1}.  
\end{align}
In the last step in line~\eqref{l:apply_combinatorix}, we used that for all  $x\in \mathbb{Z}$ the relation $\genfrac{(}{)}{0pt}{1}{0}{x} = \delta _{x,0}$ holds (a special case of Eq.~(5.1) in Ref.~\cite{GKP94}). 
From line~\eqref{l:induction4} to \eqref{l:induction5}, we used that the terms $\delta_{i,n-1}$ and $(-1)^{n+i}\delta_{i, n-1}$ cancel each other:
$\delta_{i,n-1} + (-1)^{n+i}\delta_{i, n-1} =\delta_{i, n-1}(1+(-1)^{2n-1})=0$. 

We can use the just-proven Eq.~\eqref{eq:ain2} to assemble the following
\emph{explicit} formula:
\begin{align} \label{eq:knjD3}
 k_j^n (D) \overset{\eqref{eq:knjD2},\eqref{eq:fnD2} }{=} & \nchoosej \sum_{i=0}^{j-1} a_i^{(j)}D^i
 \\ \label{eq:knjD4}
  \overset{\eqref{eq:ain2}}{=} & \nchoosej \sum_{i=0}^{j-1} \left( (-1)^j\delta_{i,0} + (-1)^{j+i+1}\genfrac{(}{)}{0pt}{1}{j}{i+1} \right)D^i
 \\  \label{eq:knjD5}
 =&  \nchoosej \left( (-1)^j  + \sum_{i=1}^j (-1)^{j+i} \genfrac{(}{)}{0pt}{1}{j}{i} D^{i-1}  \right)
 \\  \label{eq:knjD6}
 =&  \nchoosej \left( (-1)^j - (-1)^jD^{-1} + D^{-1} \sum_{i=0}^j  \genfrac{(}{)}{0pt}{1}{j}{i} D^{i}(-1)^{j-i}   \right)
 \\  \label{eq:knjD7}
=& \nchoosej \left( (-1)^j (1-D^{-1}) + D^{-1} (D-1)^j \right )
\\  \label{eq:knjD8}
=& \nchoosej \frac{(D-1)^j+(-1)^j(D-1)}{D}
\end{align} 
Until line~\eqref{eq:knjD4}, we only inserted previously established relations and simplified.
From line~\eqref{eq:knjD4} to~\eqref{eq:knjD5}, we simplified the sum and performed an index shift.
From line~\eqref{eq:knjD5} to~\eqref{eq:knjD6}, we subtracted $(-1)^{j}D^{-1}$ to make the sum start at $i=0$ instead of $i=1$. 
From line~\eqref{eq:knjD6} to~\eqref{eq:knjD7}, we used the binomial theorem.
The last step is a simplification to recover the initial claim.
This finishes the proof.
\end{proof}

\section{\label{app:QubitTables}\protect Tables of sector lengths for graph states on seven and eight qubits}

For future reference, we provide tables of the sector length distributions of all graph states up to eight qubits from the classification explained in Sec.~\ref{sec:6.1.3}. 
We produced these tables by running our C++ routine based on Eq.~\eqref{eq:sec_len_qubits}.
Table~\ref{tab:sector_7} and Tables~\ref{tab:sector_8a}-\ref{tab:sector_8c} show the sector length distributions for graph states on seven and eight qubits, respectively. 
See Figs.~4 and 5 in Ref.~\cite{HeinEisBri04} and Fig.~2 in Ref.~\cite{CaLTMoPo09} for a pictorial representation of the corresponding graphs. 
  \begin{table}
  \centering
\begin{minipage}{0.85\textwidth}
\begin{framed} 
\centering
   \begin{tabular}{|l|cccccccc|l|}  
   \hline
  Name  & $\ell^7_0$& $\ell^7_1$ & $\ell^7_2$ & $\ell^7_3$ & $\ell^7_4$ & $\ell^7_5$& $\ell^7_6$ & $\ell^7 _7$ & Family\\
   \hline
No. 20    &  1 &  0 &  21 &  0 &  35 &  0 &  7 	 &  64 & star \\
No. 21    &  1 &  0 &  11 &  10 &  15 &  20 &  37&  34 & \\
No. 22    &  1 &  0 &  9 &  0 &  35 &  24 &  19  &  40 & dandelion\\
No. 23    &  1 &  0 &  7 &  10 &  15 &  28 &  41 &  26 & \\
No. 24    &  1 &  0 &  6 &  9 &  17 &  30 &  40  &  25 & \\
No. 25    &  1 &  0 &  5 &  12 &  11 &  32 &  47 &  20 & \\
No. 26    &  1 &  0 &  5 &  4 &  27 &  32 &  31  &  28 & \\
No. 27    &  1 &  0 &  4 &  7 &  21 &  34 &  38  &  23 & \\
No. 28    &  1 &  0 &  3 &  10 &  15 &  36 &  45 &  18 & \\
No. 29    &  1 &  0 &  3 &  6 &  23 &  36 &  37  &  22 & \\
No. 30    &  1 &  0 &  2 &  9 &  17 &  38 &  44  &  17 & \\
No. 31    &  1 &  0 &  5 &  12 &  11 &  32 &  47 &  20 & \\
No. 32    &  1 &  0 &  5 &  0 &  35 &  32 &  23  &  32 & \\
No. 33    &  1 &  0 &  3 &  10 &  15 &  36 &  45 &  18 & \\
No. 34    &  1 &  0 &  3 &  6 &  23 &  36 &  37 &  22 & \\
No. 35    &  1 &  0 &  2 &  9 &  17 &  38 &  44 &  17 & line\\
No. 36    &  1 &  0 &  3 &  2 &  31 &  36 &  29 &  26 & \\
No. 37    &  1 &  0 &  2 &  5 &  25 &  38 &  36 &  21 & \\
No. 38    &  1 &  0 &  1 &  8 &  19 &  40 &  43 &  16 & \\
No. 39    &  1 &  0 &  1 &  8 &  19 &  40 &  43 &  16 & \\
No. 40    &  1 &  0 &  0 &  7 &  21 &  42 &  42 &  15 & ring \\
No. 41    &  1 &  0 &  1 &  4 &  27 &  40 &  35 &  20 & \\
No. 42    &  1 &  0 &  0 &  7 &  21 &  42 &  42 &  15 & \\
No. 43    &  1 &  0 &  0 &  7 &  21 &  42 &  42 &  15 & \\
No. 44    &  1 &  0 &  0 &  3 &  29 &  42 &  34 &  19 & \\
No. 45    &  1 &  0 &  1 &  0 &  35 &  40 &  27 &  24 & \\
   \hline
   \end{tabular} 
\caption{Sector length distributions for graph states on seven qubits. }
 \label{tab:sector_7}
\end{framed}
\end{minipage}
 \end{table}

  \begin{table}
  \centering
\begin{minipage}{0.85\textwidth}
\begin{framed} 
\centering \begin{tabular}{|l|ccccccccc|l|}  
   \hline
 Name   & $\ell^8_0$& $\ell^8_1$ & $\ell^8_2$ & $\ell^8_3$ & $\ell^8_4$ & $\ell^8_5$& $\ell^8_6$ & $\ell^8_7$& $\ell^8_8$ & Family\\
   \hline
No. 46    &  1 &  0 &  28 &  0 &  70 &  0 &  28 &  0    &  129 & star \\
No. 47    &  1 &  0 &  16 &  12 &  30 &  40 &  16 &  76 &  65 & \\
No. 48    &  1 &  0 &  13 &  0 &  55 &  0 &  103 &  0 &  84  & dandelion\\
No. 49    &  1 &  0 &  11 &  12 &  25 &  40 &  41 &  76 &  50 & \\
No. 50    &  1 &  0 &  12 &  0 &  38 &  64 &  12 &  64 &  65  &\\
No. 51    &  1 &  0 &  9 &  12 &  23 &  40 &  51 &  76 &  44  &\\
No. 52    &  1 &  0 &  8 &  16 &  18 &  32 &  64 &  80 &  37  &\\
No. 53    &  1 &  0 &  8 &  4 &  30 &  56 &  40 &  68 &  49  &\\
No. 54    &  1 &  0 &  7 &  8 &  25 &  48 &  53 &  72 &  42 & \\
No. 55    &  1 &  0 &  7 &  6 &  27 &  52 &  49 &  70 &  44 & \\
No. 56    &  1 &  0 &  5 &  14 &  17 &  36 &  75 &  78 &  30 & \\
No. 57    &  1 &  0 &  7 &  0 &  49 &  0 &  133 &  0 &  66  &\\
No. 58    &  1 &  0 &  6 &  6 &  26 &  52 &  54 &  70 &  41 & \\
No. 59    &  1 &  0 &  5 &  10 &  21 &  44 &  67 &  74 &  34&  \\
No. 60    &  1 &  0 &  5 &  6 &  25 &  52 &  59 &  70 &  38 & \\
No. 61    &  1 &  0 &  5 &  4 &  27 &  56 &  55 &  68 &  40 & \\
No. 62    &  1 &  0 &  4 &  8 &  22 &  48 &  68 &  72 &  33 & \\
No. 63    &  1 &  0 &  3 &  12 &  17 &  40 &  81 &  76 &  26&  \\
No. 64    &  1 &  0 &  4 &  12 &  18 &  40 &  76 &  76 &  29&  \\
No. 65    &  1 &  0 &  4 &  8 &  22 &  48 &  68 &  72 &  33 & \\
No. 66    &  1 &  0 &  3 &  10 &  19 &  44 &  77 &  74 &  28 & \\
No. 67    &  1 &  0 &  3 &  8 &  21 &  48 &  73 &  72 &  30 & \\
No. 68    &  1 &  0 &  2 &  10 &  18 &  44 &  82 &  74 &  25 & line\\
No. 69    &  1 &  0 &  8 &  16 &  18 &  32 &  64 &  80 &  37 & \\
No. 70    &  1 &  0 &  7 &  18 &  15 &  28 &  73 &  82 &  32 & \\
No. 71    &  1 &  0 &  8 &  0 &  50 &  0 &  128 &  0 &  69 & \\
No. 72    &  1 &  0 &  6 &  12 &  20 &  40 &  66 &  76 &  35 & \\
No. 73    &  1 &  0 &  7 &  0 &  33 &  64 &  37 &  64 &  50  &\\
No. 74    &  1 &  0 &  5 &  8 &  23 &  48 &  63 &  72 &  36  &\\
No. 75    &  1 &  0 &  4 &  12 &  18 &  40 &  76 &  76 &  29 & \\
No. 76    &  1 &  0 &  4 &  12 &  18 &  40 &  76 &  76 &  29 & \\
\hline
   \end{tabular} 
 
\caption{Sector length distributions for graph states on eight qubits (part~1 of 3). }
 \label{tab:sector_8a}
\end{framed}
\end{minipage}
 \end{table}

  \begin{table}
  \centering
\begin{minipage}{0.85\textwidth}
\begin{framed} 
\centering
   \begin{tabular}{|l|ccccccccc|l|}  
   \hline
 Name   & $\ell^8_0$& $\ell^8_1$ & $\ell^8_2$ & $\ell^8_3$ & $\ell^8_4$ & $\ell^8_5$& $\ell^8_6$ & $\ell^8_7$& $\ell^8_8$ & Family\\
   \hline
No. 77    &  1 &  0 &  5 &  4 &  27 &  56 &  55 &  68 &  40 & \\
No. 78    &  1 &  0 &  5 &  0 &  47 &  0 &  143 &  0 &  60  &\\
No. 79    &  1 &  0 &  5 &  2 &  29 &  60 &  51 &  66 &  42 & \\
No. 80    &  1 &  0 &  4 &  6 &  24 &  52 &  64 &  70 &  35 & \\
No. 81    &  1 &  0 &  3 &  10 &  19 &  44 &  77 &  74 &  28&  \\
No. 82    &  1 &  0 &  3 &  8 &  21 &  48 &  73 &  72 &  30 & \\
No. 83    &  1 &  0 &  3 &  8 &  21 &  48 &  73 &  72 &  30 & \\
No. 84    &  1 &  0 &  2 &  10 &  18 &  44 &  82 &  74 &  25&  \\
No. 85    &  1 &  0 &  3 &  6 &  23 &  52 &  69 &  70 &  32 & \\
No. 86    &  1 &  0 &  4 &  4 &  26 &  56 &  60 &  68 &  37 & \\
No. 87    &  1 &  0 &  4 &  0 &  46 &  0 &  148 &  0 &  57  &\\
No. 88    &  1 &  0 &  2 &  12 &  16 &  40 &  86 &  76 &  23&  \\
No. 89    &  1 &  0 &  3 &  6 &  23 &  52 &  69 &  70 &  32 & \\
No. 90    &  1 &  0 &  3 &  4 &  25 &  56 &  65 &  68 &  34 & \\
No. 91    &  1 &  0 &  2 &  8 &  20 &  48 &  78 &  72 &  27 & \\
No. 92    &  1 &  0 &  3 &  4 &  25 &  56 &  65 &  68 &  34 & \\
No. 93    &  1 &  0 &  2 &  8 &  20 &  48 &  78 &  72 &  27 & \\
No. 94    &  1 &  0 &  3 &  4 &  25 &  56 &  65 &  68 &  34 & \\
No. 95    &  1 &  0 &  2 &  6 &  22 &  52 &  74 &  70 &  29 & \\
No. 96    &  1 &  0 &  2 &  6 &  22 &  52 &  74 &  70 &  29 & \\
No. 97    &  1 &  0 &  1 &  10 &  17 &  44 &  87 &  74 &  22&  \\
No. 98    &  1 &  0 &  1 &  8 &  19 &  48 &  83 &  72 &  24 & \\
No. 99    &  1 &  0 &  1 &  8 &  19 &  48 &  83 &  72 &  24 & \\
No. 100   &  1 &  0 &  0 &  8 &  18 &  48 &  88 &  72 &  21 & ring \\
No. 101   &  1 &  0 &  3 &  14 &  15 &  36 &  85 &  78 &  24&  \\
No. 102   &  1 &  0 &  3 &  8 &  21 &  48 &  73 &  72 &  30 & \\
No. 103   &  1 &  0 &  2 &  8 &  20 &  48 &  78 &  72 &  27 & \\
No. 104   &  1 &  0 &  4 &  0 &  46 &  0 &  148 &  0 &  57  &\\
No. 105   &  1 &  0 &  3 &  4 &  25 &  56 &  65 &  68 &  34 & \\
No. 106   &  1 &  0 &  3 &  0 &  45 &  0 &  153 &  0 &  54  &\\
No. 107   &  1 &  0 &  4 &  0 &  30 &  64 &  52 &  64 &  41 & \\
No. 108   &  1 &  0 &  3 &  2 &  27 &  60 &  61 &  66 &  36 & \\
No. 109   &  1 &  0 &  1 &  10 &  17 &  44 &  87 &  74 &  22&  \\
No. 110   &  1 &  0 &  2 &  4 &  24 &  56 &  70 &  68 &  31 & \\
No. 111   &  1 &  0 &  2 &  4 &  24 &  56 &  70 &  68 &  31 & \\
No. 112   &  1 &  0 &  2 &  4 &  24 &  56 &  70 &  68 &  31 & \\ \hline
   \end{tabular} 
 
\caption{Sector length distributions for graph states on eight qubits (part~2 of 3). }
 \label{tab:sector_8b}
\end{framed}
\end{minipage}
 \end{table}

  \begin{table}
  \centering
\begin{minipage}{0.85\textwidth}
\begin{framed} 
\centering
 
   \begin{tabular}{|l|ccccccccc|l|}  
   \hline
  Name  & $\ell^8_0$& $\ell^8_1$ & $\ell^8_2$ & $\ell^8_3$ & $\ell^8_4$ & $\ell^8_5$& $\ell^8_6$ & $\ell^8_7$& $\ell^8_8$ & Family \\
   \hline 
No. 113   &  1 &  0 &  1 &  6 &  21 &  52 &  79 &  70 &  26 & \\
No. 114   &  1 &  0 &  1 &  6 &  21 &  52 &  79 &  70 &  26 & \\
No. 115   &  1 &  0 &  1 &  6 &  21 &  52 &  79 &  70 &  26 & \\
No. 116   &  1 &  0 &  1 &  4 &  23 &  56 &  75 &  68 &  28 & \\
No. 117   &  1 &  0 &  1 &  4 &  23 &  56 &  75 &  68 &  28 & \\
No. 118   &  1 &  0 &  0 &  8 &  18 &  48 &  88 &  72 &  21 & \\
No. 119   &  1 &  0 &  0 &  6 &  20 &  52 &  84 &  70 &  23 & \\
No. 120   &  1 &  0 &  0 &  6 &  20 &  52 &  84 &  70 &  23 & \\
No. 121   &  1 &  0 &  4 &  0 &  30 &  64 &  52 &  64 &  41 & \\
No. 122   &  1 &  0 &  1 &  10 &  17 &  44 &  87 &  74 &  22 & \\
No. 123   &  1 &  0 &  2 &  0 &  44 &  0 &  158 &  0 &  51  &\\
No. 124   &  1 &  0 &  4 &  0 &  30 &  64 &  52 &  64 &  41 & \\
No. 125   &  1 &  0 &  2 &  2 &  26 &  60 &  66 &  66 &  33 & \\
No. 126   &  1 &  0 &  1 &  6 &  21 &  52 &  79 &  70 &  26 & \\
No. 127   &  1 &  0 &  1 &  4 &  23 &  56 &  75 &  68 &  28 & \\
No. 128   &  1 &  0 &  1 &  4 &  23 &  56 &  75 &  68 &  28 & \\
No. 129   &  1 &  0 &  1 &  2 &  25 &  60 &  71 &  66 &  30 & \\
No. 130   &  1 &  0 &  0 &  6 &  20 &  52 &  84 &  70 &  23 & \\
No. 131   &  1 &  0 &  0 &  4 &  22 &  56 &  80 &  68 &  25 & \\
No. 132   &  1 &  0 &  0 &  4 &  22 &  56 &  80 &  68 &  25 & \\
No. 133   &  1 &  0 &  0 &  8 &  18 &  48 &  88 &  72 &  21 & \\
No. 134   &  1 &  0 &  3 &  0 &  45 &  0 &  153 &  0 &  54  &\\
No. 135   &  1 &  0 &  2 &  0 &  28 &  64 &  62 &  64 &  35  &\\
No. 136   &  1 &  0 &  1 &  0 &  43 &  0 &  163 &  0 &  48  &\\
No. 137   &  1 &  0 &  0 &  4 &  22 &  56 &  80 &  68 &  25 & \\
No. 138   &  1 &  0 &  0 &  2 &  24 &  60 &  76 &  66 &  27 & \\
No. 139   &  1 &  0 &  1 &  2 &  25 &  60 &  71 &  66 &  30 & \\
No. 140   &  1 &  0 &  0 &  2 &  24 &  60 &  76 &  66 &  27 & \\
No. 141   &  1 &  0 &  0 &  0 &  42 &  0 &  168 &  0 &  45  &\\
No. 142   &  1 &  0 &  0 &  0 &  42 &  0 &  168 &  0 &  45  &\\
No. 143   &  1 &  0 &  1 &  0 &  27 &  64 &  67 &  64 &  32 & \\
No. 144   &  1 &  0 &  0 &  0 &  26 &  64 &  72 &  64 &  29 & \\
No. 145   &  1 &  0 &  0 &  0 &  42 &  0 &  168 &  0 &  45  &\\
No. 146   &  1 &  0 &  0 &  0 &  26 &  64 &  72 &  64 &  29  &\\ \hline
   \end{tabular} 
\caption{Sector length distributions for graph states on eight qubits (part~3 of 3). }
 \label{tab:sector_8c}
\end{framed}
\end{minipage}
 \end{table}

\newpage

\section{\label{app:proof_choices_ring}\protect Proof of Lemma~\ref{lem:choices_ring}}
Here, we state the proof of Lemma~\ref{lem:choices_ring} from the main text. For a better readability, we repeat it now.

 \lemma\label{lem:choices_ring}
 \emph{Let $N(n,k)$ be the number of choices to place $k$ pairs of white vertices on an $n$-vertex ring graph such that any two such pairs are separated by at least one black vertex. Then, it holds $N(n,k)={\genfrac{(}{)}{0pt}{1}{n-2k-1}{k-1}} \frac{n}{k}$.}
 
 \begin{proof}
If $k$ pairs are white, there are exactly $n-2k$ black vertices on the ring graph. 
They are grouped into $k$ black chains (between any two adjacent white pairs) each consisting of $x_i\ge1$ vertices, where $i=1,\ldots,k$ is a counterclockwise enumeration of the black chains. 
The set 
\begin{align}
       L:= \left\{ (x_1,\ldots,x_k) \in \ZZ^k \ \bigg \vert \ x_i\ge1,\ \sum_{i=1}^k x_i =n-2k \right \}.
\end{align}
contains all possible tuples of lengths $x_i$ for black chains fitting on the ring graph.
We can use it to characterize all colorings with exactly $k$ white pairs which are in accordance to the rule of the puzzle from Sec.~\ref{sec:6.2}.
Given $\ell=(x_1,\ldots,x_k)\in L$, color vertex $1$ and $2$ white, vertex $3, \ldots, x_1+2$ black, and vertex $x_1+3$ and $x_1+4$ white. Continuing in a similar fashion by sequentially coloring two vertices black and $x_i$ vertices black gives rise to an allowed coloring. 
Starting with a different $\ell'\in L$ and proceeding in an analogous way yields a different allowed coloring.
However, not all allowed colorings with exactly $k$ white pairs arise in this way:
A counterclockwise shift of all colors by one vertex yields a new coloring because, e.g., vertex $n$ is black and white before and after the shift, respectively.
A second such shift again yields a new coloring as vertex 1 is now black (and was white in the two cases before). After $x_1+3$ shifts, vertex 1 is (for the first time) white again.
Since the obtained coloring is the initial (i.e., unshifted) coloring for a different tuple, $(x_2, \ldots, x_k,x_1)\in L$,  each tuple $\ell\in L$ contributes exactly $x_1(\ell)+2$ colorings which are in accordance to the rule.
Thereby, we have used the notation  $x_i(\ell):=x_i$ for an $\ell\in L$ whose explicit form is given by $\ell=(x_1,\ldots,x_k)$. 
   
These considerations show that the number of allowed coloring with exactly $k$ white pairs is given by \begin{align}\label{eq:ring_choices_1}
N(n,k)=
\sum_{\ell\in L} (x_1(\ell)+2) = 2\vert L \vert + \sum_{\ell\in L} x_1(\ell).
\end{align}
We can use the symmetry of the aforementioned shifts to evaluate this expression. 
The necessary mathematical toolbox for this is the theory of group actions, see Chapter 5.1 of Ref.~\cite{Bosch} for an introduction.
Write 
\begin{align}
\pi: \ZZ^k \longrightarrow \ZZ^k, \hspace{1em} (x_1,\ldots, x_k) \longmapsto (x_2,\ldots, x_{k},x_1)
\end{align}
for the cyclic permutation of all entries by one to the left.
Let $\mathcal {G}=\{ \pi^i \ \vert \ i\in\{1,\ldots,k\} \}$, where $\pi^2= \pi \circ \pi$ etc., be the group generated by $\pi$ and consider the group action
\begin{align}
\mathcal{G} \times L \longrightarrow L, \hspace{1em} (\pi^i, \ell) \longmapsto \pi^i(\ell).
\end{align}
Enumerating the orbits of this group action from $1$ to $m$ yields a partition of the form
\begin{align}
L = \bigcupdot _{j=1}^m L_j
\end{align}  
which we use to split up the sum in  Eq.~\eqref{eq:ring_choices_1} and obtain
\begin{align} \label{eq:ring_choices_2}
 N(n,k) =  2\vert L \vert + \sum_{j=1}^{m} \sum_{\ell\in L_j} x_1(\ell).
\end{align}
This reduces the problem to the computation of $ \sum_{\ell\in L_j} x_1(\ell)$ for a fixed orbit $L_j$.
For each orbit choose a member $\ell_j\in L_j$ and denote its stabilizer subgroup by $\mathcal{S}_j \subset \mathcal{G}$, i.e., $\mathcal{S}_j = \{\sigma \in \mathcal{G} \ \vert \ \sigma(\ell_j)=\ell_j \}$.
Furthermore, choose a representative $g_r\in \mathcal{G}$ for each coset in $\mathcal{G}/\mathcal{S}_j$. (These $g_r$ also depend on $j$.) 
This yields the parametrization
\begin{align} \label{eq:G_decomposed}
\mathcal{G} = \left\{  g_r\circ \sigma \ \Big \vert \  r\in \left\{1,\ldots, \left\vert \mathcal{G}/\mathcal{S}_j \right \vert \right\} ,\, \sigma\in \mathcal{S}_j  \right\}.
\end{align}
Note that $g_r\circ\sigma = g_{r'}\circ \sigma'$ iff $r=r'$ and $\sigma = \sigma'$.
Because of $\ell_j\in L$, we obtain
\begin{align} \label{l:action_1}
n-2k &= \sum_{i=1}^{k} x_i(\ell_j) = \sum_{i=1}^{k} x_1(\pi^i(\ell_j)) 
\\ \label{l:action_2}
&= \sum_{r=1}^{\vert \mathcal{G}/\mathcal{S}_j \vert } \sum_{\sigma \in \mathcal{S}_j} x_1( g_r( \underbrace{\sigma(\ell_j)}_{=\ell_j})) 
= \vert \mathcal{S}_j \vert \sum_{r=1}^{\vert\mathcal{G}/\mathcal{S}_j \vert} x_1(g_r(\ell_j)) 
\\ \label{l:action_5}
&= \vert \mathcal{S}_j \vert \sum_{\ell\in L_j} {x_1(\ell)}.
\end{align}
In line~\eqref{l:action_1}, we use the defining equation of elements in $L$ and that $\pi^i$ shifts the $i^\mathrm{th}$ entry of $\ell_j$ to the first entry.
To get to line~\eqref{l:action_2}, we use that the group $\mathcal {G}$ has two parametrizations, $\mathcal{G}=\{ \pi^i \ \vert \ i\in\{1,\ldots,k\} \}$ and the one given in Eq.~\eqref{eq:G_decomposed}.
Since the stabilizers $\sigma \in \mathcal{S}_j$ trivially act on $\ell_j$, all terms of the inner sum coincide such that we can factor out $\vert \mathcal{S}_j\vert$. 
To get to line~\eqref{l:action_5}, we use that the elements of the orbit $L_j$ are in 1:1-correspondence to the cosets in $\mathcal{G}/\mathcal{S}_j$.
Together with $\vert L_j\vert \times \vert \mathcal{S}_j\vert  = \vert \mathcal{G} \vert =k$, this implies
$\sum_{\ell\in L_j} x_1(\ell) = \vert L_j \vert \left(\frac{n}{k}-2\right)$.
Now, we can compute Eq.~\eqref{eq:ring_choices_1} as
\begin{align}
N(n,k)&= 2\vert L \vert + \left(\frac{n}{k} -2 \right) \sum_{j=1}^{m} \vert L_j \vert= \vert L \vert \frac{n}{k}.
\end{align}

The last information we need is the cardinality of the set $L$, i.e., the number of positive integer solutions of the equation $\sum_{i=1}^k x_i = n-2k$.
Using the method of {stars and bars}, one can show that the similar equation, $\sum_{i=1}^k x'_i = m$ has exactly $ {\genfrac{(}{)}{0pt}{1}{m+k-1}{k-1}} $ nonnegative integer solutions~\cite[Example 1.5.3]{Levin15}.
Substituting $x_i'=x_i-1$ and $m=n-3k$, yields $\vert L \vert = {\genfrac{(}{)}{0pt}{1}{n-2k-1}{k-1}}$. 
This finishes the proof.

\end{proof}


%



\end{document}